\documentclass[numberedappendix]{emulateapj}

\usepackage{color}
\def\NLOWZWV{    45}   % Number of LOWZ SN Ia in WV07 
\def\NESSEWV{    57}   % Number of ESSENCE SN Ia in WV07 
 
\def\NSNSDSSCONF{   130}   % spectro-confirmed (type=120)
\def\NSNSDSSMAYBE{    16}   % spectro-probable (type=119)
\def\NSNSDSSTOT{   146}   % sum of 120+119
\def\NSNLOWZ{    33}   % after cuts
\def\NSNSDSS{   103}   % idem
\def\NSNESSE{    56}   % idem
\def\NSNSNLS{    62}   % idem
\def\NSNHST{    34}   % idem
\def\NSNTOT{   288}   % sum of all samples
 
\def\NEPLOWZ{    52}   % after cuts 
\def\NEPSDSS{    48}   % idem 
\def\NEPESSE{    21}   % idem 
\def\NEPSNLS{    27}   % idem 
\def\NEPHST{    11}   % idem 
 
\def\NDUSTSPEC{    81}   % Type=120 (confirmed)
\def\NDUSTPROB{     6}   % Type=119 (spec-prob)
\def\NDUSTHOSTZ{    73}   % Type=105 (spec host-z)
\def\NDUST{   160}   % sum of above
\def\NDUSTMISS{    40}   % Type=104 (no spec)
\def\NDUSTEFF{    80}   % Dust Sample eff (%)
\def\NHOSTZGTZCUT{    31}   % 119+105 with z>.3
\def\NHOSTZALLZ{   110}   % 119+105, ALL Z 
\def\HOSTZCHISQDOF{107/102}   %wfit hostZ chi2/dof
\def\DUSTCHISQDOF{166/205}   %wfit DUST chi2/dof
\def\DUSTFITPROB{    98}   % prob(chi2/Ndof)
\def\HOSTZSCATTER{  0.22}   % RMS-scatter
\def\DUSTSCATTER{  0.19}   % RMS-scatter
 
\def\TAUAV{  0.334}   % AV-expon slope
\def\TAUAVERRTOT{  0.088}   % error on above
\def\TAUAVERRSTAT{  0.050}   % stat error on above
\def\TAUAVERRSYST{  0.072}   % syst error on above
\def\TAUAVERRTOTnoRV{  0.072}   % tot error sans RV-err
\def\dTAUAVdRV{  0.10}   % dTAUAV/dRV
 
\def\DELTAPEAK{ -0.24}   % peak of bifurcated Gaussian
\def\DELTASIGMINUS{  0.24}   % Gauss sigma on neg side
\def\DELTASIGPLUS{  0.48}   % Gauss sigma on pos side
 
\def\RV{  2.18}   % mean RV
\def\RVERRSTAT{  0.14}   % RV stat-error
\def\RVERRSYST{  0.48}   % RV syst-error
\def\RVERRTOT{  0.50}   % RV totol error
 
\def\AVMN{  0.358}   % <AV>
\def\AVMNERRSTAT{  0.026}   % stat error on <AV>
\def\AVMNERRSYST{  0.068}   % syst error on <AV>
\def\AVMNERRTOT{  0.072}   % total error on <AV>
 
\def\RVSYSTzcut{  0.36}   % RV syst from Zcut 
\def\RVSYSTzcuterr{  0.17}   % stat error on above
\def\RVSYSTdNdz{  0.11}   % RV syst from dN/dz(beta)
\def\RVSYSTepoch{  0.25}   % RV syst from ep-range
\def\RVSYSTSNR{  0.11}   % RV syst from SNR cut
\def\RVSYSTtwocolor{  0.08}   % RV syst, 2 colors
\def\RVSYSTcolorsmear{  0.14}   % RV syst, color model
\def\RVSYSTmlcsmodel{  0.05}   % RV syst, mlcs2k2 model
 
\def\AVMNSYSTzcut{  0.031}   % <AV> syst from Zcut 
\def\AVMNSYSTdNdz{  0.004}   % <AV> syst from dN/dz(beta)
\def\AVMNSYSTepoch{  0.045}   % <AV> syst from ep-range
\def\AVMNSYSTSNR{  0.011}   % <AV> syst from SNR cut
\def\AVMNSYSTtwocolor{  0.030}   % <AV> syst, 2 colors
\def\AVMNSYSTcolorsmear{  0.017}   % <AV> syst, color model
\def\AVMNSYSTmlcsmodel{  0.018}   % <AV> syst, MLCS model
 
\def\LOWZCHISQ           {   31.9}   % Hubble chi2, separate fits
\def\LOWZCHISQx          {   32.9}   % Hubble chi2, global fit
\def\LOWZNDOF            {    32}   % Hubble Ndof
\def\LOWZRMS             {  0.19}   % Hubble RMS, separate fits
\def\LOWZSALTDISP        {  0.15}   % SALT2 dispersion, separate fits
\def\LOWZMLCSDISP        {  0.16}   % MLCS dispersion, separate fits
 
\def\SDSSCHISQ           {   55.3}   % Hubble chi2, separate fits
\def\SDSSCHISQx          {   56.6}   % Hubble chi2, global fit
\def\SDSSNDOF            {   102}   % Hubble Ndof
\def\SDSSRMS             {  0.15}   % Hubble RMS, separate fits
\def\SDSSSALTDISP        {  0.08}   % SALT2 dispersion, separate fits
\def\SDSSMLCSDISP        {  0.09}   % MLCS dispersion, separate fits
 
\def\ESSECHISQ           {   46.8}   % Hubble chi2, separate fits
\def\ESSECHISQx          {   48.4}   % Hubble chi2, global fit
\def\ESSENDOF            {    55}   % Hubble Ndof
\def\ESSERMS             {  0.23}   % Hubble RMS, separate fits
\def\ESSESALTDISP        {  0.17}   % SALT2 dispersion, separate fits
\def\ESSEMLCSDISP        {  0.13}   % MLCS dispersion, separate fits
 
\def\SNLSCHISQ           {   63.0}   % Hubble chi2, separate fits
\def\SNLSCHISQx          {   64.6}   % Hubble chi2, global fit
\def\SNLSNDOF            {    61}   % Hubble Ndof
\def\SNLSRMS             {  0.24}   % Hubble RMS, separate fits
\def\SNLSSALTDISP        {  0.11}   % SALT2 dispersion, separate fits
\def\SNLSMLCSDISP        {  0.16}   % MLCS dispersion, separate fits
 
\def\HSTCHISQ            {   32.5}   % Hubble chi2, separate fits
\def\HSTCHISQx           {   32.4}   % Hubble chi2, global fit
\def\HSTNDOF             {    33}   % Hubble Ndof
\def\HSTRMS              {  0.28}   % Hubble RMS, separate fits
\def\HSTSALTDISP         {  0.23}   % SALT2 dispersion, separate fits
\def\HSTMLCSDISP         {  0.15}   % MLCS dispersion, separate fits

\def\WSYSTaJRKVECT{  0.013}   % JRK original vectors                    
\def\WSYSTaRVSHIFT{  0.036}   % change RV by 1 sigma                    
\def\WSYSTaXTMW{  0.021}   % change XTMW by 1 sigma                  
\def\WSYSTaFILTSHIFT{  0.007}   % shifted LOWZ filters (no color transf)  
\def\WSYSTaSIMEFFLOWZ{  0.000}   % simeff for LOWZ                         
\def\WSYSTaSIMEFFSDSS{  0.062}   % simeff for SDSS                         
\def\WSYSTaSIMEFFESSE{  0.000}   % simeff for ESSE                         
\def\WSYSTaSIMEFFSNLS{  0.000}   % simeff for SNLS                         
\def\WSYSTaLOWZCUT{  0.000}   % LOWZ CUT at .02                         
\def\WSYSTanoU{  0.310}   % deweight U-band                         
\def\WSYSTaERRUBVRI{  0.030}   % .01 mag error in U,B,V,R,I              
\def\WSYSTaERRki{  0.003}   % 1 sigma shift in color terms            
\def\WSYSTaERRABOFF{  0.004}   % 1 sigma shift in SDSS AB offsets        
\def\WSYSTaZPTSNLS{  0.000}   % 1 sigma shift in SNLS griz ZPTs         
\def\WSYSTaZPTESSE{  0.000}   % 1 sigma shift in ESSENCE R-I ZPT        
\def\WSYSTaZPTHST{  0.000}   % 1 sigma shift in HST ZPTs               
\def\OMSYSTaJRKVECT{  0.003}   % JRK original vectors                    
\def\OMSYSTaRVSHIFT{  0.007}   % change RV by 1 sigma                    
\def\OMSYSTaXTMW{  0.005}   % change XTMW by 1 sigma                  
\def\OMSYSTaFILTSHIFT{  0.001}   % shifted LOWZ filters (no color transf)  
\def\OMSYSTaSIMEFFLOWZ{  0.000}   % simeff for LOWZ                         
\def\OMSYSTaSIMEFFSDSS{  0.014}   % simeff for SDSS                         
\def\OMSYSTaSIMEFFESSE{  0.000}   % simeff for ESSE                         
\def\OMSYSTaSIMEFFSNLS{  0.000}   % simeff for SNLS                         
\def\OMSYSTaLOWZCUT{  0.000}   % LOWZ CUT at .02                         
\def\OMSYSTanoU{  0.051}   % deweight U-band                         
\def\OMSYSTaERRUBVRI{  0.006}   % .01 mag error in U,B,V,R,I              
\def\OMSYSTaERRki{  0.001}   % 1 sigma shift in color terms            
\def\OMSYSTaERRABOFF{  0.001}   % 1 sigma shift in SDSS AB offsets        
\def\OMSYSTaZPTSNLS{  0.000}   % 1 sigma shift in SNLS griz ZPTs         
\def\OMSYSTaZPTESSE{  0.000}   % 1 sigma shift in ESSENCE R-I ZPT        
\def\OMSYSTaZPTHST{  0.000}   % 1 sigma shift in HST ZPTs               
\def\MLCSWRESa{ -0.84}   % w central value
\def\MLCSWSTATERRa{  0.15}   % w stat error
\def\MLCSWSYSTERRPa{  0.08}   % w +syst error
\def\MLCSWSYSTERRMa{  0.32}   % w -syst error
\def\MLCSWTOTERRPa{  0.17}   % w +total syst
\def\MLCSWTOTERRMa{  0.35}   % w -total syst
\def\MLCSOMRESa{  0.289}   % Omega_matter
\def\MLCSOMSTATERRa{  0.033}   % stat error on O_Mat
\def\MLCSOMSYSTERRPa{  0.019}   % OM +syst error
\def\MLCSOMSYSTERRMa{  0.054}   % OM -syst error
\def\MLCSOMTOTERRPa{  0.038}   % OM +total error
\def\MLCSOMTOTERRMa{  0.064}   % OM -total error
\def\MLCSWCHISQa{  170.9}   % Hubble chi2
\def\WNDOFa{  102}   % Hubble Ndof
\def\MLCSWRMSa{  0.15}   % Hubble RMS

\def\LCDMOMa{  0.274}   % LCDM Omega_matter
\def\LCDMOMSTATERRa{  0.023}   % LCDM stat error on O_Mat
\def\LCDMOMSYSTERRa{  0.000}   % LCDM OM syst error
\def\LCDMOMTOTERRa{  0.023}   % LCDM OM total error
\def\LCDMOLa{  0.735}   % LCDM Omega_Lambda
\def\LCDMOLSTATERRa{  0.019}   % LCDM stat error on O_Lam
\def\LCDMOLSYSTERRa{  0.000}   % LCDM O_Lam syst error
\def\LCDMOLTOTERRa{  0.019}   % LCDM O_Lam total error
\def\LCDMCHISQa{   55.2}   % LCDM Hubble chi2
\def\LCDMNDOFa{  102}   % LCDM Hubble Ndof
\def\WSYSTbJRKVECT{  0.036}   % JRK original vectors                    
\def\WSYSTbRVSHIFT{  0.023}   % change RV by 1 sigma                    
\def\WSYSTbXTMW{  0.012}   % change XTMW by 1 sigma                  
\def\WSYSTbFILTSHIFT{  0.017}   % shifted LOWZ filters (no color transf)  
\def\WSYSTbSIMEFFLOWZ{  0.000}   % simeff for LOWZ                         
\def\WSYSTbSIMEFFSDSS{  0.002}   % simeff for SDSS                         
\def\WSYSTbSIMEFFESSE{  0.013}   % simeff for ESSE                         
\def\WSYSTbSIMEFFSNLS{  0.029}   % simeff for SNLS                         
\def\WSYSTbLOWZCUT{  0.000}   % LOWZ CUT at .02                         
\def\WSYSTbnoU{  0.080}   % deweight U-band                         
\def\WSYSTbERRUBVRI{  0.029}   % .01 mag error in U,B,V,R,I              
\def\WSYSTbERRki{  0.007}   % 1 sigma shift in color terms            
\def\WSYSTbERRABOFF{  0.028}   % 1 sigma shift in SDSS AB offsets        
\def\WSYSTbZPTSNLS{  0.040}   % 1 sigma shift in SNLS griz ZPTs         
\def\WSYSTbZPTESSE{  0.011}   % 1 sigma shift in ESSENCE R-I ZPT        
\def\WSYSTbZPTHST{  0.000}   % 1 sigma shift in HST ZPTs               
\def\OMSYSTbJRKVECT{  0.009}   % JRK original vectors                    
\def\OMSYSTbRVSHIFT{  0.006}   % change RV by 1 sigma                    
\def\OMSYSTbXTMW{  0.003}   % change XTMW by 1 sigma                  
\def\OMSYSTbFILTSHIFT{  0.004}   % shifted LOWZ filters (no color transf)  
\def\OMSYSTbSIMEFFLOWZ{  0.000}   % simeff for LOWZ                         
\def\OMSYSTbSIMEFFSDSS{  0.001}   % simeff for SDSS                         
\def\OMSYSTbSIMEFFESSE{  0.004}   % simeff for ESSE                         
\def\OMSYSTbSIMEFFSNLS{  0.008}   % simeff for SNLS                         
\def\OMSYSTbLOWZCUT{  0.000}   % LOWZ CUT at .02                         
\def\OMSYSTbnoU{  0.016}   % deweight U-band                         
\def\OMSYSTbERRUBVRI{  0.008}   % .01 mag error in U,B,V,R,I              
\def\OMSYSTbERRki{  0.002}   % 1 sigma shift in color terms            
\def\OMSYSTbERRABOFF{  0.007}   % 1 sigma shift in SDSS AB offsets        
\def\OMSYSTbZPTSNLS{  0.010}   % 1 sigma shift in SNLS griz ZPTs         
\def\OMSYSTbZPTESSE{  0.003}   % 1 sigma shift in ESSENCE R-I ZPT        
\def\OMSYSTbZPTHST{  0.000}   % 1 sigma shift in HST ZPTs               
\def\MLCSWRESb{ -0.71}   % w central value
\def\MLCSWSTATERRb{  0.09}   % w stat error
\def\MLCSWSYSTERRPb{  0.11}   % w +syst error
\def\MLCSWTOTERRPb{  0.14}   % w +total syst
\def\MLCSOMRESb{  0.319}   % Omega_matter
\def\MLCSOMSTATERRb{  0.025}   % stat error on O_Mat
\def\MLCSOMSYSTERRPb{  0.027}   % OM +syst error
\def\MLCSOMTOTERRPb{  0.036}   % OM +total error
\def\MLCSWCHISQb{  406.7}   % Hubble chi2
\def\WNDOFb{  220}   % Hubble Ndof
\def\MLCSWRMSb{  0.20}   % Hubble RMS

\def\LCDMOMb{  0.300}   % LCDM Omega_matter
\def\LCDMOMSTATERRb{  0.023}   % LCDM stat error on O_Mat
\def\LCDMOMSYSTERRb{  0.000}   % LCDM OM syst error
\def\LCDMOMTOTERRb{  0.023}   % LCDM OM total error
\def\LCDMOLb{  0.714}   % LCDM Omega_Lambda
\def\LCDMOLSTATERRb{  0.019}   % LCDM stat error on O_Lam
\def\LCDMOLSYSTERRb{  0.000}   % LCDM O_Lam syst error
\def\LCDMOLTOTERRb{  0.019}   % LCDM O_Lam total error
\def\LCDMCHISQb{  171.7}   % LCDM Hubble chi2
\def\LCDMNDOFb{  220}   % LCDM Hubble Ndof
\def\WSYSTcJRKVECT{  0.001}   % JRK original vectors                    
\def\WSYSTcRVSHIFT{  0.006}   % change RV by 1 sigma                    
\def\WSYSTcXTMW{  0.004}   % change XTMW by 1 sigma                  
\def\WSYSTcFILTSHIFT{  0.007}   % shifted LOWZ filters (no color transf)  
\def\WSYSTcSIMEFFLOWZ{  0.014}   % simeff for LOWZ                         
\def\WSYSTcSIMEFFSDSS{  0.072}   % simeff for SDSS                         
\def\WSYSTcSIMEFFESSE{  0.000}   % simeff for ESSE                         
\def\WSYSTcSIMEFFSNLS{  0.000}   % simeff for SNLS                         
\def\WSYSTcLOWZCUT{  0.060}   % LOWZ CUT at .02                         
\def\WSYSTcnoU{  0.310}   % deweight U-band                         
\def\WSYSTcERRUBVRI{  0.030}   % .01 mag error in U,B,V,R,I              
\def\WSYSTcERRki{  0.005}   % 1 sigma shift in color terms            
\def\WSYSTcERRABOFF{  0.030}   % 1 sigma shift in SDSS AB offsets        
\def\WSYSTcZPTSNLS{  0.000}   % 1 sigma shift in SNLS griz ZPTs         
\def\WSYSTcZPTESSE{  0.000}   % 1 sigma shift in ESSENCE R-I ZPT        
\def\WSYSTcZPTHST{  0.000}   % 1 sigma shift in HST ZPTs               
\def\OMSYSTcJRKVECT{  0.000}   % JRK original vectors                    
\def\OMSYSTcRVSHIFT{  0.002}   % change RV by 1 sigma                    
\def\OMSYSTcXTMW{  0.001}   % change XTMW by 1 sigma                  
\def\OMSYSTcFILTSHIFT{  0.001}   % shifted LOWZ filters (no color transf)  
\def\OMSYSTcSIMEFFLOWZ{  0.003}   % simeff for LOWZ                         
\def\OMSYSTcSIMEFFSDSS{  0.015}   % simeff for SDSS                         
\def\OMSYSTcSIMEFFESSE{  0.000}   % simeff for ESSE                         
\def\OMSYSTcSIMEFFSNLS{  0.000}   % simeff for SNLS                         
\def\OMSYSTcLOWZCUT{  0.014}   % LOWZ CUT at .02                         
\def\OMSYSTcnoU{  0.051}   % deweight U-band                         
\def\OMSYSTcERRUBVRI{  0.006}   % .01 mag error in U,B,V,R,I              
\def\OMSYSTcERRki{  0.001}   % 1 sigma shift in color terms            
\def\OMSYSTcERRABOFF{  0.006}   % 1 sigma shift in SDSS AB offsets        
\def\OMSYSTcZPTSNLS{  0.000}   % 1 sigma shift in SNLS griz ZPTs         
\def\OMSYSTcZPTESSE{  0.000}   % 1 sigma shift in ESSENCE R-I ZPT        
\def\OMSYSTcZPTHST{  0.000}   % 1 sigma shift in HST ZPTs               
\def\MLCSWRESc{ -0.92}   % w central value
\def\MLCSWSTATERRc{  0.13}   % w stat error
\def\MLCSWSYSTERRPc{  0.10}   % w +syst error
\def\MLCSWSYSTERRMc{  0.33}   % w -syst error
\def\MLCSWTOTERRPc{  0.16}   % w +total syst
\def\MLCSWTOTERRMc{  0.35}   % w -total syst
\def\MLCSOMRESc{  0.273}   % Omega_matter
\def\MLCSOMSTATERRc{  0.028}   % stat error on O_Mat
\def\MLCSOMSYSTERRPc{  0.023}   % OM +syst error
\def\MLCSOMSYSTERRMc{  0.056}   % OM -syst error
\def\MLCSOMTOTERRPc{  0.036}   % OM +total error
\def\MLCSOMTOTERRMc{  0.062}   % OM -total error
\def\MLCSWCHISQc{  279.5}   % Hubble chi2
\def\WNDOFc{  135}   % Hubble Ndof
\def\MLCSWRMSc{  0.16}   % Hubble RMS

\def\LCDMOMc{  0.274}   % LCDM Omega_matter
\def\LCDMOMSTATERRc{  0.023}   % LCDM stat error on O_Mat
\def\LCDMOMSYSTERRc{  0.001}   % LCDM OM syst error
\def\LCDMOMTOTERRc{  0.023}   % LCDM OM total error
\def\LCDMOLc{  0.735}   % LCDM Omega_Lambda
\def\LCDMOLSTATERRc{  0.019}   % LCDM stat error on O_Lam
\def\LCDMOLSYSTERRc{  0.006}   % LCDM O_Lam syst error
\def\LCDMOLTOTERRc{  0.019}   % LCDM O_Lam total error
\def\LCDMCHISQc{   87.1}   % LCDM Hubble chi2
\def\LCDMNDOFc{  135}   % LCDM Hubble Ndof
\def\WSYSTdJRKVECT{  0.040}   % JRK original vectors                    
\def\WSYSTdRVSHIFT{  0.007}   % change RV by 1 sigma                    
\def\WSYSTdXTMW{  0.016}   % change XTMW by 1 sigma                  
\def\WSYSTdFILTSHIFT{  0.016}   % shifted LOWZ filters (no color transf)  
\def\WSYSTdSIMEFFLOWZ{  0.008}   % simeff for LOWZ                         
\def\WSYSTdSIMEFFSDSS{  0.007}   % simeff for SDSS                         
\def\WSYSTdSIMEFFESSE{  0.012}   % simeff for ESSE                         
\def\WSYSTdSIMEFFSNLS{  0.025}   % simeff for SNLS                         
\def\WSYSTdLOWZCUT{  0.040}   % LOWZ CUT at .02                         
\def\WSYSTdnoU{  0.080}   % deweight U-band                         
\def\WSYSTdERRUBVRI{  0.021}   % .01 mag error in U,B,V,R,I              
\def\WSYSTdERRki{  0.006}   % 1 sigma shift in color terms            
\def\WSYSTdERRABOFF{  0.012}   % 1 sigma shift in SDSS AB offsets        
\def\WSYSTdZPTSNLS{  0.036}   % 1 sigma shift in SNLS griz ZPTs         
\def\WSYSTdZPTESSE{  0.015}   % 1 sigma shift in ESSENCE R-I ZPT        
\def\WSYSTdZPTHST{  0.000}   % 1 sigma shift in HST ZPTs               
\def\OMSYSTdJRKVECT{  0.010}   % JRK original vectors                    
\def\OMSYSTdRVSHIFT{  0.002}   % change RV by 1 sigma                    
\def\OMSYSTdXTMW{  0.004}   % change XTMW by 1 sigma                  
\def\OMSYSTdFILTSHIFT{  0.004}   % shifted LOWZ filters (no color transf)  
\def\OMSYSTdSIMEFFLOWZ{  0.002}   % simeff for LOWZ                         
\def\OMSYSTdSIMEFFSDSS{  0.002}   % simeff for SDSS                         
\def\OMSYSTdSIMEFFESSE{  0.002}   % simeff for ESSE                         
\def\OMSYSTdSIMEFFSNLS{  0.007}   % simeff for SNLS                         
\def\OMSYSTdLOWZCUT{  0.009}   % LOWZ CUT at .02                         
\def\OMSYSTdnoU{  0.016}   % deweight U-band                         
\def\OMSYSTdERRUBVRI{  0.005}   % .01 mag error in U,B,V,R,I              
\def\OMSYSTdERRki{  0.002}   % 1 sigma shift in color terms            
\def\OMSYSTdERRABOFF{  0.003}   % 1 sigma shift in SDSS AB offsets        
\def\OMSYSTdZPTSNLS{  0.009}   % 1 sigma shift in SNLS griz ZPTs         
\def\OMSYSTdZPTESSE{  0.004}   % 1 sigma shift in ESSENCE R-I ZPT        
\def\OMSYSTdZPTHST{  0.000}   % 1 sigma shift in HST ZPTs               
\def\MLCSWRESd{ -0.76}   % w central value
\def\MLCSWSTATERRd{  0.08}   % w stat error
\def\MLCSWSYSTERRPd{  0.11}   % w +syst error
\def\MLCSWTOTERRPd{  0.14}   % w +total syst
\def\MLCSOMRESd{  0.306}   % Omega_matter
\def\MLCSOMSTATERRd{  0.021}   % stat error on O_Mat
\def\MLCSOMSYSTERRPd{  0.026}   % OM +syst error
\def\MLCSOMTOTERRPd{  0.033}   % OM +total error
\def\MLCSWCHISQd{  517.8}   % Hubble chi2
\def\WNDOFd{  253}   % Hubble Ndof
\def\MLCSWRMSd{  0.20}   % Hubble RMS

\def\LCDMOMd{  0.302}   % LCDM Omega_matter
\def\LCDMOMSTATERRd{  0.023}   % LCDM stat error on O_Mat
\def\LCDMOMSYSTERRd{  0.001}   % LCDM OM syst error
\def\LCDMOMTOTERRd{  0.023}   % LCDM OM total error
\def\LCDMOLd{  0.713}   % LCDM Omega_Lambda
\def\LCDMOLSTATERRd{  0.019}   % LCDM stat error on O_Lam
\def\LCDMOLSYSTERRd{  0.004}   % LCDM O_Lam syst error
\def\LCDMOLTOTERRd{  0.019}   % LCDM O_Lam total error
\def\LCDMCHISQd{  203.9}   % LCDM Hubble chi2
\def\LCDMNDOFd{  253}   % LCDM Hubble Ndof
\def\WSYSTeJRKVECT{  0.026}   % JRK original vectors                    
\def\WSYSTeRVSHIFT{  0.008}   % change RV by 1 sigma                    
\def\WSYSTeXTMW{  0.022}   % change XTMW by 1 sigma                  
\def\WSYSTeFILTSHIFT{  0.013}   % shifted LOWZ filters (no color transf)  
\def\WSYSTeSIMEFFLOWZ{  0.007}   % simeff for LOWZ                         
\def\WSYSTeSIMEFFSDSS{  0.001}   % simeff for SDSS                         
\def\WSYSTeSIMEFFESSE{  0.008}   % simeff for ESSE                         
\def\WSYSTeSIMEFFSNLS{  0.017}   % simeff for SNLS                         
\def\WSYSTeLOWZCUT{  0.040}   % LOWZ CUT at .02                         
\def\WSYSTenoU{  0.080}   % deweight U-band                         
\def\WSYSTeERRUBVRI{  0.019}   % .01 mag error in U,B,V,R,I              
\def\WSYSTeERRki{  0.006}   % 1 sigma shift in color terms            
\def\WSYSTeERRABOFF{  0.013}   % 1 sigma shift in SDSS AB offsets        
\def\WSYSTeZPTSNLS{  0.024}   % 1 sigma shift in SNLS griz ZPTs         
\def\WSYSTeZPTESSE{  0.008}   % 1 sigma shift in ESSENCE R-I ZPT        
\def\WSYSTeZPTHST{  0.012}   % 1 sigma shift in HST ZPTs               
\def\OMSYSTeJRKVECT{  0.006}   % JRK original vectors                    
\def\OMSYSTeRVSHIFT{  0.003}   % change RV by 1 sigma                    
\def\OMSYSTeXTMW{  0.006}   % change XTMW by 1 sigma                  
\def\OMSYSTeFILTSHIFT{  0.003}   % shifted LOWZ filters (no color transf)  
\def\OMSYSTeSIMEFFLOWZ{  0.002}   % simeff for LOWZ                         
\def\OMSYSTeSIMEFFSDSS{  0.000}   % simeff for SDSS                         
\def\OMSYSTeSIMEFFESSE{  0.002}   % simeff for ESSE                         
\def\OMSYSTeSIMEFFSNLS{  0.004}   % simeff for SNLS                         
\def\OMSYSTeLOWZCUT{  0.009}   % LOWZ CUT at .02                         
\def\OMSYSTenoU{  0.016}   % deweight U-band                         
\def\OMSYSTeERRUBVRI{  0.005}   % .01 mag error in U,B,V,R,I              
\def\OMSYSTeERRki{  0.002}   % 1 sigma shift in color terms            
\def\OMSYSTeERRABOFF{  0.003}   % 1 sigma shift in SDSS AB offsets        
\def\OMSYSTeZPTSNLS{  0.005}   % 1 sigma shift in SNLS griz ZPTs         
\def\OMSYSTeZPTESSE{  0.002}   % 1 sigma shift in ESSENCE R-I ZPT        
\def\OMSYSTeZPTHST{  0.003}   % 1 sigma shift in HST ZPTs               
\def\MLCSWRESe{ -0.76}   % w central value
\def\MLCSWSTATERRe{  0.07}   % w stat error
\def\MLCSWSYSTERRPe{  0.11}   % w +syst error
\def\MLCSWTOTERRPe{  0.13}   % w +total syst
\def\MLCSOMRESe{  0.307}   % Omega_matter
\def\MLCSOMSTATERRe{  0.019}   % stat error on O_Mat
\def\MLCSOMSYSTERRPe{  0.023}   % OM +syst error
\def\MLCSOMTOTERRPe{  0.030}   % OM +total error
\def\MLCSWCHISQe{  568.1}   % Hubble chi2
\def\WNDOFe{  287}   % Hubble Ndof
\def\MLCSWRMSe{  0.21}   % Hubble RMS

\def\LCDMOMe{  0.312}   % LCDM Omega_matter
\def\LCDMOMSTATERRe{  0.022}   % LCDM stat error on O_Mat
\def\LCDMOMSYSTERRe{  0.001}   % LCDM OM syst error
\def\LCDMOMTOTERRe{  0.022}   % LCDM OM total error
\def\LCDMOLe{  0.705}   % LCDM Omega_Lambda
\def\LCDMOLSTATERRe{  0.018}   % LCDM stat error on O_Lam
\def\LCDMOLSYSTERRe{  0.004}   % LCDM O_Lam syst error
\def\LCDMOLTOTERRe{  0.018}   % LCDM O_Lam total error
\def\LCDMCHISQe{  237.9}   % LCDM Hubble chi2
\def\LCDMNDOFe{  287}   % LCDM Hubble Ndof
\def\WSYSTfJRKVECT{  0.043}   % JRK original vectors                    
\def\WSYSTfRVSHIFT{  0.009}   % change RV by 1 sigma                    
\def\WSYSTfXTMW{  0.023}   % change XTMW by 1 sigma                  
\def\WSYSTfFILTSHIFT{  0.014}   % shifted LOWZ filters (no color transf)  
\def\WSYSTfSIMEFFLOWZ{  0.013}   % simeff for LOWZ                         
\def\WSYSTfSIMEFFSDSS{  0.000}   % simeff for SDSS                         
\def\WSYSTfSIMEFFESSE{  0.012}   % simeff for ESSE                         
\def\WSYSTfSIMEFFSNLS{  0.026}   % simeff for SNLS                         
\def\WSYSTfLOWZCUT{  0.040}   % LOWZ CUT at .02                         
\def\WSYSTfnoU{  0.080}   % deweight U-band                         
\def\WSYSTfERRUBVRI{  0.021}   % .01 mag error in U,B,V,R,I              
\def\WSYSTfERRki{  0.006}   % 1 sigma shift in color terms            
\def\WSYSTfERRABOFF{  0.000}   % 1 sigma shift in SDSS AB offsets        
\def\WSYSTfZPTSNLS{  0.035}   % 1 sigma shift in SNLS griz ZPTs         
\def\WSYSTfZPTESSE{  0.010}   % 1 sigma shift in ESSENCE R-I ZPT        
\def\WSYSTfZPTHST{  0.000}   % 1 sigma shift in HST ZPTs               
\def\OMSYSTfJRKVECT{  0.011}   % JRK original vectors                    
\def\OMSYSTfRVSHIFT{  0.002}   % change RV by 1 sigma                    
\def\OMSYSTfXTMW{  0.006}   % change XTMW by 1 sigma                  
\def\OMSYSTfFILTSHIFT{  0.003}   % shifted LOWZ filters (no color transf)  
\def\OMSYSTfSIMEFFLOWZ{  0.003}   % simeff for LOWZ                         
\def\OMSYSTfSIMEFFSDSS{  0.000}   % simeff for SDSS                         
\def\OMSYSTfSIMEFFESSE{  0.002}   % simeff for ESSE                         
\def\OMSYSTfSIMEFFSNLS{  0.007}   % simeff for SNLS                         
\def\OMSYSTfLOWZCUT{  0.009}   % LOWZ CUT at .02                         
\def\OMSYSTfnoU{  0.016}   % deweight U-band                         
\def\OMSYSTfERRUBVRI{  0.005}   % .01 mag error in U,B,V,R,I              
\def\OMSYSTfERRki{  0.001}   % 1 sigma shift in color terms            
\def\OMSYSTfERRABOFF{  0.000}   % 1 sigma shift in SDSS AB offsets        
\def\OMSYSTfZPTSNLS{  0.008}   % 1 sigma shift in SNLS griz ZPTs         
\def\OMSYSTfZPTESSE{  0.002}   % 1 sigma shift in ESSENCE R-I ZPT        
\def\OMSYSTfZPTHST{  0.000}   % 1 sigma shift in HST ZPTs               
\def\MLCSWRESf{ -0.78}   % w central value
\def\MLCSWSTATERRf{  0.08}   % w stat error
\def\MLCSWSYSTERRPf{  0.12}   % w +syst error
\def\MLCSWTOTERRPf{  0.14}   % w +total syst
\def\MLCSOMRESf{  0.302}   % Omega_matter
\def\MLCSOMSTATERRf{  0.022}   % stat error on O_Mat
\def\MLCSOMSYSTERRPf{  0.026}   % OM +syst error
\def\MLCSOMTOTERRPf{  0.034}   % OM +total error
\def\MLCSWCHISQf{  341.9}   % Hubble chi2
\def\WNDOFf{  150}   % Hubble Ndof
\def\MLCSWRMSf{  0.23}   % Hubble RMS

\def\LCDMOMf{  0.294}   % LCDM Omega_matter
\def\LCDMOMSTATERRf{  0.023}   % LCDM stat error on O_Mat
\def\LCDMOMSYSTERRf{  0.001}   % LCDM OM syst error
\def\LCDMOMTOTERRf{  0.024}   % LCDM OM total error
\def\LCDMOLf{  0.718}   % LCDM Omega_Lambda
\def\LCDMOLSTATERRf{  0.019}   % LCDM stat error on O_Lam
\def\LCDMOLSYSTERRf{  0.004}   % LCDM O_Lam syst error
\def\LCDMOLTOTERRf{  0.019}   % LCDM O_Lam total error
\def\LCDMCHISQf{  145.8}   % LCDM Hubble chi2
\def\LCDMNDOFf{  150}   % LCDM Hubble Ndof
\def\FWCDMBAOCMBSYSTUBANDOMEGAMa{-0.020}

\def\FWCDMBAOCMBSYSTUBANDWa{-0.100}

\def\FWCDMBAOCMBSYSTUBANDOMEGAMb{0.022}

\def\FWCDMBAOCMBSYSTUBANDWb{0.104}

\def\FWCDMBAOCMBSYSTUBANDOMEGAMc{-0.024}

\def\FWCDMBAOCMBSYSTUBANDWc{-0.133}

\def\FWCDMBAOCMBSYSTUBANDOMEGAMd{0.022}

\def\FWCDMBAOCMBSYSTUBANDWd{0.104}

\def\FWCDMBAOCMBSYSTUBANDOMEGAMe{0.022}

\def\FWCDMBAOCMBSYSTUBANDWe{0.104}

\def\FWCDMBAOCMBSYSTUBANDOMEGAMf{0.022}

\def\FWCDMBAOCMBSYSTUBANDWf{0.104}

\def\FWCDMBAOCMBSYSTXTMWOMEGAMa{0.005}

\def\FWCDMBAOCMBSYSTXTMWWa{0.021}

\def\FWCDMBAOCMBSYSTXTMWOMEGAMb{0.003}

\def\FWCDMBAOCMBSYSTXTMWWb{0.012}

\def\FWCDMBAOCMBSYSTXTMWOMEGAMc{0.001}

\def\FWCDMBAOCMBSYSTXTMWWc{0.004}

\def\FWCDMBAOCMBSYSTXTMWOMEGAMd{0.004}

\def\FWCDMBAOCMBSYSTXTMWWd{0.016}

\def\FWCDMBAOCMBSYSTXTMWOMEGAMe{0.006}

\def\FWCDMBAOCMBSYSTXTMWWe{0.022}

\def\FWCDMBAOCMBSYSTXTMWOMEGAMf{0.006}

\def\FWCDMBAOCMBSYSTXTMWWf{0.023}

\def\FWCDMBAOCMBSYSTBETAZOMEGAMa{0.000}

\def\FWCDMBAOCMBSYSTBETAZWa{0.000}

\def\FWCDMBAOCMBSYSTBETAZOMEGAMb{+0.016}

\def\FWCDMBAOCMBSYSTBETAZWb{+0.073}

\def\FWCDMBAOCMBSYSTBETAZOMEGAMc{0.000}

\def\FWCDMBAOCMBSYSTBETAZWc{0.000}

\def\FWCDMBAOCMBSYSTBETAZOMEGAMd{+0.010}

\def\FWCDMBAOCMBSYSTBETAZWd{+0.045}

\def\FWCDMBAOCMBSYSTBETAZOMEGAMe{+0.002}

\def\FWCDMBAOCMBSYSTBETAZWe{+0.013}

\def\FWCDMBAOCMBSYSTBETAZOMEGAMf{+0.007}

\def\FWCDMBAOCMBSYSTBETAZWf{+0.036}

\def\FWCDMBAOCMBSYSTRTDOMEGAMa{0.002}

\def\FWCDMBAOCMBSYSTRTDWa{0.008}

\def\FWCDMBAOCMBSYSTRTDOMEGAMb{0.001}

\def\FWCDMBAOCMBSYSTRTDWb{0.005}

\def\FWCDMBAOCMBSYSTRTDOMEGAMc{0.003}

\def\FWCDMBAOCMBSYSTRTDWc{0.017}

\def\FWCDMBAOCMBSYSTRTDOMEGAMd{0.001}

\def\FWCDMBAOCMBSYSTRTDWd{0.011}

\def\FWCDMBAOCMBSYSTRTDOMEGAMe{0.002}

\def\FWCDMBAOCMBSYSTRTDWe{0.005}

\def\FWCDMBAOCMBSYSTRTDOMEGAMf{0.001}

\def\FWCDMBAOCMBSYSTRTDWf{0.005}

\def\FWCDMBAOCMBSYSTDISPOMEGAMa{0.000}

\def\FWCDMBAOCMBSYSTDISPWa{0.001}

\def\FWCDMBAOCMBSYSTDISPOMEGAMb{0.001}

\def\FWCDMBAOCMBSYSTDISPWb{0.003}

\def\FWCDMBAOCMBSYSTDISPOMEGAMc{0.000}

\def\FWCDMBAOCMBSYSTDISPWc{0.002}

\def\FWCDMBAOCMBSYSTDISPOMEGAMd{0.001}

\def\FWCDMBAOCMBSYSTDISPWd{0.006}

\def\FWCDMBAOCMBSYSTDISPOMEGAMe{0.001}

\def\FWCDMBAOCMBSYSTDISPWe{0.006}

\def\FWCDMBAOCMBSYSTDISPOMEGAMf{0.001}

\def\FWCDMBAOCMBSYSTDISPWf{0.004}

\def\FWCDMBAOCMBSYSTSImSNLSOMEGAMa{0.005}

\def\FWCDMBAOCMBSYSTSImSNLSWa{0.020}

\def\FWCDMBAOCMBSYSTSImSNLSOMEGAMb{0.003}

\def\FWCDMBAOCMBSYSTSImSNLSWb{0.011}

\def\FWCDMBAOCMBSYSTSImSNLSOMEGAMc{0.002}

\def\FWCDMBAOCMBSYSTSImSNLSWc{0.009}

\def\FWCDMBAOCMBSYSTSImSNLSOMEGAMd{0.003}

\def\FWCDMBAOCMBSYSTSImSNLSWd{0.002}

\def\FWCDMBAOCMBSYSTSImSNLSOMEGAMe{0.000}

\def\FWCDMBAOCMBSYSTSImSNLSWe{0.001}

\def\FWCDMBAOCMBSYSTSImSNLSOMEGAMf{0.003}

\def\FWCDMBAOCMBSYSTSImSNLSWf{0.012}

\def\FWCDMBAOCMBSYSTLOWZZPTOMEGAMa{0.006}

\def\FWCDMBAOCMBSYSTLOWZZPTWa{0.029}
\def\FWCDMBAOCMBSYSTLOWZZPTOMEGAMb{0.008}

\def\FWCDMBAOCMBSYSTLOWZZPTWb{0.030}
\def\FWCDMBAOCMBSYSTLOWZZPTOMEGAMc{0.005}

\def\FWCDMBAOCMBSYSTLOWZZPTWc{0.027}
\def\FWCDMBAOCMBSYSTLOWZZPTOMEGAMd{0.005}

\def\FWCDMBAOCMBSYSTLOWZZPTWd{0.022}
\def\FWCDMBAOCMBSYSTLOWZZPTOMEGAMe{0.005}

\def\FWCDMBAOCMBSYSTLOWZZPTWe{0.020}
\def\FWCDMBAOCMBSYSTLOWZZPTOMEGAMf{0.005}

\def\FWCDMBAOCMBSYSTLOWZZPTWf{0.022}
\def\FWCDMBAOCMBSYSTBESSOMEGAMa{0.000}

\def\FWCDMBAOCMBSYSTBESSWa{0.000}
\def\FWCDMBAOCMBSYSTBESSOMEGAMb{0.000}

\def\FWCDMBAOCMBSYSTBESSWb{0.000}
\def\FWCDMBAOCMBSYSTBESSOMEGAMc{0.003}

\def\FWCDMBAOCMBSYSTBESSWc{0.015}
\def\FWCDMBAOCMBSYSTBESSOMEGAMd{0.002}

\def\FWCDMBAOCMBSYSTBESSWd{0.010}
\def\FWCDMBAOCMBSYSTBESSOMEGAMe{0.001}

\def\FWCDMBAOCMBSYSTBESSWe{0.008}
\def\FWCDMBAOCMBSYSTBESSOMEGAMf{0.002}

\def\FWCDMBAOCMBSYSTBESSWf{0.013}
\def\FWCDMBAOCMBSYSTABOMEGAMa{0.004}

\def\FWCDMBAOCMBSYSTABWa{0.018}
\def\FWCDMBAOCMBSYSTABOMEGAMb{0.007}

\def\FWCDMBAOCMBSYSTABWb{0.037}
\def\FWCDMBAOCMBSYSTABOMEGAMc{0.006}

\def\FWCDMBAOCMBSYSTABWc{0.031}
\def\FWCDMBAOCMBSYSTABOMEGAMd{0.003}

\def\FWCDMBAOCMBSYSTABWd{0.015}
\def\FWCDMBAOCMBSYSTABOMEGAMe{0.003}

\def\FWCDMBAOCMBSYSTABWe{0.016}
\def\FWCDMBAOCMBSYSTABOMEGAMf{0.000}

\def\FWCDMBAOCMBSYSTABWf{0.000}
\def\FWCDMBAOCMBSYSTESSENCEZPTOMEGAMa{0.000}

\def\FWCDMBAOCMBSYSTESSENCEZPTWa{0.000}
\def\FWCDMBAOCMBSYSTESSENCEZPTOMEGAMb{0.006}

\def\FWCDMBAOCMBSYSTESSENCEZPTWb{0.035}
\def\FWCDMBAOCMBSYSTESSENCEZPTOMEGAMc{0.000}

\def\FWCDMBAOCMBSYSTESSENCEZPTWc{0.000}
\def\FWCDMBAOCMBSYSTESSENCEZPTOMEGAMd{0.006}

\def\FWCDMBAOCMBSYSTESSENCEZPTWd{0.036}
\def\FWCDMBAOCMBSYSTESSENCEZPTOMEGAMe{0.003}

\def\FWCDMBAOCMBSYSTESSENCEZPTWe{0.021}
\def\FWCDMBAOCMBSYSTESSENCEZPTOMEGAMf{0.004}

\def\FWCDMBAOCMBSYSTESSENCEZPTWf{0.025}
\def\FWCDMBAOCMBSYSTSNLSZPTOMEGAMa{0.000}

\def\FWCDMBAOCMBSYSTSNLSZPTWa{0.000}
\def\FWCDMBAOCMBSYSTSNLSZPTOMEGAMb{0.011}

\def\FWCDMBAOCMBSYSTSNLSZPTWb{0.057}
\def\FWCDMBAOCMBSYSTSNLSZPTOMEGAMc{0.000}

\def\FWCDMBAOCMBSYSTSNLSZPTWc{0.000}
\def\FWCDMBAOCMBSYSTSNLSZPTOMEGAMd{0.009}

\def\FWCDMBAOCMBSYSTSNLSZPTWd{0.046}
\def\FWCDMBAOCMBSYSTSNLSZPTOMEGAMe{0.005}

\def\FWCDMBAOCMBSYSTSNLSZPTWe{0.030}
\def\FWCDMBAOCMBSYSTSNLSZPTOMEGAMf{0.008}

\def\FWCDMBAOCMBSYSTSNLSZPTWf{0.043}
\def\FWCDMBAOCMBSYSTHSTZPTOMEGAMa{0.000}

\def\FWCDMBAOCMBSYSTHSTZPTWa{0.000}
\def\FWCDMBAOCMBSYSTHSTZPTOMEGAMb{0.000}

\def\FWCDMBAOCMBSYSTHSTZPTWb{0.000}
\def\FWCDMBAOCMBSYSTHSTZPTOMEGAMc{0.000}

\def\FWCDMBAOCMBSYSTHSTZPTWc{0.000}
\def\FWCDMBAOCMBSYSTHSTZPTOMEGAMd{0.000}

\def\FWCDMBAOCMBSYSTHSTZPTWd{0.000}
\def\FWCDMBAOCMBSYSTHSTZPTOMEGAMe{0.003}

\def\FWCDMBAOCMBSYSTHSTZPTWe{0.015}
\def\FWCDMBAOCMBSYSTHSTZPTOMEGAMf{0.000}

\def\FWCDMBAOCMBSYSTHSTZPTWf{0.000}
\def\FWCDMBAOCMBSYSTZMINOMEGAMa{0.012}

\def\FWCDMBAOCMBSYSTZMINWa{0.050}

\def\FWCDMBAOCMBSYSTZMINOMEGAMb{0.007}

\def\FWCDMBAOCMBSYSTZMINWb{0.030}

\def\FWCDMBAOCMBSYSTZMINOMEGAMc{0.012}

\def\FWCDMBAOCMBSYSTZMINWc{0.050}

\def\FWCDMBAOCMBSYSTZMINOMEGAMd{0.007}

\def\FWCDMBAOCMBSYSTZMINWd{0.030}

\def\FWCDMBAOCMBSYSTZMINOMEGAMe{0.007}

\def\FWCDMBAOCMBSYSTZMINWe{0.030}

\def\FWCDMBAOCMBSYSTZMINOMEGAMf{0.007}

\def\FWCDMBAOCMBSYSTZMINWf{0.030}

\def\FWCDMBAOCMBDISPa{0.084}
\def\FWCDMBAOCMBDISPb{0.124}
\def\FWCDMBAOCMBDISPc{0.105}
\def\FWCDMBAOCMBDISPd{0.128}
\def\FWCDMBAOCMBDISPe{0.140}
\def\FWCDMBAOCMBDISPf{0.160}
\def\FWCDMBAOCMBRMSa{0.178}
\def\FWCDMBAOCMBRMSb{0.219}
\def\FWCDMBAOCMBRMSc{0.170}
\def\FWCDMBAOCMBRMSd{0.210}
\def\FWCDMBAOCMBRMSe{0.232}
\def\FWCDMBAOCMBRMSf{0.231}

\def\FWCDMBAOCMBWaEFFSHIFT{-0.04}
\def\FWCDMBAOCMBWa{-0.87}

\def\FWCDMBAOCMBWbEFFSHIFT{-0.02}
\def\FWCDMBAOCMBWb{-0.98}

\def\FWCDMBAOCMBWcEFFSHIFT{-0.02}
\def\FWCDMBAOCMBWc{-0.92}

\def\FWCDMBAOCMBWdEFFSHIFT{ 0.00}
\def\FWCDMBAOCMBWd{-0.98}

\def\FWCDMBAOCMBWeEFFSHIFT{ 0.00}
\def\FWCDMBAOCMBWe{-0.96}

\def\FWCDMBAOCMBWfEFFSHIFT{ 0.02}
\def\FWCDMBAOCMBWf{-0.95}
\def\FWCDMBAOCMBDWSTATa{ 0.12}
\def\FWCDMBAOCMBDWSTATb{ 0.08}
\def\FWCDMBAOCMBDWSTATc{ 0.11}
\def\FWCDMBAOCMBDWSTATd{ 0.07}
\def\FWCDMBAOCMBDWSTATe{ 0.06}
\def\FWCDMBAOCMBDWSTATf{ 0.08}
\def\FWCDMBAOCMBDWSYSTPa{ 0.06}
\def\FWCDMBAOCMBDWSYSTPb{ 0.15}
\def\FWCDMBAOCMBDWSYSTPc{ 0.07}
\def\FWCDMBAOCMBDWSYSTPd{ 0.13}
\def\FWCDMBAOCMBDWSYSTPe{ 0.12}
\def\FWCDMBAOCMBDWSYSTPf{ 0.13}
\def\FWCDMBAOCMBDWSYSTMa{ 0.12}
\def\FWCDMBAOCMBDWSYSTMb{ 0.14}
\def\FWCDMBAOCMBDWSYSTMc{ 0.15}
\def\FWCDMBAOCMBDWSYSTMd{ 0.13}
\def\FWCDMBAOCMBDWSYSTMe{ 0.12}
\def\FWCDMBAOCMBDWSYSTMf{ 0.12}
\def\FWCDMBAOCMBDWTOTPa{ 0.14}
\def\FWCDMBAOCMBDWTOTPb{ 0.17}
\def\FWCDMBAOCMBDWTOTPc{ 0.13}
\def\FWCDMBAOCMBDWTOTPd{ 0.15}
\def\FWCDMBAOCMBDWTOTPe{ 0.14}
\def\FWCDMBAOCMBDWTOTPf{ 0.15}
\def\FWCDMBAOCMBDWTOTMa{ 0.17}

\def\FWCDMBAOCMBDWTOTMc{ 0.18}

\def\FWCDMBAOCMBOMEGAMaEFFSHIFT{-0.010}
\def\FWCDMBAOCMBOMEGAMa{0.281}

\def\FWCDMBAOCMBOMEGAMbEFFSHIFT{-0.007}
\def\FWCDMBAOCMBOMEGAMb{0.256}

\def\FWCDMBAOCMBOMEGAMcEFFSHIFT{-0.005}
\def\FWCDMBAOCMBOMEGAMc{0.271}

\def\FWCDMBAOCMBOMEGAMdEFFSHIFT{0.006}
\def\FWCDMBAOCMBOMEGAMd{0.264}

\def\FWCDMBAOCMBOMEGAMeEFFSHIFT{0.001}
\def\FWCDMBAOCMBOMEGAMe{0.265}

\def\FWCDMBAOCMBOMEGAMfEFFSHIFT{0.006}
\def\FWCDMBAOCMBOMEGAMf{0.267}
\def\FWCDMBAOCMBDOMEGAMSTATa{0.030}
\def\FWCDMBAOCMBDOMEGAMSTATb{0.019}
\def\FWCDMBAOCMBDOMEGAMSTATc{0.025}
\def\FWCDMBAOCMBDOMEGAMSTATd{0.017}
\def\FWCDMBAOCMBDOMEGAMSTATe{0.016}
\def\FWCDMBAOCMBDOMEGAMSTATf{0.019}
\def\FWCDMBAOCMBDOMEGAMSYSTPa{0.015}
\def\FWCDMBAOCMBDOMEGAMSYSTPb{0.033}
\def\FWCDMBAOCMBDOMEGAMSYSTPc{0.015}
\def\FWCDMBAOCMBDOMEGAMSYSTPd{0.028}
\def\FWCDMBAOCMBDOMEGAMSYSTPe{0.025}
\def\FWCDMBAOCMBDOMEGAMSYSTPf{0.027}
\def\FWCDMBAOCMBDOMEGAMSYSTMa{0.025}
\def\FWCDMBAOCMBDOMEGAMSYSTMb{0.029}
\def\FWCDMBAOCMBDOMEGAMSYSTMc{0.029}
\def\FWCDMBAOCMBDOMEGAMSYSTMd{0.027}
\def\FWCDMBAOCMBDOMEGAMSYSTMe{0.025}
\def\FWCDMBAOCMBDOMEGAMSYSTMf{0.026}
\def\FWCDMBAOCMBDOMEGAMTOTPa{0.034}
\def\FWCDMBAOCMBDOMEGAMTOTPb{0.038}
\def\FWCDMBAOCMBDOMEGAMTOTPc{0.029}
\def\FWCDMBAOCMBDOMEGAMTOTPd{0.033}
\def\FWCDMBAOCMBDOMEGAMTOTPe{0.030}
\def\FWCDMBAOCMBDOMEGAMTOTPf{0.033}
\def\FWCDMBAOCMBDOMEGAMTOTMa{0.039}

\def\FWCDMBAOCMBDOMEGAMTOTMc{0.038}

\def\FWCDMBAOCMBALPHAa{0.127}

\def\FWCDMBAOCMBALPHAb{0.123}

\def\FWCDMBAOCMBALPHAc{0.113}

\def\FWCDMBAOCMBALPHAd{0.107}

\def\FWCDMBAOCMBALPHAe{0.124}

\def\FWCDMBAOCMBALPHAf{0.106}
\def\FWCDMBAOCMBDALPHASTATa{0.017}
\def\FWCDMBAOCMBDALPHASTATb{0.015}
\def\FWCDMBAOCMBDALPHASTATc{0.014}
\def\FWCDMBAOCMBDALPHASTATd{0.013}
\def\FWCDMBAOCMBDALPHASTATe{0.014}
\def\FWCDMBAOCMBDALPHASTATf{0.019}
\def\FWCDMBAOCMBDALPHASYSTPa{0.020}
\def\FWCDMBAOCMBDALPHASYSTPb{0.021}
\def\FWCDMBAOCMBDALPHASYSTPc{0.016}
\def\FWCDMBAOCMBDALPHASYSTPd{0.020}
\def\FWCDMBAOCMBDALPHASYSTPe{0.023}
\def\FWCDMBAOCMBDALPHASYSTPf{0.023}

\def\FWCDMBAOCMBDALPHATOTPa{0.026}
\def\FWCDMBAOCMBDALPHATOTPb{0.026}
\def\FWCDMBAOCMBDALPHATOTPc{0.021}
\def\FWCDMBAOCMBDALPHATOTPd{0.024}
\def\FWCDMBAOCMBDALPHATOTPe{0.027}
\def\FWCDMBAOCMBDALPHATOTPf{0.030}

\def\FWCDMBAOCMBBETAa{ 2.52}

\def\FWCDMBAOCMBBETAb{ 2.62}

\def\FWCDMBAOCMBBETAc{ 2.50}

\def\FWCDMBAOCMBBETAd{ 2.66}

\def\FWCDMBAOCMBBETAe{ 2.64}

\def\FWCDMBAOCMBBETAf{ 2.56}
\def\FWCDMBAOCMBDBETASTATa{ 0.16}
\def\FWCDMBAOCMBDBETASTATb{ 0.13}
\def\FWCDMBAOCMBDBETASTATc{ 0.15}
\def\FWCDMBAOCMBDBETASTATd{ 0.12}
\def\FWCDMBAOCMBDBETASTATe{ 0.12}
\def\FWCDMBAOCMBDBETASTATf{ 0.17}
\def\FWCDMBAOCMBDBETASYSTPa{ 0.11}
\def\FWCDMBAOCMBDBETASYSTPb{ 0.19}
\def\FWCDMBAOCMBDBETASYSTPc{ 0.11}
\def\FWCDMBAOCMBDBETASYSTPd{ 0.19}
\def\FWCDMBAOCMBDBETASYSTPe{ 0.18}
\def\FWCDMBAOCMBDBETASYSTPf{ 0.25}

\def\FWCDMBAOCMBDBETATOTPa{ 0.19}
\def\FWCDMBAOCMBDBETATOTPb{ 0.23}
\def\FWCDMBAOCMBDBETATOTPc{ 0.19}
\def\FWCDMBAOCMBDBETATOTPd{ 0.22}
\def\FWCDMBAOCMBDBETATOTPe{ 0.22}
\def\FWCDMBAOCMBDBETATOTPf{ 0.30}

\def\LCDMBAOCMBDISPa{0.085}
\def\LCDMBAOCMBDISPb{0.123}
\def\LCDMBAOCMBDISPc{0.105}
\def\LCDMBAOCMBDISPd{0.128}
\def\LCDMBAOCMBDISPe{0.140}
\def\LCDMBAOCMBDISPf{0.160}
\def\LCDMBAOCMBRMSa{0.177}
\def\LCDMBAOCMBRMSb{0.220}
\def\LCDMBAOCMBRMSc{0.170}
\def\LCDMBAOCMBRMSd{0.210}
\def\LCDMBAOCMBRMSe{0.232}
\def\LCDMBAOCMBRMSf{0.231}

\def\LCDMBAOCMBOMEGALaEFFSHIFT{0.001}
\def\LCDMBAOCMBOMEGALa{0.734}

\def\LCDMBAOCMBOMEGALbEFFSHIFT{-0.003}
\def\LCDMBAOCMBOMEGALb{0.735}

\def\LCDMBAOCMBOMEGALcEFFSHIFT{0.001}
\def\LCDMBAOCMBOMEGALc{0.734}

\def\LCDMBAOCMBOMEGALdEFFSHIFT{-0.006}
\def\LCDMBAOCMBOMEGALd{0.734}

\def\LCDMBAOCMBOMEGALeEFFSHIFT{-0.008}
\def\LCDMBAOCMBOMEGALe{0.727}

\def\LCDMBAOCMBOMEGALfEFFSHIFT{-0.004}
\def\LCDMBAOCMBOMEGALf{0.734}
\def\LCDMBAOCMBDOMEGALSTATa{0.019}
\def\LCDMBAOCMBDOMEGALSTATb{0.017}
\def\LCDMBAOCMBDOMEGALSTATc{0.019}
\def\LCDMBAOCMBDOMEGALSTATd{0.017}
\def\LCDMBAOCMBDOMEGALSTATe{0.016}
\def\LCDMBAOCMBDOMEGALSTATf{0.017}
\def\LCDMBAOCMBDOMEGALSYSTPa{0.019}
\def\LCDMBAOCMBDOMEGALSYSTPb{0.026}
\def\LCDMBAOCMBDOMEGALSYSTPc{0.018}
\def\LCDMBAOCMBDOMEGALSYSTPd{0.020}
\def\LCDMBAOCMBDOMEGALSYSTPe{0.019}
\def\LCDMBAOCMBDOMEGALSYSTPf{0.018}

\def\LCDMBAOCMBDOMEGALTOTPa{0.027}
\def\LCDMBAOCMBDOMEGALTOTPb{0.031}
\def\LCDMBAOCMBDOMEGALTOTPc{0.026}
\def\LCDMBAOCMBDOMEGALTOTPd{0.026}
\def\LCDMBAOCMBDOMEGALTOTPe{0.025}
\def\LCDMBAOCMBDOMEGALTOTPf{0.025}

\def\LCDMBAOCMBOMEGAMaEFFSHIFT{-0.001}
\def\LCDMBAOCMBOMEGAMa{0.275}

\def\LCDMBAOCMBOMEGAMbEFFSHIFT{0.004}
\def\LCDMBAOCMBOMEGAMb{0.274}

\def\LCDMBAOCMBOMEGAMcEFFSHIFT{-0.001}
\def\LCDMBAOCMBOMEGAMc{0.275}

\def\LCDMBAOCMBOMEGAMdEFFSHIFT{0.007}
\def\LCDMBAOCMBOMEGAMd{0.275}

\def\LCDMBAOCMBOMEGAMeEFFSHIFT{0.005}
\def\LCDMBAOCMBOMEGAMe{0.279}

\def\LCDMBAOCMBOMEGAMfEFFSHIFT{0.005}
\def\LCDMBAOCMBOMEGAMf{0.275}
\def\LCDMBAOCMBDOMEGAMSTATa{0.023}
\def\LCDMBAOCMBDOMEGAMSTATb{0.021}
\def\LCDMBAOCMBDOMEGAMSTATc{0.023}
\def\LCDMBAOCMBDOMEGAMSTATd{0.020}
\def\LCDMBAOCMBDOMEGAMSTATe{0.019}
\def\LCDMBAOCMBDOMEGAMSTATf{0.021}
\def\LCDMBAOCMBDOMEGAMSYSTPa{0.014}
\def\LCDMBAOCMBDOMEGAMSYSTPb{0.021}
\def\LCDMBAOCMBDOMEGAMSYSTPc{0.013}
\def\LCDMBAOCMBDOMEGAMSYSTPd{0.020}
\def\LCDMBAOCMBDOMEGAMSYSTPe{0.017}
\def\LCDMBAOCMBDOMEGAMSYSTPf{0.016}

\def\LCDMBAOCMBDOMEGAMTOTPa{0.027}
\def\LCDMBAOCMBDOMEGAMTOTPb{0.030}
\def\LCDMBAOCMBDOMEGAMTOTPc{0.027}
\def\LCDMBAOCMBDOMEGAMTOTPd{0.029}
\def\LCDMBAOCMBDOMEGAMTOTPe{0.026}
\def\LCDMBAOCMBDOMEGAMTOTPf{0.027}

\def\LCDMBAOCMBALPHAa{0.126}

\def\LCDMBAOCMBALPHAb{0.123}

\def\LCDMBAOCMBALPHAc{0.113}

\def\LCDMBAOCMBALPHAd{0.113}

\def\LCDMBAOCMBALPHAe{0.116}

\def\LCDMBAOCMBALPHAf{0.105}
\def\LCDMBAOCMBDALPHASTATa{0.017}
\def\LCDMBAOCMBDALPHASTATb{0.015}
\def\LCDMBAOCMBDALPHASTATc{0.014}
\def\LCDMBAOCMBDALPHASTATd{0.013}
\def\LCDMBAOCMBDALPHASTATe{0.014}
\def\LCDMBAOCMBDALPHASTATf{0.019}
\def\LCDMBAOCMBDALPHASYSTPa{0.020}
\def\LCDMBAOCMBDALPHASYSTPb{0.021}
\def\LCDMBAOCMBDALPHASYSTPc{0.016}
\def\LCDMBAOCMBDALPHASYSTPd{0.019}
\def\LCDMBAOCMBDALPHASYSTPe{0.023}
\def\LCDMBAOCMBDALPHASYSTPf{0.023}

\def\LCDMBAOCMBDALPHATOTPa{0.027}
\def\LCDMBAOCMBDALPHATOTPb{0.026}
\def\LCDMBAOCMBDALPHATOTPc{0.022}
\def\LCDMBAOCMBDALPHATOTPd{0.023}
\def\LCDMBAOCMBDALPHATOTPe{0.027}
\def\LCDMBAOCMBDALPHATOTPf{0.030}

\def\LCDMBAOCMBBETAa{ 2.58}

\def\LCDMBAOCMBBETAb{ 2.64}

\def\LCDMBAOCMBBETAc{ 2.50}

\def\LCDMBAOCMBBETAd{ 2.57}

\def\LCDMBAOCMBBETAe{ 2.65}

\def\LCDMBAOCMBBETAf{ 2.46}
\def\LCDMBAOCMBDBETASTATa{ 0.15}
\def\LCDMBAOCMBDBETASTATb{ 0.12}
\def\LCDMBAOCMBDBETASTATc{ 0.15}
\def\LCDMBAOCMBDBETASTATd{ 0.12}
\def\LCDMBAOCMBDBETASTATe{ 0.12}
\def\LCDMBAOCMBDBETASTATf{ 0.17}

\def\LCDMBAOCMBDBETASYSTMa{ 0.13}
\def\LCDMBAOCMBDBETASYSTMb{ 0.18}
\def\LCDMBAOCMBDBETASYSTMc{ 0.09}
\def\LCDMBAOCMBDBETASYSTMd{ 0.17}
\def\LCDMBAOCMBDBETASYSTMe{ 0.18}
\def\LCDMBAOCMBDBETASYSTMf{ 0.19}

\def\LCDMBAOCMBDBETATOTMa{ 0.20}
\def\LCDMBAOCMBDBETATOTMb{ 0.22}
\def\LCDMBAOCMBDBETATOTMc{ 0.17}
\def\LCDMBAOCMBDBETATOTMd{ 0.20}
\def\LCDMBAOCMBDBETATOTMe{ 0.22}
\def\LCDMBAOCMBDBETATOTMf{ 0.25}
\newcommand{\DL}{d_L}
\newcommand{\LCDM}{\mbox{$\Lambda$}CDM}
\newcommand{\wCDM}{F\mbox{$w$}CDM}
\newcommand{\Kcor}{K--correction}
\newcommand{\eff}{efficiency}
\newcommand{\effs}{efficiencies}
\newcommand{\ineff}{inefficiency}

\newcommand{\spec}{spectroscopic}
\newcommand{\specy}{spectroscopically}
\newcommand{\LAMFBAR}{\bar{\lambda}_f}
\newcommand{\lamobs}{\lambda_{\rm obs}}
\newcommand{\lamrest}{\lambda_{\rm rest}}
\newcommand{\lamprime}{\lambda^{\prime}}
\newcommand{\TAUV}{\tau_{\rm V}}
\newcommand{\Trest}{T_{\rm rest}}
\newcommand{\Xhost}{X_{\rm host}}
\newcommand{\XMW}{X_{\rm MW}}
\newcommand{\zcut}{z_{\rm cut}}
\newcommand{\zcmbsym}{z_{\rm CMB}}
\newcommand{\zcmbval}{1090}
\newcommand{\RATEPOWEREQ}{\beta = 1.5 \pm 0.6}

\newcommand{\ZCUTAVTAU}{0.3}  % ZCUT to measure TAUAV and RV
\newcommand{\ZMINSYM}{z_{\rm min}}  % LOWZ ZCUT symbol for cosmolocay
\newcommand{\ZMINVAL}{0.02}     % LOWZ ZCUT value for cosmology
\newcommand{\ZBUBBLE}{0.025}    % z at anonaly, or bubble.
\newcommand{\ZCUTLOWZ}{0.02}    % same as ZMINVAL (temp while Josh edits)

\newcommand{\LOWZMUSIGNIF}{2.4}  % sigma-shift for MU(LOWZ>.025)-MU(LOWZ<.025)
\newcommand{\SDSSMUSIGNIF}{2.5}  % sigma-shift for MU(SDSS<.150)-MU(LOWZ<.025)

\newcommand{\TMINCUT}{-15}
\newcommand{\TMAXCUT}{+60} 

\newcommand{\sigmutot}{\sigma_{\mu} }
\newcommand{\sigmufit}{\sigma_{\mu}^{\rm fit} }
\newcommand{\sigmuint}{\sigma_{\mu}^{\rm int} }
\newcommand{\sigmudz}{\sigma_{\mu}^{z} }
\newcommand{\sigzspec}{\sigma_{z,spec} }
\newcommand{\sigzpec}{\sigma_{z,pec} }

\newcommand{\sigmurmsVALUE}{0.16}  % for MLCS
\newcommand{\sigzpecVALUE}{0.0012} % for both MLCS & SALT2
\newcommand{\RMSMU}{{\rm RMS}_{\mu}}
\newcommand{\MUDIF}{\Delta\mu}

\newcommand{\MLCSMUPULLSIG}{0.77}
\newcommand{\MLCSMUPULLRMS}{0.90}
\newcommand{\SALTMUPULLSIG}{0.75}
\newcommand{\SALTMUPULLRMS}{0.92}

\newcommand{\MUwCDM}{\mu_{{\rm F}w{\rm CDM}}}

\newcommand{\hostz}{host-$z$}
\newcommand{\Pfit}{ {\cal P}_{\rm fit} }

\newcommand{\simeffsurvey}{\epsilon_{\rm survey}}
\newcommand{\simeffsearch}{\epsilon_{\rm search}}
\newcommand{\simeffpipe}{\epsilon_{\rm subtr}}
\newcommand{\simeffspec}{\epsilon_{\rm spec}}
\newcommand{\simeffzspec}{\epsilon_{\rm spec}^z}
\newcommand{\simeffmagdim}{\epsilon_{\rm spec}^{\mathcal M}}
\newcommand{\simeffcuts}{\epsilon_{\rm cuts}}
\newcommand{\simeff}{\epsilon_{\rm sim}}

\newcommand{\magdim}{{\mathcal M}_{\rm dim}}

\newcommand{\effzzero}{\zeta_0}
\newcommand{\effzone}{\zeta_1}

\newcommand{\effdimzero}{m_0}
\newcommand{\effdimone}{m_1}
\newcommand{\effdimtwo}{m_2}
\newcommand{\AEFFDIM}{A_{\mathcal M}}

\newcommand{\lc}{light curve}
\newcommand{\lcs}{light curves}

\newcommand{\OM}{\Omega_{\rm M}}
\newcommand{\ODE}{\Omega_{\rm DE}}
\newcommand{\OLAM}{\Omega_{\Lambda}}
\newcommand{\OL}{\Omega_{\Lambda}}
\newcommand{\Ok}{\Omega_{\rm k}}
\newcommand{\SINSINHFUN}{{\cal S}_k}  % F(x) = sin(x), sinh(x) or x

\newcommand{\BDFULL}{{\rm BD+17}$^0$4708}
\newcommand{\BD}{{\rm BD+17}}

\newcommand{\asym}{{\it a}}
\newcommand{\bsym}{{\it b}}
\newcommand{\csym}{{\it c}}
\newcommand{\dsym}{{\it d}}
\newcommand{\esym}{{\it e}}
\newcommand{\fsym}{{\it f}}
\newcommand{\combosymlist}{{\it a-f}}

\newcommand{\ncomboword}{six}

\newcommand{\samplea}{SDSS-only}
\newcommand{\sampleb}{SDSS+ESSENCE+SNLS}
\newcommand{\samplec}{Nearby+SDSS}
\newcommand{\sampled}{Nearby+SDSS+ESSENCE+SNLS}
\newcommand{\samplee}{Nearby+SDSS+ESSENCE+SNLS+HST}
\newcommand{\samplef}{Nearby+ESSENCE+SNLS}

\newcommand{\SDSS}{SDSS-II}
\newcommand{\ZUSDSS}{0.21}
\newcommand{\ZMINSDSS}{0.04}
\newcommand{\ZMAXSDSS}{0.42}
\newcommand{\mlcs}{{\sc mlcs2k2}}

\newcommand{\Mmlcs}{M}

\newcommand{\SALTII}{{\sc salt--ii}}
\newcommand{\chisqmu}{\chi^2_{\mu}}
\newcommand{\dFrestdlam}{\frac{dF_{\rm rest}}{d\lambda}}
\newcommand{\mBstar}{m_B^*}
\newcommand{\SALTIILAMMIN}{2900}
\newcommand{\SALTIILAMMAX}{7000}

\newcommand{\HSTSEARCHFILT}{F850LP\_ACS}

\newcommand{\unc}{uncertainty}
\newcommand{\uncs}{uncertainties}

\newcommand{\AVMNSYMBOL}{\overline{A}_V}
\newcommand{\RVRESULT}{R_V = \RV \pm \RVERRSTAT_{\rm stat} \pm \RVERRSYST_{\rm syst}}
\newcommand{\AVMNRESULT}
   {\AVMNSYMBOL = \AVMN \pm \AVMNERRSTAT_{\rm stat} \pm \AVMNERRSYST_{\rm syst}}

\newcommand{\RVMW}{3.1}

\newcommand{\SIGOM}{\sigma_{\OM}}

\newcommand{\dw}{\delta w}
\newcommand{\wsystsym}{\sigma_w({\rm syst})}
\newcommand{\wstatsym}{\sigma_w({\rm stat})}
\newcommand{\wtotsym}{\sigma_w({\rm tot})}
\newcommand{\OMsystsym}{\SIGOM({\rm syst})}
\newcommand{\OMstatsym}{\SIGOM({\rm stat})}
\newcommand{\OMtotsym}{\SIGOM({\rm tot})}

\newcommand{\ABoffuerr}{0.014}
\newcommand{\ABoffgerr}{0.009}
\newcommand{\ABoffrerr}{0.009}
\newcommand{\ABoffierr}{0.009}
\newcommand{\ABoffzerr}{0.010}

\newcommand{\ABAO}{0.469}
\newcommand{\ABAOERR}{0.017}
\newcommand{\BAOCHISQ}{\chi^2_{\rm BAO}}

\newcommand{\RWMAP}{1.710}
\newcommand{\RWMAPERR}{0.019}
\newcommand{\CMBCHISQ}{\chi^2_{\rm CMB}}

\newcommand{\MLCSSDSSnoUdMU}{0.12\pm 0.02}  % shift for z > .21
\newcommand{\SALTSDSSnoUdMU}{0.079\pm 0.028}  % shift for z > .21
\newcommand{\SALTSDSSnoUdmu}{0.012\pm 0.003}  % shift for z < .21

\newcommand{\MLCSSDSSnoUdMUval}{0.12}  % shift for z > .21
\newcommand{\SALTSDSSnoUdMUval}{0.08}  % shift for z > .21

\newcommand{\SYSTUband}{Rest frame {\it U}-band}   
\newcommand{\SYSTzmin}{\mbox{$\ZMINSYM$} cut for Nearby sample}

\newcommand{\FwCDMwnoSNe}{\mbox{$w = -0.80\pm 0.20$}}    % at min chi2
\newcommand{\FwCDMOMnoSNe}{\mbox{$\OM = 0.30\pm 0.05$}}  % at min chi2

\newcommand{\FwCDMwnoSNeVal}{-0.80}    % at min chi2
\newcommand{\FwCDMOMnoSNeVal}{0.30}    % at min chi2

\newcommand{\LCDMOMnoSNe}{\mbox{$\OM   = 0.27 \pm 0.02$}}
\newcommand{\LCDMOLnoSNe}{\mbox{$\OLAM = 0.74 \pm 0.02$}}

\newcommand{\wwwNICMOS}{\tt http://www.stsci.edu/hst /nicmos/documents/handbooks/handbooks/DataHandbookv7}

\newcommand{\wwwACS}{\tt http://www.stsci.edu/hst /acs/documents/handbooks/DataHandbookv4/ACS\_longdhbcover.html}

\newcommand{\wwwMINUIT}{\tt http://wwwasdoc.web.cern.ch/wwwasdoc/minuit/minmain.html}

\newcommand{\wwwSDSS}{\tt http://www.sdss.org/}

\newcommand{\wwwCALSPEC}{\tt http://www.stsci.edu/hst/observatory/cdbs/calspec.html}

\newcommand{\wwwSNANA}{\tt http://www.sdss.org/supernova/SNANA.html}

\newcommand{\wwwTABLES}{\tt http://das.sdss.org/va/SNcosmology/sncosm09\_fits.tar.gz}
\begin{document}

\title{
First-year Sloan Digital Sky Survey-II (SDSS-II) Supernova Results: \\ 
Hubble Diagram and Cosmological Parameters
}

%\journalinfo{Accepted for publication in ApJS}
\submitted{Accepted for publication in ApJS}
\email{kessler@kicp.uchicago.edu}

%
% author list for SDSS-II 1st-year cosmology paper (sncosm09)
%

\newcommand{\NUMUCASTRO}{1}  % U.Chicago Astronomy
\newcommand{\NUMKICP}{2}     % U.Chicago KICP
\newcommand{\NUMUW}{3}       % U.Washington at Seattle
\newcommand{\NUMWAYNE}{4}    % Wayne State (Detroit)
\newcommand{\NUMFNAL}{5}     % Fermilab Center for Particle Astrophys.
\newcommand{\NUMUQ}{6}       % U. Queensland, Ausstralia
\newcommand{\NUMBOHR}{7}     % Niels Bohr Inst., Denmark
\newcommand{\NUMRUTGERS}{8}   % Rutgers University
\newcommand{\NUMNMSU}{9}     % New Mexico State U
\newcommand{\NUMPORT}{10}     % Portsmouth U, UK
\newcommand{\NUMUPENN}{11}    % U. Penn
\newcommand{\NUMCAPEMATH}{12} % U. Cape  Town,  South Africa (math)
\newcommand{\NUMSAAO}{13}     % South African Astronomical Observatory
\newcommand{\NUMTXAM}{14}     % Texas A&M U
\newcommand{\NUMTOKYOa}{15}   % Int. Physics & Math of Universe, Tokyo
\newcommand{\NUMTOKYOb}{16}   % 
\newcommand{\NUMUCB}{17}      % U.C. Berkeley
\newcommand{\NUMCFA}{18}      % CFA (for Ryan)
\newcommand{\NUMIRISH}{19}    % Notre Dame
\newcommand{\NUMLUDMAX}{20}   % Ludwig-Maximilians U, Munich
\newcommand{\NUMTOKYOc}{21}   % 
\newcommand{\NUMAPO}{22}      % Apache Point Observatory
\newcommand{\NUMTOKYOe}{23}   % 
\newcommand{\NUMGOTT}{24}     % Univ Gottingham
\newcommand{\NUMSTOCKPHYS}{25}  % Stockholm U, Physics Dept
\newcommand{\NUMOSU}{26}      % Ohio State U
\newcommand{\NUMCRTOKYO}{27}  % U.Tokyo Cosmic Ray Researcy
\newcommand{\NUMSTOCKASTRO}{28} % Stockholm U, astro
\newcommand{\NUMRIT}{29}      % Rochester Institute of Technology
\newcommand{\NUMSTSI}{30}     % Space Telescope Science Institute
\newcommand{\NUMHOPKINS}{31}  % John Hopkins U
\newcommand{\NUMKIPAC}{32}     % KICAP at Stanford
\newcommand{\NUMPENNSTATE}{33} % Penn State U, Astro Dept.
\newcommand{\NUMMcDTX}{34}     % McDondal Observatory, Texas
\newcommand{\NUMCAPEASTRO}{35} % U. Cape  Town,  South Africa (astronomy)
\newcommand{\NUMEFI}{36}      % U.Chicago EFI
\newcommand{\NUMSNT}{37}       % Center for Neighborhood Technology (Gajus)

\author{
Richard~Kessler,\altaffilmark{\NUMUCASTRO,\NUMKICP}
Andrew~C.~Becker,\altaffilmark{\NUMUW}
David~Cinabro,\altaffilmark{\NUMWAYNE}
Jake~Vanderplas,\altaffilmark{\NUMUW}
Joshua~A.~Frieman,\altaffilmark{\NUMKICP,\NUMUCASTRO,\NUMFNAL}
John~Marriner,\altaffilmark{\NUMFNAL}
%
% ===========================================
%
Tamara~M~Davis,\altaffilmark{\NUMUQ,\NUMBOHR}
Benjamin~Dilday,\altaffilmark{\NUMRUTGERS}
Jon~Holtzman,\altaffilmark{\NUMNMSU}
Saurabh~W.~Jha,\altaffilmark{\NUMRUTGERS}
Hubert~Lampeitl,\altaffilmark{\NUMPORT}
Masao~Sako,\altaffilmark{\NUMUPENN}
Mathew~Smith,\altaffilmark{\NUMPORT\NUMCAPEMATH}
Chen~Zheng,\altaffilmark{\NUMKIPAC}
Robert~C.~Nichol,\altaffilmark{\NUMPORT}
%
% ===========================================
%
Bruce~Bassett,\altaffilmark{\NUMCAPEMATH,\NUMSAAO}
Ralf~Bender,\altaffilmark{\NUMLUDMAX}
Darren~L.~Depoy,\altaffilmark{\NUMTXAM}
Mamoru~Doi,\altaffilmark{\NUMTOKYOa,\NUMTOKYOb}
Ed~Elson,\altaffilmark{\NUMCAPEMATH}
Alexei~V.~Filippenko,\altaffilmark{\NUMUCB}
Ryan~J.~Foley,\altaffilmark{\NUMUCB,\NUMCFA}
Peter~M.~Garnavich,\altaffilmark{\NUMIRISH}
Ulrich~Hopp,\altaffilmark{\NUMLUDMAX}
Yutaka~Ihara,\altaffilmark{\NUMTOKYOa,\NUMTOKYOc}
William~Ketzeback,\altaffilmark{\NUMAPO}
W.~Kollatschny,\altaffilmark{\NUMGOTT}
Kohki~Konishi,\altaffilmark{\NUMCRTOKYO}
Jennifer~L.~Marshall,\altaffilmark{\NUMTXAM}
Russet~J.~McMillan,\altaffilmark{\NUMAPO}
Gajus~Miknaitis,\altaffilmark{\NUMSNT,\NUMFNAL}
Tomoki~Morokuma,\altaffilmark{\NUMTOKYOe}
Edvard~M\"ortsell,\altaffilmark{\NUMSTOCKPHYS}
Kaike~Pan,\altaffilmark{\NUMAPO}
Jose~Luis~Prieto,\altaffilmark{\NUMOSU}
Michael~W.~Richmond,\altaffilmark{\NUMRIT}
Adam~G.~Riess,\altaffilmark{\NUMSTSI,\NUMHOPKINS}
Roger~Romani,\altaffilmark{\NUMKIPAC}
Donald~P.~Schneider,\altaffilmark{\NUMPENNSTATE}
Jesper~Sollerman,\altaffilmark{\NUMBOHR,\NUMSTOCKASTRO}
Naohiro~Takanashi,\altaffilmark{\NUMTOKYOe}
Kouichi~Tokita,\altaffilmark{\NUMTOKYOa,\NUMTOKYOc}
Kurt~van~der~Heyden,\altaffilmark{\NUMCAPEASTRO}
J.~C.~Wheeler,\altaffilmark{\NUMMcDTX}
Naoki~Yasuda,\altaffilmark{\NUMCRTOKYO}
and
Donald~York\altaffilmark{\NUMUCASTRO,\NUMEFI}
}  % end author

% ==================================
% ========== INSTITUTIONS ==========
% ==================================

\altaffiltext{\NUMUCASTRO}{
  Department of Astronomy and Astrophysics,
   The University of Chicago, 5640 South Ellis Avenue, Chicago, IL 60637
}

\altaffiltext{\NUMKICP}{
Kavli Institute for Cosmological Physics, 
The University of Chicago, 5640 South Ellis Avenue Chicago, IL 60637
}

\altaffiltext{\NUMUW}{
Department of Astronomy,
University of Washington, Box 351580, Seattle, WA 98195
}

\altaffiltext{\NUMWAYNE}{
Department of Physics and Astronomy, 
Wayne State University, Detroit, MI 48202
}

\altaffiltext{\NUMFNAL}{
Center for Particle Astrophysics, 
  Fermi National Accelerator Laboratory, P.O. Box 500, Batavia, IL 60510
}

\altaffiltext{\NUMUQ}{
School of Mathematics and Physics, University of Queensland, QLD,  
4072, Australia
}

\altaffiltext{\NUMBOHR}{
Dark Cosmology Centre, Niels Bohr Institute, 
University of Copenhagen, Juliane
Maries Vej 30, DK-2100 Copenhagen \O, Denmark
}

\altaffiltext{\NUMRUTGERS}{
Department of Physics and Astronomy, 
Rutgers University, 136 Frelinghuysen Road, Piscataway, NJ 08854
}

\altaffiltext{\NUMNMSU}{
  Department of Astronomy, MSC 4500,
   New Mexico State University, P.O. Box 30001, Las Cruces, NM 88003
}

\altaffiltext{\NUMPORT}{
Institute of Cosmology and Gravitation, Dennis Sciama Building,
Burnaby Road, University of Portsmouth, Portsmouth, PO1 3FX, UK
}

\altaffiltext{\NUMUPENN}{
Department of Physics and Astronomy,
University of Pennsylvania, 203 South 33rd Street, Philadelphia, PA  19104
}

\altaffiltext{\NUMCAPEMATH}{
Department of Mathematics and Applied Mathematics,
University of Cape Town, Rondebosch 7701, South Africa
}

\altaffiltext{\NUMSAAO}{
  South African Astronomical Observatory,
   P.O. Box 9, Observatory 7935, South Africa.
}

\altaffiltext{\NUMTXAM}{
 Department of Physics, Texas A \& M University, College Station, TX 77843
}

\altaffiltext{\NUMTOKYOa}{
 Institute of Astronomy, University of Tokyo, 2-21-1 Osawa,
 Mitaka-shi, Tokyo 181-0015, Japan
}

\altaffiltext{\NUMTOKYOb}{
 Institute for Physics and Mathematics of the Universe
 University of Tokyo, 5-1-5, Kashiwanoha, Kashiwa, Chiba, 277-8582, Japan
}

\altaffiltext{\NUMUCB}{
Departmentof Astronomy, University of California, Berkeley, CA 94720-4311
}

\altaffiltext{\NUMCFA}{
Harvard-Smithsonian Center for Astrophysics,
60 Garden Street,
Cambridge, MA 02138.
}

\altaffiltext{\NUMIRISH}{
  University of Notre Dame, 225 Nieuwland Science, Notre Dame, IN 46556-5670
}

\altaffiltext{\NUMLUDMAX}{
  Universit\"ats-Sternwarte, Ludwig-Maximilians
  Universit\"at M\"unchen, Germany
}

\altaffiltext{\NUMTOKYOc}{
 Department of Astronomy, Graduate School of Science, University of
 Tokyo, Bunkyo-ku, Tokyo 113-0033, Japan
}

\altaffiltext{\NUMAPO}{
  Apache Point Observatory, P.O. Box 59, Sunspot, NM 88349.
}

\altaffiltext{\NUMTOKYOe}{
 National Astronomical Observatory of Japan, Mitaka 181-8588, Japan
}

\altaffiltext{\NUMGOTT}{
  Institut f\"ur Astrophysik, Universit\"at G\"ottingen,
  Friedrich-Hund Platz 1, D-37077 G\"ottingen, Germany
}

\altaffiltext{\NUMSTOCKPHYS}{
Department of Physics, AlbaNova, Stockholm University, 
SE-106 91 Stockholm Sweden
}

\altaffiltext{\NUMOSU}{
  Department of Astronomy,
   Ohio State University, 140 West 18th Avenue, Columbus, OH 43210-1173.
}

\altaffiltext{\NUMCRTOKYO}{
Institute for Cosmic Ray Research,
University of Tokyo, 5-1-5, Kashiwanoha, Kashiwa, Chiba, 277-8582, Japan
}

\altaffiltext{\NUMSTOCKASTRO}{
The Oskar Klein Centre, Department of Astronomy, AlbaNova, 
Stockholm University, SE-106 91 Stockholm, Sweden
}

\altaffiltext{\NUMRIT}{
  Physics Department,
   Rochester Institute of Technology,
   85 Lomb Memorial Drive, Rochester, NY 14623-5603
}

\altaffiltext{\NUMSTSI}{
  Space Telescope Science Institute,
   3700 San Martin Drive, Baltimore, MD 21218.
}

\altaffiltext{\NUMHOPKINS}{
Department of Physics and Astronomy,
Johns Hopkins University, 3400 North Charles Street, Baltimore, MD 21218.
}

\altaffiltext{\NUMKIPAC}{
 Kavli Institute for Particle Astrophysics \& Cosmology, 
  Stanford University, Stanford, CA 94305-4060.
}

\altaffiltext{\NUMPENNSTATE}{
  Department of Astronomy and Astrophysics,
   The Pennsylvania State University,
   525 Davey Laboratory, University Park, PA 16802.
}

\altaffiltext{\NUMMcDTX}{
  Department of Astronomy,
   McDonald Observatory, University of Texas, Austin, TX 78712
}

\altaffiltext{\NUMCAPEASTRO}{
Department of Astronomy,
University of Cape Town, Private Bag X3, Rondebosch 7701, South Africa
}

\altaffiltext{\NUMEFI}{
Enrico Fermi Institute,
University of Chicago, 5640 South Ellis Avenue, Chicago, IL 60637
}

\altaffiltext{\NUMSNT}{
Center for Neighborhood Technology
2125 W. North Ave, Chicago IL 60647
}

\begin{abstract}
 
We present measurements of the Hubble diagram for 
$\NSNSDSS$ Type Ia supernovae (SNe) with redshifts 
$\ZMINSDSS < z < \ZMAXSDSS $,
discovered during the first season (Fall 2005)
of the Sloan Digital Sky Survey-II ({\SDSS}) Supernova Survey. 
These data fill in the redshift ``desert'' 
between low- and high-redshift SN~Ia surveys. 
Within the framework of the \mlcs\ light-curve fitting 
method, we use the \SDSS\ SN sample 
to infer the mean reddening parameter 
for host galaxies, $\RVRESULT$, and find that the intrinsic 
distribution of host-galaxy extinction is well fit by an 
exponential function, $P(A_V)=\exp(-A_V/\TAUV)$, 
with $\TAUV = \TAUAV \pm \TAUAVERRTOT$~mag.
We combine the \SDSS\ measurements with new distance estimates for 
published SN data from the ESSENCE survey, the Supernova Legacy Survey (SNLS), 
the Hubble Space Telescope (HST), and a compilation of nearby 
SN~Ia measurements. 
A new feature in our analysis is the use of detailed 
Monte Carlo simulations of all surveys to account for selection biases,
including those from \spec\ targeting.
Combining the SN Hubble diagram with measurements of 
baryon acoustic oscillations from the SDSS Luminous
Red Galaxy sample 
and with cosmic microwave background temperature anisotropy
measurements from WMAP, we estimate the cosmological parameters 
$w$ and $\OM$, assuming a spatially flat cosmological model 
({\wCDM}) with constant dark energy equation of state parameter, $w$. 
We also consider constraints upon $\OM$ and $\Omega_\Lambda$ 
for a cosmological constant model ({\LCDM}) 
with $w=-1$ and non-zero spatial curvature. 
For the \wCDM\ model and
the combined sample of $\NSNTOT$ SNe~Ia, we find 
$w = \MLCSWRESe \pm \MLCSWSTATERRe({\rm stat}) 
{\pm\MLCSWSYSTERRPe} ({\rm syst})$, 
$\OM= \MLCSOMRESe \pm \MLCSOMSTATERRe({\rm stat}) 
{\pm\MLCSOMSYSTERRPe}({\rm syst})$ 
using \mlcs\ and 
$w=\FWCDMBAOCMBWe \pm \FWCDMBAOCMBDWSTATe({\rm stat})
 \pm\FWCDMBAOCMBDWSYSTPe ({\rm syst})$, 
$\OM = \FWCDMBAOCMBOMEGAMe \pm \FWCDMBAOCMBDOMEGAMSTATe ({\rm stat})
  \pm \FWCDMBAOCMBDOMEGAMSYSTPe ({\rm syst})$ 
using the \SALTII\ fitter. We trace 
the discrepancy between these results to a difference 
in the rest-frame UV model combined with a different
luminosity correction from color variations; these differences 
mostly affect the distance estimates for the SNLS and HST supernovae. 
We present detailed discussions of systematic errors for both 
light-curve methods and find that they both show data-model
discrepancies in rest-frame $U$-band. 
For the \SALTII\ approach, we also see strong evidence for 
redshift-dependence of  the color-luminosity parameter ($\beta$).
Restricting the analysis to the 136 SNe~Ia in the 
Nearby+SDSS-II samples, we find much better agreement between the 
two analysis methods but with larger uncertainties:
$ w = \MLCSWRESc \pm \MLCSWSTATERRc ({\rm stat})
  {~}^{+\MLCSWSYSTERRPc}_{-\MLCSWSYSTERRMc}$ ({\rm syst}) 
for \mlcs\ and 
$ w = \FWCDMBAOCMBWc \pm \FWCDMBAOCMBDWSTATc ({\rm stat}) 
 {~}^{+\FWCDMBAOCMBDWSYSTPc}_{-\FWCDMBAOCMBDWSYSTMc}$ ({\rm syst}) 
for \SALTII.

\end{abstract}

%% \tableofcontents  % for internal editing only

% #########################################
\section{Introduction}
\label{sec:intro}
%
% Introduction for SDSS-II SN Cosmology I paper

Ten years ago, measurements of the Hubble diagram of 
Type Ia supernovae (SNe) provided the first direct evidence for 
cosmic acceleration \citep{Riess_98,Perlmutter99}.  
In the intervening decade,
dedicated SN surveys have brought tremendous improvements in 
both the quantity and quality of SN~Ia data, and SNe~Ia remain 
the method of choice for precise relative distance determination 
over cosmological scales 
\citep[e.g.,][]{Leibundgut_01,filippenko05}. 
We now have in hand large, homogeneously selected samples of 
SNe~Ia with relatively dense time-sampling in multiple passbands 
at redshifts $z \gtrsim 0.3$, most recently from the 
ESSENCE project \citep{Miknaitis_07,WV07} 
and Supernova Legacy (SNLS) Survey \citep{Astier06}, 
augmented by smaller samples from the Hubble Space Telescope (HST) 
that extend to higher redshift 
\citep{Garnavich98,Knop03,Riess_04,Riess_06}.  
These data have confirmed and sharpened the
evidence for accelerated expansion. Cosmic acceleration
is most commonly attributed to a new energy-density
component known as dark energy  
(for a review, see \citet*{FTH08}).  
The recent SN measurements, in combination with measurements 
of the baryon acoustic oscillation (BAO) feature in 
galaxy clustering and of the cosmic microwave background (CMB) 
anisotropy, have provided increasingly 
precise constraints on the density, $\ODE$, 
and equation of state parameter, $w$, of dark energy.

Despite these advances, a number of concerns remain about the 
robustness of current SN cosmology constraints. 
The SN~Ia Hubble diagram is
constructed from combining low- and high-redshift 
SN samples that have been observed with a variety of 
telescopes, instruments, and photometric passbands. 
Photometric offsets between these samples are highly 
degenerate with changes in cosmological parameters, 
and these offsets could be hidden in part because there 
is a gap or ``redshift desert'' between the low-redshift 
($z\lesssim 0.1$) SNe, found with small-aperture, wide-field telescopes, 
and the high-redshift ($z \gtrsim 0.3$) SNe discovered by large-aperture 
telescopes with relatively narrow fields. 
In addition, the low-redshift 
SN measurements that are used both to anchor the Hubble 
diagram and to train SN distance estimators were 
themselves compiled from combinations of several
surveys using different telescopes, instruments, and selection criteria.  
Increasing the robustness of the cosmological results 
calls for larger supernova samples with 
continuous redshift coverage of the Hubble diagram; it also 
necessitates {\it high-quality} data, with homogeneously selected, 
densely sampled, multi-band SN light curves 
and well-understood photometric calibration.

The Sloan Digital Sky Survey-II Supernova Survey 
({\SDSS} SN Survey)
\citep{Frieman07}, 
one of the three components of the \SDSS\ project, 
was designed to address both the paucity of SN~Ia data at 
intermediate redshifts and the systematic limitations of 
previous SN~Ia samples, thereby leading to more robust 
constraints upon the properties of the dark energy.
Over the course of three three-month seasons, 
the \SDSS\ SN Survey
discovered and measured well-sampled, 
multi-band light curves for roughly 500 spectroscopically confirmed 
SNe~Ia in the redshift range \hbox{$0.01 \lesssim z \lesssim 0.45$}. 
This data set fills in the redshift desert and for the first time 
includes both low- and high-redshift SN measurements in a single survey. 
The survey takes advantage of the extensive database of
reference images, object catalogs, and photometric calibration
previously obtained by the SDSS
(for a description of the SDSS, see \citet{York_00}).

In this paper, we present the Hubble diagram based on \specy\ 
confirmed SNe~Ia from the first full season (Fall 2005) of the 
\SDSS\ SN Survey.
To derive cosmological results, we include information from BAO 
\citep{Eisenstein05} and CMB measurements \citep{Komatsu2008},
and we also combine our data with our own analysis of public 
SN~Ia data sets at lower and higher redshifts. 
We fit the SN~Ia \lcs\ with two models, 
{\mlcs} \citep{Jha07} and {\SALTII} \citep{Guy07}. 
We use the publicly available {\SALTII}
software with minor modifications, 
but we have made a number of improvements to the implementation
of the \mlcs\ method, as described in \S \ref{sec:anal}.

Two companion papers
explore related analyses with the same SN data sets. 
\cite{Lampeitl09} combine the \SDSS\ SN data with 
different BAO constraints and with measurements 
of redshift-space distortions and of the Integrated Sachs-Wolfe 
effect to derive joint constraints on dark energy 
from low-redshift ($z<0.4$) measurements only; they also explore 
the consistency of the SN and BAO distance scales. 
\cite{Sollerman09} use SN, BAO, and CMB measurements 
to constrain cosmological models 
with a time-varying dark energy equation of state parameter 
as well as more exotic models for cosmic acceleration. In all three papers,
we use a consistent analysis of the SN data. Differences in cosmological 
inferences are attributable to differences in (a) the SN data 
included, (b) the other cosmological data sets included, and (c) the 
cosmological model space considered.

The outline of the paper is as follows.
In \S \ref{sec:survey}, we briefly describe the operation 
and  data processing for the \SDSS\ SN Survey, 
which have been more extensively described in \citet{Sako08}. 
In \S \ref{sec:spec-photo}, we summarize the \spec\ analysis 
leading to final redshift and SN type determinations \citep{Zheng08}
and the photometric analysis leading to final 
supernova flux measurements \citep{Holtz08} for \SDSS\ SNe. 
In \S~\ref{sec:sample}, we present the SN samples and 
selection criteria applied to the light-curve data.
We describe and compare the {\mlcs} and {\SALTII} methods 
in  \S~\ref{sec:anal}. 
In \S \ref{sec:sim} we describe detailed Monte Carlo simulations 
for the \SDSS\ SN Survey and other SN data sets that 
we use to determine survey efficiencies and their dependences
on SN luminosity, extinction, and redshift. 
Modeling of the survey efficiencies is needed to  
correct for selection biases that affect SN distance estimates. 
In \S \ref{sec:dust} we use a larger spectroscopic+photometric 
\SDSS\ SN sample to determine host-galaxy dust properties that 
are used in the \mlcs\ fits.  
In particular, we present new measurements of the mean dust parameter, 
$R_V$, and of the extinction ($A_V$) distribution.
In \S \ref{sec:wfit} we describe the cosmological 
likelihood analysis, which combines 
the SN~Ia Hubble diagram with BAO and CMB measurements.
In \S \ref{sec:syst}, we present a detailed study of 
systematic errors, showing how uncertainties in model parameters 
and in calibrations impact the results. 
In \S \ref{sec:results} we discuss the supernova Hubble diagrams
using the \mlcs\ and \SALTII\ fitters  
and derive constraints on cosmological parameters. We provide a 
detailed comparison of the \mlcs\ and \SALTII\ results
in \S~\ref{sec:results_compare}, and 
we conclude in \S~\ref{sec:conclude}.
Appendices provide details on the methods for 
warping the SN~Ia spectral template for {\Kcor s}, 
modeling the filter passbands for the nearby SN~Ia sample,
determining the magnitudes of the primary photometric standard stars, 
extracting the distribution of host-galaxy dust extinction
from the \SDSS\ sample, and estimating error contours that include
systematic \uncs. They also include discussion of the scatter in 
the \SDSS\ Hubble diagram and of the translation of the \SALTII\ 
model into the \mlcs\ framework.

% #########################################
\section{SDSS-II Supernova Survey}
\label{sec:survey}
%%
%% Section 2 of SDSS SN cosmology: SURVEY
%%

The scientific goals, operation, and basic data processing 
for the \SDSS\ SN Survey are described in \citet{Frieman07},
and details of the SN search algorithms and \spec\ observations
are given in \citet{Sako08}.
Here we provide a brief summary of the Fall 2005 campaign, in 
order to set the context for the data analysis. 

The \SDSS\ Supernova Survey primary instrument was the 
SDSS CCD camera \citep{Gunn_98} mounted on a
dedicated 2.5-m telescope \citep{SDSS_telescope}
at Apache Point Observatory (APO), New Mexico.  
The camera obtains, nearly simultaneously, 
images in five broad optical bands: $ugriz$ \citep{Fukugita_96}. 
The camera was used in time-delay-and-integrate 
(TDI, or drift scan) mode, which provides efficient sky coverage.
The Supernova Survey scanned at the normal (sidereal) SDSS 
survey rate, which yielded 55-s integrated exposures in each 
passband; the instrument covered the sky at a rate of approximately 
20 square degrees per hour.

On most of the usable observing nights in 
the period 1 September through 30 November 2005, 
the \SDSS\ SN Survey
scanned a region (designated stripe~82) centered on the celestial equator
in the Southern Galactic hemisphere that is 
2.5$^{\circ}$ wide and runs between right ascensions of
20$^{\rm hr}$ and 4$^{\rm hr}$, 
covering a total area of~300~sq.~deg. 
Due to gaps between the CCD columns, on a given night slightly more than
half of the declination range of the stripe was imaged; on succeeding
nights, the survey alternated between the northern (N) and southern (S)
declination strips of stripe 82
(see \citet{Stoughton_02} for a description of the 
SDSS observing geometry).
Accounting for CCD gaps, bad weather, nearly full Moon, 
and other observing programs, a given region was imaged 
on average every four to five nights under a variety of conditions. 
This relatively high cadence enabled us to obtain
well-sampled light curves, typically starting well before peak light.
%On average, there are 16 $ugriz$ epochs per confirmed SN, 
%9 of which satisfy the detection threshold described below.

At the end of each night of imaging, the SN data were processed 
using a dedicated 20-CPU computing cluster at APO. 
Images were processed through the PHOTO photometric reduction 
pipeline to produce corrected 
$u,g,r,i,z$ frames \citep*{Lupton_01,Ivezic_04},
each with an astrometric solution \citep{Pier_03}, 
point-spread-function map, and zero-point. 
A co-added template image, consisting of typically 8 
stacked images taken in previous years, was matched to the 
new image and subtracted from it.
Subtracted $gri$ images were searched for pixel clusters with an 
excess flux (roughly $3 \sigma$) above the noise in the subtracted 
image, and a position and total PSF flux were assigned for each 
significant detection.
We positionally matched detections in multiple passbands: 
{\it objects} are detections in at least two of the three $gri$ 
passbands with a displacement of less than $0.8\arcsec$
between detections in each filter. 
This displacement cut was chosen to ensure high efficiency
for objects with low signal-to-noise.
The $g$ and $r$ exposures 
of a given object were taken five minutes apart, enabling many 
fast asteroids to be removed by the 0.8\arcsec\ requirement. 
Finally, a catalog of $10^5$  previously detected variables 
(mainly stars and AGN) and 4~million stars ($r<21.5$)
was used to reject detections within 1\arcsec\ of any object 
in the catalog; nearly 40,000 such detections 
were automatically vetoed during the Fall 2005 survey.

During the season, $20''\times 20''$
cut-outs of the resulting $\sim 140,000$ 
object images were visually scanned by 
humans,\footnote{During the 2006 season
we implemented more aggressive software cuts that 
reduced the number of objects scanned by over an order of magnitude.
} % end footnote
typically within 24 hours of when the data were obtained.
The human scanning was done to eliminate objects that 
were clearly not supernovae, 
such as unsubtracted diffraction spikes, other subtraction artifacts, 
%% unmasked saturated stars, 
and obvious asteroids. 
To monitor the software pipelines and human scanning efficiency,
`fake' supernovae were inserted on top of galaxies
in the images during processing.
Approximately 11,400 of the objects were tagged by a scanner 
as a possible supernova {\it candidate}. 
Nearly 60\% of the candidates appeared only once during the survey;
most of these are likely slow-moving solar system objects. 
After a night of observations, each candidate \lc\ (in $g,r,i$) 
was updated and compared with 
a set of supernova \lc\ templates that include SNe~Ia 
as a function of redshift, intrinsic luminosity, and extinction, 
as well as non-Ia SN types. 
Light curves that matched best to a SN~Ia template 
(at any reasonable redshift, luminosity, and extinction)
were preferentially scheduled for \spec\ follow-up observations. 
Candidates with $r$-band magnitude $r \la 20.5$
were given highest priority for follow-up, 
regardless of photometric SN type; for SNe~Ia, 
this magnitude cut corresponds roughly to redshifts $z < 0.15$.
For fainter SN~Ia candidates, \spec\ priority was given to candidates 
with the best chance of acquiring a useful spectrum.
In order of importance, the prioritization criteria were:
(i) SN is well-separated ($\gtrsim 1''$) from the core of its host galaxy,
(ii) reasonable SN/galaxy brightness contrast based on visual inspection, and
(iii) SN host-galaxy is relatively red (early-type). 
In most cases, a detection in at least two epochs was required before 
a spectrum was obtained.

%This pixel-level simulation accounted for variations in
%the intrinsic supernova brightness,
%seeing, and non-photometric conditions.

Spectra of supernova candidates and, where possible,
their host galaxies were obtained in Sept.-Dec. 2005 
with a number of telescopes \citep{Frieman07,Zheng08}:
the Hobby-Eberly 9.2-meter at McDonald Observatory,
the Astrophysical Research Consortium 3.5-meter at APO,
the Subaru 8.2-meter on Mauna Kea,
the Hiltner 2.4-meter at MDM Observatory,
the 4.2-meter William Herschel Telescope on La Palma,
%
%the Magellan 6.5-meter at Las Campanas,
%the 11-meter South African Large Telescope,
%the 3.6-meter Telescopio Nazionale Galileo on La Palma
%
and the Keck 10-meter on Mauna Kea. 
Approximately 90\% of the SN~Ia candidates that were \specy\
observed were confirmed as SNe~Ia.

As noted below (\S \ref{subsec:typez}), 
$\NSNSDSSTOT$ \specy\ observed candidates from 2005 
were classified as definitive or possible SNe~Ia 
based on analysis of their spectra. For these candidates,
there are a total of 
more than 2000 photometric epochs, 
where each epoch corresponds to a measurement 
(not necessarily a detection)
in the $ugriz$ passbands within a time window of
$\TMINCUT$ days to $\TMAXCUT$ days relative to peak brightness
in the supernova rest-frame.
About half of the epochs were recorded in ``photometric''
conditions, defined as no moon, PSF less than 1.7\arcsec,
and no clouds as indicated by the SDSS cloud camera, which
monitors the sky at $10\mu m$ \citep{Hogg_01}.
Another 30\% of the measurements were recorded in
non-photometric (but moonless) conditions.
The remaining 20\% of the measurements were taken
with the moon above the horizon.

% #########################################
\section{SDSS SN Spectroscopic and Photometric Reduction}
\label{sec:spec-photo}
%
% spectroscopic and photometric analysis
%

For each supernova candidate found during the survey, the 
on-mountain software pipeline described in \S~\ref{sec:survey}
delivered preliminary photometric measurements.
Similarly, \spec\ observations were reduced in near-real time
so that 
estimates of SN type and redshift could be made.
%used for subsequent \spec\ targeting. 
Although these initial 
measurements were sufficient for discovering and confirming 
SNe, for the final analysis and sample selection 
we require more accurate photometry \citep{Holtz08}
and a more uniform \spec\ analysis \citep{Zheng08}.
This section briefly describes these techniques.

% --------------------------------------------------
  \subsection{Supernova Typing and Redshift Determination }
  \label{subsec:typez}
% --------------------------------------------------

After the finish of the Fall 2005 season, 
all of the supernova
spectra were processed
with IRAF \citep*{IRAF}.
Classification 
of the reduced SN spectra was aided by 
the IRAF package  {\tt rvsao.xcsao}
\citep{rvsao.xcsao}, 
which cross-correlates the spectra
with libraries of SN spectral templates and 
searches 
for significant peaks.
Details of this analysis are described in \citet{Zheng08}.
About half of the supernova spectra had an excellent template
match, while the other half required more
human judgment for the SN typing.
Based on this analysis, $\NSNSDSSCONF$ candidates 
were classified as confirmed SNe~Ia and
$\NSNSDSSMAYBE$ candidates were classified
as probable SNe~Ia.

%
% Oct 10, 2008: update z-source numbers using SDSS_HOLTZ08
%

\newcommand{\NHOSTZDB}{29}          % DR4 hostZ's 
\newcommand{\NHOSTZDBovp}{2?}       % DR4 hostZ's that overlap SDSS hostz
\newcommand{\NHOSTZSDSSIInoDB}{82}  % SDSS-II hostZ's not in DR4
\newcommand{\NHOSTZSDSSII}{1??}     % SDSS-II hostZ's, total
\newcommand{\NSNZSDSSII}{34}        % SN z from spectral features
\newcommand{\FRACHOSTZ}{77}         % percentage with hostz
\newcommand{\FRACSNZ}{23}           % percentage with SN-z
\newcommand{\HOSTZDBNAME}{DR4}

For {\NHOSTZDB} of these $\NSNSDSSTOT$ candidates,
we have used the SDSS host-galaxy \spec\ redshift
as reported in the SDSS {\HOSTZDBNAME} database;
typical redshift \uncs\ are 1-2$\times 10^{-4}$.
For SN 2005hj, a host-galaxy \spec\ redshift
and its \unc\ were obtained by \citet{Quimby2007}. 
For {\NHOSTZSDSSIInoDB} of the candidates that do not
have a host \spec\ redshift in the DR4 database,
we use the redshift from host-galaxy spectral features
obtained with our own \spec\ observations. 
The redshift precision in those cases is estimated to be 0.0005, 
the rms difference between our host-galaxy redshifts
and those measured by the SDSS spectroscopic survey
({\HOSTZDBNAME})
for a sample in which both redshifts are available.
For the remaining {\NSNZSDSSII} candidates,
our redshift estimate is based on \spec\ features 
of the supernovae, with an estimated uncertainty of 0.005, 
the rms spread between the SN redshifts
and host-galaxy redshifts.
In summary, $\FRACHOSTZ$\% of the \specy\ confirmed and probable SNe~Ia
have \spec\ redshifts determined from host-galaxy features,
while the rest have redshifts based on
SN spectral features.
The redshifts are determined in the heliocentric frame 
and then transformed to the CMB frame as described in
\S~\ref{sec:wfit}.

The redshift distribution for the $\NSNSDSSCONF$
confirmed SNe~Ia from the 2005 season is shown below in
Fig.~\ref{fig:sncuts_sdss}e.
The relative deficit of confirmed
SNe at redshifts between $0.15$ and $0.25$ is due to the
finite spectroscopic resources that were available for the 
Fall 2005 campaign and to the relative priorities given to 
low- and high-redshift candidates for the different 
telescopes \citep{Sako08}.
Subsequently, host-galaxy redshifts have been obtained for
most of the ``missing'' SN~Ia candidates with SN~Ia-like light curves
in this redshift range.
These photometrically identified (but spectroscopically unconfirmed) 
candidates with 
host-galaxy redshifts are used in the determination of
host-galaxy dust properties (\S~\ref{sec:dust}),
but we do not include them in the Hubble diagram for this analysis.
Compared to the Fall 2005 season, \spec\ observations during
the 2006 and 2007 seasons
were more complete around redshifts  $z\sim 0.2$.

% --------------------------------------------------
   \subsection{Supernova Photometry}
   \label{subsec:SMP}
% --------------------------------------------------

To achieve precise and reliable SN photometry, we developed 
a new technique called ``Scene Model Photometry'' (SMP)
that optimizes the determination of supernova and host-galaxy fluxes.
This method and the Fall 2005 SN photometry results are 
described in detail in \citet{Holtz08}.

The basic approach of SMP is to simultaneously model the ensemble of survey 
images covering a SN location as a time-varying point
source (the SN) and sky background plus 
time-independent galaxy background and nearby calibration stars, 
all convolved with a time-varying PSF.
The calibration stars are taken from the SDSS catalog for 
stripe 82 produced by \citet{Ivezic_07}.
The fitted parameters are supernova position,
supernova flux for each epoch and passband,
and the host galaxy intensity distribution in each passband.
The galaxy model for each passband is a $20 \times 20$ grid 
(with a grid-scale set by the CCD pixel scale, 
$0.4\arcsec \times 0.4\arcsec$) 
in sky coordinates, and each of the $400 \times 5 = 2000$  
galaxy intensities is an independent fit parameter.
As there is no pixel re-sampling or image convolution, 
the procedure yields correct statistical error estimates. 
\citet{Holtz08} describes the rigorous tests that were carried out 
to validate the accuracy of SMP photometry and of the 
error estimates. 

\begin{figure*}  %  [b]
\centering
\epsscale{.28}
\plotone{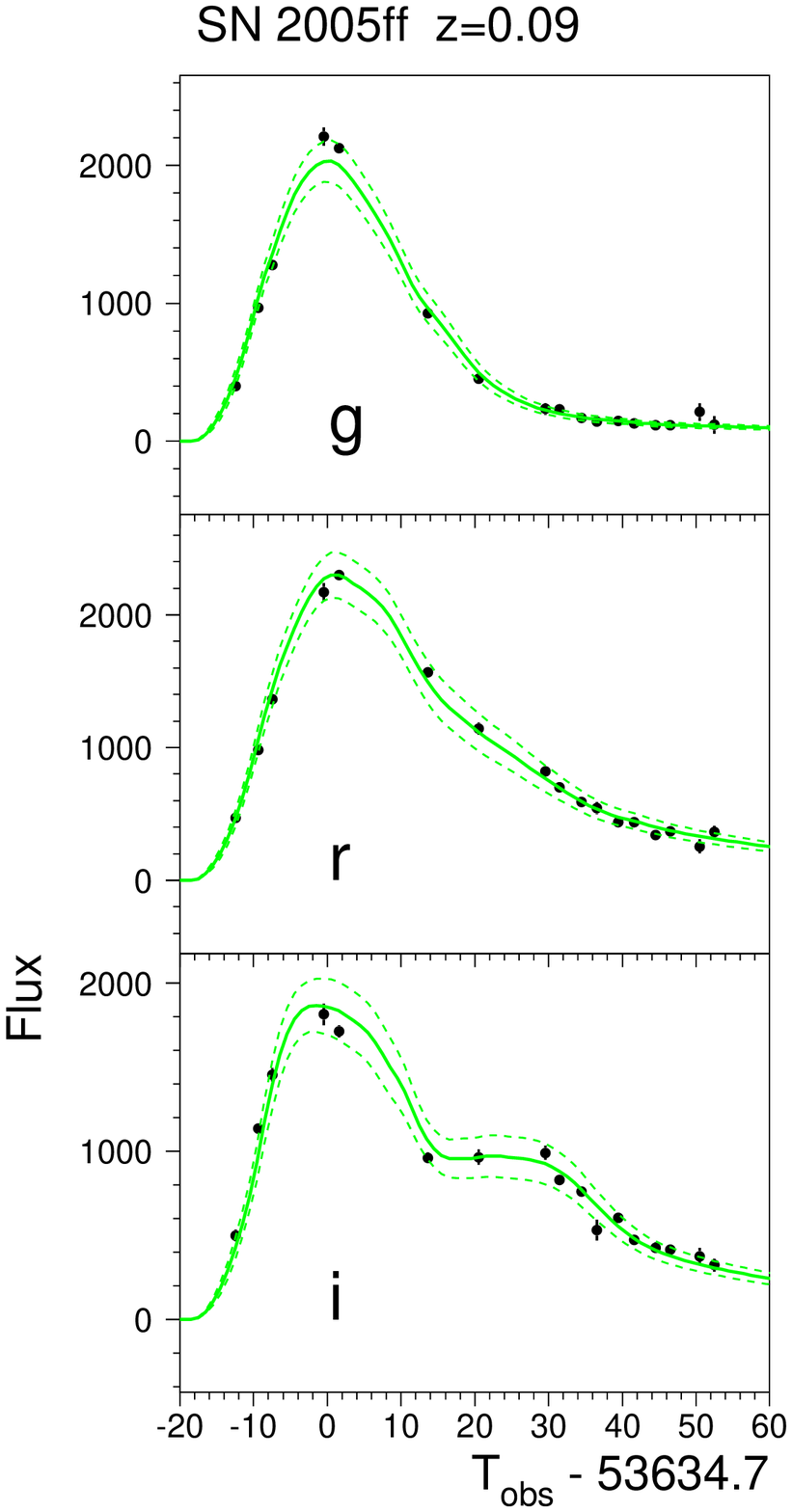}
\plotone{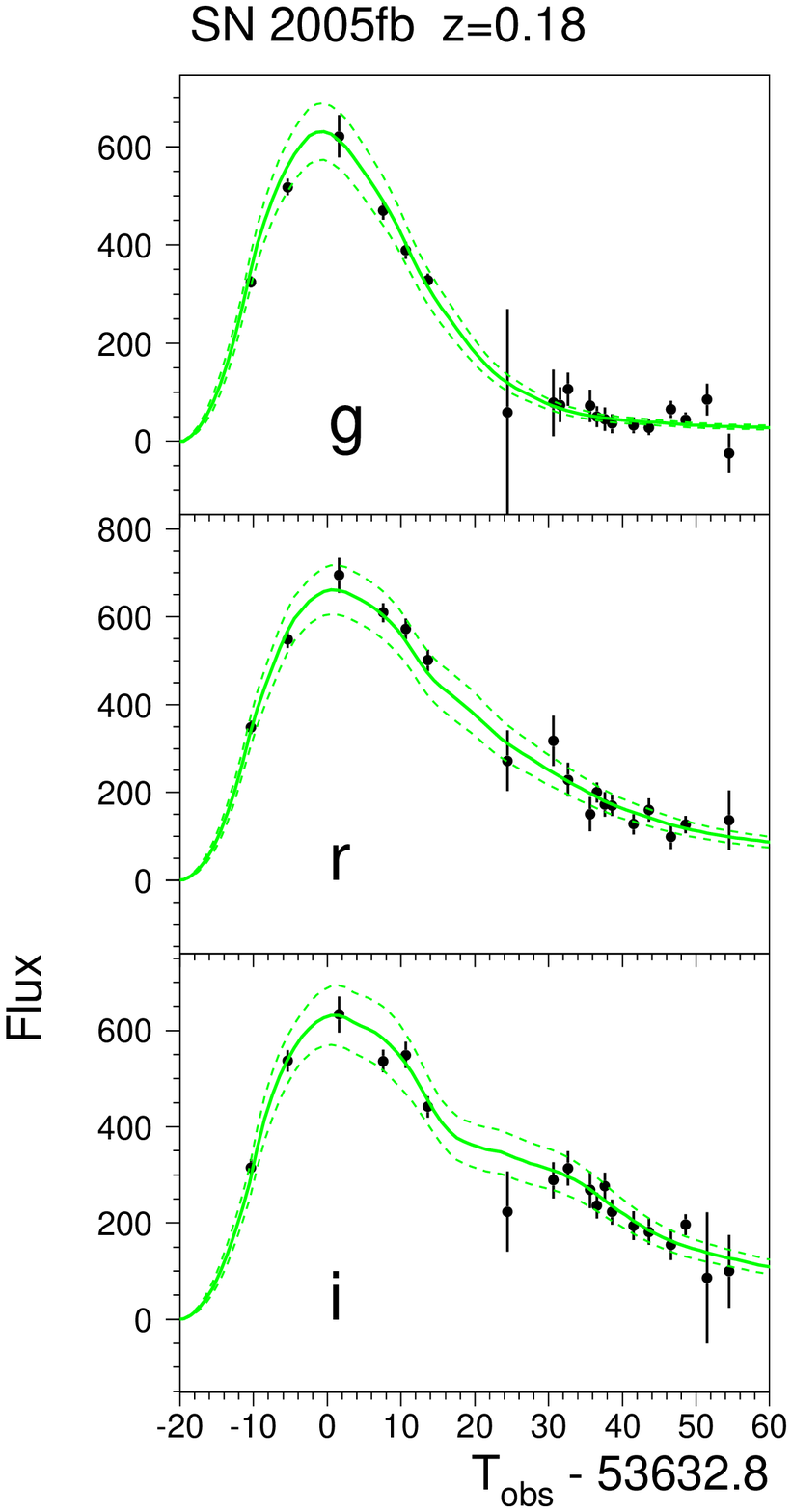}
\plotone{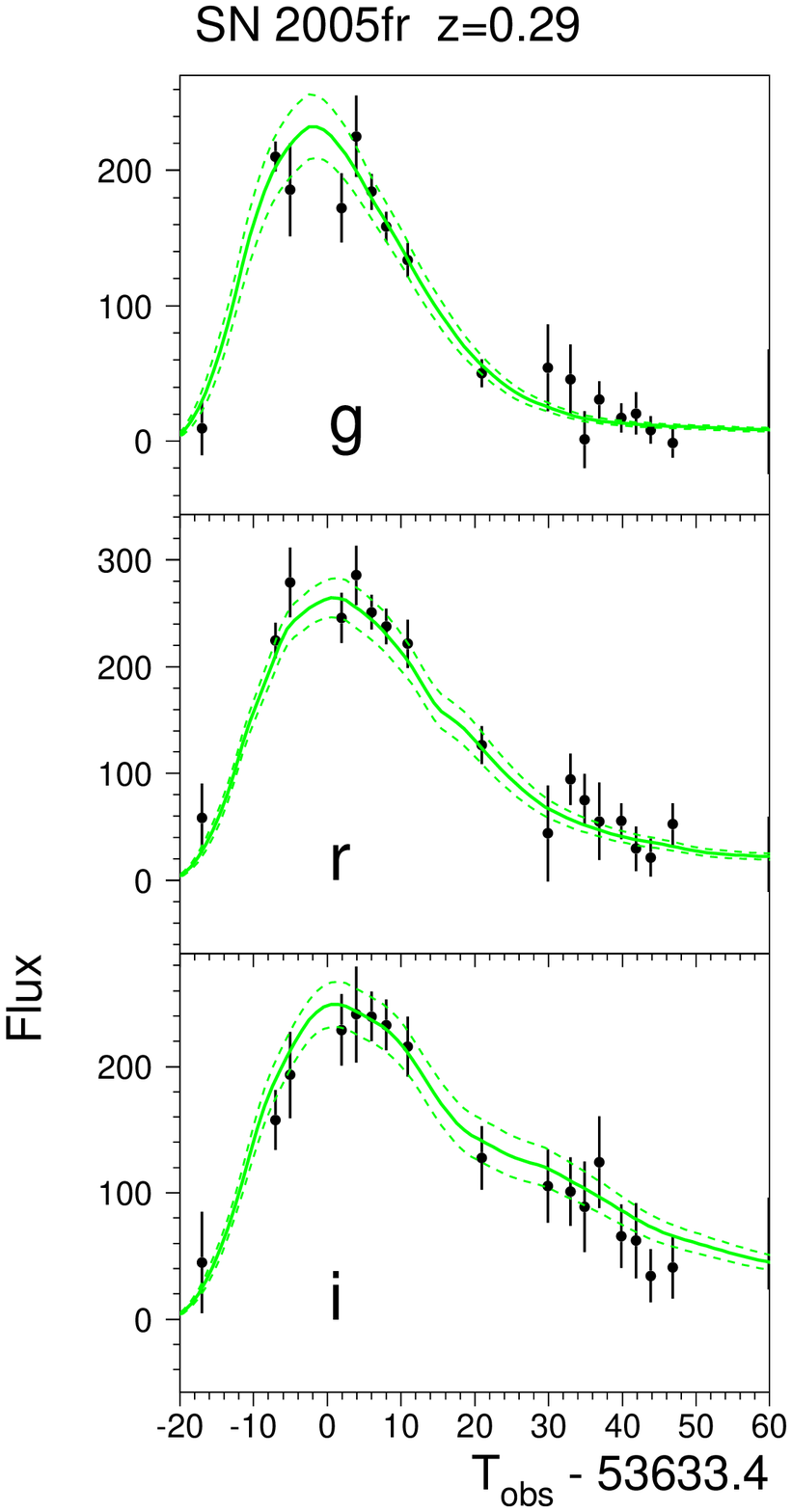}
\plotone{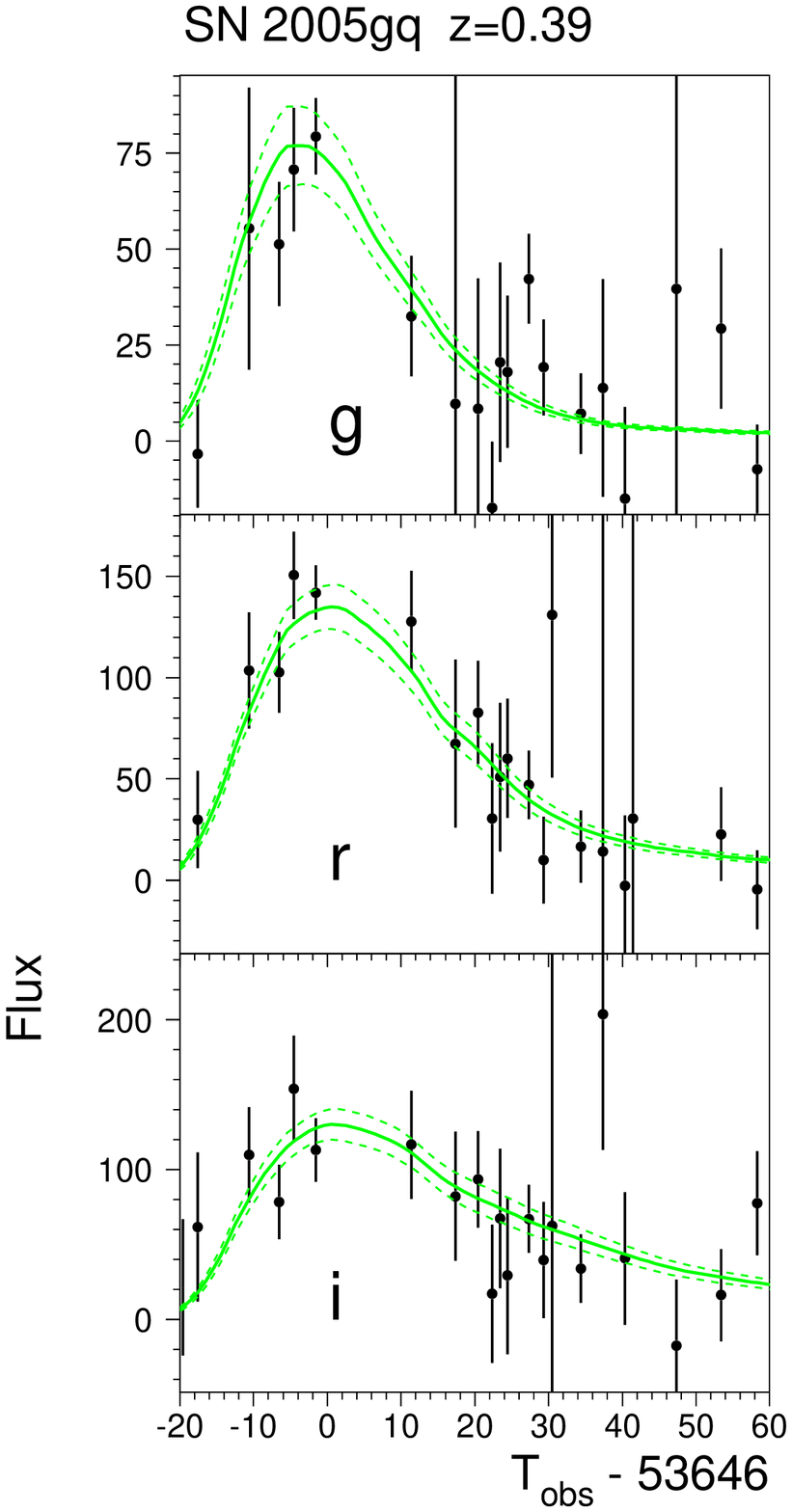}
  \caption{
	Light curves for four \SDSS\ SNe~Ia at
	different redshifts:
	SN 2005ff at $z = 0.09$,
	SN 2005fb at $z = 0.18$,
 	SN 2005fr at $z = 0.29$,
	and
 	SN 2005gq at $z = 0.39$.
 	The passbands are SDSS $g$ (top), $r$ (middle), and $i$ (bottom).
 	Points are the SMP flux measurements 
	(${\rm flux} = 10^{(11-0.4m)}$, where $m$ is the SN magnitude)
	with $\pm 1~\sigma$
	photometric errors indicated.  
	Solid curves show the best-fit \mlcs\ model fits 
	(see \S \ref{subsec:MLCS2k2}),
	and dashed curves give the $\pm 1~\sigma$ error bands
  	on the model fits.
	The Modified Julian Date (MJD) under each set of light curves
	is the fitted time of peak brightness for rest-frame $B$-band. 
     }  % end caption
  \label{fig:mlcsfit_sdss}
\end{figure*}

Although we have obtained additional imaging on other 
telescopes for a subsample of the confirmed SNe~Ia,
only photometry from the SDSS~2.5~m telescope is used in this analysis.
Fig.~\ref{fig:mlcsfit_sdss} shows four representative \SDSS\ 
SN~Ia light curves processed through SMP and provides an indication of 
the typical sampling cadence and signal-to-noise as a function of 
redshift.

The fluxes and magnitudes returned by SMP are in the native 
SDSS system \citep{Ivezic_07}.
The SDSS photometric system is nominally on the AB system,
but the native flux in each filter differs from that of a 
true AB system by a small amount. 
AB-magnitudes are obtained by adding 
the AB-offsets in Table~\ref{tb:ABoff}
to the native magnitudes.
The offsets are determined by comparing photometric 
measurements of the HST standard solar analogs
P3330E, P177D, and P041C with synthetic magnitudes 
based on the published HST spectra \citep*{Bohlin06} 
and SDSS filter bandpasses.
Since the standard stars are too bright to be measured directly
with the SDSS 2.5~m telescope, the measurements are
taken with the 0.5-meter SDSS Photometric Telescope (the PT)
and transformed to the native system of the SDSS telescope.
The technique of transferring the PT magnitudes to
the native SDSS system is identical to that used to obtain the SDSS
photometric calibration \citep*{Tucker_06}. 
The \unc\ in the AB offsets is estimated to be
0.003, 0.004, 0.004, 0.007, 0.010~mag (for $u,g,r,i,z$)
based on the internal consistency of the three standard solar analogs.
The \uncs\ given in Table~\ref{tb:ABoff} are larger, 
since they also account for the $\sim 10$~\AA\ 
\uncs\ in the central wavelengths 
(given in the same Table) 
of the SDSS filters.

\begin{table}[!h]
\caption{
  AB offsets and central wavelength \uncs\ for the SDSS filters.
	} % end caption
\begin{center}
\leavevmode
\begin{tabular}{ccc}
\tableline\tableline
                & AB offset (mag) and           & \unc\ (\AA) on     \\
   SDSS filter  & its {\unc}\tablenotemark{a}   & central wavelength \\
\tableline  % ------------------------ 
  $u$  & $-0.037 \pm  \ABoffuerr$  & 8  \\ 
  $g$  & $+0.024 \pm  \ABoffgerr$  & 7  \\
  $r$  & $+0.005 \pm  \ABoffrerr$  & 16 \\
  $i$  & $+0.018 \pm  \ABoffierr$  & 25 \\
  $z$  & $+0.016 \pm  \ABoffzerr$  & 38 \\
\tableline  % ------------------------
\end{tabular}
\end{center}
\tablenotetext{a}{
	Errors account for \uncs\ in the
	central wavelengths of the SDSS filters.
	}
  \label{tb:ABoff}
\end{table}

% ##############################################################
% set clearpage for preprint format,
% but comment this out for emulateapj format.
% \clearpage

\section{Supernova Sample Selection}
\label{sec:sample}
%
% SN Sample Selection section
%

In this section, we describe the light-curve selection criteria 
used to define the SN~Ia samples.
To minimize systematic errors associated with analysis methods 
and assumptions, we perform a nearly uniform analysis on data from \SDSS, 
the published data from 
ESSENCE (\citet{WV07}; hereafter WV07),
SNLS \citep{Astier06}, HST \citep{Riess_06},
and a nearby SN~Ia sample 
collected over a decade from several surveys and a number of telescopes 
(\citet*{Jha07}; hereafter JRK07).
Although these data samples are analyzed in a homogeneous fashion,
we present more details about the \SDSS\  analysis
since these data are presented here for the first time and, 
more importantly, because we use the \SDSS\ sample in \S~\ref{sec:dust} 
to make inferences about the SN~Ia population that we apply 
to {\it all} the data samples.

Light curves with good time sampling and good signal-to-noise are 
needed to yield reliable distance estimates. We therefore 
apply stringent selection cuts
to all five photometric data samples used in this analysis. 
The cuts are also chosen to define samples whose selection 
functions can be reliably modeled with the Monte Carlo 
simulations described in \S \ref{sec:sim}. 
In future analyses the cuts will be further refined
based on studies with simulated samples.

We first present the selection cuts we have applied and then discuss 
briefly the rationale for each of them.
Defining $\Trest$ as the rest-frame time, such that $\Trest=0$ 
corresponds to peak brightness in rest-frame $B$-band according to {\mlcs}, 
we select for inclusion in the cosmology analysis SN~Ia light curves that 
satisfy the following criteria:
\begin{enumerate}
  \item For \SDSS, ESSENCE, SNLS, and HST, 
 	at least one measurement is required 
        before peak brightness ($\Trest < 0$~days);  
        for the nearby sample,
        at least one measurement is required with $\Trest < +5$~days.
	The requirement on the nearby sample is relaxed,
	because nearly half the sample would be rejected
	by the more stringent cut of $\Trest < 0$~days.
  \item at least one measurement with $\Trest > +10$~days.
  \item at least five measurements with $\TMINCUT < \Trest < \TMAXCUT$ days.
  \item at least one measurement with signal-to-noise ratio (SNR) 
        above 5 for: each of SDSS $g$, $r$, and $i$; 
        both SNLS $r$ and $i$ (no requirement on $g,z$);
        HST {F814W\_WFPC2} and at least one other HST passband.
      	For the ESSENCE sample, we adopt the
	cuts from WV07:
	at least one measurement at $\Trest < +4$~days that
	has SNR$>5$, 
	at least one measurement at $\Trest > +9$~days that
	has SNR$>5$,
	and at least 8 total measurements with SNR$>5$.
        Since the nearby SN~Ia sample includes only events 
        with high SNR, no SNR requirement is needed for that sample.
  \item $\Pfit > 0.001$, where $\Pfit$ is the \mlcs\ light-curve 
        fit probability based on the $\chi^2$ per 
        degree of freedom (see \S \ref{subsec:MLCS2k2}).
  \item $z > \ZMINSYM = 0.02$, 
	which only affects the nearby SN~Ia sample.
\end{enumerate}
For all the data samples we only include unambiguous spectroscopically 
confirmed SNe~Ia; in particular, 
for the \SDSS\ sample, we do not include 
the $\NSNSDSSMAYBE$ \specy\ probable SNe~Ia 
(see \S \ref{subsec:typez}). 
Moreover, for the \SDSS\ sample, 
we use only $g,r,i$ photometry in the analysis,
and we reject $\sim 4$\% of the epochs for which the 
Scene Model Photometry pipeline 
(\S \ref{subsec:SMP})
did not return a reliable flux estimate.

For the first three requirements in the list above, 
a ``measurement'' corresponds 
to a recorded photometric measurement in a 
single passband and can have any signal-to-noise value, 
i.e., a significant detection is not necessary.
These requirements collectively ensure that the time sampling of the light curve is 
sufficient 
to yield a robust light-curve model fit, with coverage before and 
after peak light so that the epoch of peak light can be reliably estimated. 
To illustrate the motivation for requiring a measurement before peak light 
(which was not explicitly required in either the 
\citet{Astier06} or WV07 analyses), 
consider SN~g133 in the ESSENCE sample, which has no measurements
before peak and is therefore rejected in our analysis. 
Compared to the published values in WV07,
our \mlcs\ fitted time of maximum brightness is 10 days earlier,
and our fitted distance modulus is 0.4 magnitudes smaller. 
The fourth requirement, on SNR, similarly puts a floor on the quality 
of the light-curve data. The fifth requirement, on \mlcs\ 
light-curve fit probability $\Pfit$, is designed 
to remove obviously peculiar SNe in an objective fashion. 
This cut removes the previously identified peculiar SNe~Ia in the 
\SDSS\ sample, 
2005hk \citep{Phillips_07,Chornock_06}, 
2005gj \citep{SNF_06,Prieto_07}, and 
\SDSS\ SN 7017 (which is similar to 2005gj).
In the nearby sample, it rejects the following peculiar SNe:
1992bg, 1995bd, 1998de, 1999aa, 1999gd, 2001ay, 2001bt, 2002bf, and 2002cx.

The sixth selection criterion, corresponding to $c\ZMINSYM= 6000$~km/sec,
removes objects from the nearby sample for which the typical 
galaxy peculiar velocity, $v_{\rm pec} \sim 300$ km/sec, 
is a non-negligible fraction of the Hubble recession velocity. 
In principle, this cut on redshift could be replaced by a 
redshift- and position-dependent weighting covariance factor 
that includes the effects of both random and correlated peculiar  
velocities \citep{Hui_06,Cooray_06}. In this analysis, we follow 
recent practice and simply impose a lower redshift bound, 
but this approach raises the issue of how to select $\ZMINSYM$.
\citet{Astier06} and WV07 used $\ZMINSYM = 0.015$. 
However, using {\mlcs}, JRK07 found that the 
Hubble parameter inferred from the lowest-redshift SNe, with 
$z  \lesssim \ZBUBBLE$, 
is systematically higher than that obtained using more 
distant ($\ZBUBBLE < z < 0.1$) nearby SNe, consistent with 
an earlier result of \citet{Zehavi98}. 
JRK07 also noted that varying $\ZMINSYM$ from 0.008 to 0.027 
changes the dark energy equation of state by $\delta w \sim 0.2$ for 
the nearby SN~Ia sample in combination with a simulated 
ESSENCE sample. As a consequence, 
\citet{Riess_04,Riess_06}
used $\ZMINSYM = 0.023$ 
($cz_{\rm min}=7000$ km/sec), i.e., they only 
included SNe beyond the so-called ``Hubble bubble''. 
On the other hand, \citet{Conley2007} found that the Hubble bubble 
is not significant when the {\SALTII} fitter is used.   
As discussed in \S \ref{subsec:syst_mlcs}, we 
find that the best-fit value of $w$ is sensitive to the choice of 
$\ZMINSYM$ whether we use {\mlcs} or {\SALTII}.  
Varying $\ZMINSYM$, we find that  $\ZMINSYM = \ZCUTLOWZ$ corresponds 
to the middle of the range of $w$ variations for the \mlcs\ method.
For the \SALTII\ method, $w$ varies rapidly with $\ZMINSYM$ 
near $\ZMINSYM \sim 0.015$,
and is more stable when $\ZMINSYM \gtrsim \ZCUTLOWZ$.  
On this basis, we choose $\ZMINSYM  = \ZCUTLOWZ$
for both light-curve fitting methods and include the effects of 
varying $\ZMINSYM$ in the systematic error budget.

% ----------------------------------------------------
%
% REJECTED SN from ESSENCE & SNLS (updated Feb 2, 2008)
%   SNLS:  6 lost from Trestmax, 1 lost from Tmin, 1 lost from Pfit<.001
%   ESSE:  3 lost from Trestmax, 1 lost from Tmin
%   TOTAL: 9 lost from Trestmax, 2 lost from Tmin, 1 lost from Pfit<.001
%
% ----------------------------------------------------

For the \SDSS\ sample of $\NSNSDSSCONF$ \specy\ confirmed SNe~Ia from 
the Fall 2005 season, 
$\NSNSDSS$ satisfy these selection criteria. 
The cut-rejection 
statistics are as follows:
3 are photometrically peculiar SNe~Ia
  that fail the $\Pfit$ requirement;
9   have no measurement before peak brightness---most of these 
   were discovered early in the survey season; 
11  have no measurement with $\Trest > +10$~days---most of these were 
    discovered late in the survey season or were at the high-redshift 
    end of the distribution; and
4 SNe~Ia in the high-redshift tail, $z \sim 0.4$, 
  fail the SNR requirement.

With the selection criteria defined above,  
the number of SN~Ia events used for fitting is 
shown in Table~\ref{tb:NCUTS} for each sample;
a total of $\NSNTOT$ SNe~Ia are included in the fiducial analysis 
(in systematic error tests, e.g., 
varying $\ZMINSYM$, this number fluctuates by a small amount).
Table~\ref{tb:NCUTS} also shows 
the average number of measurements per SN~Ia for each
sample, where a measurement 
is an observation in a single passband  
in the rest-frame time interval $\TMINCUT$ to $\TMAXCUT$ days.
The average number of measurements is about 50 for both the 
nearby and \SDSS\ samples, in the twenties for 
ESSENCE and SNLS, and $\NEPHST$  for HST. We note that 
our selection requirements are more restrictive than those applied 
in previous analyses. 
WV07 included 60 out of 105 spectroscopically confirmed 
ESSENCE SNe~Ia for their \mlcs\ 
analysis\footnote{
	WV07 include 60 ESSENCE SNe~Ia for the analysis 
	that includes the SNLS sample; for their analysis of
	Nearby+ESSENCE (excluding SNLS) they
	require $z<0.67$, resulting in 57 ESSENCE SNe~Ia.
}, 
while our cuts select $\NSNESSE$.
WV07 selected 45 SNe~Ia from the nearby sample ($z>0.015$), 
while we include $\NSNLOWZ$ ($z>0.02)$;
the difference is mainly due to the different
redshift cuts. \citet{Astier06} included 71 SNLS SNe~Ia, 
while we retain $\NSNSNLS$ from the same sample.

\begin{table}[!h]
%\centering
\caption{  
  	Redshift range, number of SNe passing selection cuts, 
	and mean number of measurements for each SN sample.
     	} % end caption
\begin{center}
\leavevmode
\begin{tabular}{lccc}
\tableline\tableline % --------------------------------------------
   sample          & redshift &                  &             \\
 (obs passbands)   & range    
        &  $N_{\rm SN}$\tablenotemark{a}
        &  $\langle N_{\rm meas}\rangle$\tablenotemark{b} \\
\tableline % -------------------------------------------------
 Nearby ($UBVRI$)  & 0.02 -- 0.10  &  $\NSNLOWZ$  & $\NEPLOWZ$  \\
 \SDSS\ ($gri$)    & $\ZMINSDSS$ -- $\ZMAXSDSS$
                                  &  $\NSNSDSS$  & $\NEPSDSS$  \\
 ESSENCE ($RI$)    & 0.16 -- 0.69  &  $\NSNESSE$  & $\NEPESSE$  \\
 SNLS ($griz$)     & 0.25 -- 1.01  &  $\NSNSNLS$  & $\NEPSNLS$  \\
 HST (F110W, F160W, 
                   & 0.21 -- 1.55  &  $\NSNHST$   & $\NEPHST$   \\
F606W, F775W, F850LP)  & & & \\
\tableline % -------------------------------------------------
\end{tabular}
\end{center}
\tablenotetext{a}{Number of SNe~Ia passing cuts.}
\tablenotetext{b}{Average number of measurements per SN~Ia,
    in the interval $\TMINCUT < \Trest < \TMAXCUT$ days.}
   \label{tb:NCUTS}
\end{table}

\begin{figure}[hb!]
  \epsscale{1.1}
  \plotone{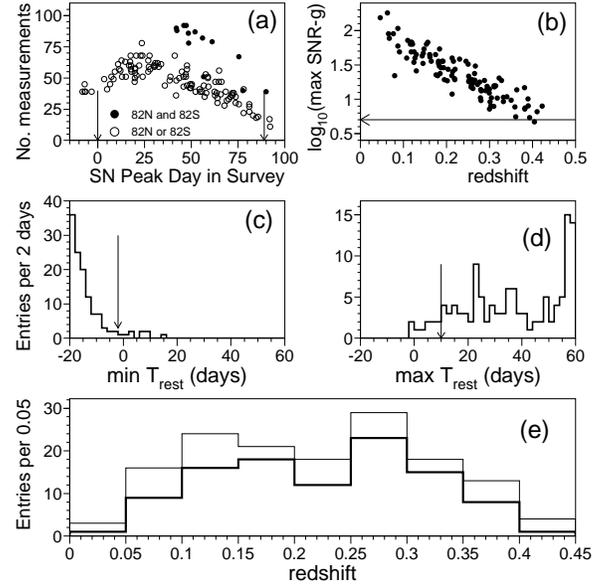}
  \caption{
       For the \specy\ confirmed SN~Ia sample from the \SDSS\ 2005 season, 
       distributions are shown for:
       (a) number of $gri$ measurements 
       with $\TMINCUT < \Trest < \TMAXCUT$~days 
       as a function of day in the survey season when the SN
       reached peak luminosity. 
       Vertical arrows show the start (Sept. 1) and end dates (Nov. 30) 
       of the survey season.
       SNe that lie in the overlap region of strips 82N and 82S (solid dots)
       tend to have more measurements;
       (b) $\log_{10}$ of maximum $g$-band SNR versus redshift; 
       (c) time of first measurement relative to peak light 
         in rest-frame $B$;  
       (d) time of last measurement (not necessarily detection)
           relative to peak light --
            the pile-up near 60 days is from SNe that have measurements
            past 60 days; 
       (e) redshifts before (130, thin line)
            and after (103, thick line) selection cuts are applied. 
       	The arrows in panels (b), (c), and (d) indicate the selection cuts.
     }
  \label{fig:sncuts_sdss}
\end{figure}

\begin{figure}[hb]
  \epsscale{1.1}
  \plotone{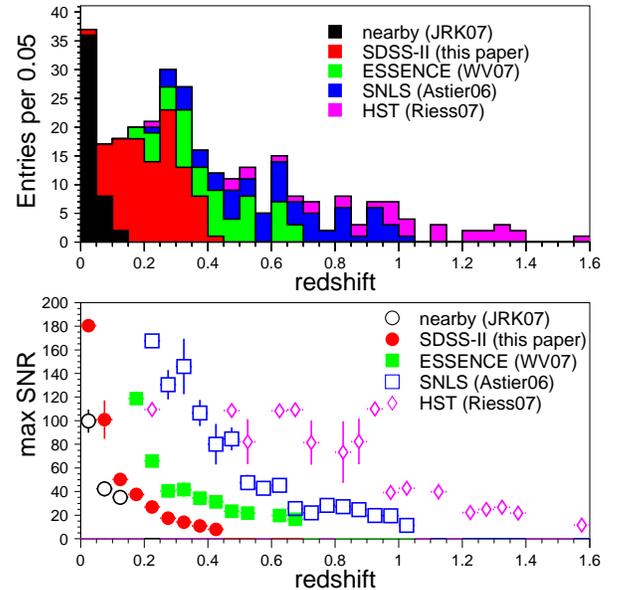}
  \caption{
     {\it Top panel:} summed redshift distribution for the 
     five SN~Ia samples indicated in the legend. 
     {\it Bottom panel:} maximum observed signal-to-noise ratio 
     (among all passbands) as a function of redshift, 
     averaged in bins of width $\Delta z=0.05$;   
     error bars indicate the rms spread within each bin. 
     All selection requirements have been applied.
     }
  \label{fig:zsamples}
\end{figure}

Figure~\ref{fig:sncuts_sdss} shows distributions in the \SDSS\ 
sample---before selection cuts---for some 
of the variables used in sample selection, 
as well as the \SDSS\ SN~Ia redshift distribution
before and after selection cuts are applied.
Figure~\ref{fig:zsamples} shows the redshift 
distribution for all five samples, along with the average of the maximum
observed signal-to-noise as a function of redshift.

% ####################################################

\section{Light-Curve Analysis}
\label{sec:anal}
% ======================================
%
%  Description of Light-curve fitting
%  *** TOP ***
%
% 
% ======================================

In this section, we describe our methods of analyzing supernova light
curves and extracting distance estimates. 
The two light-curve fitting methods we employ, {\mlcs} and {\SALTII},
reflect different assumptions about the nature
of color variations in SNe~Ia, different approaches to training 
the models using pre-existing data, and different ways of determining 
model parameters.

% ---------------------------------------------------------
\subsection{\mlcs\ Fitting Method}
   \label{subsec:MLCS2k2}
% ---------------------------------------------------------

The Multicolor Light Curve Shape method, known as {\mlcs} in its
current incarnation (JRK07), has been in use for more than a decade;
the original MLCS version \citep{Riess_98} was used by the High-z
Supernova Team in the discovery of cosmic acceleration.
For each supernova, \mlcs\ returns an estimated distance modulus
and its uncertainty;
the redshift and distance modulus for each SN are inputs 
to the cosmology fit discussed in \S~\ref{sec:wfit}.

\mlcs\ describes the variation among SN~Ia light curves with a 
single parameter ($\Delta$).  
Excess color variations relative to the one-parameter model are 
assumed to be the result of extinction by dust in the host galaxy 
and in the Milky way.  
The \mlcs\ model magnitude is given by
\begin{eqnarray}
  m_{\rm model}^{e,f} & = & 
    {\Mmlcs}^{e,f'} + p^{e,f'}\Delta + q^{e,f'}\Delta^2 
   \nonumber \\
     &  +  &
    \Xhost^{e,f'} +   K^e_{ff'} + \mu + \XMW^{e,f}~,
   \label{eq:MLCS2k2model}
\end{eqnarray}
where $e$ is an epoch index that runs over the observations,
$f$ are observer-frame filter indices,
$f' = UBVRI$ are the rest-frame filters for which
the model is defined,
$\Delta$ is the \mlcs\ shape-luminosity parameter that accounts for the
correlation between peak luminosity and the shape/duration of the 
light curve, $\Xhost$ is the host-galaxy extinction,
$\XMW$ is the Milky Way extinction,
$K_{ff'}$ is the \Kcor\ between rest-frame  and observer-frame filters,
and $\mu$ is the distance modulus, which satisfies 
$\mu = 5 \log_{10}(\DL/10~pc)$, where $\DL$ is the luminosity distance.
We use this model for SN epochs in the rest-frame time range 
$\TMINCUT < \Trest < \TMAXCUT$~days relative to rest-frame $B$-band maximum.
Observer-frame passbands are included that satisfy 
$3200 < \LAMFBAR /(1+z) < 9500$~\AA,
where $\LAMFBAR$ is the mean wavelength of the filter passband,
and $z$ is the redshift of the SN~Ia.
To account for larger model \uncs\ in the restframe-UV region,
a \Kcor\ \unc\ of $0.0006\times (3500-\LAMFBAR)$~mag is added
in quadrature to the model error
for $\LAMFBAR < 3500$~\AA.

In the {\mlcs} model, the shape-luminosity parameter $\Delta$ 
describes the intrinsic SN color dependence on brightness, 
and $\Xhost$ describes SN color variations
from reddening (extinction) by dust in the host galaxy,
which is assumed to behave in a manner similar to dust 
in the Milky Way. In particular, the extinction is 
described by 
the parametrization of \citet*{ccm89} (hereafter CCM89), 
$\Xhost^{e,f'}=\zeta^{e,f'}(a^{f'}+b^{f'}/R_V)A_V$, where 
$A_V$ is the extinction in magnitudes in $V$-band, 
$a^V=1$, $b^V=0$, and the relative extinction in other 
passbands is determined by the parameter $R_V$, 
the ratio of $V$-band extinction to color-excess, 
$R_V = A_V/E(B-V)$.  
For the Milky Way, the value of $R_V$ averaged over a 
number of lines of sight is $R_V  = \RVMW$; 
this global value has been adopted in previous SN analyses using {\mlcs}.
For the galaxies that host SNe~Ia,
we instead adopt $R_V = \RV \pm \RVERRTOT$,
as derived in \S~\ref{subsec:RV} from the \SDSS\ SN data.

The coefficients ${\Mmlcs}^{e,f'}$, $p^{e,f'}$, and $q^{e,f'}$ 
are model vectors that have been evaluated using nearly 100
well-observed low-redshift SNe as a training set. 
$\Mmlcs^{e,f'}$ is the absolute magnitude for a SN~Ia with 
$\Delta=0$.
Assuming a Hubble parameter $h=H_0/100$ km/sec/Mpc $=0.65$,
the resulting absolute magnitudes at peak brightness are
$-20.00$, $-19.54$, $-19.46$, $-19.45$, $-19.18$~mag
for $U,B,V,R,I$, respectively.
The $p$ and $q$ vectors translate the shape-luminosity parameter $\Delta$
into a change in the SN~Ia absolute magnitude.
The $p^{0,f'}$ values (at peak brightness)
vary among passbands from 0.6 to 0.8,
and the $q^{0,f'}$ vary from 0.1 to 0.9; therefore, intrinsically
faint (bright) SNe have positive (negative) values of $\Delta$.

We use model vectors based on the procedure outlined in JRK07,
but with two notable differences. First, the vectors have
been re-evaluated based on our determination of the  
dust parameter, $R_V = \RV$. 
Since most of the nearby objects used in the training have
low extinction, retraining with a different value of $R_V$ 
has little effect on the vectors and therefore on the cosmological results.
%with $w$-shifts of a few hundredths.
Note that the insensitivity of the \mlcs\ training to the
value of $R_V$ does not imply that the estimated distances
for high-redshift SNe are insensitive to the value of $R_V$,
especially since the latter samples include highly 
extinguished SNe. 
The impact of $R_V$ on the cosmology results 
is presented in \S \ref{sec:syst}.

The second change in the model vectors from JRK07 
involves 
adjustments to the $\Mmlcs^{e,f'}$  that were developed during 
the course of the WV07 analysis of the ESSENCE data. 
For the model training with the nearby SN~Ia sample, 
it was assumed that the observed $A_V$ distribution 
has the functional form of an exponential distribution
convolved with a Gaussian centered at $A_V=0$. However,
the \mlcs\ training process resulted in a convolution Gaussian
that is not centered at zero; adjustments
were made in the vectors such that the Gaussian
is centered at zero. The main caveat in this procedure
is that the selection \eff\ for the nearby sample is small and 
unknown, and therefore it is not straightforward to model the 
observed $A_V$ distribution in terms of an underlying population.
The $\Mmlcs^{e,f'}$  adjustments for $UBVRI$
depend only on the passband and are independent of epoch.
For the model vectors determined with $R_V = 2.2$, 
the magnitude adjustments relative to the values in JRK07 are
\begin{equation}
  \delta \Mmlcs^{UBVRI} = +0.050,+0.020, 0.0, -0.002, -0.033.
   \label{eq:Mtweak}
\end{equation}

\Kcor s transform the \mlcs\ SN rest-frame Landolt-system
magnitudes to the magnitudes of a redshifted SN in an observed 
passband.
\Kcor s are computed following the prescription of \citet*{Nugent2002},
which requires a SN spectrum at each epoch, 
the spectrum of a reference star, 
and the reference star magnitude in each passband. 
As explained in Appendix~\ref{app:kcor},
we use a single template spectrum for each SN epoch and warp it to match  
the colors of the SN model.  
Since the Landolt photometry is not associated with a precisely 
defined set of filters, 
we use the standard $UBVRI$ and $B_X$ filters 
defined by \citet{Bessell90} and apply a color transformation to 
obtain photometry in the Landolt system.
That procedure is detailed in Appendix~\ref{app:lowzfilters}.
In place of the traditional primary reference star Vega,
we choose \BDFULL\ \citep{OkeGunn83} as our primary reference because
it has been measured by Landolt, it has a precise 
HST STIS spectrum \citep{CALSPEC07}
and it is the primary reference for SDSS photometry.  
We have also carried out the analysis with Vega 
as the primary reference and include the difference as a systematic error.
The primary magnitudes for each filter system are 
given in Table~\ref{tb:primary_mags} of Appendix \ref{app:primarymags}
for both \BD\ and Vega.

A light-curve fit determines the likelihood function 
$\cal L$ of the observed magnitudes or fluxes as a 
function of four model parameters for each SN~Ia: 
(i) time of peak luminosity in rest-frame $B$-band, $t_0$, 
(ii) shape-luminosity parameter, $\Delta$,
(iii) host-galaxy extinction at central wavelength of 
     rest-frame $V$-band, $A_V$, and
(iv) the distance modulus, $\mu$.
The redshift ($z$) is accurately determined from the \spec\
analysis, so it is not included as a fit parameter; 
the redshift uncertainty is included in the 
cosmology analysis (\S~\ref{sec:wfit}).
For each supernova, the log of the posterior probability $P_{\rm post}$,
or $\chi^2$ statistic, is given by 
\begin{eqnarray}
   \chi^2 & = & -2\ln P_{\rm post}(t_0,\Delta,A_V,\mu | {\rm data})
       \nonumber \\
          & = & -2\ln {\cal L}({\rm data}|t_0,\Delta,A_V,\mu) 
                -2\ln P_{\rm prior}(z,A_V,\Delta)~,  \nonumber \\
          & &
   \label{eq:mlcs_chi2def}
\end{eqnarray}
where $P_{\rm prior}$ is a Bayesian prior (see below), 
and the log-likelihood is given by
\begin{equation}
    -2\ln {\cal L} = \left\{ 
    \sum_i \frac{\left[F_i^{\rm data} - F_i^{\rm model}(t_0,\Delta,A_V,\mu)\right]^2}
              {  \sigma_{i,{\rm stat}}^2 + \sigma_{i,{\rm model}}^2}
     \right\} ~.
\end{equation}
Here the index $i$ runs over all measured epochs and observer-frame 
passbands, and $F_i^{\rm data}$ is the observed flux for measurement $i$. 
The statistical measurement uncertainty, $\sigma_{\rm stat}$, 
is estimated from the Scene Model Photometry 
as described in \S~\ref{subsec:SMP}.
For the model uncertainty, $\sigma_{\rm model}$, we use 
the diagonal elements of the \mlcs\ covariance matrix,
which are estimated from the spread in the training sample of SNe.
For example, at the epoch of peak brightness ($t_0$) 
these model errors are 
$0.11$, $0.07$, $0.08$, $0.10$, $0.11$~mag
%% RV=RVMW: {$0.12,0.10,0.06,0.06,0.10$}~mag
for $U,B,V,R,I$, respectively;
the \uncs\ increase monotonically with time away from $t_0$.
As explained below in the list of modifications, we do not use 
the off-diagonal \mlcs\ correlations in this analysis.

Since $A_V$ is a physical parameter that is always positive,
and since it is not well constrained if the peak signal-to-noise 
is low or if the observations do not span a large wavelength range, 
the \mlcs\ fit includes a Bayesian prior on the extinction.
The prior forbids negative values of $A_V$ 
and encodes information about the distribution
of extinction in SN host galaxies as well as the
selection \eff\ of the survey.
Since there is degeneracy between the inferred values of $A_V$ and $\mu$, 
the prior  leads to reduced scatter in the Hubble diagram. 
For the nearby SN sample, which has high peak signal-to-noise for 
all objects, the prior has no impact on the Hubble scatter;
for the other samples we employ, 
the prior reduces the Hubble scatter by a factor of 1.3--2.
For this analysis, the prior is defined to be
\begin{eqnarray}
  P_{\rm prior}(z,A_V,\Delta) & = & 
      P(A_V) P(\Delta) \times  \nonumber \\
     & &
      \simeffsearch(z,A_V,\Delta) 
      \simeffcuts(z,A_V,\Delta),
   \label{eq:prior_master}
\end{eqnarray}
where $P(A_V)$ and $P(\Delta)$ are the underlying SN~Ia population 
distributions of $A_V$ and $\Delta$, 
and we assume that these distributions
are independent of redshift. We determine them from
\SDSS\ SN data in \S~\ref{sec:dust}. For SNe passing 
the selection cuts, the parameter $\Delta$ 
is typically precisely determined by the light-curve fit, so 
the prior on $\Delta$ does not have a significant impact 
on the inferred parameters. 
The functions $\simeffsearch$ and $\simeffcuts$ are
survey-dependent efficiency factors associated with the 
survey selection functions and with the sample selection cuts.
The efficiencies are determined from Monte Carlo simulations 
in conjunction with the observed data distributions for each 
survey in \S~\ref{sec:sim}. 
Tests with high-statistics simulations have verified that the 
prior in Eq.~\ref{eq:prior_master} leads to 
unbiased results for cosmological parameters.

In the \mlcs\ fit, 
the estimated value and \unc\ for each 
model parameter, 
e.g., the distance modulus $\mu$, are obtained by 
marginalizing the posterior (Eq.~\ref{eq:mlcs_chi2def})
over the three other parameters and taking the mean and rms
of the resulting one-dimensional probability distribution. 
In the marginalization integrals, we use 11 bins in each 
of the parameters; this choice is dictated 
by the computational time required for the large number of 
systematics tests 
(see \S~\ref{subsec:syst_mlcs}). We have compared results 
with 11 and 15 integration bins and find excellent agreement.

\begin{figure}[hb]
  \epsscale{1.00}
  \plotone{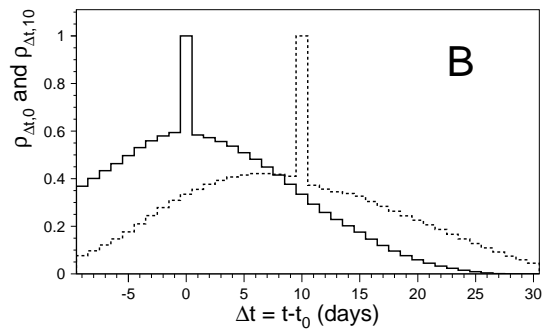}
  \caption{
	For the \mlcs\ model,
	correlation coefficient $\rho_{\Delta t,0}$ between 
	$B$-band epoch at peak brightness ($t_0$) and time
	$\Delta t = t - t_0$, where 
	$\rho_{\Delta t,0} \equiv
	  {\rm cov}(\Delta t,0)/\sigma_{\Delta t}\sigma_0$,
	as a function of $\Delta t$, and correlation coefficient
	$\rho_{t-t_{10},t_{10}}$ between epoch at 10 days past peak 
	($t_{10}$) and time $t-t_{10}$.
	The spikes at 0 and 10 days correspond to the requirements
	$\rho_{0,0} = 1$ and $\rho_{10,10} = 1$.
     }
  \label{fig:mlcs_corr}
\end{figure}

To implement the \mlcs\ method, we have written a new version 
of the fitting package with several modifications from JRK07:
\begin{enumerate}
  \item	We fit in calibrated flux instead of magnitudes. 
	In previous analyses using {\mlcs}, the fits were carried
	out using magnitudes, and data with SNR$<5$ were typically
	excluded in order to avoid ill-defined magnitudes associated
	with negative flux measurements.
	The SNR cut results in a biased determination of the
	shape-luminosity parameter $\Delta$ 
	and therefore of the distance modulus $\mu$.
	Fitting in flux enables a proper treatment of errors
	for all measurements and results in a
	negligible bias in $\Delta$ and $\mu$, as determined from
	a simulation.
	This change is crucial for our analysis, since $\sim 40$\%
	of the \SDSS\ SN measurements
	(with $\TMINCUT < \Trest < \TMAXCUT$~days)
	have SNR$<5$.
 \item	We have made two improvements to the treatment of \Kcor s.
  	First, we use the updated SN~Ia spectral templates 
	from \citet*{Hsiao07}, which result in better consistency
	between the data and the best-fit \mlcs\ model
 	for observer-frame filters that map onto rest-frame $R$-band.
	Second, we have improved the spectral warping used for
	{\Kcor s} as explained in Appendix~\ref{app:kcor}.
 \item	The \mlcs\ model includes off-diagonal 
	covariances in the model magnitudes to account for 
	brightness correlations between different epochs and passbands; 
	in this analysis, we ignore the off-diagonal covariances
     	for two reasons. The primary reason is that the \mlcs\ model
     	covariances appear to display unphysical behavior.  
     	The correlation coefficient 
	$\rho_{ij} \equiv cov(i,j)/\sigma_i\sigma_j$ between epochs 
	$i$ and $j$ decreases discontinuously from unity at $t_i=t_j$
	(Fig.~\ref{fig:mlcs_corr}):  
	the correlation between epochs separated by
	only one day is weak, $0.2 < \rho_{t,t+1} < 0.8$,
 	and thus does not penalize (via $\chi^2$)
   	random variations of $\sim 0.1$~mag over one-day time-scales.
	The observed smoothness of high-quality SN~Ia
	light-curve data rules out such large intrinsic fluctuations,
 	suggesting that random instrumental noise may have been included
	in the model covariance matrix. 
	The impact of the off-diagonal covariances on determination 
	of the cosmological
	parameters ($w$ and $\OM$) from the SN data is much smaller 
	than the statistical \uncs. 
	Second, 
	there is a subtle limitation when measurements at the
	same epoch in two observer-frame passbands $f_1, f_2$ are
  	matched onto the same rest-frame filter $f'$ using
  	$\lamrest = \lamobs/(1+z)$ for each passband. 
	In the \mlcs\ model, there is an artificial 100\% correlation 
	between the two rest-frame model magnitudes.
   	This feature arises for the observed
	$ugriz$ filters used by \SDSS\ and SNLS,
  	but does not appear for the Bessell filters used in the
   	nearby and ESSENCE samples.
 \item	We have extensively modified the prior
	(Eq.~\ref{eq:prior_master}).
	The \mlcs\ prior in JRK07 is intended to reflect
	the true distribution of $A_V$. In analyzing the ESSENCE 
	data, WV07 used a different $A_V$ prior and multiplied 
	it by a simulated \eff\ that depends upon
  	extinction, intrinsic luminosity, and redshift. 
	In our analysis we use more detailed Monte Carlo
	simulations (\S~\ref{sec:sim}) of each data sample to
	estimate the survey efficiencies that are incorporated
	into the priors,
	and we use the \SDSS\ SN data sample to determine
	the underlying $A_V$ distribution.
 \item 	In {\mlcs}, the reddening parameter $R_V$ is treated as 
	a fixed global parameter. In JRK07 and  WV07,
	$R_V$ was set to the average Milky Way value of $\RVMW$.
	In our analysis, we use $R_V= \RV \pm \RVERRTOT$
	as empirically determined from the \SDSS\ SN sample
	(\S~\ref{sec:dust}).
\end{enumerate}

\begin{figure}[hb]
  \epsscale{1.10}
  \plotone{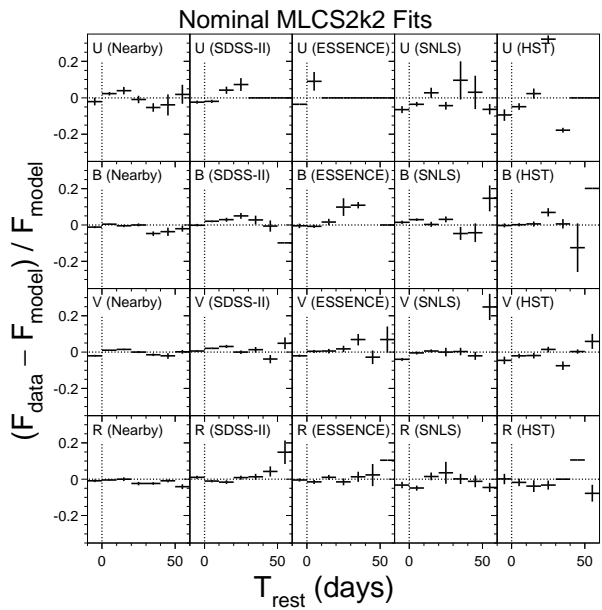}
  \caption{
	Data-model fractional residuals as a function of rest-frame
	epoch in 5-day bins for \mlcs\ light-curve fits. 
	The rest-frame passband and SN sample are indicated on each plot.
	Measurements with SNR$<6$ are excluded,
	and error bars indicate the rms spread.
	For SNLS, the residuals are shown only for SNe with $z<0.5$
     	as explained later in \S \ref{subsec:Uanom_SALT2}.
	${\rm F}_{\rm data}$ (${\rm F}_{\rm model}$)
	is the SN flux from the data (best-fit \mlcs\ model).
	Vertical dashed lines indicate epoch of peak brightness
	($\Trest=0$); horizontal dashed lines indicate
	${\rm F}_{\rm data} = {\rm F}_{\rm model}$. 
        }
  \label{fig:lcresid_MLCS}
\end{figure}

\begin{figure}[ht]
\centering
  \epsscale{1.00}
\plottwo{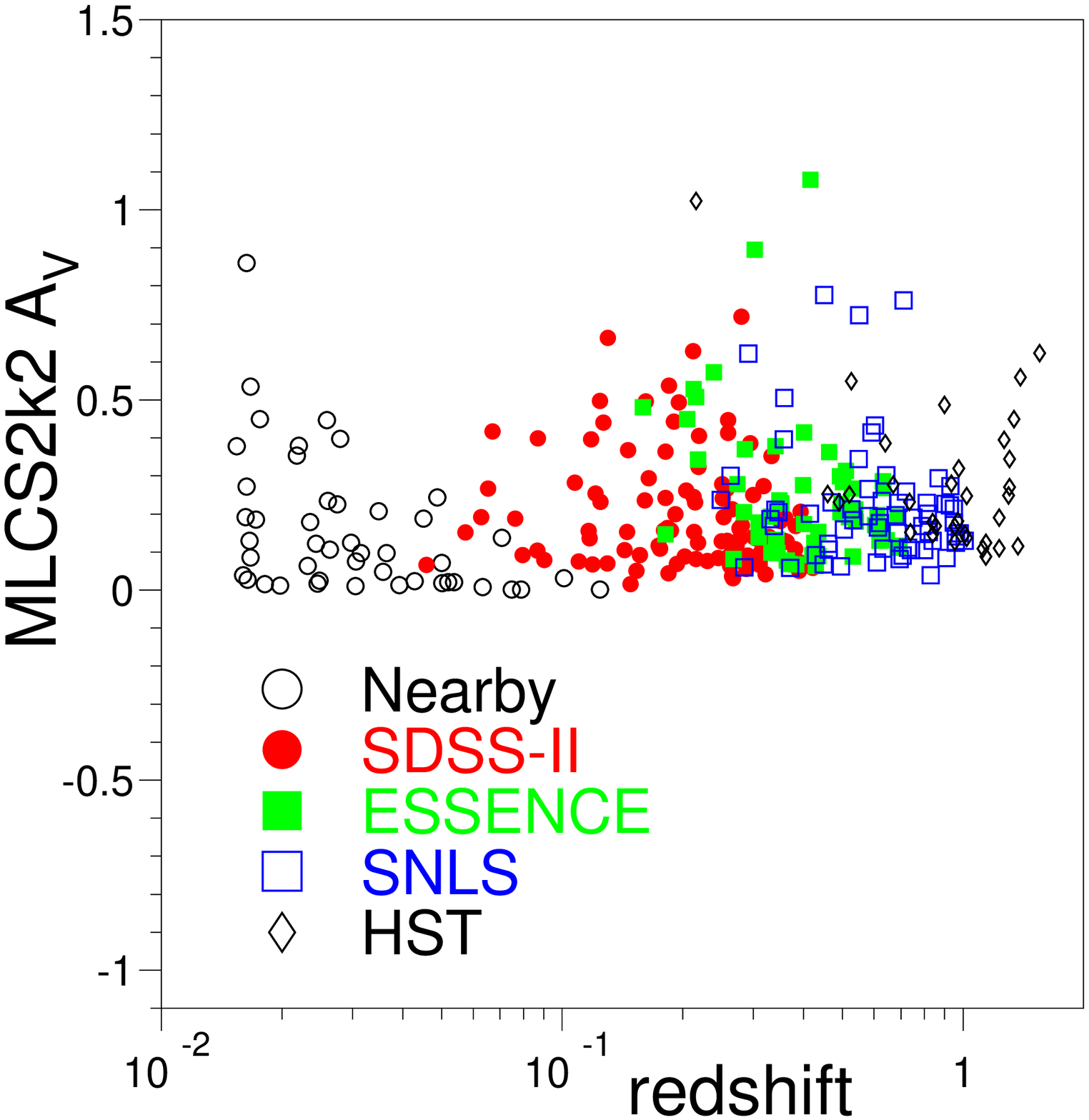}{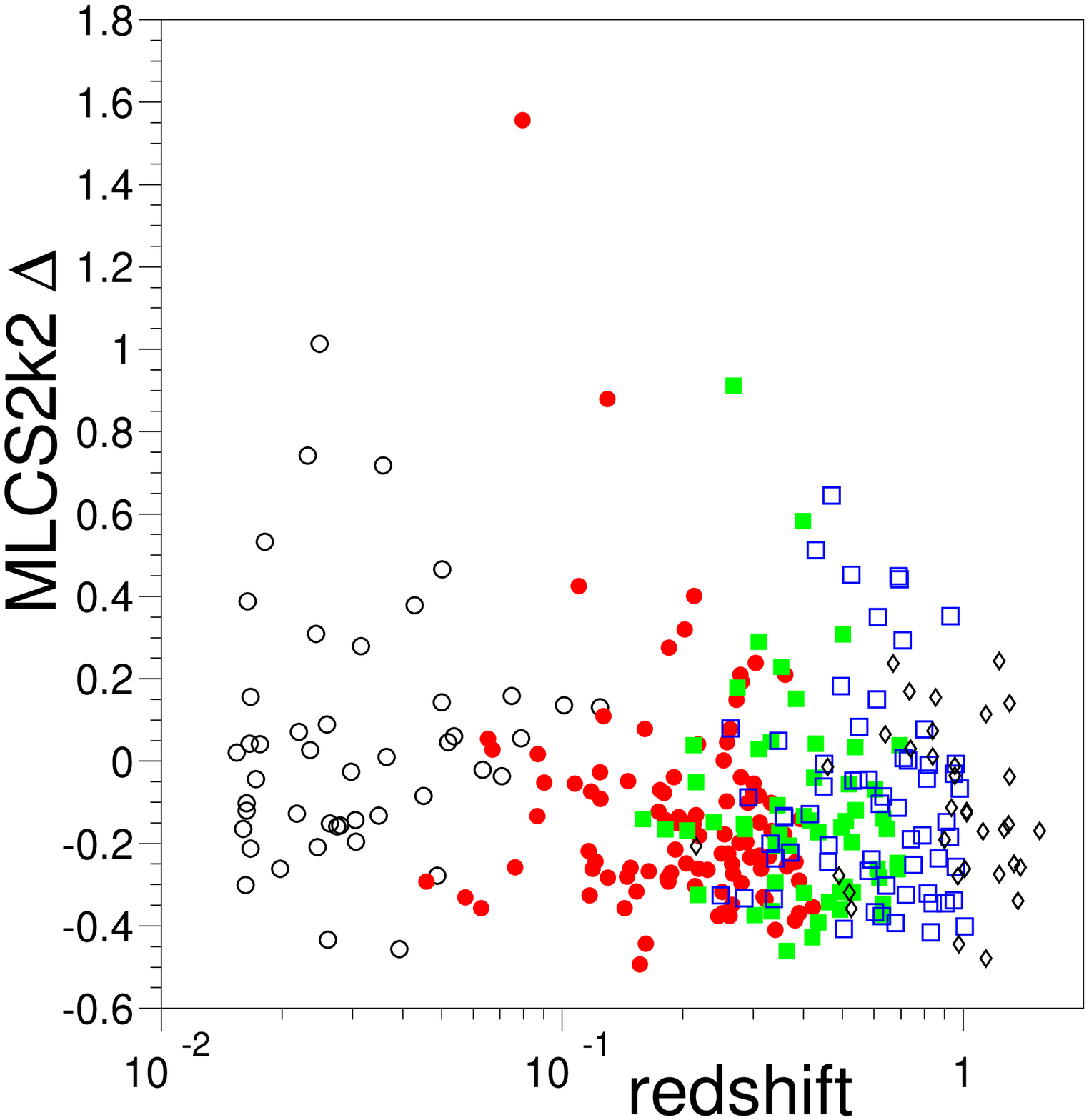}
\plottwo{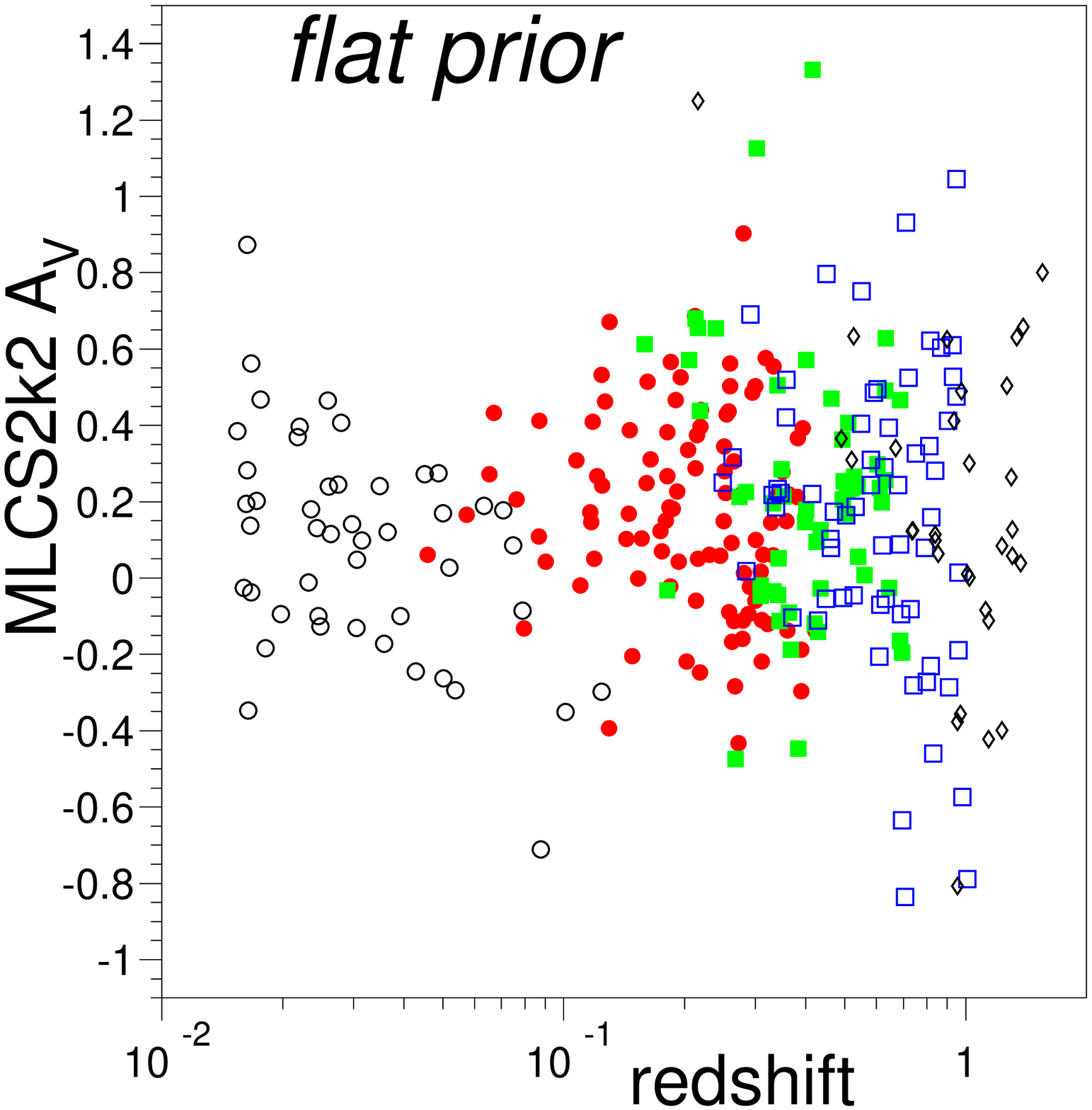}{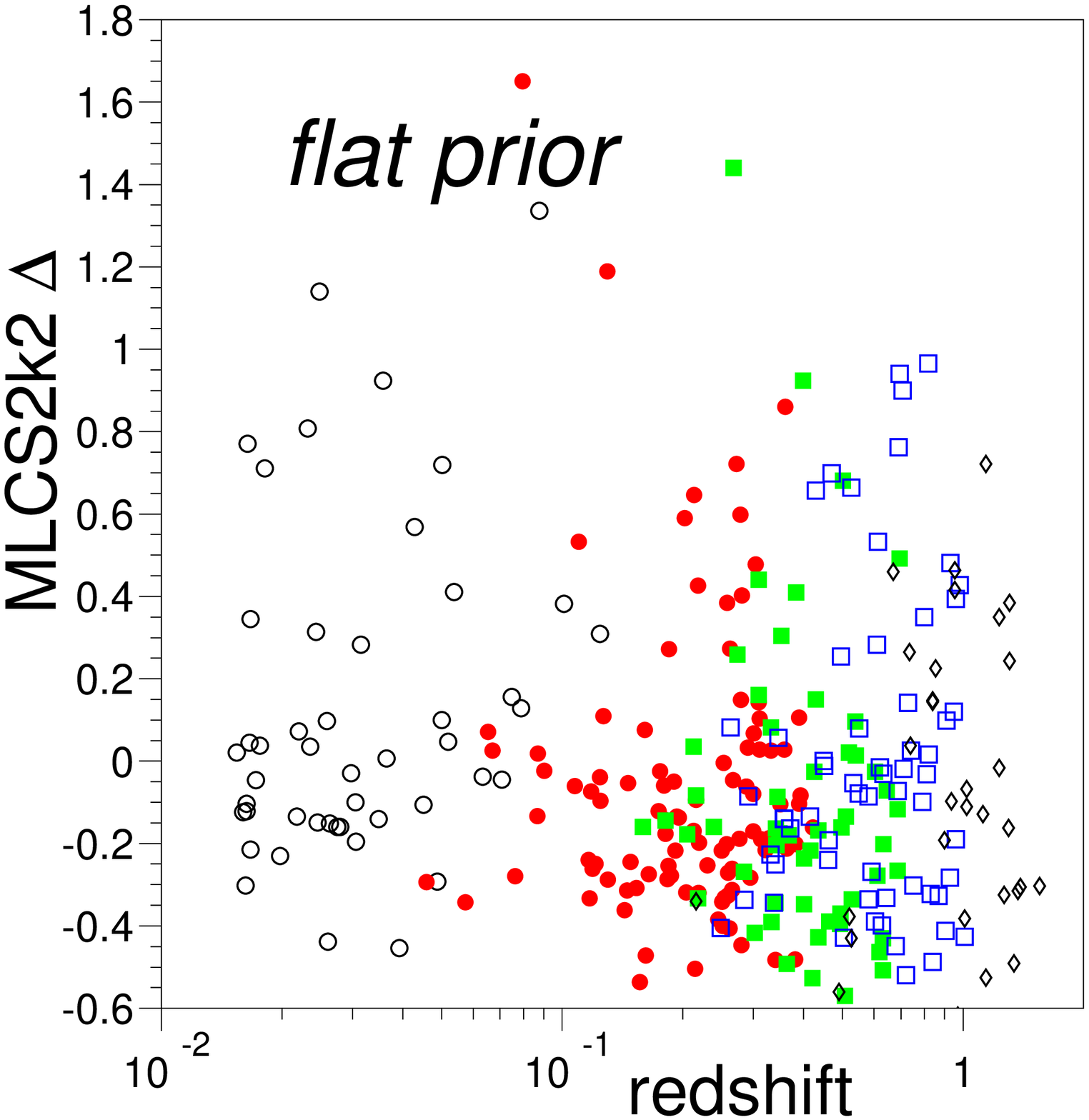}
  \caption{
	Left panels: {\mlcs} fitted dust extinction values $A_V$
	vs. redshift, for the different SN~Ia samples indicated
	on the plot.
	Right panels: fitted $\Delta$ vs. redshift.
	Upper panels are from fit with nominal prior;
	lower panels are from fit with flat prior.
      }   % end caption
  \label{fig:AVzDeltaz}
\end{figure}

Some example fits for \SDSS\ SN light curves using the modified 
version of \mlcs\ are shown above in Fig.~\ref{fig:mlcsfit_sdss}.
Figure \ref{fig:lcresid_MLCS} shows the average fractional residuals 
between the \mlcs\ model light curves and the data
for the SNe in each survey and for each rest-frame $UBVR$ passband.
The overall data-model agreement is good, except for
some late-time epochs and $U$-band.
The $U$-band residuals are discussed later in more detail
(\S~\ref{subsec:Uanom_MLCS}).
Fig.~\ref{fig:AVzDeltaz} shows the fit parameters $A_V$ and $\Delta$ 
vs. redshift for SNe in the different surveys.
The impact of the prior requiring $A_V>0$ is immediately
evident in the top-left panel. 
If we split each SN sample at its median redshift, the average $A_V$
for the lower-redshift SNe is larger than for the higher-redshift SNe;
this $A_V$-difference is 0.1~mag for the nearby sample,
and $\sim 0.05$~mag for the other SN samples. 
The prior discussed above accounts for this redshift-dependent shift.
The right panel in Fig.~\ref{fig:AVzDeltaz} 
shows the fitted $A_V$ vs. redshift using a flat prior,
$P(A_V) = P(\Delta) = 1$ in Eq.~\ref{eq:prior_master}.
Although there are many SNe with $A_V<0$, in 
\S~\ref{subsec:AV} we show that an underlying extinction distribution
with $A_V>0$, combined with measurement \uncs,
is consistent with the ``negative-$A_V$'' distribution
obtained from fitting with a flat prior.
Since $\Delta$ is well constrained by the light-curve fits, 
the $\Delta$ distribution with a flat prior is very similar
to that using the nominal prior.

We have checked the results of the modified \mlcs\ fitter
with the distance estimates derived by WV07 for the ESSENCE
and nearby SN samples. For this comparison, we use the WV07
extinction prior and {\eff}, as described
in their Eqs.~2 and 3.
The WV07 \eff\ function accounts for missing SNe at high redshift
and for the bias arising from 
using only measurements with SNR $>5$; for comparison, we therefore
use the same SNR cut.
The modified fitter is run in a mode that
replicates the original \mlcs\ fitter, with two exceptions:
First, as noted above, we fit in flux instead of magnitude. 
Second, the K-corrections use the average spectral template of 
\citet*{Hsiao07}, while WV07 used a library
of spectra and interpolated {\Kcor s} to the desired epoch.
To compare our ``nearby+ESSENCE'' analysis with WV07,
we fit \lcs\ for the $\NLOWZWV$ nearby SNe~Ia 
($0.015<z<0.1$) and $\NESSEWV$ ESSENCE SNe~Ia analyzed by WV07.
The rms scatter between our fitted distance moduli ($\mu$)
and those from WV07 is 0.03~mag and 0.05~mag
for the nearby and ESSENCE samples, respectively.
Our marginalized value for the dark energy equation of
state parameter $w$, using the SDSS BAO prior
(see \S~\ref{sec:wfit}), agrees to within 0.01 with the
result of WV07.

We stress that this comparison with WV07 is a consistency
check of our version of {\mlcs} relative to previous versions.
When we analyze the present SN samples with {\mlcs}
(\S~\ref{sec:results}), our different prior and
{\mlcs} model parameter values result in cosmological parameter
estimates that differ significantly from those of WV07,
as discussed in \S~\ref{subsec:WV07_compare}.

% ---------------------------------------------------------

\subsection{{\SALTII\ Fitting Method}}
\label{subsec:SALTII}
% TOP
% DESCRIPTION of SALT2 LIGHT CURVE FIT
%

The \SALTII\ light curve fitting method \citep{Guy07}
has been developed by the SNLS collaboration. The \SALTII\ 
model employs a two--dimensional surface in time and wavelength
that describes the temporal evolution of the 
rest--frame spectral energy distribution (SED) for SNe~Ia. 
The temporal resolution of the model  is 1~day, and the
wavelength resolution is 10~{\AA},
allowing accurate synthesis of model fluxes to compare 
with photometric data.
The model is created from a combination of 
photometric light curves and hundreds of SN~Ia spectra.
When there are measurement gaps in the spectral surface,
the unmeasured regions of the SED are determined
from interpolations of the measured regions.
The photometric data is mostly from the nearby sample (JRK07)
but also includes higher--redshift data ($z > 0.1$) to
better constrain the rest--frame ultraviolet behavior of the model.
For a complete list of SN light curves and spectra used
for training, see Table~2 in \citet*{Guy07}.

In \SALTII, the rest-frame flux at wavelength $\lambda$ 
and time $t$ ($t = 0$ at B--band maximum) is modeled by
\begin{eqnarray}
   {\dFrestdlam}(t,\lambda)  & =  &
   x_0 \times [M_0(t,\lambda) + x_1 \times M_1(t,\lambda)] 
   \nonumber \\
      &  \times  & \exp[c \times CL(\lambda)]~.
  \label{eq:SALTII_flux_rest} 
\end{eqnarray}
$M_0(t,\lambda)$, $M_1(t,\lambda)$, and $CL(\lambda)$ are determined
from the training process described in \cite{Guy07}. 
The $M_0$ surface represents the average spectral sequence, 
and is very similar to the sequence of average spectral templates 
\citep{Hsiao07} that we use for the \mlcs\ \Kcor s.
$M_1$ is the first moment of variability about this average, 
accounting for the well-known correlation of both  
peak brightness and color with light-curve shape,
and $x_1$ is the stretch parameter, the analog of the 
\mlcs\ $\Delta$ parameter.
$CL(\lambda)$ is the mean color correction term, and 
$c$ is a measure of SN~Ia color. Although the color 
variation is not explicitly attributed to dust extinction, 
in the optical region $CL(\lambda)$ is 
reasonably well approximated by the CCM89 extinction law 
with $R_V \sim 2$. In the $UV$ region, $CL(\lambda)$ 
exceeds the CCM89 extinction by about 0.07~mag.

The spectral-time surfaces are defined for rest-frame times 
$-20 < \Trest < +50$~days relative to the time of 
maximum brightness, and for rest-frame wavelengths that span 
2000 to 9200~{\AA}. 
We use the \SALTII\ spectral surfaces obtained 
from retraining the model using Bessell-filter shifts based
on HST standards, as discussed in Appendix~\ref{app:lowzfilters} 
(Table~\ref{tb:lamshift}),  
but otherwise using the same technique and data
as described in \cite{Guy07}.
The $UBVRI$ magnitudes for the primary reference Vega are taken from 
\citet{Fukugita_96}: these are 
0.02, 0.03, 0.03, 0.03, 0.024, respectively,
and are slightly different from those used 
to crosscheck the \mlcs\ method.
Although the wavelength coverage of the spectral surface
is rather broad, the \SALTII\ model includes only those 
observer-frame passbands for which 
$\SALTIILAMMIN < \LAMFBAR/(1+z) < \SALTIILAMMAX$~{\AA},
where $\LAMFBAR$ is the mean wavelength of the filter
and $z$ is the SN~Ia redshift.

To compare with photometric SN data, the observer-frame flux in 
passband $f$ is calculated as 
\begin{equation}
   F^f_{\rm obs}(t) = 
    (1+z)\int d\lamprime \left [ \lamprime
        {dF_{\rm rest}\over d\lamprime}(t,\lamprime) 
        T^f(\lamprime (1+z))   
     \right ],
  \label{eq:SALTII_flux_obs}    
\end{equation}
where $T^f(\lambda)$  defines the transmission 
curve 
of observer-frame passband $f$.
For the \SDSS, ESSENCE, SNLS, and HST samples, $T^f(\lambda)$ 
is provided by each survey. For the nearby sample,
$T^f(\lambda)$ is given by the \citet*{Bessell90} 
$UBVRI$ filter response curves, with wavelength shifts
as described in Appendix~\ref{app:lowzfilters}
and listed in Table~\ref{tb:lamshift}.

The model \unc\ accounts for the covariance between 
$M_0(t,\lambda)$ and $M_1(t,\lambda)$ 
at the same epoch and wavelength.
Although spectral covariances between different epochs and
wavelengths are not considered,
the model does account for covariances between integrated fluxes
at different epochs within the same filter.
Each SN~Ia light curve is fit separately using 
Eqs. \ref{eq:SALTII_flux_rest} and \ref{eq:SALTII_flux_obs} 
to determine the parameters $x_0$, $x_1$, and $c$. 
However, the \SALTII\ light-curve fit does not yield an 
independent distance modulus estimate for each SN. 
As discussed in \S~\ref{subsec:wfit_SALTII}, the 
distance moduli are determined 
as part of 
a global fit to an ensemble of SN light curves 
in which cosmological parameters 
and global SN properties 
are also determined. 
The \SALTII\ fits do not include informative priors on the fit 
parameters or the effects of selection efficiencies.  
We correct the \SALTII\ results for selection 
biases using a Monte Carlo simulation 
(see \S \ref{sec:sim} and \ref{subsec:syst_salt2}).

\begin{figure}[hb]
  \epsscale{1.10}
  \plotone{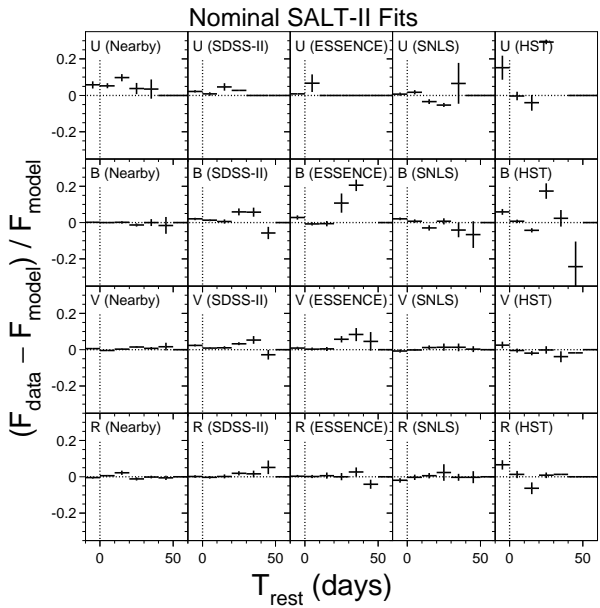}
  \caption{
        Data-model fractional residuals as a function of rest-frame
        epoch in 5-day bins, for \SALTII\ light-curve fits.
        The rest-frame passband and SN sample are indicated on each plot.
        Measurements with SNR$<6$ are excluded. 
	Note the discrepancy for the $U$-band residuals 
	in the nearby sample.
	For SNLS, the residuals are shown only for SN with $z<0.5$
     	as explained later in \S~\ref{subsec:Uanom_SALT2}.
        ${\rm F}_{\rm data}$ (${\rm F}_{\rm model}$)
        is the SN flux from the data (best-fit \SALTII\ model).
        Vertical dashed lines indicate epoch of peak brightness
        ($\Trest=0$); horizontal dashed lines indicate
        ${\rm F}_{\rm data} = {\rm F}_{\rm model}$.
        }
  \label{fig:lcresid_SALT2}
\end{figure}

\begin{figure}[ht]
\centering
  \epsscale{1.10}
\plottwo{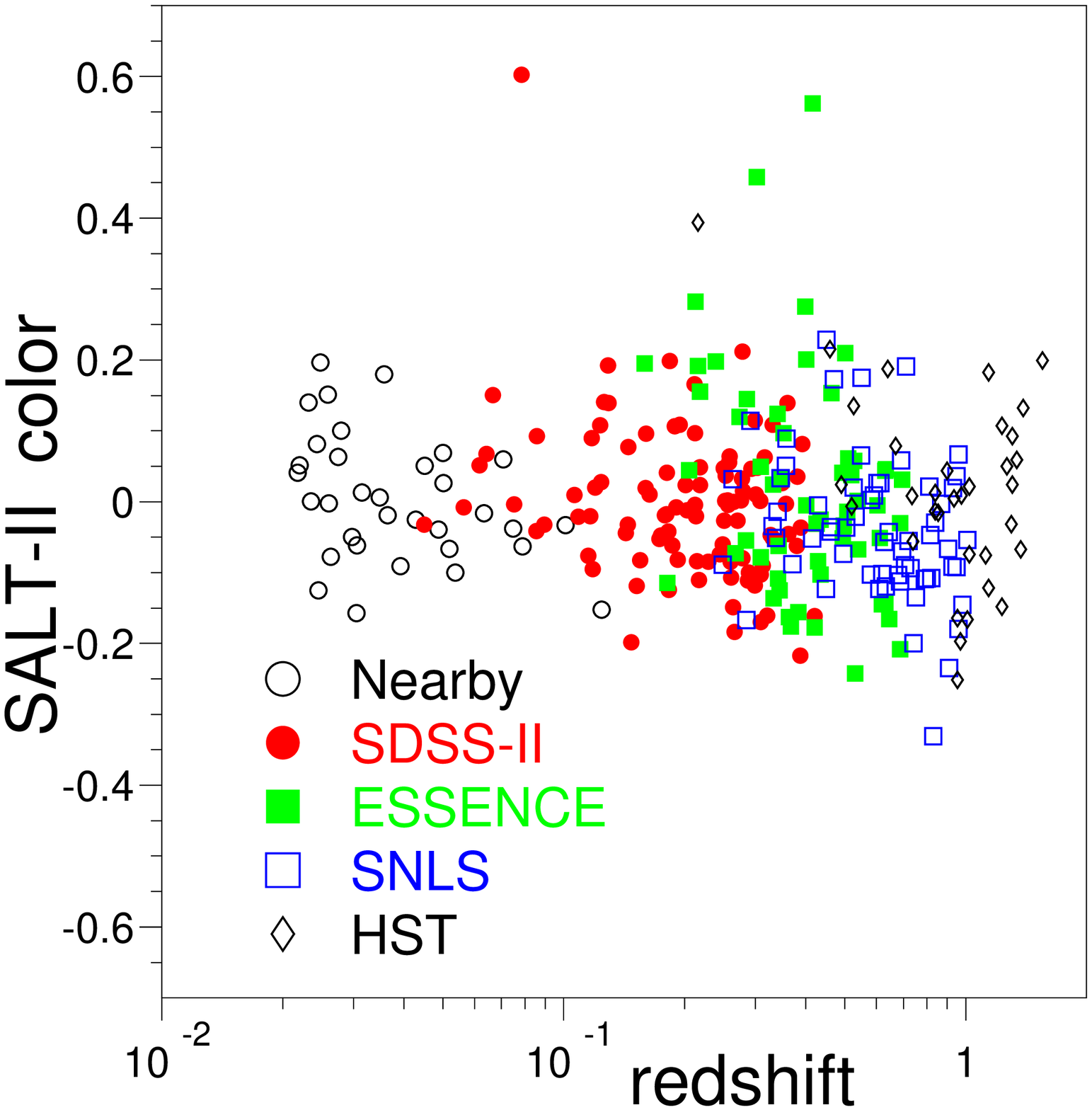}{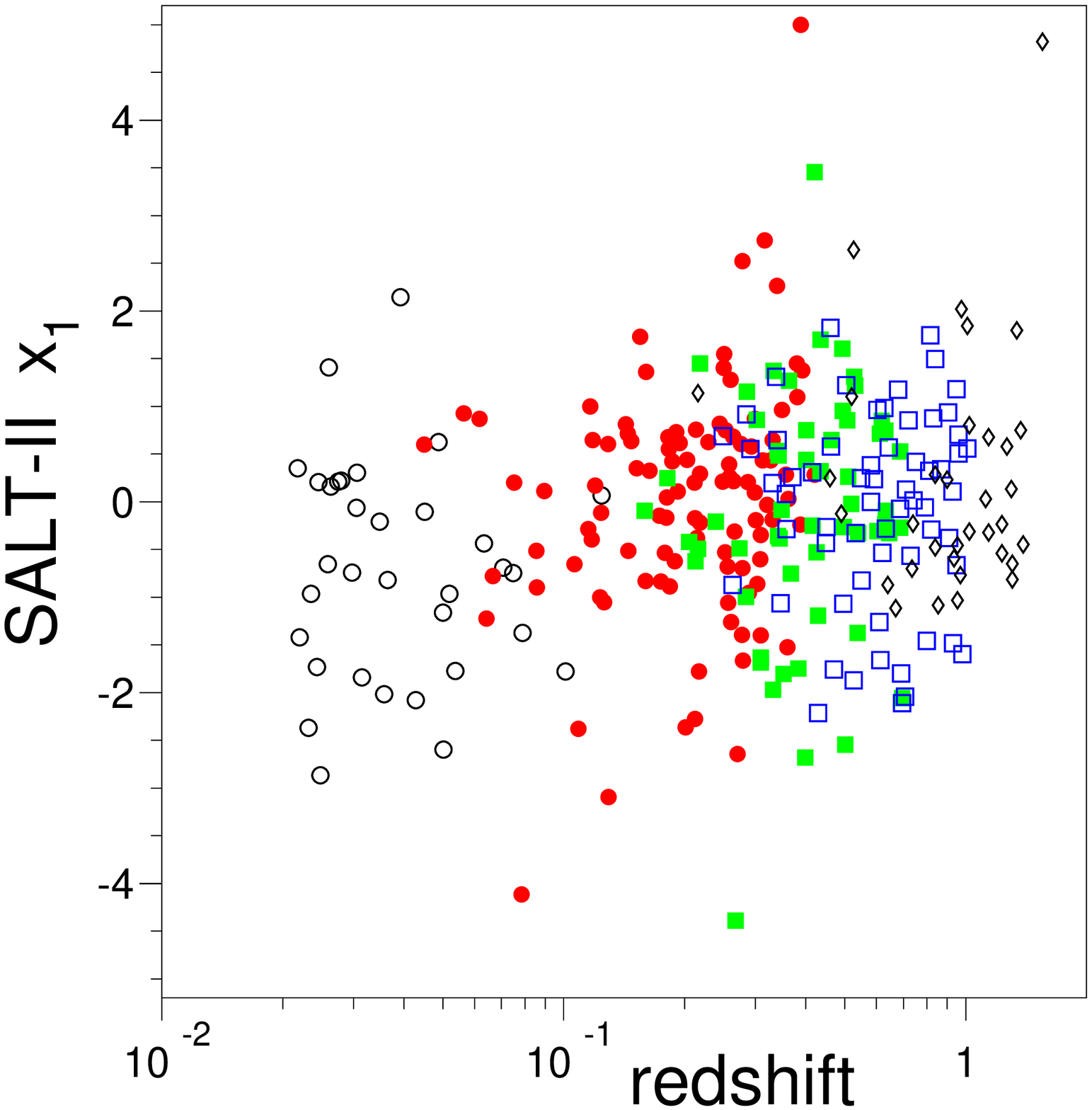}
  \caption{
 	Left panel: {\SALTII} fitted color values ($c$) 
	vs. redshift, for the SN~Ia samples indicated on the plot.
 	Right panel: fitted stretch parameter values, $x_1$, 
	vs. redshift. 
      }
  \label{fig:x1cvsz}
\end{figure}

In most cases, the \SALTII\ light-curve fits are qualitatively very 
similar to the \mlcs\ fits on a per-object basis.
The average rest-frame \lc\ residuals for the \SALTII\ fits are shown in 
Fig.~\ref{fig:lcresid_SALT2} for each survey 
for filters $UBVR$; note that the $U$-band residuals 
for the nearby SNe show some discrepancy, as will be discussed later.
Fig.~\ref{fig:x1cvsz} shows the fitted values for the 
color parameter $c$ and stretch parameter $x_1$ 
versus redshift for SNe in the different surveys.

As a crosscheck on our use of the public \SALTII\ code,
we have compared our fits of the 71 SNe~Ia from 
\citet{Astier06} to fits done
by the developer (J. Guy, private communication) and find good agreement.
The mean difference in the color ($c$) is
$0.003\pm 0.003$, with an rms dispersion of 0.02,
and the mean difference in the shape-luminosity parameter ($x_1$) is
$-0.014 \pm 0.021$, with an rms dispersion of 0.18.
The slight differences are attributable to the use of different
versions of the code and of the error (dispersion) map.

% ---------------------------------------------------------

% ---------------------------------------------------------

\subsection{Comparison of \mlcs\ and \SALTII\ Fitters}
\label{subsec:compare}
% ==================================================%
% End of light curve fitting section.
% Compare MLCS & SALT2 methods conceptually ... 
% no comparisons of cosmology results.
%
% ===============================================

We end this section by briefly comparing and contrasting the
\SALTII\ and \mlcs\ methods.
The \mlcs\ rest-frame model for the intrinsic SN brightness
is defined in discrete $UBVRI$ passbands
corresponding to the Landolt system. 
For each SN \lc, a composite SN spectrum is warped based on the 
model fit to the observed SN colors at each epoch,  and the warped spectrum is
used to perform the {\Kcor s} needed to transform the rest-frame 
model to the observer-frame fluxes.
The \SALTII\ model uses 
a composite SN spectrum that depends on both the epoch and 
intrinsic luminosity as well as an epoch-independent color term.
This spectrum is used to model the rest-frame fluxes.
Both the \SALTII\ and \mlcs\ models are trained using nearby 
SN~Ia data, but \SALTII\ training also includes higher--redshift data
that reduces the dependence on the nearby SN sample and 
provides better constraints on the 
rest--frame ultraviolet regions of the spectrum.

The \SALTII\ parameters $x_1$ and $c$ are analogous to the \mlcs\ 
parameters $\Delta$ and $A_V$. The parameters $x_1$ and $\Delta$ 
are essentially equivalent in describing the correlation between 
SN light-curve shape and brightness, but $c$ and $A_V$
have different meanings. 
\mlcs\ assumes that all intrinsic SN color variations are captured 
in the model by the light-curve shape-luminosity 
correlation and that any additional observed color 
variation is due to reddening by host-galaxy dust.
The color term ($c$) in \SALTII\
describes the excess color (red or blue) of a SN 
relative to that of a fiducial SN 
with fixed stretch parameter $x_1$. The excess color could be 
from host-galaxy extinction, from 
variations in SN color that are independent of $x_1$,
or from other effects, and \SALTII\ does not 
attempt
to separate these effects. 
\SALTII\ uses $c$ to reduce the scatter
in the Hubble diagram in a manner analogous to the
use of $x_1$.
The global \SALTII\ parameter $\beta$,  defined below
in \S~\ref{subsec:wfit_SALTII},
is the analog of the global \mlcs\ dust parameter $R_B = R_V+1$;
one expects $\beta \simeq R_B$ if 
excess color variation is purely due to host-galaxy extinction.
The \SALTII\ $\beta$ parameter is determined from the
global fit to the Hubble diagram for the entire SN~Ia sample under analysis;
we determine the \mlcs\ $R_V$ parameter by modeling the 
observed colors of a specific
subset of the SN data (\S~\ref{subsec:RV}).

Concerning correlations among model parameters,
\mlcs\ and \SALTII\ treat different aspects.
The \mlcs\ model includes covariances
between different epochs and passbands,
but we have excluded the off-diagonal covariances
as explained in \S~\ref{subsec:MLCS2k2}.
The \SALTII\ model includes covariances between
integrated fluxes at different epochs within the 
same passband, but covariances between passbands
are not considered.
\SALTII\ also includes 
the covariance between the spectral surfaces 
$M_0(t,\lambda)$ and $M_1(t,\lambda)$ 
at each epoch and wavelength bin
(Eq.~\ref{eq:SALTII_flux_rest}), 
but it does not include covariances between different 
epochs and passbands.

Within the \mlcs\ framework, each light-curve
fit yields an estimated distance modulus along with its 
estimated error, independent of cosmological assumptions. 
By contrast, in \SALTII\  the distance modulus estimate for 
a given  supernova is based on a global fit to 
the ensemble of
supernovae within a parametrized cosmological model 
(see \S~\ref{subsec:wfit_SALTII}). 
A result of this global minimization in \SALTII\ is that
a distance modulus bias in a particular redshift range,
such as could arise from including a poorly calibrated filter,
will induce a bias over the entire redshift range of the sample.
For the determination of cosmological parameters,
this tends to reduce the sensitivity 
to systematic problems and hence can lead to smaller
systematic \uncs. However, 
this reduced sensitivity
can also make biases more difficult to identify.
An explicit example of this is
described in \S~\ref{subsec:Uanom_SALT2}.

Fitting with \mlcs\ usually incorporates a Bayesian prior 
(Eqs.~\ref{eq:mlcs_chi2def}-\ref{eq:prior_master})
that reduces the scatter in the Hubble diagram by
incorporating information about the underlying $A_V$ 
distribution and the survey \effs.
The prior, and the resulting Hubble scatter, do not depend on
cosmological parameters. Because of the assumption that 
excess color variation is due to extinction by dust,
the prior excludes values of $A_V < 0$.  In {\mlcs}, 
SNe with very blue apparent colors (bluer than the template) 
are assigned $A_V \simeq 0$, and the data-model color discrepancy 
is attributed to fluctuations. In {\SALTII}, apparently blue SNe 
are assigned negative colors ($c<0$) 
that result in larger luminosities and distance moduli
compared to \mlcs.

Within the \SALTII\ framework, 
scatter in the Hubble diagram is explicitly minimized
by simultaneously adjusting global SN parameters 
along with the cosmological parameters;
this minimization is described in \S~\ref{subsec:wfit_SALTII}.
In contrast to \mlcs, the \SALTII\ Hubble scatter depends
on the cosmological parameters, and there is
no mechanism to account for the survey \eff\ directly in the fits.
To correct for biases related to the survey efficiencies
(\S~\ref{subsec:wfit_SALTII}),
we use the Monte Carlo simulations 
described in \S~\ref{sec:sim}.

The \mlcs\ and \SALTII\ \lc\ fit-residuals
can be visually compared in 
Figures~\ref{fig:lcresid_MLCS} and \ref{fig:lcresid_SALT2};
the data and models
are consistent for rest-frame passbands $BVR$,
but there are discrepancies for $U$-band in both cases.
We address this issue in more detail in 
\S \ref{subsec:Uanom_MLCS} and \ref{subsec:Uanom_SALT2}, and 
we compare the \mlcs\ and \SALTII\ results explicitly in 
\S~\ref{sec:results_compare}.

% ---------------------------------------------------------

% ##############################################################

\section{Monte Carlo Simulation: Determining the Selection Efficiency}
\label{sec:sim}
%
% TOP: describe simulation and spectroscopic efficiency
%
All surveys suffer from incompleteness and selection effects of 
various kinds. Supernovae that are intrinsically subluminous or 
highly extinguished by dust have less chance of being included 
in a flux-limited sample than more typical SNe. 
In addition, with limited \spec\ resources, 
higher priority may be given to SN
candidates with the best chances of yielding  
reliable identifications, e.g., by focusing on events that 
appear well separated from the host galaxy or for which the
host has either low surface brightness or early-type colors 
and morphology that suggest low dust content. These selection 
effects become more pronounced at the high-redshift end of a survey, 
where only the brightest, unextinguished
SNe will satisfy selection cuts. 
If SN~Ia brightness were a 
perfectly standardizable distance indicator, 
such selection effects would not be an issue for cosmological analysis. 
However, intrinsic variations in SN brightness, 
photometric errors, and uncertainties in estimating 
host-galaxy dust extinction lead to 
significant uncertainties and possible biases 
in distance estimates, particularly for SNe observed
with low signal-to-noise. 
In order to extract unbiased cosmological parameter estimates, 
biases must either be reduced to an acceptably 
small level by the analysis procedure or else 
a correction scheme must be  adopted.

We have developed detailed Monte Carlo simulations of the different 
SN surveys in order to determine the survey selection (or efficiency) 
functions and their impacts on SN distance estimates for both 
\mlcs\ and {\SALTII}.
The simulations also enable us to verify the estimates of 
systematic errors due to uncertainties in the light-curve 
model parameters.
The simulated \eff\ is a major component in the \mlcs\ 
fit prior discussed above in \S~\ref{subsec:MLCS2k2}. 
Determining the host-galaxy extinction dependence of the \eff\
is critical for the \mlcs\ method, because the extinction
is often poorly determined from the data.
For the \SALTII\ method, the simulation and efficiency play
no direct role in the fitting, but they enable us to 
estimate and correct for 
biases in the cosmological parameters
as desribed in \S~\ref{subsec:wfit_SALTII}.

Ideally, survey simulations would be based on artificial SNe~Ia embedded
into survey images, as was done during the \SDSS\ SN survey to monitor
the \eff\ of the search pipelines (see \S~\ref{sec:survey}). 
We do not have access to the images for the other surveys, 
and full image-level simulations would require a large amount of
computing to perform the many variations that are needed for the analysis. 
We have instead developed a fast light-curve 
simulation\footnote{The simulation, along with the
light curve fitters described in \S~\ref{sec:anal}, 
are publicly available in a software package called {\tt SNANA}: 
{\wwwSNANA}
} % end footnote
that is based upon actual survey conditions
and that therefore accounts for non-photometric conditions and
varying time intervals between observations due to bad weather.
At each survey epoch and sky location, the simulation uses 
the measured point spread function (PSF), zero point, CCD gain, 
and sky background to determine the noise and to convert
the simulated model magnitudes into CCD counts.
The simulation also incorporates a model for host-galaxy light 
and dust extinction.
We have obtained the necessary observational information 
for the \SDSS, ESSENCE, SNLS, and HST surveys to carry out these 
detailed simulations.
The nearby SN~Ia sample is a heterogeneous sample collected
over many years by different
observers and telescopes,
and we do not have the information 
needed to make detailed simulations of this sample.

Here we describe the simulation within the context of the
\mlcs\ light-curve model and comment on the differences
needed to simulate light curves in the \SALTII\ model. 
We select a random SN redshift from a power-law
distribution, $dN/dz \sim (1+z)^\beta$, with $\RATEPOWEREQ$, 
as determined by our recent analysis of the 
SN~Ia rate \citep*{Dilday08}.
A SN~Ia luminosity parameter $\Delta$ and host-galaxy 
extinction $A_V$ are selected from underlying distributions 
that we have inferred from our data (\S~\ref{subsec:AV}).
The \mlcs\ model is used to 
convert $\Delta$ into rest-frame $UBVRI$ magnitudes.
These generated SN~Ia magnitudes are increased according 
to the selected $A_V$ and the CCM89 extinction law 
using $R_V = \RV$, as determined in \S~\ref{subsec:RV}. 
The reddened 
$UBVRI$ magnitudes are K--corrected into 
observer-frame magnitudes.
A random sky coordinate is selected from the survey area,
and Milky Way extinction is applied based on the maps of  
\citet*{Schlegel_98}.
A random date for peak brightness is selected from 
the survey time frame, 
and all observed epochs at the selected sky coordinate
are identified from the actual survey observations. 
For each observation epoch, the measured survey zero point is used to convert
the simulated magnitude into a simulated flux.
For simulations based on the \SALTII\ model, the \mlcs\ parameters $\Delta$
and $A_V$ are simply replaced by the corresponding \SALTII\ 
parameters ($x_1,c$), drawn from empirical distributions.

The simulated noise for each epoch and filter includes 
Poisson fluctuations from the SN~Ia (signal) flux, sky background, 
CCD read noise, and host-galaxy background.
The signal noise is based on the number of CCD photoelectrons
calculated from the simulated flux.
The sky background is computed from the measured 
sky background per pixel, which is summed
over an effective aperture   
based on the measured PSF at that survey epoch and sky coordinate.
For SN redshifts $z_{\rm SN} < 0.4$, noise from the host galaxy is 
simulated by associating the SN with a host from the 
SDSS galaxy photometric redshift catalog \citep{Oyaizu_07},
randomly selected such that $z_{\rm gal} \sim z_{\rm SN}$.
From the SDSS DR5 \citep{SDSS_DR5} 
{\tt photoPrimary} database \citep{Stoughton_02},
we use the fitted exponential surface brightness profile in $r$-band
as a probability distribution from which the SN position within the 
galaxy is randomly selected, i.e., we assume that the SN~Ia 
rate within a galaxy is proportional to the local 
$r$-band luminosity. 
The host-galaxy background is computed by integrating the 
exponential galaxy model within the same effective aperture
that is used for the sky noise. The exponential profile is not 
appropriate for early-type galaxies, but this model is meant 
only as an estimate of the range of host-galaxy 
background light expected. The host-galaxy noise exceeds the 
sky noise 
for only $\sim 10$\% of the simulated SNe~Ia with $z_{SN} < 0.4$.  
For redshifts greater than 0.4, the lack of simulated 
host-galaxy noise is not significant, because the sky noise 
is dominant at these higher redshifts

There remain two important aspects of the simulation 
that are less well-defined and therefore more difficult to model:
(i) intrinsic variations in SN~Ia properties, beyond the shape-luminosity 
correlation, that lead to (so far) irreducible 
scatter in the Hubble diagram; and 
(ii) search-related inefficiencies beyond those due to photometric 
signal-to-noise and selection cuts, e.g., those associated with 
\spec\ selection.
Below, we describe our modeling of
these features in the simulation.

% --------------------------------------------
\subsection{Simulating Variations of Intrinsic SN Brightness}
\label{subec:hubblescat}
% --------------------------------------------

\newcommand{\PEAKMLCSSIGMA}{\sigma_{f'}^0}

Using the {\mlcs}-based simulation described above, 
the resulting scatter in the Hubble diagram for \SDSS\ SNe 
at $z<0.15$ is only 0.06 magnitudes, 
well below the observed scatter of $\sim 0.15$~mag. 
To make the model more realistic, we introduce {\it intrinsic} 
fluctuations in the simulated luminosity.
The models for intrinsic fluctuations 
described below are empirically determined
to match the observed Hubble scatter and are not based 
on a physical model.
% such as a variable viewing angle from
% asymmetric SN explosions \citep{Wang03}.

We have implemented two models of intrinsic SN variations.
The default method we use for {\mlcs},
called ``color-smearing,'' 
introduces an independent fluctuation in each passband,
and the fluctuation is the same for all epochs within each passband.
A random number $r_{f'}$ from a unit-variance Gaussian distribution 
is chosen for each rest-frame passband $f'$. 
A magnitude fluctuation,
$\delta m_{f'} = r_{f'} \PEAKMLCSSIGMA$,
is added to the generated magnitude at all epochs, 
where $\PEAKMLCSSIGMA$ is the magnitude uncertainty at peak brightness
given by the \mlcs\ model in passband $f'$.
In this method, the intrinsic model colors are randomly varied
by typically $\sim 0.1$~mag.
Since the simulated color variations are the same 
at all epochs, this model does not respect the Lira law 
\citep{Phillips_99}, 
the empirical observation that intrinsic SN~Ia colors 
have smaller variations at
epochs later than about 2 months after explosion;
this defficiency has a negligible impact on our analysis because
our requirements of good \lc\ coverage make our simulated 
\effs\ insensitive to the magnitudes at such late epochs.

The second model of intrinsic variation, which we use 
for the \SALTII\ method, and as a crosscheck for the \mlcs\ method,
is called ``coherent luminosity smearing:''
a coherent random magnitude shift, typically $\sim 0.15$~magnitudes,
is added to all epochs and passbands.
In the coherent smearing method, 
the intrinsic model colors are not varied.

A caveat in our implementation of intrinsic luminosity variations
is that the \mlcs\ simulation and fitter use different models of
intrinsic fluctuations and covariances. 
Although the \mlcs\ model includes a full covariance matrix,
we argued in \S~\ref{subsec:MLCS2k2} that these covariances 
do not accurately reflect intrinsic correlations. 
Since we use only the diagonal
elements of the \mlcs\ covariance matrix in the fitter, 
a literal translation for the simulation would be to implement a random 
intrinsic fluctuation of $\sim 0.1$~mag independently for each 
epoch and passband.
Since the observed smoothness of high-quality SN~Ia light curve 
data rules out such large intrinsic epoch-to-epoch fluctuations, 
we have chosen the above methods to simulate 
intrinsically smooth light curves. 
In future training of SN~Ia \lcs, it will be desirable 
to extract a model of intrinsic fluctuations and covariances
that can be used consistently in both the light-curve fitter and the simulation.

% --------------------------------------------
\subsection{Simulation of Survey Search Efficiency}
\label{subec:searcheff}
% --------------------------------------------

The final step in the simulation is to model losses related to
the SN search. In particular, we need to account for SNe~Ia that 
would have passed the light-curve selection cuts of \S~\ref{sec:sample}
had they been identified, but that were missed due to inefficiencies 
in the SN search.

Search-related losses come from the following sources:
(a) the image-differencing pipeline can fail to detect objects 
    with very low signal-to-noise as well as objects with nearby 
    artifacts such as a diffraction spike; 
(b) humans tasked with evaluating objects detected in subtracted 
    images may not correctly identify them as possible SN candidates; 
(c) software used in \spec\ targeting to fit light curves and  
    photometrically classify SN candidates identified by humans 
    may not correctly classify all SNe~Ia; and 
(d) due to limited resources for \spec\ 
    observations,
    not all photometrically identified SN~Ia 
    candidates will be targeted \specy\ or result in a
    spectrum with sufficient signal-to-noise to confirm 
    the SN type and determine its redshift. 

We define the overall survey efficiency, or survey selection function, 
as $\simeffsurvey=\simeffsearch \times \simeffcuts$, 
where the search efficiency is further decomposed as 
$\simeffsearch = \simeffpipe \times \simeffspec$. 
Here, $\simeffpipe$ describes the net search efficiency 
of the image-subtraction pipeline corresponding
to step (a) above.
The term $\simeffspec$ describes the combination of 
steps (b), (c), and (d), 
which depends in part on human judgment for each SN,
and $\simeffcuts$ is the factor 
associated with the final selection cuts described in \S~\ref{sec:sample}. 
These decompositions of \eff\ components are convenient because,
if sufficient information about the search is available, 
then $\simeffcuts$ and $\simeffpipe$ can be reliably simulated. 
By contrast, it is usually impossible to directly simulate 
$\simeffspec$ because it involves complex decision-making under 
varying circumstances by many people involved in \spec\
observations.
However, if the \spec\ \eff\ is nearly 100\% below some redshift 
for a given survey, then we can model $\simeffspec$ 
at higher redshifts by comparing observed 
distributions of SNe properties (including redshift)
to simulated distributions expected for a \specy\ complete survey. 
In summary, our philosophy is to make as detailed a model of the 
survey efficiency components as the data and survey information allow, 
and to use data-simulation comparisons to constrain the other components.

The different components of the \eff\ are likely to be
correlated, so that the overall \eff\ is not really a 
simple product as defined above. 
As an extreme example, if both the search and selection cuts 
required only that the maximum signal-to-noise ratio
was greater than 5, then $\simeffsearch = \simeffcuts$,
and the combined \eff\ would be the same (not the product of the two).
As discussed below for the \SDSS\ survey, 
we can simulate the combined effect of the 
image-subtraction \eff\ ($\simeffpipe$)
and selection cuts ($\simeffcuts$).
To simplify notation we write the combined \eff\ as a product,
$\simeffpipe \times \simeffcuts$, but it should be understood
that this product really refers to the combined \eff\ 
taking into account all correlations.
The $\simeffspec$ term is treated differently in that
it is defined as an independent \eff\ function that multiplies 
the other \eff\ terms.

% it is defined to be 
% $\simeffspec \equiv \simeffsurvey/(\simeffpipe \times \simeffcuts)$.

For the \SDSS, ESSENCE, SNLS, and HST samples, we have obtained detailed
observing conditions, and the simulation accurately
describes the effects of sample selection criteria, $\simeffcuts$.
For the \SDSS, we determine $\simeffpipe$ using 
information from the fake SNe~Ia that were inserted into
the images during the survey (see below). 
For the non-SDSS samples, we do not have access to the 
pixel-level data, so we set $\simeffpipe = 1$
and absorb the image-subtraction pipeline \eff\ into the \spec\ {\eff}, 
$\simeffspec = \simeffsearch$.
For the nearby sample, we do not even have the information needed
to determine the impact of the selection cuts; we therefore
absorb all sources of \ineff\ into the \spec\ \eff,
$\simeffspec = \simeffsurvey$.

As noted above, for the \SDSS, fake SNe~Ia are used to infer 
$\simeffpipe$ as a function of signal-to-noise ratio (SNR), as 
shown in Fig.~7 of \citet*{Dilday08}.   
These efficiency curves are used by the 
light-curve simulation to probabilistically determine which measurements 
in which passbands result in detections; 
a single-epoch detection in any two of the three ($gri$) passbands 
is considered to be an {\it object}.
An object found at two or more epochs results in a
supernova {\it candidate} and is considered to be discovered
by the image-subtraction pipeline.
For the \SDSS, the image-subtraction pipeline \eff\ ($\simeffpipe$)
is complete up to a redshift of $z\sim 0.2$ 
and then drops gradually to about 60\% at $z \simeq 0.4$.

For the HST simulation, we use the single-epoch 
search efficiency as a function of magnitude,
as described in \citet{Strolger_05} and \citet{Strolger_06}.
The search was done with the \HSTSEARCHFILT\ passband 
(mean wavelength: 9070~\AA); it is fully efficient
down to mag~$\sim 23$, and the efficiency drops to half
at mag~$\sim 26$. 
Incorporating this single-epoch efficiency profile into our simulation 
and requiring any one detection to discover a supernova, 
the simulated \eff\ for finding SNe~Ia is 100\% up to $z\sim 1$, 
and drops to about 65\% at $z=1.6$.

% ------------------------------------------------------
\subsubsection{Modeling Spectroscopic Efficiency}
\label{subsec:speceff}
% ------------------------------------------------------

\begin{figure}[h]
\centering
\epsscale{1.1}
\plottwo{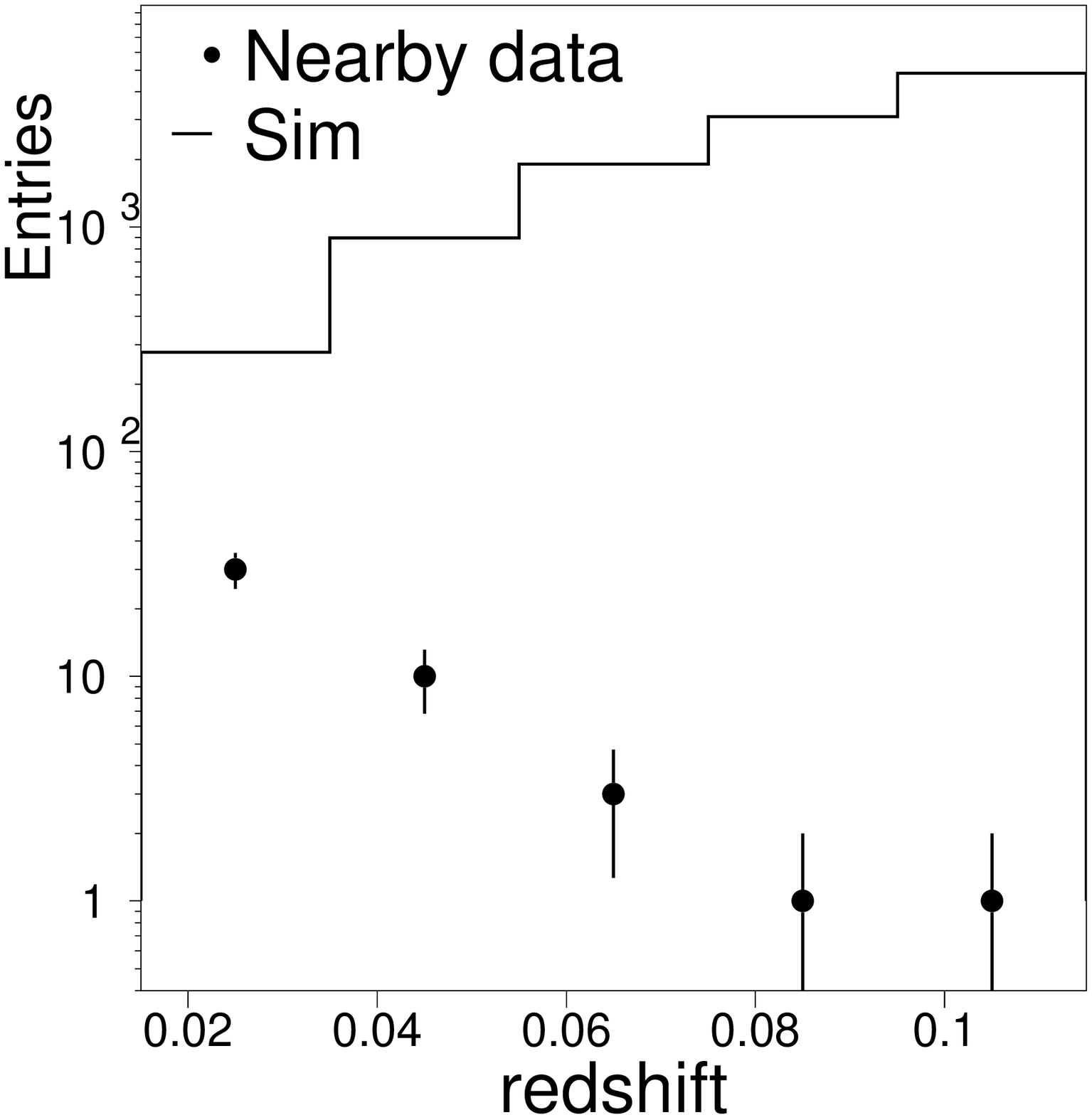}{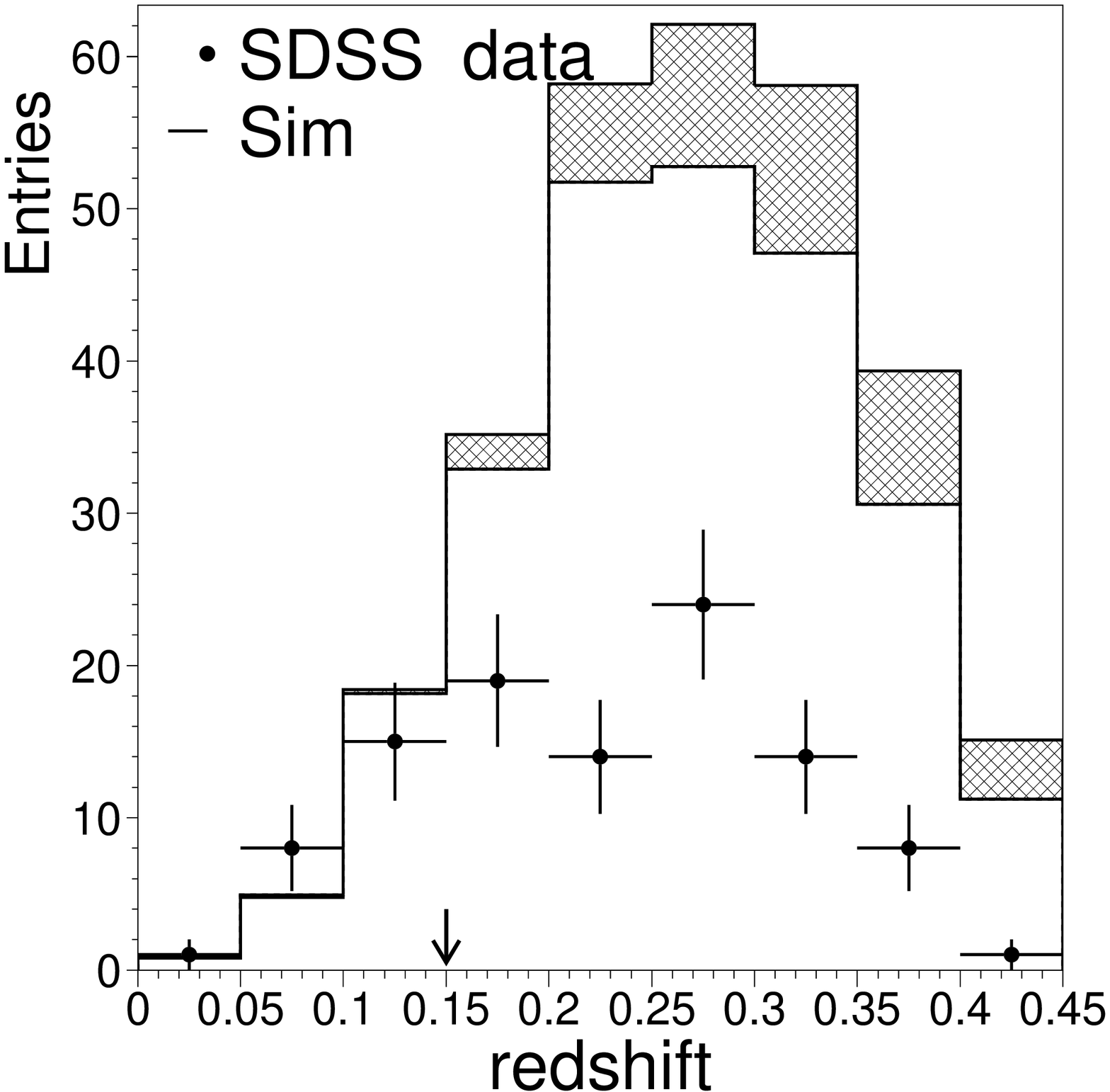}
\plottwo{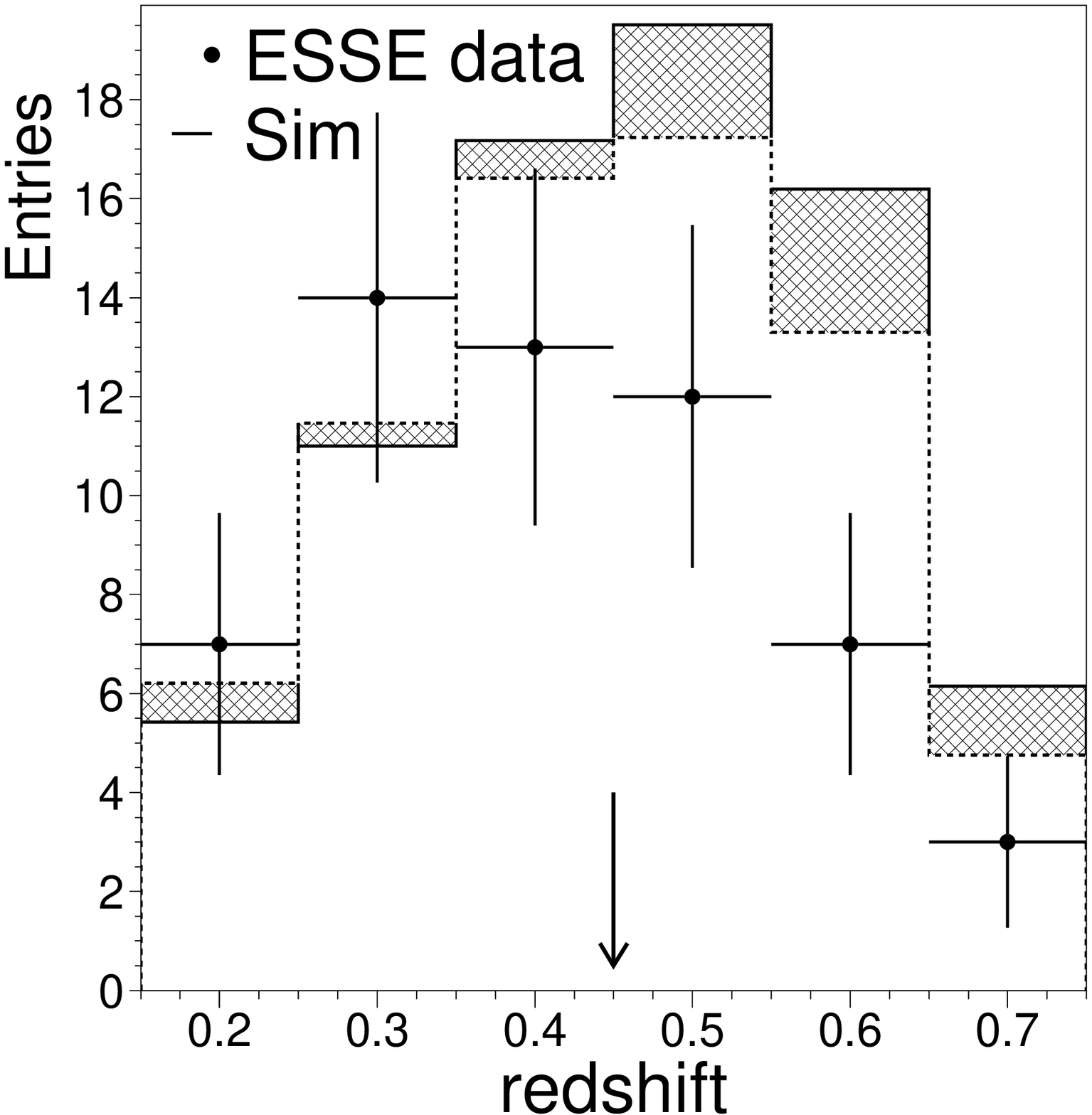}{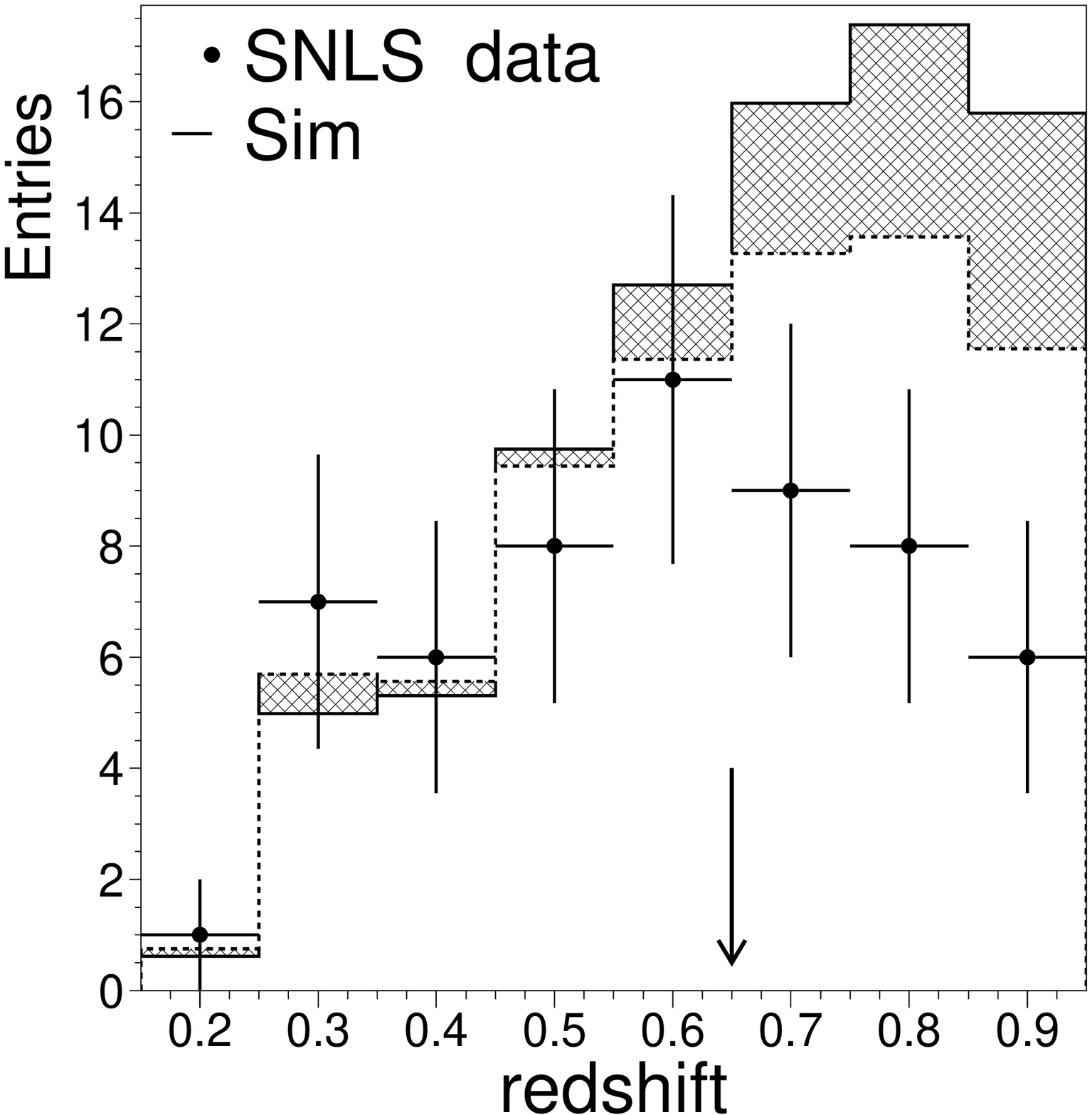}
  \caption{
	Comparison of redshift distributions for data (dots) and 
	simulations (histograms) for the 
	nearby \citep{Jha07}, \SDSS\ (2005 data only), 
	ESSENCE \citep{WV07}, and SNLS \citep{Astier06} samples.
 	The nearby sample is shown on a logarithmic scale, 
	while the other vertical scales are linear.
 	The simulations include losses from selection requirements
 	(\S~\ref{sec:sample}), and, for the \SDSS, losses from
	the image-subtraction pipeline.
     	The shaded regions show uncertainties in the simulated 
	distributions due to uncertainty in the SN~Ia rate vs.
 	redshift ($\RATEPOWEREQ$ for $dN/dz \sim (1+z)^\beta$).
 	Except for the nearby sample, the simulated distributions 
	are scaled such that the integrated number of simulated 
	SNe left of the cutoff redshifts 
	(indicated by vertical arrows) match the data. 
	Vertical error bars show the statistical \unc;
	horizontal bars show the bin-size. 
      }  % end caption
  \label{fig:ovzdatasim}
\end{figure}

Although selection effects associated with the photometry can be 
simulated based on available information, modeling 
the \spec\ \eff\ for each survey is more of a challenge.
The SN redshift distributions $N(z)$ for the non-HST data and simulations
are shown in Fig.~\ref{fig:ovzdatasim}.
With the exception of the nearby SN sample, 
the simulations include losses from selection requirements
(\S~\ref{sec:sample}); for \SDSS, losses from
the image-subtraction pipeline are also included.
For the nearby sample, the simulation is based on
\SDSS\ observing conditions and is chosen to be 100\% efficient, 
since we do not have the information to 
model the effects of photometric selection requirements in that case;
the simulated redshift distribution in narrow bins of 
redshift therefore scales as $N(z) \sim z^2$ (ignoring evolution 
of the SN~Ia rate at low redshift).

Below some cutoff redshift, 
$\zcut$, the \spec\ \eff\ is assumed to be $\sim 100$\%; 
for $z > \zcut$ the data-simulation discrepancy in 
$N(z)$ is taken to be due to \spec\ \ineff. 
For the \SDSS\ and SNLS samples, the cutoff redshifts are estimated 
to be 0.15 and 0.65, respectively, 
based on the levels of completeness given 
in their SN~Ia rate measurements \citep{Dilday08,Neill06}. 
For the ESSENCE sample, we estimate $\zcut \sim 0.45$ based 
on visual inspection of the redshift histogram in 
Fig.~\ref{fig:ovzdatasim}.
For the nearby ($z<0.1$) sample, there is no redshift below  
which the spectroscopic completeness is assumed to be 100\% (see 
discussion below). All 
four of these samples show significant \spec\ \ineff\
at the upper ends of their redshift ranges.

\begin{figure}[h]
\epsscale{1.1}
\centering
\plotone{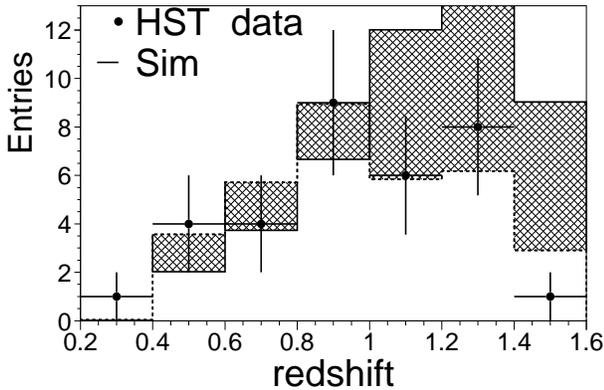}
  \caption{
	Redshift distribution for HST sample \citep{Riess_06};
	data are shown by dots and the simulation by a histogram.
	The shaded region shows the \unc\ from the redshift-dependence 
	of the SN~Ia rate (see Fig.~\ref{fig:ovzdatasim} caption), 
	but with the rate uncertainty doubled for $z>1$.
      }
  \label{fig:ovzdatasim_HST}
\end{figure}

The data-simulation redshift comparison for the HST sample
is shown in Fig.~\ref{fig:ovzdatasim_HST}. 
Although the uncertainty in the SN~Ia rate at these 
high redshifts is large, there is no evidence for significant 
\spec\ \ineff: 
the data-simulation redshift comparison
is consistent with the claim that there are no significant
losses for $z<1.4$ \citep{Riess_06}.
We therefore assume that $\simeffspec = 1$ for the HST survey, 
and this sample is not included in the efficiency discussion below.

The \spec\ \eff\  as a function of redshift is defined to be 
$\simeffspec \equiv \simeffsurvey/(\simeffpipe\times \simeffcuts)$,
which corresponds to the ratios of the data and simulation 
histograms in Fig.~\ref{fig:ovzdatasim}. For each sample, 
this \eff\ can be fitted by an exponential function,
\begin{eqnarray}
   \simeffspec(z)  & = & 
     \effzzero \exp[-(z-\zcut)/\effzone] 
     ~~~~~{\rm for~}z > \zcut  
          \label{eq:effz} \\
    \simeffspec(z) & = & 1    ~~~~~{\rm for~}z < \zcut ~, \nonumber
\end{eqnarray}
where $\effzzero$ and $\effzone$ are 
determined from the fit.
The discontinuity at $\zcut$ is motivated by the {\SDSS} survey, 
for which our \spec\ observation strategy targeted $z<0.15$ 
candidates with very high priority. 
We have no evidence that the other surveys should have 
discontinuities in their redshift distributions,
but the model above provides a reasonably accurate representation for 
the redshift distributions of the ESSENCE \& SNLS samples as well. 
This functional form
is adopted for computational convenience in generating 
large Monte Carlo samples.

Although the redshift dependence of the \spec\ \eff\
has now been estimated, we know that $\simeffspec(z)$ depends in 
reality on a variety of factors, not simply redshift, and we must 
model its dependence on those factors in order to properly model 
the selection function and associated biases. As noted above, since 
spectroscopic targeting is a complex process, 
the underlying mechanism determining $\simeffspec$ 
is not easily characterized. 
The simplest possibility would be that $\simeffspec$
depends purely on redshift: in this case, 
$\simeffzspec \equiv \simeffspec(z)$ 
would be given by Eq.~\ref{eq:effz},
and there would be no impact of \spec\ \eff\ 
on the survey bias, since redshift is precisely 
measured for each SN in the samples we consider. However, this 
model is not {\it a priori} likely: it is more probable that 
$\simeffspec$ depends explicitly on both redshift and apparent  
SN~Ia brightness, the well-known Malmquist bias. 

To consider this alternative, 
we define a  ``magnitude dimming'' parameter $\magdim$ 
as the difference between the simulated rest-frame magnitude 
and the magnitude of the brightest possible SN~Ia at peak light in 
rest-frame $V$ band (for {\mlcs}) or in rest-frame $B$ band 
(for {\SALTII}).
For the \mlcs\ and \SALTII\ models, $\magdim$ is given by
\begin{eqnarray}
   \magdim({\rm MLCS2k2}) & = &  A_V +    
       \Theta [ p^{0,V}(\Delta - \Delta_{\rm ref})  \nonumber \\
            &  + &  
                 q^{0,V}(\Delta^2 - \Delta^2_{\rm ref}) ]  
     \label{eq:magdim_mlcs} \\
   \magdim({\rm SALT\!-\!II}) & = &
       \beta(c - c_{\rm ref}) - \alpha(x_1 - x_{1,{\rm ref}}).
     \label{eq:magdim_SALTII}    
\end{eqnarray}
For \mlcs, $A_V$ is the host-galaxy extinction in $V$-band,
$p^{0,V}$ and $q^{0,V}$ are model parameters at the epoch 
of peak brightness in $V$-band (see Eq.~\ref{eq:MLCS2k2model}), 
and $\Delta_{\rm ref} = -0.3$ corresponds to nearly the most 
intrinsically luminous SN~Ia in the training sample.
For the \SDSS, ESSENCE, and SNLS samples, $\Theta=1$;
for the nearby sample, we set $\Theta=0$, because the 
$\Delta$ distribution for the nearby sample appears unbiased 
relative to that of the underlying SN~Ia population (see 
\S \ref{subsec:AV})
while its $A_V$ distribution is clearly biased.
For \SALTII, $x_1$ and $c$ describe the light-curve shape and color
for each SN~Ia (see \S~\ref{subsec:SALTII}), 
$x_{1,{\rm ref}}=2.6$ and $c_{\rm ref} = -0.26$ correspond to the
brightest SN~Ia, and $\alpha,\beta$ are global SN~Ia parameters 
determined in the cosmology fit (see \S~\ref{subsec:wfit_SALTII}).

Using the magnitude dimming parameter, 
we model the \spec\ inefficiency due to apparent magnitude as 
\begin{equation}
   \simeffmagdim = \exp[-\magdim/m(z)]~,
   \label{eq:simeffmagdim}
\end{equation}
where $m(z)$ is an exponential slope function
determined by numerically solving
\begin{eqnarray}
    & & \int d\magdim ~ N_{\rm SIM}(z,\magdim) \exp[-\magdim/m(z)] 
   \nonumber \\
     & = & N_{\rm DATA}(z)~.
   \label{eq:magdim_slope}
\end{eqnarray}
Here, $N_{\rm SIM}(z,\magdim)$ is the number of simulated events
in a two-dimensional bin of redshift and simulated $\magdim$,
and $N_{\rm DATA}(z)$ is the number of data events in 
the redshift bin. 
%The integral in each redshift bin is 
%evaluated as a sum over simulated $\magdim$ bins.
In practice,  $m(z)$ is evaluated in discrete redshift bins
for $z > \zcut$ and fit to an exponential function
of redshift,
\begin{equation}
   m(z)  =  \effdimzero \exp[-(z-\zcut)/\effdimone] + \effdimtwo  
   \label{eq:effdim} 
\end{equation}
where $\effdimzero$, $\effdimone$, and $\effdimtwo$ are 
determined in the fit, and  $m(z) = \infty$ for $z < \zcut$.
% also given in Table~\ref{tb:effz}. 
Since $\magdim$ is defined for $V$-band in the \mlcs\ model 
and for $B$-band in the \SALTII\ model, the associated 
$\effdimzero$, $\effdimone$ parameters are different. 
The parameter $\effdimtwo$ is non-zero only for the nearby sample.
Qualitatively, Eq. \ref{eq:simeffmagdim} says that SNe 
that are intrinsically faint (large positive $\Delta$) or 
extinguished 
(large $A_V$) will be under-represented 
at the high-redshift end of a sample.

\begin{figure}[h]
\epsscale{1.2}
\centering
\plotone{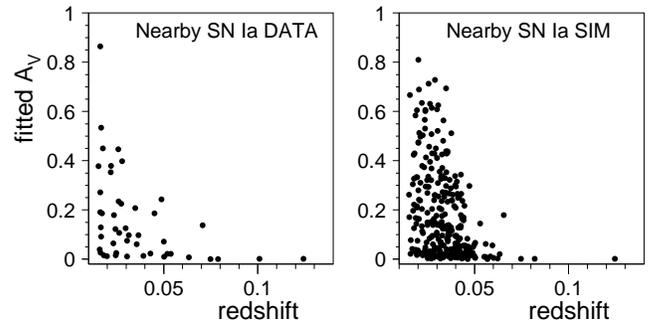}
  \caption{
  	Left panel: \mlcs\ fitted extinction $A_V$ versus redshift 
	for the nearby SN~Ia sample \citep{Jha07}. 
	Right panel: fitted $A_V$ distribution for the simulation 
	using the parametrized \eff\ function of 
	Eqs. \ref{eq:simeffmagdim} and \ref{eq:simspec}
	with $\AEFFDIM = 1$.
	Light-curve fits were made using 
	the \mlcs\ prior of Eq. \ref{eq:prior_master} and the 
	exponential $A_V$ distribution of \S \ref{subsec:RV}.
      }
  \label{fig:nearbyavz}
\end{figure}

In principle, we now have two models for the 
\spec\ \eff: one, denoted $\simeffzspec$, and explicitly given 
in Eq. \ref{eq:effz}, depends solely on redshift;
the other depends explicitly on apparent brightness and implicitly on 
redshift and is also constrained to match Eq. \ref{eq:effz}.
While the latter model corresponds to expectations for Malmquist bias,
the ``redshift-only'' model $\simeffzspec$ may be better suited to modeling \spec\ 
selection that assigns lower priority to SN candidates 
near the cores of host galaxies, since SN-host angular separation tends to 
decrease with redshift. 
Since both models satisfy Eq.~\ref{eq:effz}, these two \eff\ models
can be linearly combined into a more general model:
\begin{equation}
  \simeffspec(z,\magdim) = 
       (1-\AEFFDIM)\simeffzspec + \AEFFDIM\simeffmagdim ~.
  \label{eq:simspec}
\end{equation}  
To break this model degeneracy and determine the coefficient $\AEFFDIM$,  
we compare the mean fitted extinction $\bar{A}_V(z)$ versus redshift 
for the data and the simulation after applying all efficiency factors, 
and we minimize the $\chi^2$ for the data-simulation difference. 
%% The results are shown in Table~\ref{tb:effz}. 
For the SDSS-II, $\AEFFDIM = 0.8\pm 0.2$, indicating
that most of the search \ineff\ is related to SN~Ia apparent brightness.
For the ESSENCE survey, $\AEFFDIM = 0\pm 0.2$, indicating
that most of the search \ineff\ is simply a function of redshift.
For the SNLS sample, the best-fit $\AEFFDIM = 1$;
however, since the data-simulation $\chi^2$ for $\AEFFDIM=0$ 
is only 0.3 larger than the minimum $\chi^2$ at $\AEFFDIM=1$, 
there is essentially no information on $\AEFFDIM$. 
The large \unc\ on $\AEFFDIM$ is in part due to the
relatively small spectroscopic \ineff\ for the SNLS.
For the nearby sample, there is no redshift range
for which the sample is 100\% efficient, and we therefore
need an additional constraint to determine the
\eff\ parametrization.
Noting that the mean fitted extinction drops rapidly with redshift 
for the nearby sample (see Fig~\ref{fig:nearbyavz}),
we assume that $\AEFFDIM=1$. 
Fig.~\ref{fig:nearbyavz} shows that the simulated \eff\ 
works well in reproducing the strong extinction gradient
in the nearby sample.

\begin{figure}[h]
  \epsscale{1.00}
  \plotone{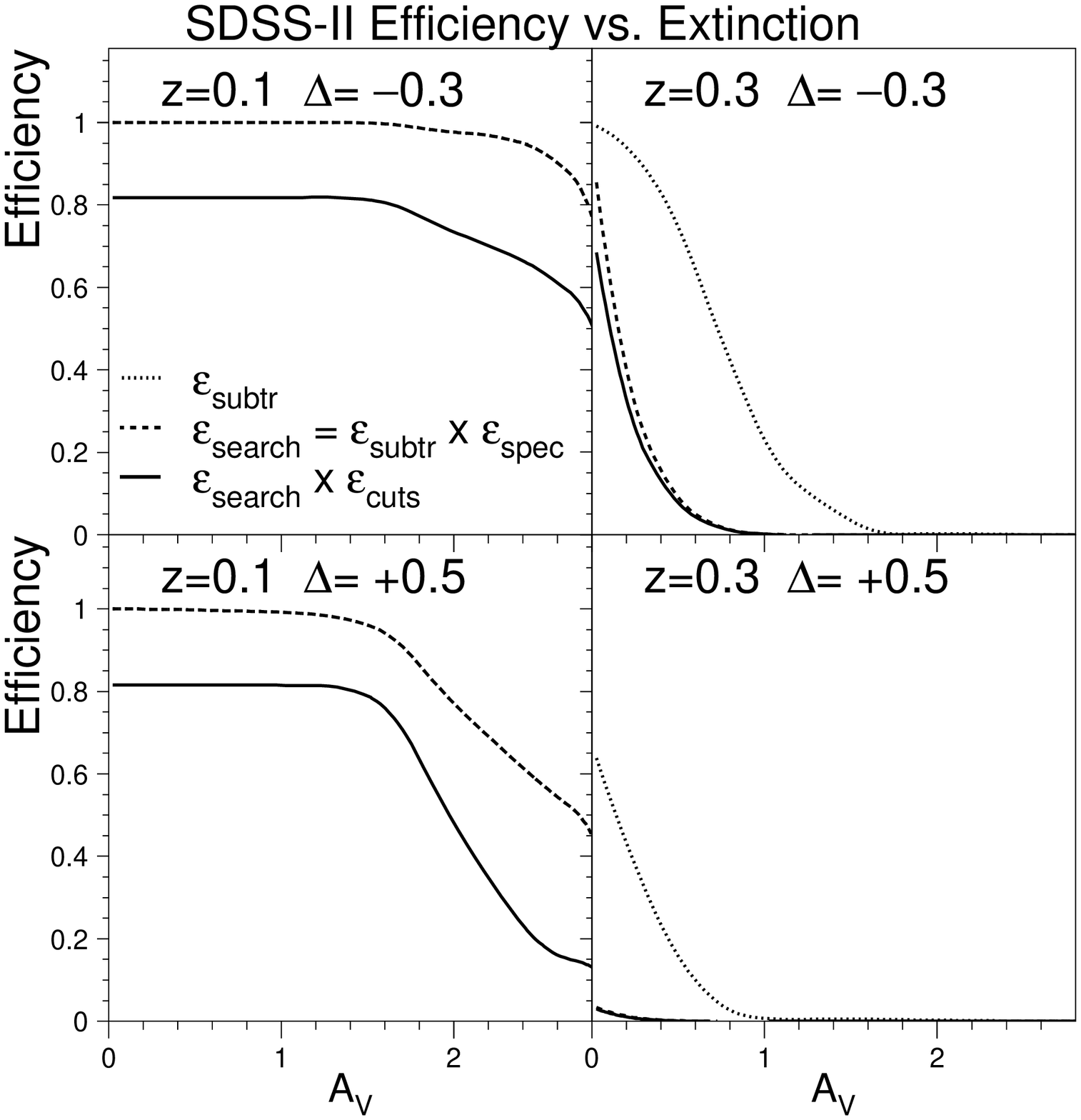}
  \caption{
	Estimated efficiency versus extinction for the SDSS-II SN sample.
  	Each panel corresponds to a different combination of
  	redshift ($z = 0.1,~0.3$) and intrinsic SN~Ia brightness 
  	($\Delta = -0.3,+0.5 \to$ bright, faint).
  	The curves correspond to different stages of the
    	efficiency:  
    	after the image-subtraction pipeline, $\simeffpipe$ (dotted);
    	after spectroscopic confirmation,
   	$\simeffsearch = \simeffpipe\times\simeffspec$ (dashed); 
    	and after selection cuts, 
   	$\simeffsurvey=\simeffsearch\times \simeffcuts$ (solid).      
  	For $z=0.1$, the spectroscopic efficiency is 100\%, so the 
  	search (dashed) and image-subtraction (dotted) curves are the same.
  	In all cases, $\simeffcuts \sim 0.8$
 	at low extinction; this loss is mainly due
  	to the requirement of good light-curve coverage, i.e., SNe that 
  	peak very early or very late in the observing season do not have 
  	adequately sampled light curves to satisfy the selection criteria.
     }
  \label{fig:prior_eff}
\end{figure}

\begin{figure}[h]
\centering
\epsscale{.36}
\plotone{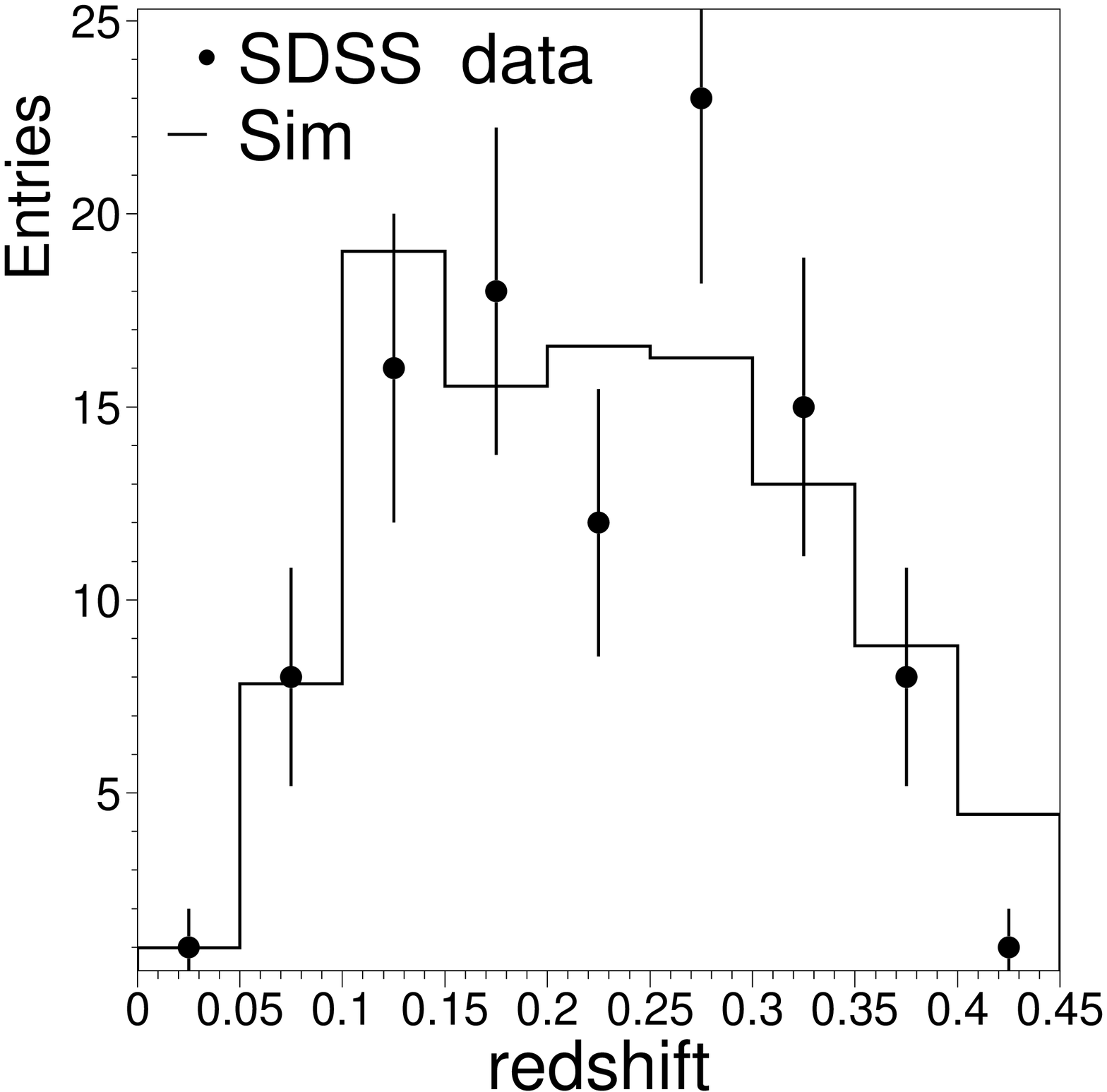}
\plotone{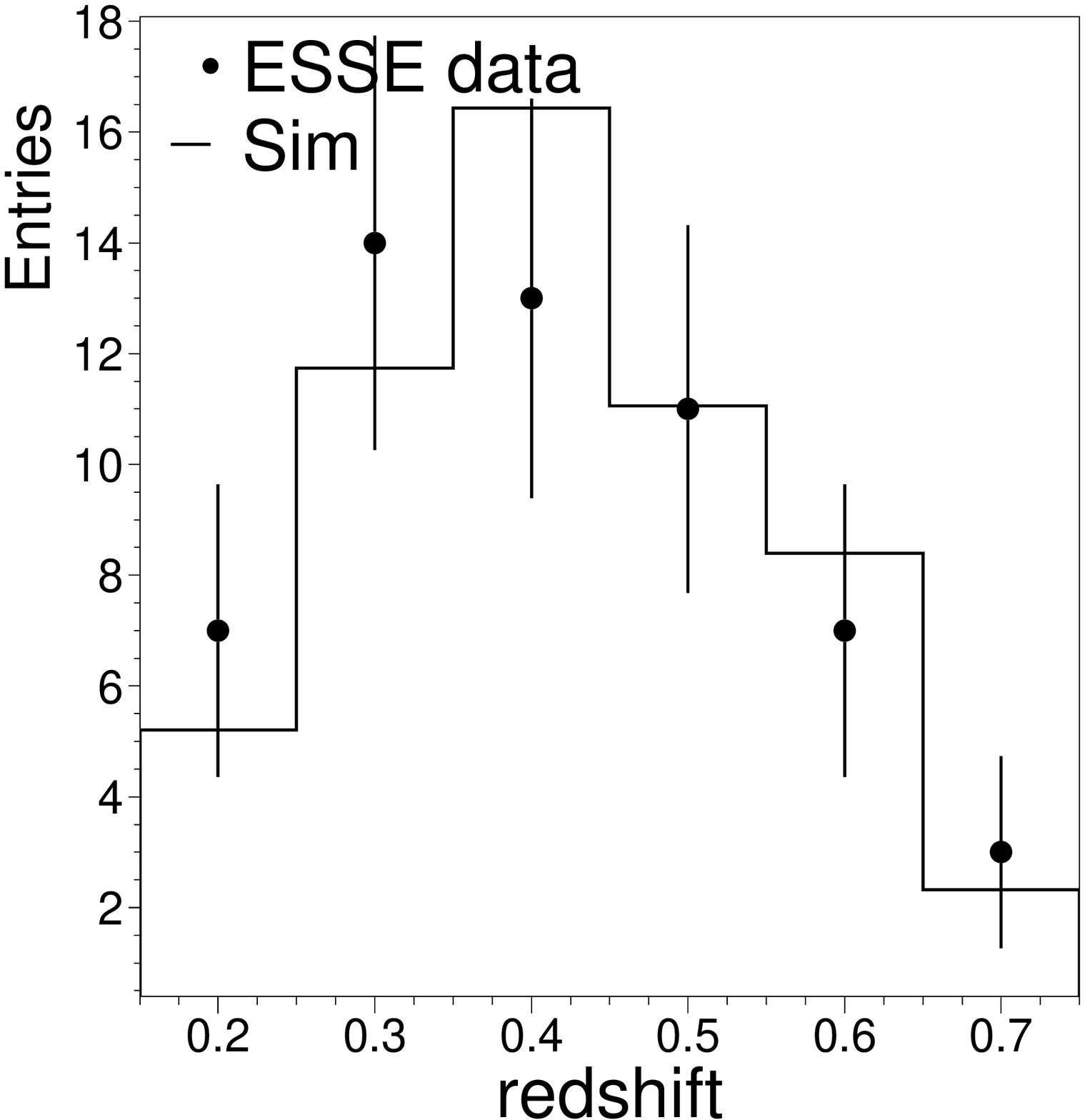}
\plotone{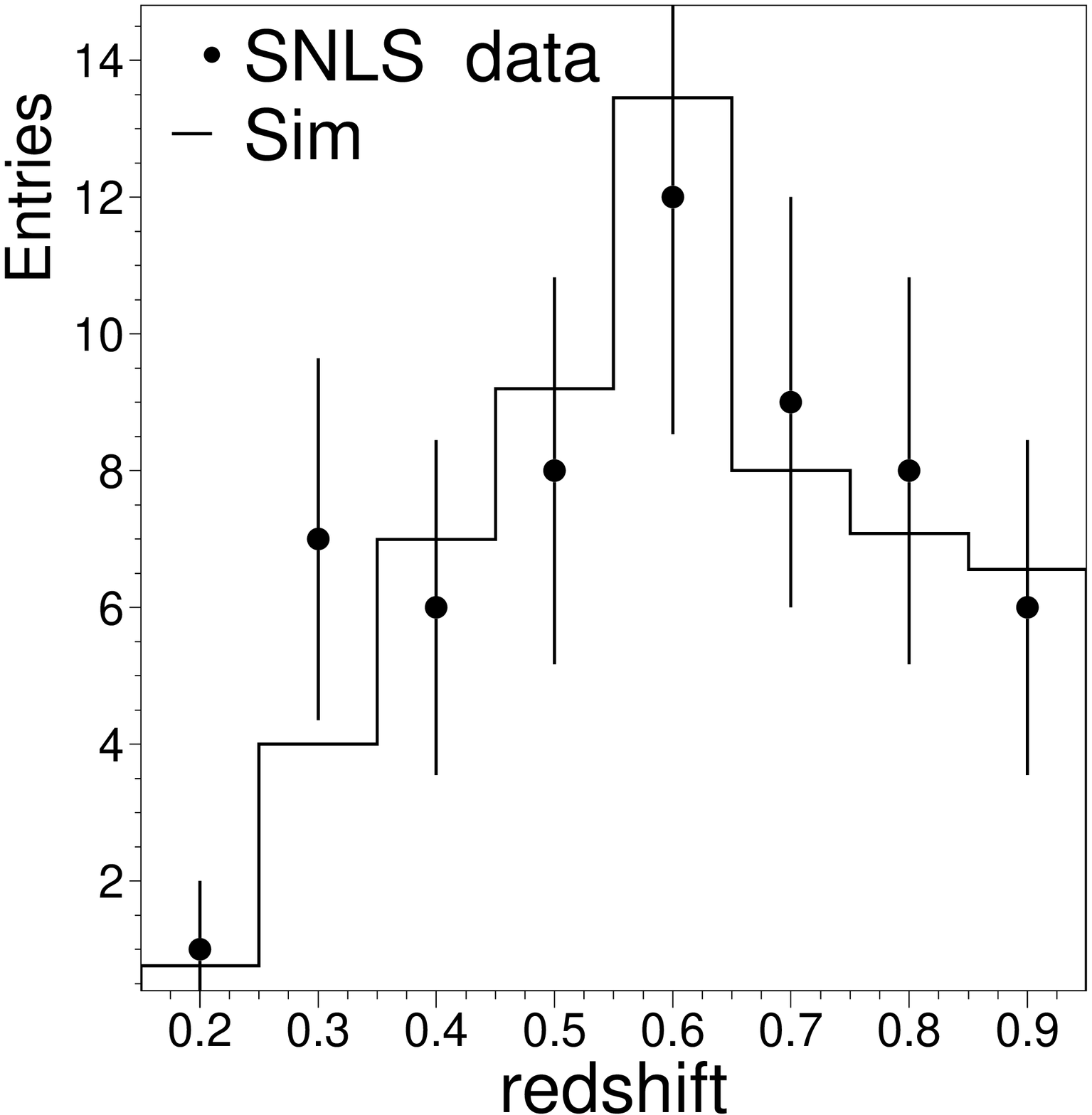}
  \caption{
	Comparison of redshift distributions for data (dots) and 
 	simulations (histograms), 
	for the \SDSS, ESSENCE, and SNLS samples 
	after applying the selection requirements 
   	in \S~\ref{sec:sample}.
     	The simulations include all known effects, 
	{\it including} the \spec\ \ineff.
     	Each simulated distribution is scaled such that the 
	total number of SNe matches the data. 
  	The SDSS data-simulation discrepancy for $z > 0.4$ is 
	an artifact of our simple modeling of
	$\simeffspec$ (Eq.~\ref{eq:effdim}).	
      }
  \label{fig:ovzdatasim2}
\end{figure}

\begin{figure*}  % [h]
\centering
\epsscale{.28}
\plotone{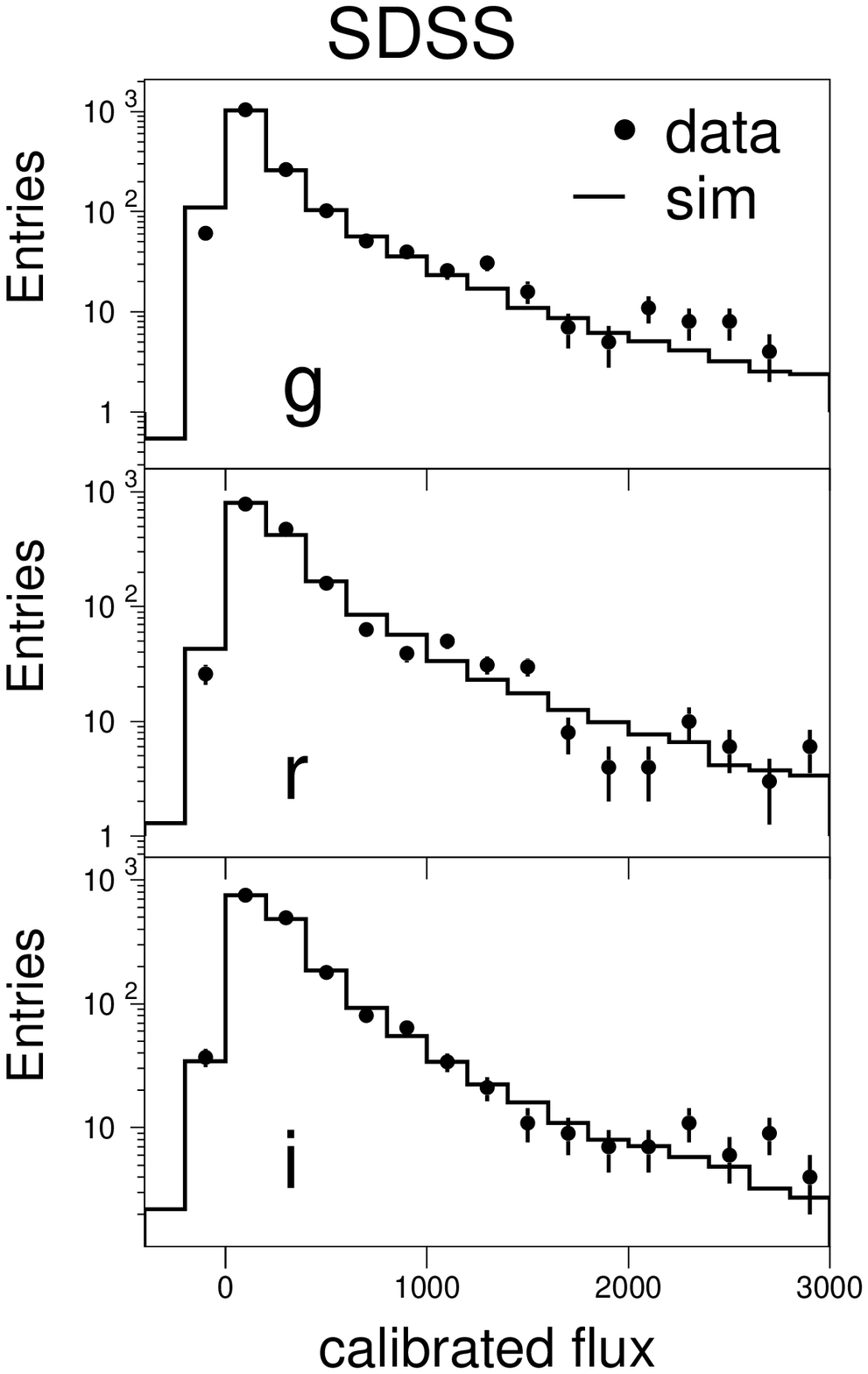}
\plotone{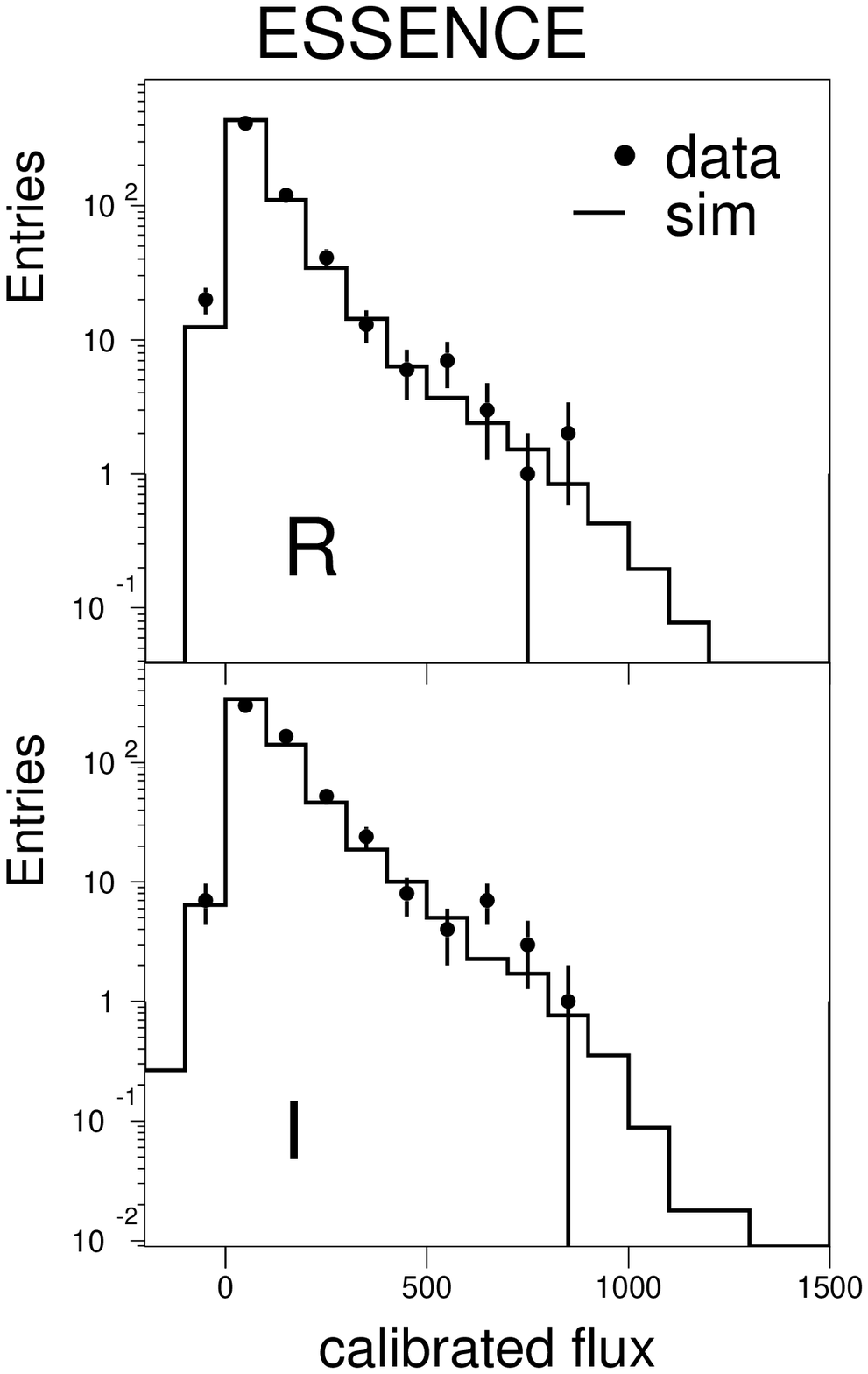}
\plotone{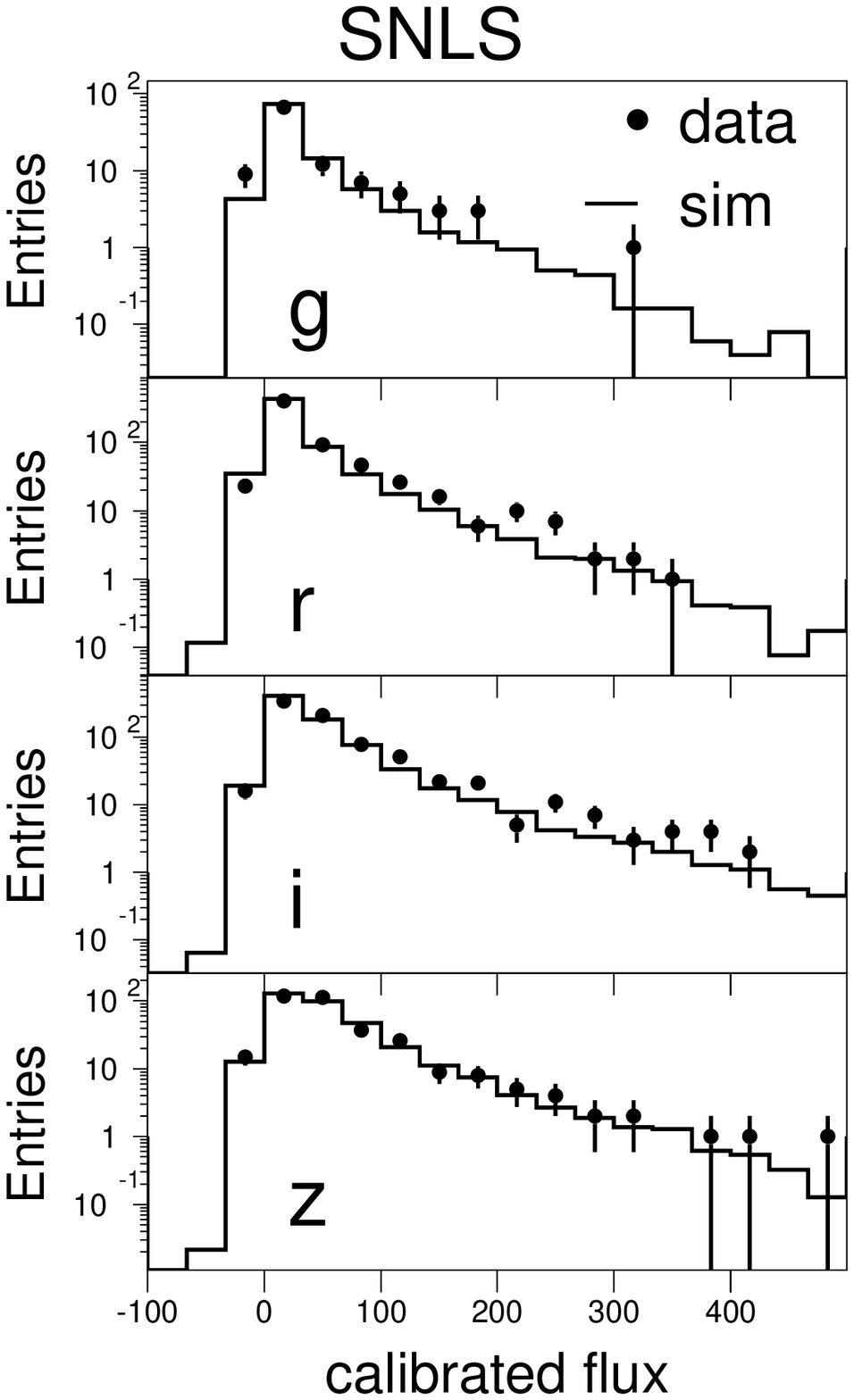}
\plotone{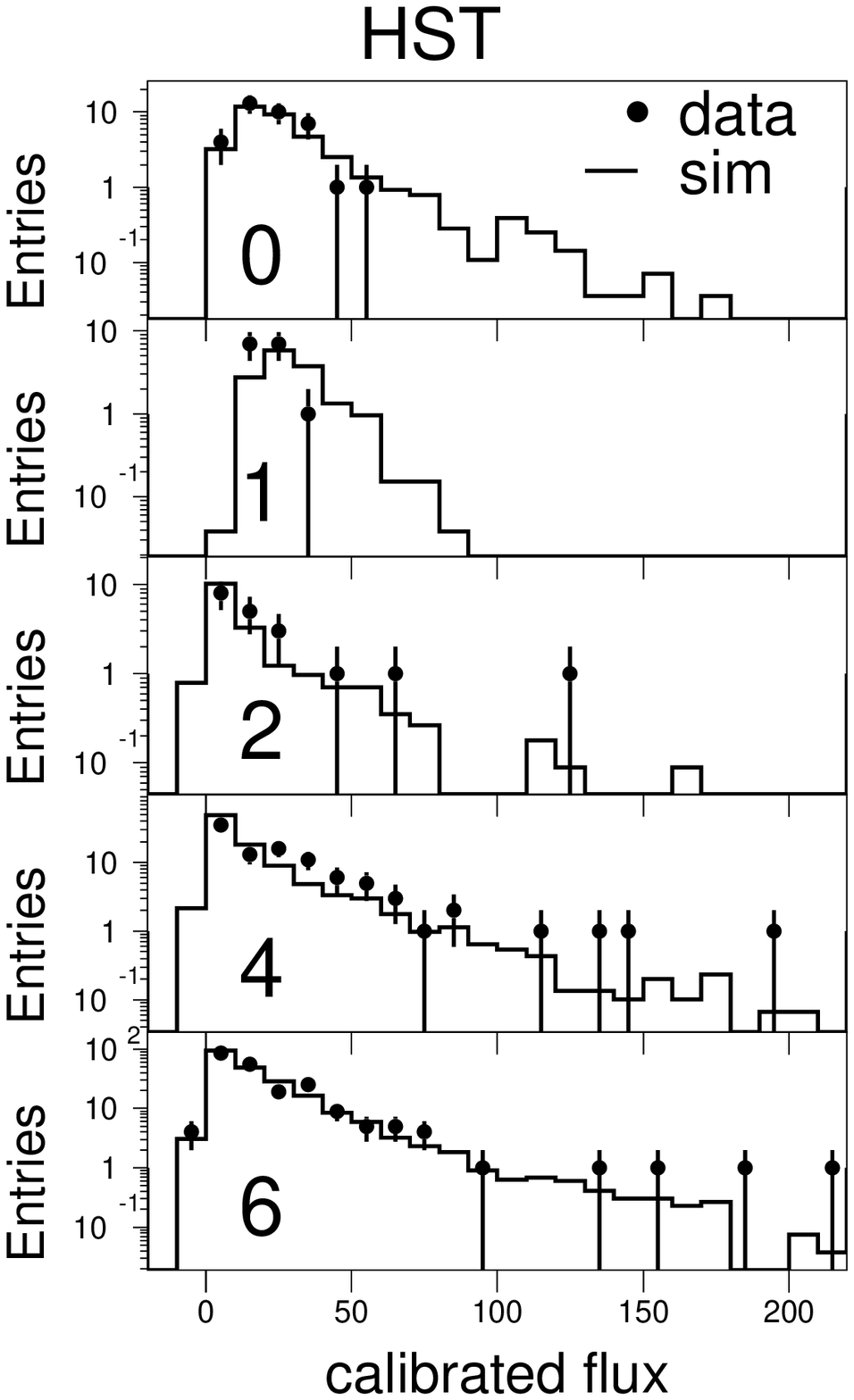}
  \caption{
     	Comparison of flux distributions for data (dots) and 
     	simulations (histograms) for 
	\SDSS, ESSENCE, SNLS, and HST.
     	The observer-frame passband is indicated on each plot
	(HST filter symbols are
	$0=$NIC-F110W, $1=$NIC-F160W, $2=$ACS-F606W, 
	$4=$ACS-F775W, $6=$ACS-F850LP).
     	Each simulated distribution is scaled to have
     	the same total number of entries as the data;
	bins with zero entries are not plotted.
	The calibrated flux is $10^{-0.4\cdot {\rm mag} + 11}$.
	As in Fig.~\ref{fig:ovzdatasim2}, the simulations include 
	all the survey inefficiencies. 
      }
  \label{fig:ovdatasim_flux}
\end{figure*}

As an illustration,
the inferred efficiencies for the \SDSS\ SN sample are shown in 
Fig.~\ref{fig:prior_eff} as a function of $A_V$, for different 
values of the redshift and shape-luminosity parameter $\Delta$. 
The high quality of the simulation for the 
\SDSS, ESSENCE, SNLS, and HST samples is illustrated in  
Figs.~\ref{fig:ovzdatasim2}-\ref{fig:ovdatasim_flux}, 
where we compare the observed redshift and flux distributions 
to simulations that include the efficiency functions derived above. 
The redshift comparisons have excellent $\chi^2/dof$ as a result
of how the \spec\ \eff\ is determined. For the flux comparisons,
the $\chi^2/dof$ vary from 1 to a few; the larger $\chi^2$ values
are due to a few notable discrepancies in some of the flux bins.

We have also compared the distributions of the number of epochs,
earliest and latest epochs, noise, and peak colors,
and find good data-simulation agreement in these distributions as well.

% ##############################################################

\section{Estimation of Host-Galaxy Dust Properties}
\label{sec:dust}
%
% TOP:  Determination of RV & AV-distribution
%

The Monte Carlo simulations used in the previous section to 
model survey selection functions rely on knowledge 
of host-galaxy dust properties, 
in particular on the underlying distribution of extinction, 
$P(A_V)$, and on the mean reddening law parameter, $R_V$. 
Moreover, \mlcs\ distance estimates rely directly on our 
knowledge of these two quantities, which are assumed to be 
independent of redshift.
In this section, we describe how we determine these
dust properties from the \SDSS\ SN sample.
These global dust properties are used in the \mlcs\
fitting prior for all SN samples.

% ------------------------------------------
\subsection{ \SDSS\ ``Dust'' Sample}
\label{subsec:dustsample}
% ------------------------------------------

The results of \S~\ref{subsec:speceff} indicate that current
SN samples suffer from significant \spec\ selection effects
(see Fig.~\ref{fig:ovzdatasim}).
Since \spec\ selection is likely to be biased against highly
extinguished SNe, use of purely spectroscopic SN samples may 
lead to biased estimates of the distribution of host-galaxy 
dust properties and thereby to potentially biased distance 
estimates when a dust-distribution prior is applied. To address this
issue, we use a nearly complete set of \SDSS\ SNe~Ia
to measure dust properties.
For \SDSS\ events with SN~Ia-like light curves that were {\it not} 
\specy\ confirmed as SNe~Ia,
we have embarked upon a program to obtain host-galaxy spectra 
and measure spectroscopic redshifts;
we call these photometric SN~Ia candidates.
Based on distributions of the \mlcs\ fit parameters
as well as visual inspection of the \lc\ fits,
we find that the requirement of a good SN~Ia light-curve fit
(see cut 5 in \S~\ref{sec:sample}) to a well-sampled light curve
is a good substitute in identifying SNe~Ia when
a confirming SN spectrum is lacking.

To identify photometric SN~Ia candidates, we start with 
all 4100 candidate events that were detected by the on-mountain 
frame-subtraction pipeline on two or more epochs in the 2005 observing 
season (see \S \ref{sec:survey}). We process these candidate light curves 
through Scene Model Photometry (\S~\ref{subsec:SMP}), 
fit them with the \mlcs\ method, and prioritize them for 
host-galaxy spectroscopy based on the quality of the light curve 
and the fit. 
We have obtained host-galaxy redshifts for the 
majority of candidates for which the host-galaxy
$r$-band magnitude satisfies $r \la 20$ and  
for a large subsample of fainter hosts as well.
Adding this ``{\hostz}'' photometric SN~Ia sample to the 
\specy\ confirmed and probable SN~Ia sample,
the combined sample appears to be nearly spectroscopically
complete to $z \simeq 0.3$, when compared with the simulated sample.
For $z<0.3$, after the selection cuts of \S \ref{sec:sample} are 
applied, the combined sample comprises 
	$\NDUSTSPEC$  confirmed SNe~Ia,
	$\NDUSTPROB$  probable SNe~Ia, and 
	$\NDUSTHOSTZ$  \hostz\ SNe~Ia.
We refer to these $\NDUST$ SNe~Ia as the \SDSS\ dust sample.
To illustrate the importance of including the \hostz\ subset,
we note that the average fitted extinction ($A_V$) is
about 0.2 for the \specy\ confirmed sample and almost 0.4 for the
\hostz\ sample: ignoring the \hostz\ subset would clearly 
lead to biased results for the distribution of host-galaxy dust properties.
On the other hand, 
giving highest priority to bright host galaxies for the {\hostz} follow-up 
program 
may preferentially select hosts of more extinguished SNe.

\begin{figure}[h]
  \epsscale{1.1}
  \plotone{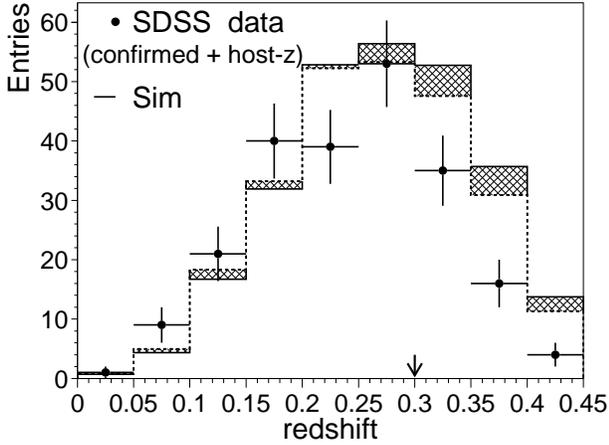}
  \caption{
 	Redshift distribution for the \SDSS\ dust sample (dots), 
	which includes confirmed, probable, and {\hostz} photometric 
	SN~Ia events,
 	and for the simulation (histogram), which includes losses 
	from the image-subtraction pipeline and selection cuts.
	Shaded region reflects simulated uncertainty in the SN~Ia 
	rate as explained in Fig.~\ref{fig:ovzdatasim} caption.
 	Vertical arrow indicates $z<\ZCUTAVTAU$ requirement to select
 	the dust sample used for the determination of host-galaxy 
	dust properties. 
     }
  \label{fig:ovzdust}
\end{figure}

To illustrate our understanding of the \eff,
Fig.~\ref{fig:ovzdust} shows the redshift distribution
for the dust sample, compared with the simulation
that includes known losses from the image-subtraction
pipeline and selection cuts but does not include losses
related to \spec\ 
observations.
This comparison shows that the dust sample is 
indeed nearly complete for redshifts $z < \ZCUTAVTAU$. 
We can obtain an independent estimate of the 
dust-sample completeness by counting the number of photometric 
SN~Ia candidates that pass our selection cuts, 
that have a photometric redshift 
(based on the host galaxy or the SN light curve)
less than $z_{\rm phot}=0.3$, and that have not yet been targeted for
host-galaxy spectroscopy. 
There are nearly $\NDUSTMISS$ such events, 
indicating a completeness, after selection cuts, 
of about $\NDUSTEFF$\% for the dust sample at $z<0.3$.

% ------------------------------------------------------

\subsection{Determination of Dust Reddening Parameter $R_V$}
\label{subsec:RV}

Although 
$R_V$ varies along sight-lines through the Milky Way and likely 
varies with local environment within galaxies, we 
follow standard practice in treating
it as a global parameter, 
because most current SN data are not adequate to determine 
it on an object-by-object basis. 
After briefly reviewing previous determinations of $R_V$, 
we describe the method we have 
developed to determine $R_V$ and its results.

Measurements of dust properties in the Milky Way have favored an average 
value of $R_V \sim \RVMW$ \citep{FitzMassa2008} along sight lines 
with moderate to substantial extinction, and this has been used 
as a canonical value in the literature.
However, studies of stellar colors in the direction of several 
thin cirrus clouds in the Milky Way indicate a value of 
$R_V \sim 2$ for those relatively low-extinction 
environments \citep{SzomoruGahathakura98}. 
For galaxies that host SNe~Ia, two methods have been commonly used 
to determine $R_V$. 
The first infers an average or global $R_V$ value 
based on statistical averages of optical light-curve colors.
This method has been applied to the nearby SN sample
\citep{Phillips_99,Altavilla_04,Reindl_05,Riess_96b,Nobili2007},
resulting in $R_V$ values in the range of $2-3$, 
somewhat lower than the Milky Way average value.
The second method uses SNe~Ia  that have densely sampled, high 
signal-to-noise optical and NIR photometry, for which $R_V$ can 
be estimated for individual events: 
this method has resulted in $R_V \sim 1-2$
\citep{Kris07,Kris06,ER06,XWang08}, 
significantly below the canonical Milky Way value.
Calculations and simulations have shown that multiple scattering 
by circumstellar dust can lead to lower values for the reddening 
parameter inferred from SNe~Ia, $R_V \sim 1.5 - 2.5$
\citep{Wang05,Goobar2008}.

Within the framework of the \mlcs\ light-curve model,
we have used the \SDSS\ dust sample and a variant of the average 
color method to make a new determination of $R_V$. 
Previous measurements with this method
were based on samples with
large and unknown selection effects;
our determination of $R_V$ is based upon a SN~Ia
sample with a well-understood selection \eff.
We assume that reddening is due to dust extinction
with a wavelength dependence described by the CCM89 model  
and that the \mlcs\ model parameters 
(${\Mmlcs},p,q$ in Eq.~\ref{eq:MLCS2k2model}) accurately describe 
the SN~Ia brightness and colors.
The measurement of $R_V$ is based upon comparing the average colors
as a function of SN epoch of the \SDSS\ data to those of the simulation. 
By using the nearly complete $z<0.3$ dust sample, we minimize potential 
selection bias against extinguished SNe~Ia; this sample also has 
a selection \eff\ that is well described by the simulation.

\begin{figure}
  \epsscale{1.0}
  \plotone{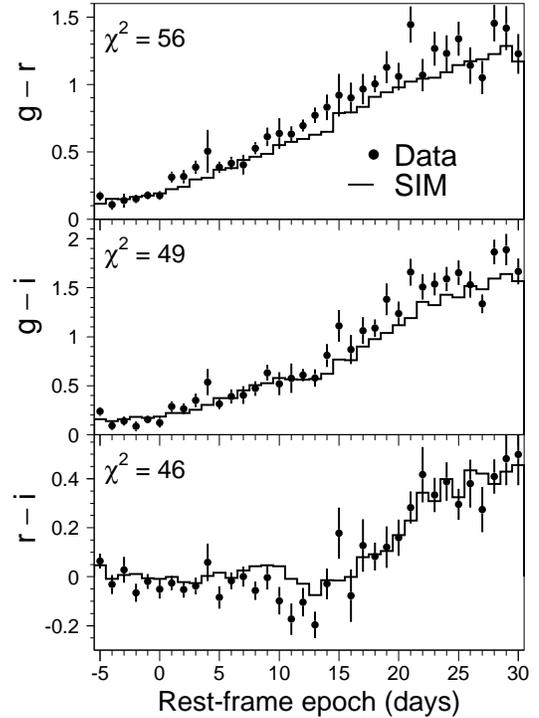}
  \caption{ 
	Mean SN~Ia colors $\langle g-r \rangle$, $\langle g-i \rangle$, 
	and $\langle r-i \rangle$ vs. rest-frame epoch, 
	for the \SDSS\ SN dust sample. Data are shown by the
	filled circles; the histogram overlay shows the simulation
	using the best-fit values of $R_V$ and $\AVMNSYMBOL$.
      	The data-simulation $\chi^2$ (for 36 degrees of freedom) 
	is indicated on each panel.
     }
  \label{fig:ovdatasim_color}
\end{figure}

For the dust sample, we compute three mean observed SN~Ia colors, 
$g-r$, $r-i$, and $g-i$, in one-day bins in rest-frame epoch, 
as shown in Fig.~\ref{fig:ovdatasim_color}. 
Although the third color ($g-i$) is redundant in most cases,
it provides information in the few cases in which the
$r$-band measurement is not available.
The rest-frame time is relative to the time of peak brightness in 
$B$-band as determined by the \mlcs\ light-curve fit 
(see \S~\ref{subsec:MLCS2k2}).
While the light-curve fits include all observations
regardless of signal-to-noise,
the estimates of the mean observed colors include only those
observations for which the signal-to-noise ratio is greater than 4.
Since the $g,r,i$ measurements are taken simultaneously in \SDSS,
the colors are determined directly from the data without the need to 
interpolate in time.

% define a few things needed for the RV-chi2.
\newcommand{\gminusr}{\langle g-r\rangle_e}
\newcommand{\gminusrPeak}{\langle g-r\rangle_0}
\newcommand{\sigmagrdata}{\sigma_{\gminusr}^{\rm data}}
\newcommand{\sigmagrsim}{\sigma_{\gminusr}^{\rm sim}}
\newcommand{\siggrmodel}{\sigma_{\gminusr}^{\rm model}}
\newcommand{\siggrmodelPeak}{\sigma_{\gminusrPeak}^{\rm model}}
\newcommand{\Wgr}{W_{\gminusr}^{\rm model}} % chi2 weight from model error

We compare these color-versus-epoch measurements 
with those of a grid of simulated samples, 
where each sample is 
generated with a different value of $R_V$ and $\AVMNSYMBOL$, 
and where $\AVMNSYMBOL$ describes the generated $A_V$ 
distribution, with $P(A_V) = \exp(-A_V/\AVMNSYMBOL)$. 
The $R_V$ and $\AVMNSYMBOL$ grid sizes are 0.2 and 0.05, respectively.
The simulated and data samples are subject to the same
selection cuts and fit with \mlcs\ in the same way. 
We determine the best-fit values of $R_V$ and $\AVMNSYMBOL$ by
minimizing the following $\chi^2$ statistic between the 
data and the grid of simulations, 
\begin{eqnarray}
  \chi^2 & = & \sum_e 
  \left[\frac{ [ {\gminusr}^{\rm data} - {\gminusr}^{\rm sim}) ]^2 }
                  { [{\sigmagrdata}]^2 + [{\sigmagrsim}]^2}
  \right]
  \left[\frac{\siggrmodelPeak}{\siggrmodel}\right]^2
    \nonumber \\
         & + & \sum_e (g,r \to r,i)
           +   \sum_e (g,r \to g,i)~,
   \label{eq:colorchi2}
\end{eqnarray}
where the epoch-index $e$
runs over 1-day bins with $-5 < \Trest^i < 30$ days, 
the data averages $\langle \rangle^{data}_e$ are taken over all 
dust-sample SNe and epochs $e$ surviving the cuts above, 
and the measured colors in the simulated samples 
depend upon the input values of $R_V$ and $\AVMNSYMBOL$.
Each color uncertainty, e.g., $\sigma_{(g-r)_e}$,  
is estimated as ${\rm rms}/\sqrt{N_e}$,
where $N_e$ is the number of color measurements 
(summed over all SNe) at epoch $e$. 
The second term in brackets is a weighting to account for the 
\mlcs\ model \unc;
$\siggrmodelPeak$ is the minimum model \unc\ at the epoch of 
peak brightness, and
$\siggrmodel$ is the model \unc\ at epoch $e$.
The model-\unc\ ratio is therefore unity at $\Trest=0$ 
and decreases as the model \unc\ increases 
for epochs away from peak brightness, so that epochs with large 
errors are downweighted.
We have tested this method with 100 simulated mock data samples
as described in \S~\ref{sec:sim};
the input values of $R_V$ and $\AVMNSYMBOL$ are recovered,
and the statistical \uncs, although they are smaller than the grid sizes,
match the spread in recovered values.

Using the \SDSS\ data and simulations, we find 
\begin{eqnarray}
   {\RVRESULT}   \label{eq:RVresult}  \\
   {\AVMNRESULT} \label{eq:AVMNresult}
\end{eqnarray}
and a correlation coefficient of 0.17.
The relatively small correlation coefficient confirms that 
the method is independently sensitive to $R_V$ and $\AVMNSYMBOL$.
Figure~\ref{fig:ovdatasim_color} compares the average observed 
colors with those of the simulation using the
best-fit values in Eq.~\ref{eq:RVresult}.
The $\chi^2$ values are somewhat larger than expected,
particularly for $g-r$. There is also a notable
data-simulation discrepancy for epochs past about 10~days
for the $g-r$ and $g-i$ colors, 
although these late-time epochs carry less weight in the $\chi^2$
due to the increasing model errors.

To estimate the systematic uncertainties in this measurement,
we have varied aspects of the procedure and determined 
their effects on the recovered $R_V$ and $\AVMNSYMBOL$.
In particular, we lowered the redshift cutoff for the dust sample 
to $z<0.25$ (nominal  is 0.3), 
varied the simulated redshift distribution of the underlying 
SN~Ia population, $dN/dz = \alpha(1+z)^{\beta}$,
within the correlated 1$\sigma$ errors on $\alpha$ and $\beta$ 
from \cite{Dilday08},
varied the minimum epoch from $-10$ to $0$ days (nominal is $-5$ days),
varied the maximum epoch from $+25$ to $+35$ days 
(nominal is $+30$ days), 
varied the minimum signal-to-noise between 2 and 8 (nominal is 4),
excluded each of the three colors $g-r$, $g-i$, and $r-i$ 
individually from the $\chi^2$ minimization,
and ignored the \mlcs\ model-\unc\ terms in the $\chi^2$ statistic. 
We also ran the simulation without intrinsic color fluctuations 
(see \S \ref{subec:hubblescat}). 
The changes due to each of these variations to the nominal results 
for $R_V$ and $\AVMNSYMBOL$ were added in quadrature 
to obtain an estimate of the total systematic uncertainty.  
Table~\ref{tb:colors} summarizes the contributions to the 
systematic uncertainties. 
The largest source of \unc\ comes from reducing the redshift-range
from 0.3 to 0.25, which changes $R_V$ by 
$\RVSYSTzcut \pm \RVSYSTzcuterr$,
and the error reflects the uncorrelated \unc.
Although this $R_V$-shift is marginally consistent with a random
fluctuation, we have included the shift as a systematic error.

As a crosscheck of how the inferred $R_V$ depends on the
assumed exponential form of the extinction distribution, 
$P(A_V)$, we have repeated the procedure using different
distributions for $A_V$ in the simulation. In particular, we  
use a flat, truncated $A_V$ distribution, 
$P(A_V) = 1$ for $A_V \le 2\AVMNSYMBOL$, and 
$P(A_V) = 0$ when $A_V > 2\AVMNSYMBOL$;
this results in a negligible change in the inferred  
values of $R_V$ and $\AVMNSYMBOL$. 
Thus, the inferred value of $R_V$ appears to be insensitive 
to the assumed $A_V$ distribution.

\begin{table}[hb!]
\centering
\caption{  
	Summary of uncertainties for the determination of
   	$R_V$ and $\AVMNSYMBOL$.
        }
\begin{tabular}{lcc}
\tableline\tableline % --------------------------------------------
source       & $\sigma(R_V)$            & $\sigma(\AVMNSYMBOL)$ \\
\tableline % -------------------------------------------------
statistical                        
    & $\RVERRSTAT$    & $\AVMNERRSTAT$ \\
redshift range  $0.25-0.3$         
    & $\RVSYSTzcut$   & $\AVMNSYSTzcut$ \\
assumed $dN/dz:$ $\beta = 0.9-2.1$  
    & $\RVSYSTdNdz$   & $\AVMNSYSTdNdz$ \\
epoch range     
    & $\RVSYSTepoch$   & $\AVMNSYSTepoch$ \\
signal to noise $>2,8$             
    & $\RVSYSTSNR$   & $\AVMNSYSTSNR$ \\
exclude one color                  
    & $\RVSYSTtwocolor$   & $\AVMNSYSTtwocolor$ \\
ignore \mlcs\ model \unc\ 
    & $\RVSYSTmlcsmodel$   & $\AVMNSYSTmlcsmodel$ \\
remove color smearing model        
    & $\RVSYSTcolorsmear$   & $\AVMNSYSTcolorsmear$ \\
\tableline % -------------------------------------------------
Total Systematic  & $\RVERRSYST$ & $\AVMNERRSYST$ \\
\tableline % -------------------------------------------------
\tableline % -------------------------------------------------
Total Uncertainty & $\RVERRTOT$ & $\AVMNERRTOT$ \\
\tableline % -------------------------------------------------
\end{tabular}
   \label{tb:colors}
\end{table}

% ------------------------------------------------------
\subsubsection{The \SALTII\ Approach to Measuring $R_V$}
\label{subsec:minRV}

The \SALTII\ analog of $R_V$ (called $\beta$, 
see Eq. \ref{eq:MUSALTII} below)
is determined by a very different method that involves 
minimizing the residual scatter in the Hubble diagram. 
For comparison, we have tried a similar procedure within 
the {\mlcs} framework, 
selecting the global value of $R_V$ that minimizes the scatter
in the Hubble diagram. For this test, we select \specy\
confirmed SNe~Ia from the \SDSS\ sample that satisfy the
light-curve criteria of \S~\ref{sec:sample} and that have 
redshifts $z<0.15$; the resulting sample includes 24 SNe~Ia. 
We concentrate on this low-redshift sample because it is  
nearly complete and because the measurement \unc\ on each  
distance modulus estimate is well below the intrinsic scatter. 
The resulting Hubble scatter-minimized $R_V$ value is about 1.7, 
roughly $0.5$
below our nominal result in Eq.~\ref{eq:RVresult}. 
This result holds whether we use the default prior
(Eq.~\ref{eq:prior_master}) or a flat prior in the light-curve fits.
When we apply this procedure to the larger
\SDSS\ dust sample (\S~\ref{subsec:dustsample}) 
that extends to $z=0.3$,  
the Hubble scatter-minimized $R_V$ value is slightly smaller. 
\citet{Hicken09b} have applied the same approach to the nearby CfA3
sample \citep{Hicken09a} and find that $R_V=1.7$ minimizes the Hubble 
scatter, consistent with our results for the \SDSS\ samples.

To test this approach to extracting $R_V$, we have implemented 
it on the simulated \SDSS\ SN sample (\S~\ref{sec:sim}). 
We generate a set of light curves with fixed $R_V$ and 
fit them with \mlcs\ in the same way that
the data are fit. 
For the default color-smearing model of intrinsic SN luminosity 
variation (see \S~\ref{subec:hubblescat}), 
the Hubble scatter-minimized $R_V$ extracted from this process 
is biased low by 0.5 with respect to the input value. 
This bias is consistent with the difference we see between 
the scatter-minimized $R_V$ and the $R_V$ we infer from the mean 
colors of the \SDSS\ dust sample.
For the alternative ``coherent luminosity smearing'' model, however,
which results in no intrinsic color variations, 
the scatter-minimized $R_V$ is unbiased.

While the result above is suggestive,
an  important caveat is that we have not evaluated 
the uncorrelated systematic \unc\ on the difference
between our nominal $R_V$ extraction and the 
scatter-minimized $R_V$.
Our study suggests, however, that the \SALTII\ $\beta$ parameter
could be biased low if  the color-smearing 
model is a reasonable description of intrinsic SN~Ia 
luminosity variation.

% ----------------------------------------------

\subsection{
	Determination of the Underlying Distributions of 
	$A_V$ and $\Delta$ }
\label{subsec:AV}
% ----------------------------------------------

The \mlcs\ simulation and the prior used in the \mlcs\ fitting method 
require knowledge of the underlying distribution for the 
$V$-band extinction due to host-galaxy dust, $P(A_V)$.
The distribution is also needed for the shape-luminosity 
parameter $\Delta$, although the latter has less impact 
on the results, because $\Delta$ is better-determined 
by the light-curve data for each SN. 
We determine these distributions 
using the Bayesian unfolding method of \citet{Agostini_95},
which essentially uses
underlying {\it trial} distributions in the simulation to 
make predictions for the observed distributions of fitted 
$A_V$ and $\Delta$. 
To limit selection biases in this procedure,
we use the \SDSS\ dust sample of \S~\ref{subsec:dustsample}. 

We fit the \SDSS\ light curves with {\mlcs}
using a flat prior on $A_V$, 
i.e., the fitted $A_V$ value is allowed to be negative. In the fit, 
the extinction as a function of wavelength is given 
by the  CCM89 model with $R_V = \RV$, as derived in \S~\ref{subsec:RV}.
The underlying distributions for $A_V$ and $\Delta$ are determined  
such that when they are input into the simulation, 
the fitted distributions from the simulated \lcs\ match the  
distributions from the \mlcs\ fits to the \SDSS\ data.
Technical descriptions of this procedure and of the assumed 
underlying distributions are given in
Appendix~\ref{app:unfold}.

\begin{figure}[h]
  \centering
  \epsscale{1.15}
  \plottwo{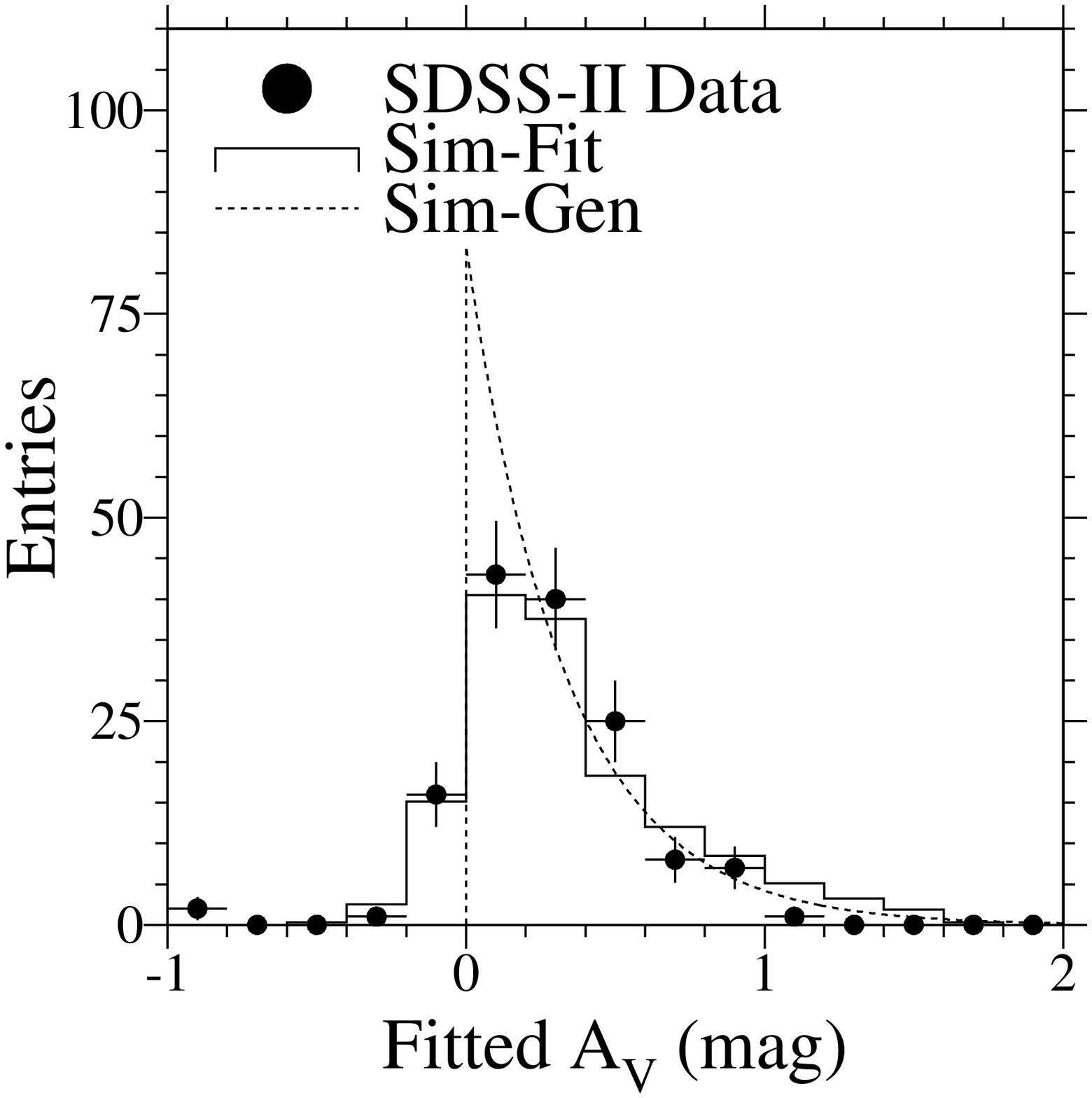}{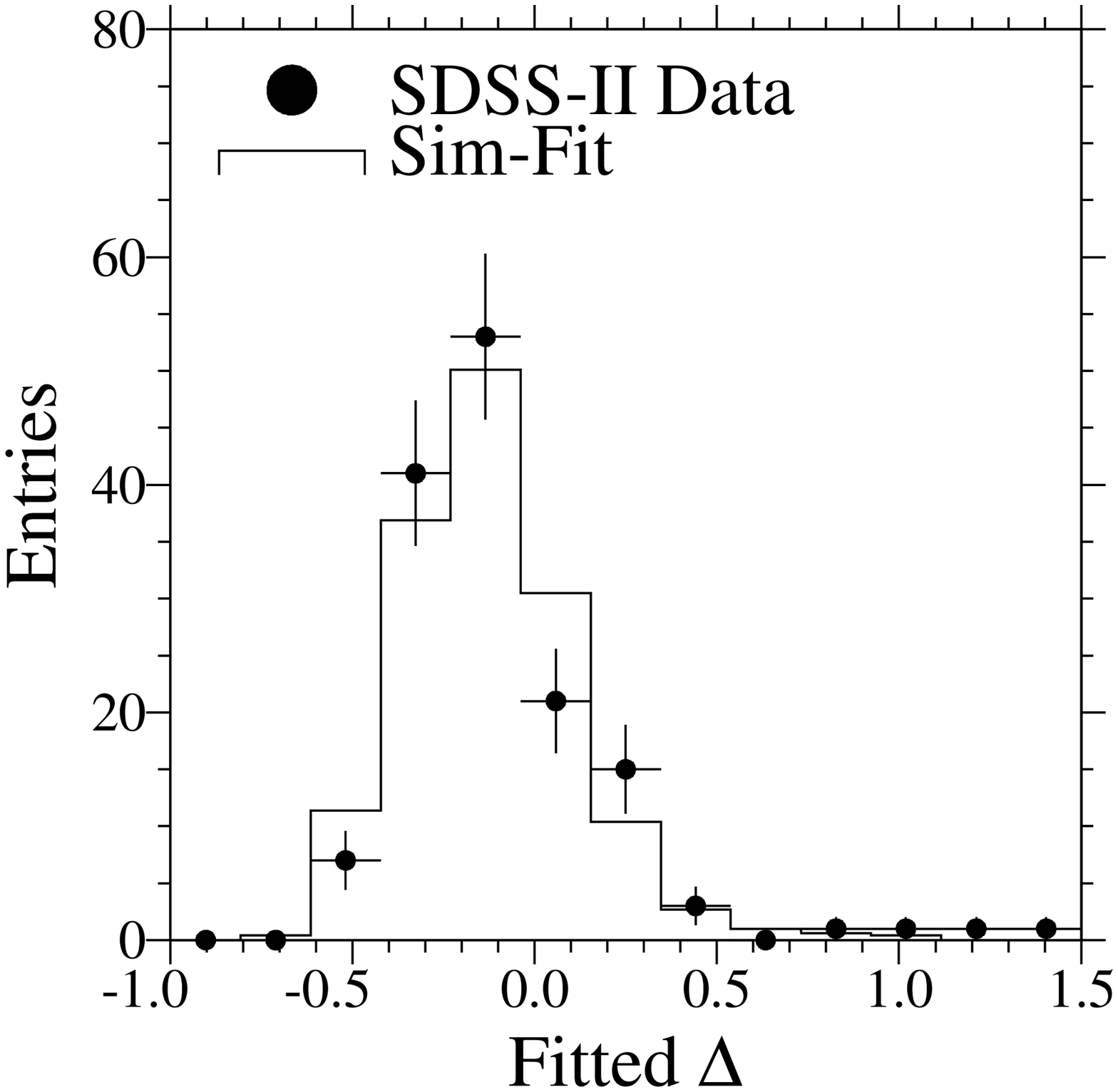}
  \caption{
     Fitted $A_V$ (left) and $\Delta$ (right) distributions,
     where the fits use the \mlcs\ model with a flat $A_V$ prior.
     The measured \SDSS\ dust sample distributions are shown by dots; 
     the simulations are shown by histograms.
     The underlying (generated) $A_V$ distribution
     is shown by the dashed curve in the left panel.     
     }
  \label{fig:ovdatasim_avdelta}
\end{figure}

The results for the data and Monte Carlo 
simulation are shown in Fig.~\ref{fig:ovdatasim_avdelta}. 
Although the generated $A_V$ distribution includes only 
non-negative values of $A_V$, there is a tail of fitted negative 
values that is 
well-described by the simulation and arises from photometric 
errors and intrinsic SN color variations.
Our procedure determines the underlying $A_V$ distribution 
without assuming a functional form for $P(A_V)$, 
but it turns out that this distribution is well described 
by an exponential, $P(A_V) =\exp(-A_V/\TAUV)$, with 
\begin{eqnarray}
  \TAUV & = & [\TAUAV \pm \TAUAVERRTOTnoRV] + \dTAUAVdRV\times(R_V-\RV)
                \label{eq:TAUV1} \\
        & = & \TAUAV \pm \TAUAVERRTOT ~. 
                \label{eq:TAUV2}
\end{eqnarray}
The \unc\ in $\TAUV$ includes statistical and systematic
\uncs\ as described in Appendix~\ref{app:unfold}.
The $R_V$-dependence is shown explicitly in Eq.~\ref{eq:TAUV1},
along with the internal measurement error. 
Eq.~\ref{eq:TAUV2} shows the total \unc\, including 
the \unc\ on $R_V$.
The agreement between $\TAUV$ and $\AVMNSYMBOL$ (from \S~\ref{subsec:RV})
is well within the expected dispersion 
based on simulated tests.

\begin{figure}[h]
  \epsscale{1.15}
  \plotone{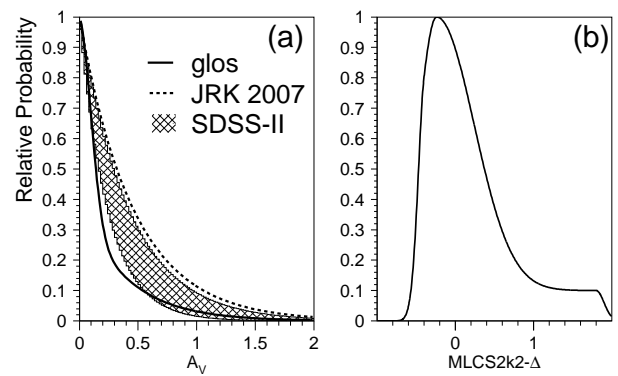}
  \caption{
	(a) Underlying $A_V$ distributions, $P(A_V)$: dotted is from JRK07, 
	solid is the glos distribution from WV07, 
	and the hatched region shows the distribution used in this paper, 
	derived from the \SDSS\ SN dust sample;
	(b) Underlying $\Delta$ distribution, $P(\Delta)$, 
	used in the \mlcs\ prior, also derived from the \SDSS\ dust sample.
     }
  \label{fig:prior_true}
\end{figure}

The underlying $A_V$ distribution, including the uncertainties,
is shown as the hatched region in Fig.~\ref{fig:prior_true}a;
we use this $P(A_V)$ as part of the \mlcs\ prior
in Eq. \ref{eq:prior_master}.
Our inferred $A_V$ distribution is marginally consistent with that of JRK07,
who derived a prior of exponential form 
from the nearby sample, with $\tau_{\rm V}^{JRK}=0.46$
(dashed curve in Fig.~\ref{fig:prior_true}a).
Our $A_V$ distribution is also somewhat consistent 
with the ``galactic line-of-sight'' (glos) prior,
based on theoretical considerations 
\citep{Hatano98,Commins04,Riello05},
that was used by WV07 for the ESSENCE analysis
(solid curve in Fig.~\ref{fig:prior_true}a).

The underlying distribution of $\Delta$ is described by
an asymmetric Gaussian with mean $\Delta_0=\DELTAPEAK$,
and Gaussian widths of
$\sigma_{-} = \DELTASIGMINUS$ for $\Delta<0$ and
$\sigma_{+} = \DELTASIGPLUS $ for $\Delta>0$
(see Appendix \ref{app:unfold} for {\uncs}).
As shown in the right panel of Fig.~\ref{fig:ovdatasim_avdelta},
when input into the simulation this distribution leads to a distribution 
of fitted $\Delta$ that is in good agreement with the observed 
distribution of fitted $\Delta$ from the \SDSS\ dust sample. 
The underlying $\Delta$ distribution, $P(\Delta)$, is shown in 
Fig.~\ref{fig:prior_true}b. 
In addition to the asymmetric Gaussian,
we have added a tail to the $\Delta$ distribution at positive $\Delta$, 
allowing for underluminous SNe where the underlying
distribution is poorly measured due to small-number statistics;
when used in the \mlcs\ prior of Eq. \ref{eq:prior_master},
this tail ensures that the fitter is not heavily
biased against underluminous SNe.
Since $\Delta$ is well determined from the light-curve shape,
the fitted value of $\Delta$ has little sensitivity 
to the functional form of $P(\Delta)$. 
To check the effect of the arbitrary tail in $P(\Delta)$,
we have run the fits with amplitude of the tail region multiplied 
by half and by two; in both cases the rms-variation in fitted 
$\Delta$ is $0.01$.

\begin{figure}[h]
  \epsscale{1.1}
  \plotone{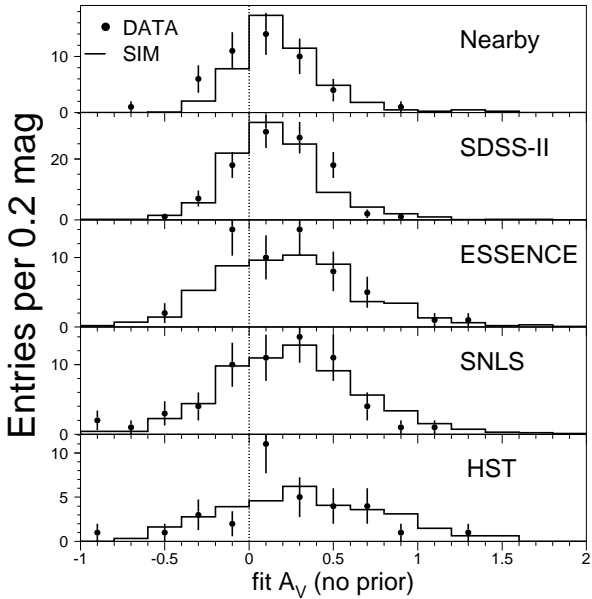}
  \caption{
	Comparisons of fitted $A_V$ 
	distributions, using a flat prior, for data and 
	simulations. Simulations were generated using the 
	underlying $A_V$ distribution derived from the 
	\SDSS\ dust sample, with $A_V \geq 0$. 
	SN samples used in the cosmology analysis
	are indicated on each panel. 
     }
  \label{fig:ovdatasim_avnoprior}
\end{figure}

To check that the derived extinction distribution reflects that 
of the global SN population rather than that of a (possibly biased) 
\SDSS\ sub-sample, we compare the fitted $A_V$ distributions for 
the data and for simulations generated with the underlying 
$A_V$ and $\Delta$ distributions derived above for each SN 
sample in Fig.~\ref{fig:ovdatasim_avnoprior}.
Here, the fits are carried out with a flat $A_V$ prior, allowing $A_V<0$.
The overall agreement is good for all samples. 
This test illustrates that the inferred negative extinction values 
(when an uninformative prior is used) are consistent with being artifacts 
of the combination of low signal-to-noise and intrinsic color fluctuations, 
since the simulation is generated with $A_V \geq 0$. 
Note that the procedure used to determine the
underlying $A_V$ distribution 
ensures that the data and simulation agree well 
for positive $A_V$
for the \SDSS\ sample, 
but parameters have not been adjusted 
to match the distribution for the other samples, or for
negative $A_V$.

% ##############################################################

\section{Fitting the Hubble Diagram for Cosmological Parameters}
\label{sec:wfit}

The estimation of cosmological parameters from supernovae 
is based on measurements of the luminosity distance, $\DL$, 
as a function of redshift. For a Friedmann-Robertson-Walker 
cosmological model, assuming the stress-energy comprises 
non-relativistic matter (M) and dark energy (DE) with constant 
equation of state parameter $w=p/\rho c^2$, 
the luminosity distance depends on four parameters:
$w$, the matter density $\OM$, the dark energy density $\ODE$, 
and the Hubble parameter $H_0$:
\begin{eqnarray}
 & & \DL(z;w,\OM,\ODE,H_0) =  \nonumber \\
 & &(1+z)|\Ok|^{-1/2}~\SINSINHFUN
   \left[
       \frac{c|\Ok|^{1/2}}{H_0} \int_{0}^{z} {dz' \over E(z')}
   \right] ~,
  \label{eq:dl}
\end{eqnarray}
where the curvature density $\Ok \equiv 1 -\OM-\ODE$, 
and the function 
$\SINSINHFUN(x)=\sin(x)$ for $\Ok <0$, 
$\SINSINHFUN(x)=\sinh(x)$ for $\Ok >0$, and 
$\SINSINHFUN(x)=x$ for a flat Universe with $\Ok=0$.
The dimensionless expansion rate is given by
\begin{eqnarray}
 & & E(z) = H(z)/H_0 =  \nonumber \\
 & & \left[\OM(1+z)^3 + \ODE(1+z)^{3(1+w)} +\Ok(1+z)^2\right]^{1/2}.
%%    \nonumber \\
   \label{eq:Ez}
\end{eqnarray}

For this analysis, we assume that the dark energy equation of state 
parameter $w$ does not evolve in time, mainly because current 
data do not yield precise constraints on the time derivative 
of $w$. For constraints on time-varying $w$, see \cite{Sollerman09}. 
Loosely, one can interpret the constraints we derive on constant 
$w$ as constraints on the mean value of $w(z)$ over the redshift 
interval $0\lesssim z \lesssim 0.4$. More precisely, 
one can interpret these constraints as providing information on 
one parameter in 
a two-parameter model that describes the evolution of $w$. That is, 
if the evolution of $w$ 
is described by a linear two-parameter model, e.g., by
$w(z)=w_0+w_a z/(1+z)$, then there is a pivot redshift $z_p$
at which the measurements of $w_0$ and $w_a$ are uncorrelated and 
the error in $w_p \equiv w(z_p)$ reaches a minimum. 
For the combined SN~Ia data sets considered in this paper, 
$z_p \approx 0.25$, and the constraints we derive on constant 
$w$ are equivalent to constraints on $w_p$.

We will study constraints on two lower-dimensional 
subspaces in the above family of cosmological models. 
For the first, which we denote \LCDM,
we consider models in which the dark energy is vacuum energy, 
i.e., the cosmological constant $\Lambda$, with $w=-1$, 
but we allow non-zero spatial curvature $\Ok$; 
the parameters of interest are $\OM$ and $\OLAM$.
In the second case, denoted \wCDM, we assume spatial flatness, 
$\Ok=0$, but allow $w$ to differ from $-1$; 
here the parameters of interest are $\OM$ and $w$. 
The rationale for considering the \LCDM\ model is that 
one can consider the cosmological constant as a ``null hypothesis'' 
for dark energy, so it is worth exploring whether it provides 
a reasonable description of the data. The rationale for the \wCDM\ 
model is that the WMAP data from the cosmic microwave background (CMB)
anisotropy constrain the spatial curvature to be very small.

In deriving parameter estimates, we will combine the SN data with
two independently measured constraints.
The first is from the measurement of 
baryon acoustic oscillation (BAO) feature
in the SDSS Luminous Red Galaxy (LRG) sample by \citet{Eisenstein05}.
The BAO measurements constrain several different parameters
\citep{Eisenstein05,Percival07}, depending on whether and how
information from the CMB is used; we explore this in more
detail in \cite{Sollerman09} and \cite{Lampeitl09}. Here, we follow
\citet{Astier06} and WV07 in using the combination of angular
diameter distance, Hubble parameter, and $\OM$ given by \citet{Eisenstein05},
\begin{eqnarray}
  A(z_1 & & ;~w,\OM,\ODE) =  {\sqrt{\OM} \over E(z_1)^{1/3}} \times
    \nonumber \\
& &     \left[
         \frac{1}{z_1\sqrt{|\Ok|}} 
          \SINSINHFUN \left(|\Ok|^{1/2}
          \int_0^{z_1} \frac{dz'}{E(z')}\right)
    \right]^{2/3} ~,
    \label{eq:ABAO}  
\end{eqnarray}
where the SDSS LRG BAO constraint is given by
\begin{equation}
   \BAOCHISQ =  [(A(z_1;~w,\OM,\ODE) - \ABAO) / \ABAOERR]^2 ~,
      \label{eq:BAO_prior}
\end{equation}
with the effective LRG redshift $z_1=0.35$. 
The second constraint is from the five-year results of the
Wilkinson Microwave Anistropy Probe \citep*{Komatsu2008}. 
We use the shift parameter, 
\begin{eqnarray}
  R(\zcmbsym;~ w,\OM,\ODE) & & = 
 \nonumber \\
  \sqrt{\frac{\OM}{|\Ok|}}
 \SINSINHFUN && \left(
        |\Ok|^{1/2} \int_0^{\zcmbsym}\frac{dz'}{E(z')} 
            \right),
   \label{eq:RCMB} 
\end{eqnarray}
with the constraint
\begin{equation}
   \CMBCHISQ =  [(R(\zcmbsym;~w,\OM,\ODE) - \RWMAP) / \RWMAPERR]^2 ~,
      \label{eq:CMB_prior}   
\end{equation}
where $\zcmbsym$ is the redshift of decoupling of the CMB. Although 
$\zcmbsym$ depends upon $\OM$ and upon the baryon density $\Omega_{\rm B}$ 
at the $\sim 2$ percent level, we fix this redshift 
to the WMAP5 maximum-likelihood 
value of $\zcmbsym = \zcmbval$ \citep*{Komatsu2008}.

Minimizing $\BAOCHISQ+\CMBCHISQ$ without using SNe,
the best-fit cosmological parameters for the \wCDM\ model
are \FwCDMwnoSNe\ and 
\FwCDMOMnoSNe;\footnote{If one marginalizes over $\OM$,
the marginalized value of $w$ is 0.09 smaller.
When SN constraints are included,
the difference between best-fit and marginalized
parameter values is much smaller.
} % end footnote
for  the \LCDM\ model, \LCDMOMnoSNe\ and \LCDMOLnoSNe.
Combining these measurements with the SN samples results in
improved constraints for the \wCDM\ model 
but has little impact for the \LCDM\ model.

We have analyzed several different combinations of the five 
SN~Ia data sets mentioned at the beginning of \S \ref{sec:anal}: 
\SDSS, nearby (low-redshift), ESSENCE, SNLS, and HST.
In \S~\ref{sec:syst} and \S~\ref{sec:results}, 
we present systematic \uncs\  and results for the 
\ncomboword\ sample combinations ({\combosymlist})
shown in Table~\ref{tb:sample_combos}.
\begin{table}[hb]
\centering
\caption{  
    Sample-combinations used to extract cosmological parameters.
     }
\begin{tabular}{cl}
\tableline\tableline
            & sample-combination \\
\tableline  % ------------------------
  ({\asym}) & \samplea  \\
  ({\bsym}) & \sampleb  \\
  ({\csym}) & \samplec  \\
  ({\dsym}) & \sampled  \\
  ({\esym}) & \samplee  \\
  ({\fsym}) & \samplef  \\
\tableline  % ------------------------
\end{tabular}
  \label{tb:sample_combos}  
\end{table}
Combination (a) includes only the 
\SDSS\ SN data, without lower- or  higher-redshift SNe. 
For combination (b), the nearby SNe~Ia are again excluded, 
leaving the \SDSS\ SN sample to serve as the low-redshift
anchor for the Hubble diagram. 
Combination (f) excludes the \SDSS\ sample, so that we can directly 
compare with previously published results such as those of WV07. 
In combination (c), the \SDSS\ serves as the `high-redshift' sample. 
Combination (d) includes 
all four  ground-based samples; 
combination (e) includes all five 
samples and is used for our nominal analysis.

% -------------------------------------------
\subsection{Fitting Distances with \mlcs\ }
\label{subsec:wfit_mlcs}

As described in \S \ref{subsec:MLCS2k2}, \mlcs\ provides an 
estimate of the distance modulus, $\mu$, 
for each supernova. Cosmological parameter estimates are 
derived by minimizing the following $\chi^2$ statistic 
($=-2\ln$ of the posterior probability) for the 
SN~Ia sample over a grid of model parameter values, 
\begin{eqnarray}
   \chisqmu & = & 
    \left\{ 
       \sum_i \frac{ [\mu_i - \mu(z_i;w,\OM,\ODE,H_0) ]^2}
                   {{\sigmutot}^2}
      \right\} 
          \nonumber \\
  &  + & \BAOCHISQ + \CMBCHISQ ~,
   \label{eq:mlcs_chisqmu}
\end{eqnarray}
where $\mu_i$ is the distance modulus estimated from 
the \mlcs\ fit for the $i$'th supernova, $z_i$ is its 
\specy\ determined redshift,
and $\mu(z_i;w,\OM,\ODE,H_0) = 5\log(\DL/10 ~{\rm pc})$ 
is computed from Eq.~\ref{eq:dl}.
The $\BAOCHISQ$ and $\CMBCHISQ$ terms in Eq.~\ref{eq:mlcs_chisqmu} 
incorporate the prior information from the SDSS LRG BAO measurement
(Eq.~\ref{eq:BAO_prior}) and the WMAP CMB measurement
(Eq.~\ref{eq:CMB_prior}).
When reporting values and errors on individual cosmological parameters, 
we use the values at the $\chisqmu$-minimum, marginalized only
over $H_0$ (due to the degeneracy between the Hubble parameter 
and the fiducial peak rest-frame model magnitude). In 
determining $w$, marginalizing over $\Omega_M$ shifts the 
value from the $\chisqmu$-minimum by 0.025 for the SDSS-only sample 
and by $\sim 0.01$
for the other sample combinations.

In Eq. \ref{eq:mlcs_chisqmu}, the distance-modulus 
uncertainty is given by
\begin{equation}
  {\sigmutot}^2 = 
  (\sigmufit)^2 + 
  (\sigmuint)^2 + 
  (\sigmudz)^2,   \label{eq:sigmudef}
\end{equation}
where $\sigmufit$ is the statistical uncertainty reported by \mlcs,
$\sigmuint = \sigmurmsVALUE$ is the additional (intrinsic) 
error added so that the $\chi^2$ per degree of freedom is 
unity for the Hubble diagram constructed from the nearby SN~Ia sample 
(\S~\ref{subsec:results_mlcs})
and $\sigmudz$ is calculated from the redshift uncertainty
as described below.

The redshift uncertainty contains two components:
from \spec\ measurements and from peculiar motions of the host galaxy.
The first source of uncertainty, $\sigzspec$, is from the uncertain 
redshift determination, either from the spectrum of the SN or from 
its host galaxy.  For \SDSS\ SNe we use $\sigzspec=0.0005$ 
for host-galaxy-based redshifts, 
$\sigzspec=0.005$ for SN-based redshifts, 
and the reported SDSS redshift errors for host  
redshifts from the SDSS spectroscopic galaxy survey
(\S~\ref{subsec:typez}).
For the ESSENCE and SNLS data, we use the estimated  
redshift errors reported in their public data tables. 
For the nearby sample, redshift measurement errors were usually 
not reported; we take 50 km/sec as a conservative estimate 
(M. Hamuy, private communication).
The second source of redshift uncertainty, $\sigzpec$, 
arises from peculiar velocities of and within host galaxies.
We take $\sigzpec = \sigzpecVALUE$, the quadrature sum of  
typical galaxy peculiar velocities of 
300~km/sec and typical internal motions of 200~km/sec.
For most of the \SDSS\ SNe, for which 
the redshift is obtained from the host galaxy,
the contribution from internal motions is overestimated
because these spectra are averaged over the internal galaxy motions.
The final redshift uncertainty is defined to be
$\sigma_z^2 = \sigzspec^2 + \sigzpec^2$.
To simplify the treatment of redshift errors, we project these
uncertainties onto distance modulus using the expression for the
distance-redshift relation for an empty universe, 
\begin{equation}
  \sigmudz =  \sigma_z \left(\frac{5}{\ln 10}\right) \frac{1+z}{z(1+z/2)} ~.
  \label{eq:sigmudz}
\end{equation}
Using a different cosmological model to compute 
$\sigmudz$ from $\sigma_z$ leads to negligible
changes in the Hubble diagram fits.

Galaxy peculiar velocities are not random, as the above treatment 
assumes, but are spatially correlated, since they are induced 
by the gravitational effects of large-scale structure. SN observations 
are affected by the peculiar velocities of both the host galaxies and 
of the Milky Way. We have accounted for the Milky Way peculiar velocity 
by correcting for the CMB dipole: all supernova redshifts have been 
transformed into the comoving frame of the CMB. This is particularly important for
the {\SDSS}, because the equatorial stripe comes within $7^0$ of the
CMB dipole direction, and the redshift correction 
from the heliocentric frame is negative along the entire
stripe. We use
$(1+z) = (1+z_{\sun})/(1-\vec{v}_0\cdot \hat{n})$, where 
$z_{\sun}$ is the heliocentric redshift as described in
\S~\ref{subsec:typez},
$\hat{n}$ is the unit vector pointing from earth to the SN,
and $\vec{v}_0$ is the CMB velocity vector, with a magnitude of 
$371$~km/sec and direction given by Galactic coordinates
$l = 264.14^0,~ b = +48.26^0$ \citep{Fixsen96}. The CMB-frame 
redshifts for the non-\SDSS\ samples are taken from the literature.

In transforming to the CMB 
frame and making no other corrections for velocity correlations, we 
are implicitly assuming that the peculiar velocities of the host 
galaxies are uncorrelated with that of the Milky Way and approximately 
uncorrelated with each other. 
That 
assumption
is a good approximation for the \SDSS\ and higher-redshift SN samples, 
which cover large spatial volumes and are distant from the Milky Way. 
It is not a good approximation for the nearby SN sample
on both theoretical \citep{Hui_06,Cooray_06} and observational 
\citep{Neill07} grounds. 
Not including velocity correlations
means that the low-redshift supernovae are over-weighted in 
the $\chi^2_\mu$ statistic and that SN-derived cosmological parameter
errors in the literature have been underestimated. However,  
inclusion of the \SDSS\ SN sample significantly reduces the impact 
of the uncertainties due to low-redshift peculiar velocities.

% -------------------------------------------
\subsection{Fitting Distances with \SALTII\ }
\label{subsec:wfit_SALTII}

In the first stage of the \SALTII\ analysis framework, 
the photometry for each supernova light curve is fit to an 
empirical model (\S~\ref{subsec:SALTII}) to determine
a shape-luminosity parameter ($x_1$), a color parameter ($c$),
and an overall flux normalization ($x_0$).
For the \SALTII\ Hubble diagram analysis,
the reference $B$-band magnitude is
%
%%\begin{equation}
$
 \mBstar = -2.5\log 
  \left[   
     x_0 \int d\lamprime M_0(t,\lamprime) T^B(\lamprime)
  \right]
$.
The flux-integral is the same as Eq.~\ref{eq:SALTII_flux_obs}
with $z = c = x_1 = 0$ and $f=B$.
The fitted parameter $x_0$ depends on the
luminosity distance and the SN brightness.

To estimate cosmological parameters, the above parameters are related
to the distance modulus for each supernova by the expression
\begin{equation}
  \mu_i = {\mBstar}_i - M + {\alpha} \cdot x_{1,i} -  
     {\beta} \cdot c_i ~.
   \label{eq:MUSALTII}
\end{equation}
The global parameters $M$, $\alpha$, and $\beta$ 
describe the SN~Ia population and are estimated simultaneously 
with the cosmological parameters by carrying out a 
$\chi^2$ minimization using an expression analogous to
Eq. \ref{eq:mlcs_chisqmu}. 
The minimization and error estimation are performed using 
the program {\sc minuit}.\footnote{\wwwMINUIT}
The expression for the distance modulus \unc, $\sigmutot$, 
is similar to that in Eq.~\ref{eq:sigmudef} for {\mlcs}, 
and the redshift uncertainty is treated identically. 
The intrinsic dispersion ($\sigmuint$) is determined 
separately for each sample combination by setting the global 
best-fit $\chi^2$ for that sample combination to unity, in contrast 
to the \mlcs\ method for which we determine $\sigmuint$ solely from 
the nearby sample. In addition, the \SALTII\ expression for $\chisqmu$ 
includes a covariance matrix to account for correlations between
the parameters $x_1$, $c$, and $x_0$ (\S~\ref{subsec:syst_salt2}).
In determining cosmological parameters, the Hubble parameter 
is marginalized over, but the parameters $\alpha$ and $\beta$ are not.

If the color corrections were due only to extinction by  
host-galaxy dust,  the \SALTII\ parameter $\beta$ would 
be equal to the \citet{ccm89} extinction parameter, 
$R_B$ \citep{Conley2007}. 
Further,  if host-galaxy extinction were similar to the 
mean extinction in the Milky Way,
one would expect $\beta \simeq R_B = R_V+1 \simeq 4$. However, 
in {\SALTII}, $\beta$ is left as a free parameter that is 
determined in the global fit to the Hubble diagram.

\newcommand{\simwcorr}{\delta w}
\newcommand{\simOMcorr}{\delta {\OM}}

To account for selection bias,
we have applied the \SALTII\ fitting
method to simulated sample combinations {\asym}-{\fsym}. 
The Monte Carlo simulations are generated from the 
{\SALTII} model 
using parameters based on our analysis of the data
(\S~\ref {subsec:results_salt2}).
In particular,
$\alpha = 0.13$ and $\beta = 2.56$,
and the simulated distributions of $c$ and $x_1$ 
are described by Gaussians with 
$\bar{c} =  0.04$, $\sigma_c = 0.13$, 
$\bar{x_1}=-0.13$, $\sigma_{x_1} = 1.24$.
These distributions were estimated directly from the 
light-curve fits rather than from simulating underlying 
$c$ and $x_1$ distributions such that the observed and 
simulated distributions match. 
The intrinsic luminosity is randomly varied using
a coherent luminosity smearing factor of 0.12~mag,
and the \spec\ \eff\ is based on the \SALTII\ parameters
described in \S~\ref{subsec:speceff}.

To isolate a potential \SALTII\ bias from a shift due to the
CMB and BAO priors, the simulated SN sample is generated with the cosmological
parameter values obtained from fitting without any SNe, 
$w^0 = \FwCDMwnoSNeVal$ and  $\OM^0 = \FwCDMOMnoSNeVal$.
The bias in the cosmological parameters is estimated to be
$w-w^0$ and $\OM-\OM^0$, where $w$ and $\OM$ are obtained
from the analysis of the simulated SN sample combined
with the BAO+CMB priors.
To increase the significance of the measured bias,
500 data-sized samples were simulated and analyzed; 
the resulting statistical \unc\ on the $w$-bias 
is typically 0.004.
The average bias for each sample combination and cosmological 
model is shown in Table \ref{tb:SALT2_simcor}. For the 
results shown below in Tables \ref{tb:results_salt2_FWCDM_BAOCMB} and 
\ref{tb:results_salt2_LCDM_BAOCMB}, 
these bias corrections have been added to the cosmological parameters.

\begin{table}[hb]
\caption{
	\SALTII\ corrections to cosmological parameters
	based on simulations.
     }
\begin{center}
\leavevmode
\begin{tabular}{l | cc | rr }
\tableline\tableline
  Sample      & \multicolumn{2}{c|}{\wCDM}   & \multicolumn{2}{c}{\LCDM} \\
  combination\tablenotemark{1} 
              &  $\simwcorr$  & $\simOMcorr$ & $\delta \OLAM$ & $\simOMcorr$ \\
\tableline  % ------------------------
 (\asym)  & 
$\FWCDMBAOCMBWaEFFSHIFT$     & $\FWCDMBAOCMBOMEGAMaEFFSHIFT$  &
$\LCDMBAOCMBOMEGALaEFFSHIFT$ & $\LCDMBAOCMBOMEGAMaEFFSHIFT$   \\
% --------
 (\bsym)  & 
$\FWCDMBAOCMBWbEFFSHIFT$     & $\FWCDMBAOCMBOMEGAMbEFFSHIFT$  &
$\LCDMBAOCMBOMEGALbEFFSHIFT$ & $\LCDMBAOCMBOMEGAMbEFFSHIFT$   \\
% --------
 (\csym) & 
$\FWCDMBAOCMBWcEFFSHIFT$     & $\FWCDMBAOCMBOMEGAMcEFFSHIFT$  &
$\LCDMBAOCMBOMEGALcEFFSHIFT$ & $\LCDMBAOCMBOMEGAMcEFFSHIFT$   \\
% --------
 (\dsym)  & 
$\FWCDMBAOCMBWdEFFSHIFT$     & $\FWCDMBAOCMBOMEGAMdEFFSHIFT$  &
$\LCDMBAOCMBOMEGALdEFFSHIFT$ & $\LCDMBAOCMBOMEGAMdEFFSHIFT$   \\
% --------
 (\esym)  & 
$\FWCDMBAOCMBWeEFFSHIFT$     & $\FWCDMBAOCMBOMEGAMeEFFSHIFT$  &
$\LCDMBAOCMBOMEGALeEFFSHIFT$ & $\LCDMBAOCMBOMEGAMeEFFSHIFT$   \\
% --------
 (\fsym) & 
$\FWCDMBAOCMBWfEFFSHIFT$     & $\FWCDMBAOCMBOMEGAMfEFFSHIFT$  &
$\LCDMBAOCMBOMEGALfEFFSHIFT$ & $\LCDMBAOCMBOMEGAMfEFFSHIFT$   \\
% --------
\tableline  % ------------------------
\end{tabular}
\end{center}
\tablenotetext{1}{{\asym}-{\fsym} are defined in 
                    Table~\ref{tb:sample_combos}.}
\label{tb:SALT2_simcor}
\end{table}

% #########################################

\section{Systematic Uncertainties}
\label{sec:syst}

The likelihood analysis described in the previous section accounts
for the impact of statistical errors on the determination of
cosmological parameters.
Here we give a detailed description of systematic \uncs\
on the cosmological parameters $w$ and $\OM$ 
within the context of the \wCDM\ model. 
For the \LCDM\ model, the systematic \uncs\ are evaluated
in a similar fashion and are summarized along with the results
in \S~\ref{sec:results}.
For both \mlcs\ and {\SALTII}
and for each of the \ncomboword\ sample combinations 
({\combosymlist}) in Table \ref{tb:sample_combos},
we have carried out several dozen systematic tests
by varying analysis parameters and methods.
The resulting variations in the Hubble diagram
and shifts in the parameter $w$ (denoted $\dw$ below)
are used to assess the systematic \unc.
To help gauge the significance of systematic
shifts observed in the data,
the same systematic parameter and method variations have also been applied
to the analysis of Monte Carlo simulations generated with 
$w=-1$ and $\OM=0.3$,
using the efficiencies determined in \S~\ref{sec:sim}.
There are four main categories of systematic {\uncs}:
(i) uncertainties in 
SN~Ia model parameters obtained from the training procedure for the
light-curve fitter; 
(ii) uncertainties in reddening from host-galaxy dust and in
intrinsic SN color variations;
(iii) errors in survey selection efficiencies and in associated
selection biases; and
(iv) uncertainties in relative photometric calibration between the nearby
sample and the higher-redshift surveys.

We discuss the impact of these uncertainties in the 
context of the \mlcs\ method in \S \ref{subsec:syst_mlcs} and 
for the {\SALTII} method in \S \ref{subsec:syst_salt2}. 
Summaries of the systematic \uncs\ for the \ncomboword\ 
sample combinations listed in Table~\ref{tb:sample_combos} 
are presented in Tables \ref{tb:wsyst_mlcs} and \ref{tb:omsyst_mlcs} 
for \mlcs\ and in 
Tables \ref{tb:wsyst_salt2_FWCDM_BAOCMB} and 
\ref{tb:omegamsyst_salt2_FWCDM_BAOCMB} 
for {\SALTII}. For the flat \wCDM\ model, 
these tables show systematic errors in the 
dark energy equation of state parameter $w$ and 
matter density parameter $\OM$, including 
the BAO and CMB priors discussed in \S \ref{sec:wfit}.
A more detailed discussion of one major source of 
systematic uncertainty, associated with data-model 
discrepancies in the rest-frame $U$-band, 
is postponed to \S \ref{subsec:Uanom_MLCS} and
\ref{subsec:Uanom_SALT2}.

% ------------------------------------
\subsection{Systematic Uncertainties with \mlcs\ }
\label{subsec:syst_mlcs}

% ===========================================
\bigskip \noindent {\bf \SYSTUband} \\
As discussed in detail in \S \ref{subsec:Uanom_MLCS}, there are  
systematic discrepancies between the \mlcs\ rest-frame 
$U$-band model and the light-curve data for all but the 
nearby SN~Ia sample. We have therefore carried out a test in 
which the observer-frame filter corresponding to rest-frame 
$U$-band is excluded from the light-curve fits and the 
resulting distance estimates. 
Figure~\ref{fig:noUtest_MLCS} shows 
that for the {\samplea} sample, the exclusion of $U$-band 
causes a mean systematic shift in estimated distance modulus of 
$\MLCSSDSSnoUdMU$~mag for $z > \ZUSDSS$. 
The resulting tilt in the Hubble diagram leads to 
a systematic change of $-\WSYSTanoU$ in the equation of state 
parameter $w$ for both the \samplea\ ({\asym}) and 
\samplec\ ({\csym}) sample combinations; 
we include this shift as an asymmetric systematic \unc,
as indicated by the minus signs in the first row of
Tables~\ref{tb:wsyst_mlcs}-\ref{tb:omsyst_mlcs}.

For the other sample combinations (\bsym,\dsym,\esym,{\fsym}), 
which include the ESSENCE \& SNLS samples, the exclusion of 
$U$-band results in $w$-shifts of 0.04 to 0.08.
Based on tests with simulations, we cannot distinguish these shifts 
from random fluctuations;
we therefore include the largest shift, $\dw = 0.08$, as a
symmetric systematic \unc\ for these four sample combinations.

% ===========================================
\bigskip \noindent {\bf Minimum Redshift for Nearby SN sample}\\
\newcommand{\mucalc}{\mbox{$\mu_{\rm calc}(z)$}}
\newcommand{\mufit}{\mbox{$\mu_{\rm fit}(z)$}}
As discussed in \S \ref{sec:sample}, the choice of minimum
redshift $\ZMINSYM$ for the Nearby sample is complicated 
by the so-called ``Hubble bubble,'' 
a jump in estimated SN~Ia distance modulus around 
$z \sim \ZBUBBLE$ 
within the Nearby sample (JRK07). 
This shift is shown in the residual Hubble diagram in 
Fig.~\ref{fig:lowz2sdss}, which compares the 
fitted SN distance modulus in redshift bins, \mufit, 
with the calculated distance modulus ({\mucalc}) for a 
cosmological model with $w=-1$ and $\OM=0.3$.
Since \mucalc\ is a smooth function of redshift,
the jump between the two left-most points in the plot reflects 
a $\simeq 0.12$ mag discontinuity in SN distance modulus 
({\mufit}) at $z \simeq \ZBUBBLE$.
The significance of the shift in $\mu$ between SNe at 
$z < \ZBUBBLE$ and $z > \ZBUBBLE$ is $\LOWZMUSIGNIF\sigma$.
JRK07 found a comparable shift, corresponding to 0.14 mag. 
To further explore this issue, we also show the same residuals 
for the lower-redshift portion of the \SDSS\ sample. 
The fitted \SDSS\ SN distances are consistent with those of the 
$z> \ZBUBBLE$ subset of the nearby SNe~Ia 
and less
consistent with the $z< \ZBUBBLE$ subset
($\sim \SDSSMUSIGNIF\sigma$ discrepancy).

The Hubble bubble may be 
a real feature due to a local, large-scale void, 
an artifact of selection biases in the nearby SN sample,
or an artifact of the light-curve fitter assumptions
about host-galaxy dust and color variations.
Regardless of its interpretation,
this feature 
has been used as an argument to discard 
SN~Ia measurements at $z< \ZBUBBLE$ from cosmological fits 
\citep[e.g.,][]{Riess_06}. 
We have decided to make a more agnostic choice for $\ZMINSYM$. 
We carry out cosmological fits using two sample combinations, 
\samplec\ ($\csym$) and \sampled\ ({\dsym}),
and vary $\ZMINSYM$ from 0.015 to 0.03; 
for {\mlcs}, the resulting variation in $w$ is shown
in the first two panels of Fig.~\ref{fig:zminscan}. 
For both sample combinations, the approximate midpoint of the 
$w$-variation occurs at $\ZMINSYM = \ZMINVAL$, 
which we therefore take as the nominal choice. 
As $\ZMINSYM$ is varied around this value, 
the $w$-variations are approximately
$\pm \WSYSTcLOWZCUT$ for sample combination ({\csym}) and 
$\pm \WSYSTdLOWZCUT$ for ({\dsym}); 
we include these variations as systematic uncertainties in 
Table~\ref{tb:wsyst_mlcs}, with associated results for $\OM$ in 
Table~\ref{tb:omsyst_mlcs}.

\begin{figure}[hb]
\centering
 \epsscale{1.1}
 \plotone{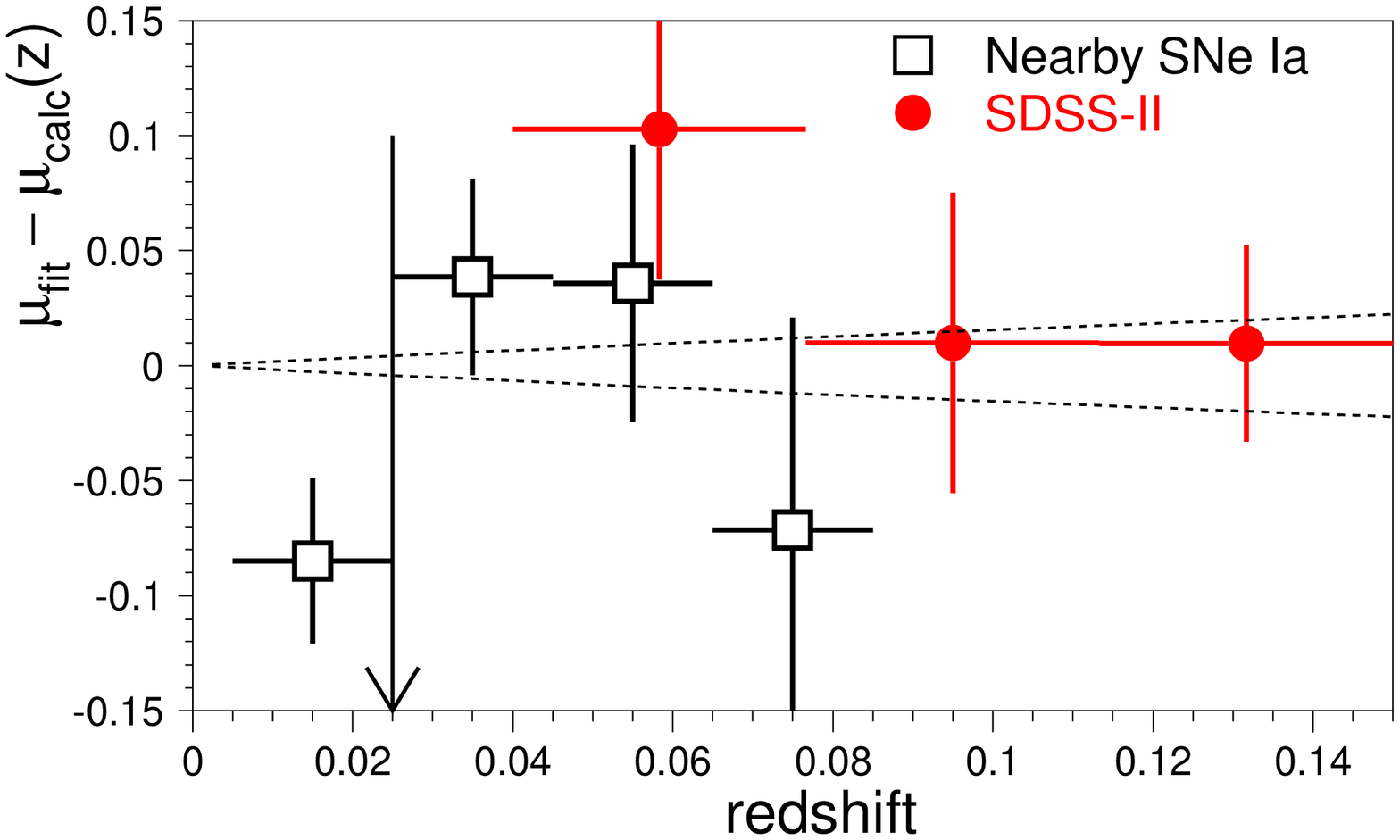}
  \caption{
	Difference between fitted
	 {\mlcs} SN distance modulus 
	for low-redshift SNe, \mufit, and the calculated distance modulus 
	{\mucalc} for an $w=-1$, $\OM=0.3$ model.
	The residuals are averaged in redshift bins
	of width 0.02 for the Nearby sample (black squares) 
	and in bins of width 0.037 for the nearer portion of the 
	\SDSS\ sample (red circles). 
	Errors are computed from $0.16/\sqrt{N}$, where 0.16 is 
	the typical magnitude dispersion per SN and 
	$N$ is the number of SNe in a given redshift bin.
	The dotted lines indicate the \unc\ in \mucalc\
	resulting from an \unc\ in $w$ of 0.15.
	The vertical arrow at $z=\ZBUBBLE$ indicates the 
	redshift associated with the Hubble bubble in JRK07.
     }
  \label{fig:lowz2sdss}
\end{figure}

\begin{figure}[hb]
\centering
 \epsscale{1.1}
\plotone{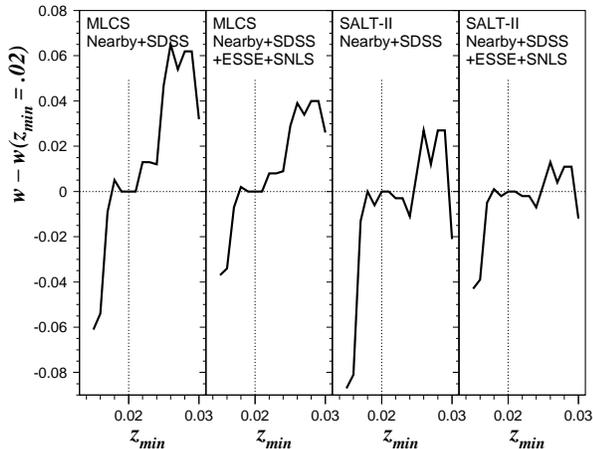}
  \caption{
	Variation in dark energy equation of state 
	parameter $w$ with  
	minimum redshift $\ZMINSYM$ as the latter is varied 
	around the nominal value of $\ZMINSYM = \ZMINVAL$ 
	(shown by dotted vertical line).
	The fitting method (\mlcs\ or {\SALTII}) and sample combination
	are indicated in each panel.
     }
  \label{fig:zminscan}
\end{figure}

% ===========================================
\bigskip \noindent {\bf \mlcs\ SN~Ia Model Parameters} \\
The \mlcs\ model vectors $M$, $p$, and $q$ in Eq. \ref{eq:MLCS2k2model}  
were determined by training the light-curve fitter  
on a nearby SN~Ia sample, as described in JRK07. As discussed 
in \S \ref{subsec:MLCS2k2}, 
our analysis uses a set of model vectors that includes the adjustments to the
$\Mmlcs^{e,f'}$  values given in Eq.~\ref{eq:Mtweak}. 
To estimate the sensitivity of the results to these adjustments,
we have also carried out the cosmology analysis 
using the \mlcs\ model vectors without those adjustments.
These two sets of model vectors result in 
values for $w$ that differ by $\dw = 0.01$ to $0.04$,
depending on the sample combination.

% ===========================================
\bigskip \noindent {\bf Galactic Extinction} \\
The wavelength-dependence of the Milky Way Galactic extinction 
is expressed as
$A(\lambda) = A_V\times ( a(\lambda)  + b(\lambda) /R_V)$,
where $R_V = \RVMW$, $A_V = R_V\times E(B-V)$,
$E(B-V)$ is the color excess,
and the functions $a(\lambda)$  and $b(\lambda)$
are defined by \citet*{ccm89}.
To estimate the systematic \unc\ for each SN, we coherently
decrease the color excess by $0.01 + 0.16\times E(B-V)$~mag
(but requiring non-negative color excess)
relative to the nominal value in \cite*{Schlegel_98}.
The color-excess \unc\ of 0.01~mag is based on optical-versus-IR 
discrepancies \citep{Burstein_03}.
The resulting $w$-\uncs\ are 0.01--0.02.
Note that we have not varied the functions
$a(\lambda)$  and $b(\lambda)$, and therefore
this \unc\ may be underestimated.
The mean Galactic extinctions in the $r$-band are
0.20 (nearby), 0.14 (SDSS), 0.07 (ESSENCE), 0.05 (SNLS)
and 0.01~mag (HST).

% ===========================================
\bigskip \noindent {\bf Dust Parameter $R_V$} \\
As described in \S~\ref{subsec:dustsample}, 
the SDSS dust sample was used to determine the 
dust reddening parameter,
$R_V = \RV \pm \RVERRTOT$ (\S~\ref{subsec:RV}),
and the extinction distribution exponential slope,
$\TAUV = \TAUAV \pm \TAUAVERRTOT$ (\S~\ref{subsec:dustsample}).
Propagating the correlated \uncs\ on $R_V$ and $\TAUV$,
the \unc\ on $w$ is $\dw = \WSYSTaRVSHIFT$ for the \samplea\ 
sample. For the combined data samples, $\dw \la 0.03$.
This color-related  \unc\ appears rather small, considering 
that the issue of color variations has been a major source
of \unc\ in previous studies such as WV07.

% ===========================================
\bigskip \noindent {\bf Simulated Spectroscopic Efficiency} \\
The simulated \eff\ is part of the \mlcs\ fitting prior,
as indicated by the terms $\simeffsearch$ and $\simeffcuts$
in Eq.~\ref{eq:prior_master}. 
The main effect of an error in determining the \spec\ \eff\ is 
to introduce a bias in the \mlcs\ prior (Eq.~\ref{eq:prior_master}).
As discussed in \S \ref{sec:sim}, the largest \unc\ in simulating the 
\eff\ is related to modeling of the incompleteness of the \spec\
observations,
$\simeffspec$.

For the nearby, SDSS, and SNLS samples, $\simeffspec$
depends mainly on the $\magdim$ component, 
i.e., $\AEFFDIM \simeq 1$ in \S~\ref{subsec:speceff}.
We explore the systematic error due to spectroscopic efficiency 
modeling for these samples by repeating the  
analysis with $\simeffspec$ set to unity, an extreme 
variation from the fiducial efficiency model of \S \ref{subsec:speceff}. 
Since the ESSENCE sample favors a purely redshift-dependent 
$\simeffspec$ ($\AEFFDIM = 0$ in Eq.~\ref{eq:simspec}),
setting $\simeffspec=1$ has no impact in that case; we instead evaluate 
the systematic error for that sample by replacing
the purely $z$-dependent $\simeffspec$ with a purely $\magdim$-dependent
$\simeffspec$, i.e., by setting $\AEFFDIM = 1$.
In these tests, the \eff\ due to photometric selection cuts 
($\simeffcuts$) is included in the fitting prior 
as well as the image-subtraction \eff\ ($\simeffpipe$)
for the SDSS sample.

\begin{figure}[h!]
\centering
 \epsscale{1.1}
\plotone{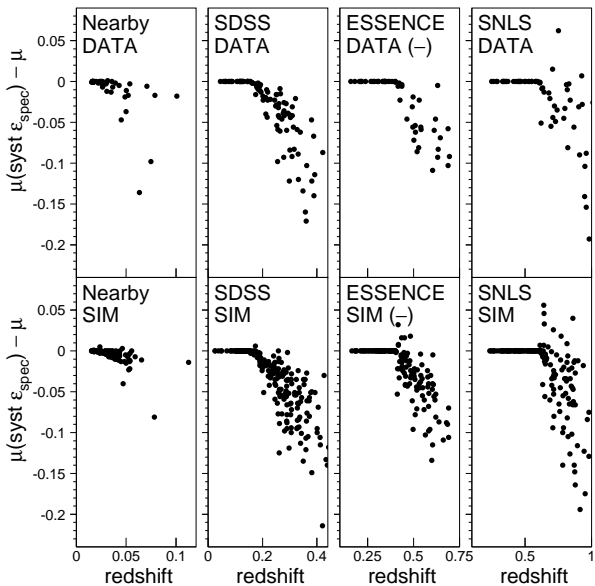}
  \caption{
	Systematic impact of uncertainty in the \spec\ \eff\ model 
	for different samples.  {\it Upper panels:} 
	data samples; {\it lower panels:} simulated samples. 
	For each sample, the difference in distance modulus between 
	using an extreme spectroscopic efficiency model 
	($\mu({\rm syst}~ \epsilon_{\rm spec})$) and using the 
	fiducial efficiency model ($\mu$) is shown vs. redshift. 
	For the nearby, SDSS, and SNLS data sets, the extreme model 
	assumes no \spec\ incompleteness ($\simeffspec=1$). 
	For ESSENCE, the extreme model assumes pure Malmquist-induced 
	incompleteness, and the negative of the $\mu$-difference 
	is plotted instead in order to better compare with the 
	results for the other samples. 
	The simulated samples have $2.5$ times more points than the 
	data samples for enhanced statistics.
  	The differences in $\mu$ appear only for $z > \zcut$ as 
	explained in \S~\ref{subsec:speceff}.
     }
  \label{fig:effspec_dmu}
\end{figure}

For the above systematic changes in $\simeffspec$, 
the corresponding changes in distance modulus vs. redshift
are shown in Fig.~\ref{fig:effspec_dmu} 
for both the data and the Monte Carlo simulations. 
One sees clearly that incorporating 
a non-trivial spectroscopic efficiency model has a significant systematic 
impact on distance estimates for the more distant SNe in each sample. 
Moreover, the data-simulation agreement is good, adding confidence
in our implementation of the \spec\ \eff\ model in the fitting prior.

From Fig.~\ref{fig:ovzdatasim}, 
there is no doubt that $\simeffspec$ is significantly 
less than one and that our model for $\simeffspec$ 
is closer to reality than the extreme 
assumption 
of setting $\simeffspec=1$. 
To be conservative, we use 
half of the absolute-value difference in $w$ between the fiducial 
and extreme efficiency models as our estimate of the  
systematic error associated with uncertainty in the efficiency model. 
The largest \unc\ related to the simulated \eff\ occurs for 
sample combinations 
({\asym}) and ({\csym}); the $w$-\uncs\ (half the shifts) are
$\dw = \WSYSTaSIMEFFSDSS$ for the {\samplea} sample 
and
$\dw = \WSYSTcSIMEFFSDSS$ for the {\samplec}  
sample combination 
({\csym}). 
We also note that including vs. not including the simulated
\eff\ in the nearby SN sample changes $w$ by a few hundredths 
for the other sample combinations that include the Nearby sample,
half of which is included as a systematic \unc.

% ===========================================
\bigskip \noindent {\bf Calibration of Primary Reference Star, {\BD} } \\
As discussed in Appendix~\ref{app:lowzfilters}, the $UBVRI$ 
magnitudes of the primary spectrophotometric reference star, 
\BD, are used in the {\Kcor s} 
to relate the flux calibration of the nearby SN sample 
to that of the higher-redshift samples.
To evaluate the \unc\ in the \BD\ magnitudes,
we compare the consistency of the synthetic magnitudes 
of the HST spectra with the SDSS photometric magnitudes.
We first compute $u,g,r,i,z$ zeropoint offsets as the difference
between synthetic \BD\ magnitudes computed from the
HST-measured spectrum \citep{Bohlin06} 
and the magnitudes measured by \citet{Landolt07b}.
These offsets are then compared to those based on
three solar analogs (P3330E, P177D, P041C).
The differences between the zeropoints offsets are 
$-0.004$, $-0.013$, $0.005$, $0.010$, $0.012$~mag
for $u,g,r,i,z$, respectively.
We therefore assume a 1\% \unc\ in each of the 
$U,B,V,R,I$ magnitudes for \BD. 
To propagate this error, we change
each {\BD} magnitude by 1\% independently in each
passband; for each change, all of the SN light curves are re-fit 
with the \mlcs\ model and cosmological parameters are extracted.
The resulting five independent $w$-shifts are then added in quadrature
and included as a systematic \unc.
The resulting $w$-\uncs\ are 0.02 -- 0.03.

As a crosscheck, we have replaced {\BD} with Vega as the primary 
reference star. 
The resulting changes in $w$ are consistent with the changes
from applying 1\% shifts in the $U,B,V,R,I$ magnitudes of {\BD}. 
The differences between \BD\ and Vega are therefore not 
included as a systematic \unc.

% ===========================================
\bigskip \noindent {\bf Landolt-Bessell Color Transformations} \\
As discussed in Appendix~\ref{app:lowzfilters}, we use 
color transformations to transform between the
Landolt network and the filter system defined by \citet*{Bessell90}.
Table~\ref{tb:color} shows the color terms ($k_i$) between
Landolt magnitudes and synthetic magnitudes using
the \citet*{Bessell90} filters. We vary 
each color term independently by one standard deviation,
and the corresponding changes in $w$ are added in quadrature.
The resulting change is $\dw < 0.01$ for all sample
combinations.

% ===========================================
\bigskip \noindent {\bf Shifted Bessell Filters instead 
    of Color Transformation} \\
As an alternative to the Landolt-Bessell color transformation, 
we consider the approach of \citet*{Astier06}, 
using a modified set of $UBVRI$ filter responses in which the
shapes of the \citet{Bessell90} response curves are held fixed, 
but the central wavelength of each filter passband is shifted 
so that the color terms (Eq.~\ref{eq:kdef}) are zero.
The corresponding wavelength shifts are given in 
Table~\ref{tb:lamshift} under the column ``HST standards.''
Using these wavelength shifts and no color transformations in place 
of the color transformation method of Appendix \ref{app:lowzfilters},
$w$ shifts by  
$\dw = {\WSYSTaFILTSHIFT}$ for the {\samplea} sample.
For the other sample combinations,
$\dw \sim 0.01$.

% ===========================================
\bigskip \noindent {\bf SDSS AB Offsets} \\
As discussed in \S \ref{subsec:SMP}, the \uncs\ in the SDSS AB offsets are 
$\ABoffgerr$, $\ABoffrerr$, $\ABoffierr$~mag
for $g,r,i$, respectively (see Table~\ref{tb:ABoff}).
These \uncs\ include zeropoint \uncs\ of 0.004, 0.004, 0.007~mag
based solely on the consistency of the solar analogs and 
also account for \uncs\ in the
central wavelengths of the SDSS filters:
$\delta \lambda \simeq 7$, $16$, $25$~{\AA} 
for $g,r,i$, respectively.
We independently vary each AB offset 
by one standard deviation,
and the resulting changes in $w$ are added in quadrature.
The resulting change in $w$ is a few hundredths 
for the sample combinations that include \SDSS\ SNe.
Note that the $w$-\unc\ for \samplea\ 
($\dw = \WSYSTaERRABOFF$) is much smaller than that
for the \samplec\ combination ($\dw = \WSYSTcERRABOFF$).
For \samplea, the effect of changing the AB offsets
is mainly to shift all of the distance moduli by the same
amount; the corresponding change in
cosmological parameters is small.
For \samplec, the \SDSS\ distances are shifted relative
to those of the nearby sample, and the abrupt feature 
in the Hubble diagram results in a larger change in the
cosmological parameters.

% ===========================================
\bigskip \noindent {\bf ESSENCE $R-I$ Color Zeropoint} \\
WV07 report an \unc\ of $0.02$~mag in their $R-I$ color zeropoint,
resulting in a systematic \unc\ of $\dw = 0.04$  for their analysis of
the nearby+ESSENCE data sets. 
As a crosscheck, we propagate their $R-I$ \unc\
into our analysis and find $\dw = 0.05$ for the same 
nearby+ESSENCE combination, in reasonable agreement with WV07.
We apply their $R-I$ \unc\ to our sample combinations
and find that the \unc\ in
$w$ is much smaller, $\dw \sim 0.01$ for the combinations
(\bsym,\dsym,{\esym},{\fsym}) that include the ESSENCE sample.
Note that these sample combinations also 
include the SNLS sample, suggesting that the impact of the ESSENCE 
$R-I$ \unc\ is reduced by the presence of another 
high-redshift data sample.

% ===========================================
\bigskip \noindent {\bf SNLS $g,r,i,z$  Zeropoints} \\
\citet{Astier06} report zeropoint \uncs\ of 
$0.01$, $0.01$, $0.01$, $0.03$ for their $g,r,i,z$ passbands,
respectively. In their analysis of the nearby+SNLS combination,
they vary each zeropoint independently and add the corresponding
$w$-shifts in quadrature: the resulting systematic is $\dw = 0.05$.
As a crosscheck, we propagate their zeropoint \uncs\ into
our analysis and also find $\dw = 0.05$
for the same nearby+SNLS combination.
We apply their zeropoint \uncs\ to our sample combinations
and find slightly smaller \uncs\
for the combinations (\bsym,\dsym,{\esym},{\fsym}) 
that include the SNLS sample.

% ================================================
\bigskip \noindent {\bf HST Zeropoints}\\
The HST zeropoint \uncs\ are 0.02~mag for the F110W and F160W 
filters\footnote{See HST Handbook for NICMOS, {\wwwNICMOS}
} % end footnote
(mean wavelengths are $1.14~\mu m$ and $1.61\mu m$, respectively),
and 0.01~mag for the optical 
filters.\footnote{See HST Handbook for ACS,  {\wwwACS}} % end footnote
For the analysis of the five-sample combination that 
includes HST ({\esym}),
the resulting $w$-\unc\ is $\dw = \WSYSTeZPTHST$.

The cosmological parameter shifts due to all of these 
sources of systematic error are added in quadrature 
to derive total systematic error estimates $\delta w$ 
and $\delta \OM$ for the {\mlcs} analysis of the \wCDM\ model; 
these are given in 
Tables~\ref{tb:wsyst_mlcs} and \ref{tb:omsyst_mlcs}.

%
% Systematic error table for w (MLCS)
%

\begin{table*}
\centering
\caption{  
   	Systematic \uncs\ in dark energy equation of state parameter 
	$w$ for the {\mlcs} analysis of the \wCDM\ model, 
	including the BAO+CMB prior.    
	Negative values indicate asymmetric \uncs.
    }
\begin{tabular}{l | cccccc }
\tableline\tableline
% -------------------------
                &  \multicolumn{6}{c}{{\unc} on $w$ for sample:}  \\
source of {\unc}
      & {\asym}
      & {\bsym}
      & {\csym}
      & {\dsym}
      & {\esym}
      & {\fsym}
         \\
\tableline
 \SYSTUband\
       & $-\WSYSTanoU$
       & \WSYSTbnoU
       & $-\WSYSTcnoU$
       & \WSYSTdnoU
       & \WSYSTenoU
       & \WSYSTfnoU
    \\
  \SYSTzmin\
       & \WSYSTaLOWZCUT
       & \WSYSTbLOWZCUT
       & \WSYSTcLOWZCUT
       & \WSYSTdLOWZCUT
       & \WSYSTeLOWZCUT
       & \WSYSTfLOWZCUT
    \\
 \mlcs\ SN~Ia Model Parameters
       & \WSYSTaJRKVECT 
       & \WSYSTbJRKVECT 
       & \WSYSTcJRKVECT 
       & \WSYSTdJRKVECT 
       & \WSYSTeJRKVECT 
       & \WSYSTfJRKVECT   
  \\
 Galactic Extinction
       & \WSYSTaXTMW
       & \WSYSTbXTMW
       & \WSYSTcXTMW
       & \WSYSTdXTMW
       & \WSYSTeXTMW
       & \WSYSTfXTMW
    \\
\hline  ~~~~~~~ FORM OF PRIOR   & & & & & & \\
 correlated $1\sigma$ changes $R_V$ and $\TAUV$
       & \WSYSTaRVSHIFT
       & \WSYSTbRVSHIFT  
       & \WSYSTcRVSHIFT
       & \WSYSTdRVSHIFT
       & \WSYSTeRVSHIFT
       & \WSYSTfRVSHIFT
    \\
 simulated efficiency for nearby SN~Ia
       & \WSYSTaSIMEFFLOWZ
       & \WSYSTbSIMEFFLOWZ
       & \WSYSTcSIMEFFLOWZ
       & \WSYSTdSIMEFFLOWZ
       & \WSYSTeSIMEFFLOWZ
       & \WSYSTfSIMEFFLOWZ
    \\
 \spec\ efficiency for SDSS
       & \WSYSTaSIMEFFSDSS
       & \WSYSTbSIMEFFSDSS
       & \WSYSTcSIMEFFSDSS
       & \WSYSTdSIMEFFSDSS
       & \WSYSTeSIMEFFSDSS
       & \WSYSTfSIMEFFSDSS
    \\
 \spec\ efficiency for ESSENCE
       & \WSYSTaSIMEFFESSE
       & \WSYSTbSIMEFFESSE
       & \WSYSTcSIMEFFESSE
       & \WSYSTdSIMEFFESSE
       & \WSYSTeSIMEFFESSE
       & \WSYSTfSIMEFFESSE
    \\
 \spec\ efficiency for SNLS
       & \WSYSTaSIMEFFSNLS
       & \WSYSTbSIMEFFSNLS
       & \WSYSTcSIMEFFSNLS
       & \WSYSTdSIMEFFSNLS
       & \WSYSTeSIMEFFSNLS
       & \WSYSTfSIMEFFSNLS
    \\
\hline  ~~~~~~~ CALIBRATION   & & & & & & \\
  0.01~mag errors in $U,B,V,R,I$
       & \WSYSTaERRUBVRI
       & \WSYSTbERRUBVRI
       & \WSYSTcERRUBVRI
       & \WSYSTdERRUBVRI
       & \WSYSTeERRUBVRI
       & \WSYSTfERRUBVRI
    \\
 shifted Bessell filters
       & \WSYSTaFILTSHIFT
       & \WSYSTbFILTSHIFT
       & \WSYSTcFILTSHIFT
       & \WSYSTdFILTSHIFT
       & \WSYSTeFILTSHIFT
       & \WSYSTfFILTSHIFT
    \\
 vary $k_i$ color terms
       & \WSYSTaERRki
       & \WSYSTbERRki
       & \WSYSTcERRki
       & \WSYSTdERRki
       & \WSYSTeERRki
       & \WSYSTfERRki
    \\
 vary SDSS AB offsets for $g,r,i$
       & \WSYSTaERRABOFF
       & \WSYSTbERRABOFF
       & \WSYSTcERRABOFF
       & \WSYSTdERRABOFF
       & \WSYSTeERRABOFF
       & \WSYSTfERRABOFF
    \\
 vary ESSENCE $R-I$ color zeropoint
       & \WSYSTaZPTESSE
       & \WSYSTbZPTESSE
       & \WSYSTcZPTESSE
       & \WSYSTdZPTESSE
       & \WSYSTeZPTESSE
       & \WSYSTfZPTESSE
    \\
 vary SNLS $g,r,i,z$ zeropoints
       & \WSYSTaZPTSNLS
       & \WSYSTbZPTSNLS
       & \WSYSTcZPTSNLS
       & \WSYSTdZPTSNLS
       & \WSYSTeZPTSNLS
       & \WSYSTfZPTSNLS
    \\
 vary HST zeropoints
       & \WSYSTaZPTHST
       & \WSYSTbZPTHST
       & \WSYSTcZPTHST
       & \WSYSTdZPTHST
       & \WSYSTeZPTHST
       & \WSYSTfZPTHST
    \\
\tableline
 & & & & & & \\
 Total  
    & \large{${}^{+\MLCSWSYSTERRPa}_{-\MLCSWSYSTERRMa}$}
    & \MLCSWSYSTERRPb
    & \large{${}^{+\MLCSWSYSTERRPc}_{-\MLCSWSYSTERRMc}$}
    & \MLCSWSYSTERRPd
    & \MLCSWSYSTERRPe
    & \MLCSWSYSTERRPf
      \\ 
 & & & & & & \\
\tableline
\end{tabular}
\tablenotetext{\asym}{ \samplea } 
\tablenotetext{\bsym}{ \sampleb }
\tablenotetext{\csym}{ \samplec } 
\tablenotetext{\dsym}{ \sampled }
\tablenotetext{\esym}{ \samplee }
\tablenotetext{\fsym}{ \samplef }
\label{tb:wsyst_mlcs} 
\end{table*}

%
% Systematic error table for OM (MLCS)
%

\begin{table*}
\centering
\caption{  
   	Systematic \uncs\ in matter density parameter $\OM$ for 
	the {\mlcs} analysis of the \wCDM\ model, 
	including the BAO+CMB prior.
	Negative values indicate asymmetric \uncs.
    }
\begin{tabular}{l | cccccc}
\tableline\tableline
% -------------------------
                &  \multicolumn{6}{c}{{\unc} on $\OM$ for sample:}  \\
source of {\unc}
      & {\asym}
      & {\bsym}
      & {\csym}
      & {\dsym}
      & {\esym}
      & {\fsym}
         \\
\tableline
 \SYSTUband\
       & $-\OMSYSTanoU$
       & \OMSYSTbnoU
       & $-\OMSYSTcnoU$
       & \OMSYSTdnoU
       & \OMSYSTenoU
       & \OMSYSTfnoU
    \\
  \SYSTzmin\
       & \OMSYSTaLOWZCUT
       & \OMSYSTbLOWZCUT
       & \OMSYSTcLOWZCUT
       & \OMSYSTdLOWZCUT
       & \OMSYSTeLOWZCUT
       & \OMSYSTfLOWZCUT
    \\
 \mlcs\ SN~Ia Model Parameters
       & \OMSYSTaJRKVECT 
       & \OMSYSTbJRKVECT 
       & \OMSYSTcJRKVECT 
       & \OMSYSTdJRKVECT 
       & \OMSYSTeJRKVECT 
       & \OMSYSTfJRKVECT 
    \\
 Galactic Extinction
       & \OMSYSTaXTMW
       & \OMSYSTbXTMW
       & \OMSYSTcXTMW
       & \OMSYSTdXTMW
       & \OMSYSTeXTMW
       & \OMSYSTfXTMW
    \\
\hline  ~~~~~~~ FORM OF PRIOR   & & & & & & \\
 correlated $1\sigma$ changes $R_V$ and $\TAUV$
       & \OMSYSTaRVSHIFT
       & \OMSYSTbRVSHIFT  
       & \OMSYSTcRVSHIFT
       & \OMSYSTdRVSHIFT
       & \OMSYSTeRVSHIFT
       & \OMSYSTfRVSHIFT
    \\
 simulated efficiency for nearby SN~Ia
       & \OMSYSTaSIMEFFLOWZ
       & \OMSYSTbSIMEFFLOWZ
       & \OMSYSTcSIMEFFLOWZ
       & \OMSYSTdSIMEFFLOWZ
       & \OMSYSTeSIMEFFLOWZ
       & \OMSYSTfSIMEFFLOWZ
    \\
 \spec\ efficiency for SDSS
       & \OMSYSTaSIMEFFSDSS
       & \OMSYSTbSIMEFFSDSS
       & \OMSYSTcSIMEFFSDSS
       & \OMSYSTdSIMEFFSDSS
       & \OMSYSTeSIMEFFSDSS
       & \OMSYSTfSIMEFFSDSS
    \\
 \spec\ efficiency for ESSENCE
       & \OMSYSTaSIMEFFESSE
       & \OMSYSTbSIMEFFESSE
       & \OMSYSTcSIMEFFESSE
       & \OMSYSTdSIMEFFESSE
       & \OMSYSTeSIMEFFESSE
       & \OMSYSTfSIMEFFESSE
    \\
 \spec\ efficiency for SNLS
       & \OMSYSTaSIMEFFSNLS
       & \OMSYSTbSIMEFFSNLS
       & \OMSYSTcSIMEFFSNLS
       & \OMSYSTdSIMEFFSNLS
       & \OMSYSTeSIMEFFSNLS
       & \OMSYSTfSIMEFFSNLS
    \\
% -------------------------------
\hline  ~~~~~~~ CALIBRATION   & & & & & & \\
  0.01~mag errors in $U,B,V,R,I$
       & \OMSYSTaERRUBVRI
       & \OMSYSTbERRUBVRI
       & \OMSYSTcERRUBVRI
       & \OMSYSTdERRUBVRI
       & \OMSYSTeERRUBVRI
       & \OMSYSTfERRUBVRI
    \\
 shifted Bessell filters
       & \OMSYSTaFILTSHIFT
       & \OMSYSTbFILTSHIFT
       & \OMSYSTcFILTSHIFT
       & \OMSYSTdFILTSHIFT
       & \OMSYSTeFILTSHIFT
       & \OMSYSTfFILTSHIFT
    \\
 vary $k_i$ color terms
       & \OMSYSTaERRki
       & \OMSYSTbERRki
       & \OMSYSTcERRki
       & \OMSYSTdERRki
       & \OMSYSTeERRki
       & \OMSYSTfERRki
    \\
 vary SDSS AB offsets for $g,r,i$
       & \OMSYSTaERRABOFF
       & \OMSYSTbERRABOFF
       & \OMSYSTcERRABOFF
       & \OMSYSTdERRABOFF
       & \OMSYSTeERRABOFF
       & \OMSYSTfERRABOFF
    \\
 vary ESSENCE $R-I$ color zeropoint
       & \OMSYSTaZPTESSE
       & \OMSYSTbZPTESSE
       & \OMSYSTcZPTESSE
       & \OMSYSTdZPTESSE
       & \OMSYSTeZPTESSE
       & \OMSYSTfZPTESSE
    \\
 vary SNLS $g,r,i,z$ zeropoints
       & \OMSYSTaZPTSNLS
       & \OMSYSTbZPTSNLS
       & \OMSYSTcZPTSNLS
       & \OMSYSTdZPTSNLS
       & \OMSYSTeZPTSNLS
       & \OMSYSTfZPTSNLS
    \\
 vary HST zeropoints
       & \OMSYSTaZPTHST
       & \OMSYSTbZPTHST
       & \OMSYSTcZPTHST
       & \OMSYSTdZPTHST
       & \OMSYSTeZPTHST
       & \OMSYSTfZPTHST
    \\
\tableline
  & & & & & & \\
 Total                     
    & \large{${}^{+\MLCSOMSYSTERRPa}_{-\MLCSOMSYSTERRMa}$}
    & \MLCSOMSYSTERRPb
    & \large{${}^{+\MLCSOMSYSTERRPc}_{-\MLCSOMSYSTERRMc}$}
    & \MLCSOMSYSTERRPd
    & \MLCSOMSYSTERRPe
    & \MLCSOMSYSTERRPf
      \\ 
  & & & & & & \\
\tableline
\end{tabular}
%
%\tablenotetext{\asym}{ \samplea } 
%\tablenotetext{\bsym}{ \sampleb }
%\tablenotetext{\csym}{ \samplec } 
%\tablenotetext{\dsym}{ \sampled }
%\tablenotetext{\esym}{ \samplee }
%\tablenotetext{\fsym}{ \samplef }
%
\label{tb:omsyst_mlcs} 
\end{table*}

% ------------------------------------

\subsection{Systematics Uncertainties with {\SALTII}}
\label{subsec:syst_salt2}
%
% SALT2 Systematic errors on w.
%

To examine systematic \uncs\ in the context of the \SALTII\ model, 
we undertake an analysis similar to that carried out for \mlcs.
We also determine systematic \uncs\ for the \SALTII\
parameters $\alpha$ and $\beta$ that enter Eq. \ref{eq:MUSALTII}.
Since the \SALTII\ training software is not available
for public use, with one exception we make approximations in cases 
where re-training of the spectral surfaces is needed.
In such cases 
we either use the nominal \SALTII\ surfaces
and propagate changes only in the light-curve fits,
or we use the \uncs\ based on the \mlcs\ analysis.
The systematic \uncs\ in $w$ and $\OM$ for the 
different sample combinations are given in 
Tables \ref{tb:wsyst_salt2_FWCDM_BAOCMB} 
and \ref{tb:omegamsyst_salt2_FWCDM_BAOCMB}.

% =============================================
\bigskip \noindent {\bf \SYSTUband} \\
As discussed in detail in \S~\ref{subsec:Uanom_SALT2}, 
there are systematic discrepancies between the \SALTII\ 
rest-frame $U$-band model and the observer-frame 
$U$-band light-curve data for the nearby SN~Ia sample. 
As was noted in \S~\ref{subsec:syst_mlcs}, a related 
issue is seen for {\mlcs}.
We carry out a test of the \SALTII\ fits in which the observer-frame 
filter corresponding to rest-frame $U$-band is excluded from the 
fits. Figure~\ref{fig:noUtest_SALT2} shows that for the \samplea\ 
sample, the exclusion of $U$-band results in a redshift-dependent 
change in the distance modulus. This results in a $w$-shift of 
$\FWCDMBAOCMBSYSTUBANDWa$ for the \samplea\ sample and 
$\FWCDMBAOCMBSYSTUBANDWc$ for the {\samplec} combination;
these shifts are included as asymmetric systematic \uncs.

For the other sample combinations, which include the 
higher-redshift ESSENCE \& SNLS samples (\bsym,\dsym,\esym,{\fsym}), 
the exclusion of rest-frame $U$-band results in 
$w$-shifts of .04 to 0.09.
Based on tests with simulations, we cannot distinguish these shifts 
from random fluctuations;
we therefore include the largest shift, $\dw = 0.09$, as a
symmetric systematic \unc\ for these four sample combinations.
If observer-frame $U$-band is excluded from the nearby sample,
while rest-frame $U$-band is included in the higher-redshift
samples, the change in $w$ is no more than 0.01.

% =============================================
\bigskip \noindent {\bf Minimum Redshift for Nearby SN sample } \\
The dependence of the \SALTII\ results for $w$ on the $\ZMINSYM$ cut is shown
in the third and fourth panels of Fig.~\ref{fig:zminscan}. 
For $\ZMINSYM \gtrsim 0.018$, the inferred value of $w$ is fairly insensitive 
to the value of $\ZMINSYM$.
However, when $\ZMINSYM$ is reduced to 0.015, 
$w$ changes by almost 0.09 for the \samplec\ combination,
and by 0.04 -- 0.05 for the combinations that include
the higher-redshift (ESSENCE \& SNLS) samples.
To account for these variations, we have
assigned a systematic \unc\ of $\dw = 0.05$ for the 
\samplec\ combination and $\dw = 0.03$ for the other 
sample combinations that include the nearby SN~Ia sample.

% ===========================================

\noindent {\bf Galactic Extinction} \\
The systematic \unc\ from Galactic extinction is
described in \S~\ref{subsec:syst_mlcs},
resulting in $\dw \sim 0.02$.

% =============================================
\bigskip \noindent {\bf \SALTII\ Training with \SDSS\ Data} \\
Here we examine the \SALTII\ training process that produces 
the spectral surfaces $M_0(t,\lambda)$ and $M_1(t,\lambda)$ 
in Eq. \ref{eq:SALTII_flux_rest}.  
Because \SDSS\ SN probes a relatively unexplored range of SN 
redshifts, the rest--frame behavior of the \SDSS\
light curves may not be as well described by the \SALTII\ model 
as that of other SN samples that were   
used in the training of the model.
To quantify this issue, J. Guy has retrained the \SALTII\ spectral surfaces
twice, first including the light curves of the SDSS \spec\ sample 
and second including those of the SDSS dust sample 
(\S~\ref{subsec:dustsample}).
Evaluating cosmological parameters obtained with 
each retrained set of spectral surfaces and comparing the 
results with those from the standard \SALTII\ training,
we include the larger of the two $w$-shifts as a systematic \unc.
For the \samplea\ and \samplec\ ({\asym} \& {\csym}) sample combinations,
the \unc\ is $\dw \sim 0.02$.
For the other combinations, $\dw \sim 0.01$.

% =============================================
\bigskip \noindent {\bf \SALTII\ Dispersions} \\
Recall from \S~\ref{subsec:SALTII} that the spectral
surfaces, $M_0(t,\lambda)$ and $M_1(t,\lambda)$,
were retrained using the Bessell filter shifts
based on HST standards (Table~\ref{tb:lamshift}).
The model dispersions around these surfaces, however,
were not determined in the retraining, and we
therefore use the model dispersions from \citet*{Guy07}.
To allow for the resulting uncertainty in the dispersions, 
we assign a systematic \unc\ of half the difference
between using and ignoring the dispersions.
The resulting \uncs\ are $\dw \sim 0.01$ to 0.02
for combinations that include the higher-redshift
ESSENCE \& SNLS samples.
For the \samplea\ and \samplec\ sample combinations, 
the $w$-\unc\ is negligible.

% =============================================
\bigskip\noindent {\bf $\beta$-Variation with Redshift} \\
If the \SALTII\ SN parameters ($\alpha$,$\beta$,$M$) are allowed 
to vary independently in redshift bins, while the cosmological
parameters are fixed, we find a strong redshift-dependence 
of $\beta$ for $z>0.6$ (see \S~\ref{subsec:SALT2z}).
To estimate the corresponding systematic \unc,
the Hubble diagram fits have been redone
allowing $\alpha$, $\beta$, and $M$ to vary with redshift
using a simple model in which each SN parameter is constant
for $z<0.6$, and is then allowed to vary linearly with
redshift for $z>0.6$.
Compared to the nominal \SALTII\ model with 
redshift-independent parameters, the largest 
change, $\dw = \FWCDMBAOCMBSYSTBETAZWb$,
occurs for \sampleb\ ({\bsym}) in which
the nearby sample is excluded.
For sample-combinations {\dsym} and {\fsym},
$\dw \sim 0.04$, and for the full SN set ({\esym})
that includes the HST, $\dw \sim 0.01$.
These $w$-shifts are included as asymmetric
systematic \uncs.

% =============================================

\bigskip \noindent {\bf Simulated Bias Correction} \\
For \SALTII, we have determined bias corrections 
from simulations, as described in \S~\ref{subsec:wfit_SALTII} 
(see Table~\ref{tb:SALT2_simcor}).
We include half the $w$-shift as a systematic \unc.
The largest \unc\ is $\dw = \FWCDMBAOCMBSYSTSImSNLSWa$
for \samplea.

% =============================================
\bigskip \noindent {\bf Calibration of Primary Reference Star, Vega} \\
We assume uncertainties of 0.01 magnitudes in the calibration of 
$U,B,V,R,I$ for the primary reference, Vega. 
Since a full accounting of this effect would require 
another retraining of the \SALTII\ surfaces, 
we instead adopt the \uncs\ derived from the \mlcs\ analysis
(Table~\ref{tb:wsyst_mlcs}).
In \citet*{Astier06}, the corresponding \unc\ is $\dw = 0.024$
for the nearby+SNLS combination;
as a crosscheck, we have evaluated this \unc\ 
for the same sample combination and find good agreement, 
$\dw \simeq 0.021$.
For the sample combinations analyzed here,
the resulting $w$-\uncs\ are 0.02 -- 0.03.

% =============================================
\bigskip \noindent {\bf Calibration: 
     Shifted Bessell Filters for Nearby Data} \\
As discussed in \S~\ref{subsec:SALTII},
we use the \citet*{Bessell90} filter responses with
wavelength shifts given in Table~\ref{tb:lamshift}
of Appendix~\ref{app:lowzfilters}.
Since these shifts differ from those in \citet*{Astier06}, 
we use the difference in cosmological results derived from both sets
of wavelength shifts to define an additional systematic \unc\ on $w$.
This \unc\ is $\dw \sim 0.02$ for sample combinations that
include the nearby sample.

% =============================================
\bigskip \noindent {\bf Calibration: 
	Zeropoint offsets for SDSS, ESSENCE, SNLS, HST} \\
Zeropoint \uncs\ for the SDSS, ESSENCE, SNLS, and HST bandpasses
are propagated in the same manner
as for the \mlcs\ method (\S~\ref{subsec:syst_mlcs}).
The SNLS zeropoints are varied in the fit,
but not in the training
of the spectral surface,
and therefore these
$w$-\uncs\ might be slightly overestimated.
The other survey samples were not used in the training,
and therefore varying the zeropoints in the fit
is sufficient to estimate the systematic error.
Note that the $w$-\unc\ from the SDSS AB offsets is smaller
for \samplea\ than for \samplec,
as explained in \S~\ref{subsec:syst_mlcs}.

The parameter shifts due to all of these systematic errors are 
added in quadrature to derive total systematic error estimates  
in Tables \ref{tb:wsyst_salt2_FWCDM_BAOCMB} and 
\ref{tb:omegamsyst_salt2_FWCDM_BAOCMB}.

\begin{table*}
\centering
\caption{
     	Systematic uncertainties in $w$ for the \SALTII\ analysis 
     	of the \wCDM\ model, including the BAO+CMB prior.
	$+/-$ values indicate asymmetric \uncs.
      }
\label{tb:wsyst_salt2_FWCDM_BAOCMB}
\begin{tabular}{l | cccccc}
\tableline\tableline
% -------------------------
                   &  \multicolumn{5}{c}{Uncertainty on $w$ for Sample:}  \\
 Source of Uncertainty
      & {{\asym}}
      & {{\bsym}}
      & {{\csym}}
      & {{\dsym}}
      & {{\esym}}
      & {{\fsym}}
         \\
\tableline
\SYSTUband
   & $\FWCDMBAOCMBSYSTUBANDWa$
   & $\FWCDMBAOCMBSYSTUBANDWb$
   & $\FWCDMBAOCMBSYSTUBANDWc$
   & $\FWCDMBAOCMBSYSTUBANDWd$
   & $\FWCDMBAOCMBSYSTUBANDWe$
   & $\FWCDMBAOCMBSYSTUBANDWf$
\\
\SYSTzmin
   & $\FWCDMBAOCMBSYSTZMINWa$
   & $\FWCDMBAOCMBSYSTZMINWb$
   & $\FWCDMBAOCMBSYSTZMINWc$
   & $\FWCDMBAOCMBSYSTZMINWd$
   & $\FWCDMBAOCMBSYSTZMINWe$
   & $\FWCDMBAOCMBSYSTZMINWf$
\\
Galactic Extinction
   & $\FWCDMBAOCMBSYSTXTMWWa$
   & $\FWCDMBAOCMBSYSTXTMWWb$
   & $\FWCDMBAOCMBSYSTXTMWWc$
   & $\FWCDMBAOCMBSYSTXTMWWd$
   & $\FWCDMBAOCMBSYSTXTMWWe$
   & $\FWCDMBAOCMBSYSTXTMWWf$
\\
\hline  \SALTII\ SN Ia MODEL PARAMETERS    & & & & & \\
 retraining : include SDSS data
   & $\FWCDMBAOCMBSYSTRTDWa$
   & $\FWCDMBAOCMBSYSTRTDWb$
   & $\FWCDMBAOCMBSYSTRTDWc$
   & $\FWCDMBAOCMBSYSTRTDWd$
   & $\FWCDMBAOCMBSYSTRTDWe$
   & $\FWCDMBAOCMBSYSTRTDWf$
\\
 dispersions of SALT-II surfaces
   & $\FWCDMBAOCMBSYSTDISPWa$
   & $\FWCDMBAOCMBSYSTDISPWb$
   & $\FWCDMBAOCMBSYSTDISPWc$
   & $\FWCDMBAOCMBSYSTDISPWd$
   & $\FWCDMBAOCMBSYSTDISPWe$
   & $\FWCDMBAOCMBSYSTDISPWf$
\\
 $\beta$-variation with redshift
   & $\FWCDMBAOCMBSYSTBETAZWa$
   & $\FWCDMBAOCMBSYSTBETAZWb$
   & $\FWCDMBAOCMBSYSTBETAZWc$
   & $\FWCDMBAOCMBSYSTBETAZWd$
   & $\FWCDMBAOCMBSYSTBETAZWe$
   & $\FWCDMBAOCMBSYSTBETAZWf$
\\
\hline  SELECTION EFFICIENCY    & & & & & \\
 simulated bias
   & $\FWCDMBAOCMBSYSTSImSNLSWa$
   & $\FWCDMBAOCMBSYSTSImSNLSWb$
   & $\FWCDMBAOCMBSYSTSImSNLSWc$
   & $\FWCDMBAOCMBSYSTSImSNLSWd$
   & $\FWCDMBAOCMBSYSTSImSNLSWe$
   & $\FWCDMBAOCMBSYSTSImSNLSWf$
\\
\hline  CALIBRATION    & & & & & \\
 0.01~mag errors in $U,B,V,R,I$
   & $\FWCDMBAOCMBSYSTLOWZZPTWa$
   & $\FWCDMBAOCMBSYSTLOWZZPTWb$
   & $\FWCDMBAOCMBSYSTLOWZZPTWc$
   & $\FWCDMBAOCMBSYSTLOWZZPTWd$
   & $\FWCDMBAOCMBSYSTLOWZZPTWe$
   & $\FWCDMBAOCMBSYSTLOWZZPTWf$
\\
 shifted Bessel90 filters
   & $\FWCDMBAOCMBSYSTBESSWa$
   & $\FWCDMBAOCMBSYSTBESSWb$
   & $\FWCDMBAOCMBSYSTBESSWc$
   & $\FWCDMBAOCMBSYSTBESSWd$
   & $\FWCDMBAOCMBSYSTBESSWe$
   & $\FWCDMBAOCMBSYSTBESSWf$
\\
 vary SDSS AB offsets for $g,r,i$
   & $\FWCDMBAOCMBSYSTABWa$
   & $\FWCDMBAOCMBSYSTABWb$
   & $\FWCDMBAOCMBSYSTABWc$
   & $\FWCDMBAOCMBSYSTABWd$
   & $\FWCDMBAOCMBSYSTABWe$
   & $\FWCDMBAOCMBSYSTABWf$
\\
 vary ESSENCE $R-I$ color zeropoint
   & $\FWCDMBAOCMBSYSTESSENCEZPTWa$
   & $\FWCDMBAOCMBSYSTESSENCEZPTWb$
   & $\FWCDMBAOCMBSYSTESSENCEZPTWc$
   & $\FWCDMBAOCMBSYSTESSENCEZPTWd$
   & $\FWCDMBAOCMBSYSTESSENCEZPTWe$
   & $\FWCDMBAOCMBSYSTESSENCEZPTWf$
\\
 vary SNLS $g,r,i,z$ zeropoints
   & $\FWCDMBAOCMBSYSTSNLSZPTWa$
   & $\FWCDMBAOCMBSYSTSNLSZPTWb$
   & $\FWCDMBAOCMBSYSTSNLSZPTWc$
   & $\FWCDMBAOCMBSYSTSNLSZPTWd$
   & $\FWCDMBAOCMBSYSTSNLSZPTWe$
   & $\FWCDMBAOCMBSYSTSNLSZPTWf$
\\
 vary HST zeropoints
   & $\FWCDMBAOCMBSYSTHSTZPTWa$
   & $\FWCDMBAOCMBSYSTHSTZPTWb$
   & $\FWCDMBAOCMBSYSTHSTZPTWc$
   & $\FWCDMBAOCMBSYSTHSTZPTWd$
   & $\FWCDMBAOCMBSYSTHSTZPTWe$
   & $\FWCDMBAOCMBSYSTHSTZPTWf$
\\
\hline 
 & & & & & & \\
  Total
  & \Large{${}^{+\FWCDMBAOCMBDWSYSTPa}_{-\FWCDMBAOCMBDWSYSTMa}$}
  & \Large{${}^{+\FWCDMBAOCMBDWSYSTPb}_{-\FWCDMBAOCMBDWSYSTMb}$}
  & \Large{${}^{+\FWCDMBAOCMBDWSYSTPc}_{-\FWCDMBAOCMBDWSYSTMc}$}
  & \Large{${}^{+\FWCDMBAOCMBDWSYSTPd}_{-\FWCDMBAOCMBDWSYSTMd}$}
  & \Large{${}^{+\FWCDMBAOCMBDWSYSTPe}_{-\FWCDMBAOCMBDWSYSTMe}$}
  & \Large{${}^{+\FWCDMBAOCMBDWSYSTPf}_{-\FWCDMBAOCMBDWSYSTMf}$}
\\
 & & & & & & \\
\tableline
\tableline
\end{tabular}
\tablenotetext{{\asym}}{ \samplea }
\tablenotetext{{\bsym}}{ \sampleb }
\tablenotetext{{\csym}}{ \samplec }
\tablenotetext{{\dsym}}{ \sampled }
\tablenotetext{{\esym}}{ \samplee }
\tablenotetext{{\fsym}}{ \samplef }
\end{table*}

\begin{table*}
\centering
\caption{
     	Systematic uncertainties in $\OM$ for the  \SALTII\ 
	analysis of the \wCDM\ model, including the BAO+CMB prior.
	$+/-$ values indicate asymmetric \uncs.
      }
\label{tb:omegamsyst_salt2_FWCDM_BAOCMB}
\begin{tabular}{l | cccccc}
\tableline\tableline
% -------------------------
                   &  \multicolumn{5}{c}{Uncertainty on $\OM$ for Sample:}  \\
 Source of Uncertainty
      & {{\asym}}
      & {{\bsym}}
      & {{\csym}}
      & {{\dsym}}
      & {{\esym}}
      & {{\fsym}}
         \\
\tableline
\SYSTUband
  & $\FWCDMBAOCMBSYSTUBANDOMEGAMa$
  & $\FWCDMBAOCMBSYSTUBANDOMEGAMb$
  & $\FWCDMBAOCMBSYSTUBANDOMEGAMc$
  & $\FWCDMBAOCMBSYSTUBANDOMEGAMd$
  & $\FWCDMBAOCMBSYSTUBANDOMEGAMe$
  & $\FWCDMBAOCMBSYSTUBANDOMEGAMf$
\\
\SYSTzmin
  & $\FWCDMBAOCMBSYSTZMINOMEGAMa$
  & $\FWCDMBAOCMBSYSTZMINOMEGAMb$
  & $\FWCDMBAOCMBSYSTZMINOMEGAMc$
  & $\FWCDMBAOCMBSYSTZMINOMEGAMd$
  & $\FWCDMBAOCMBSYSTZMINOMEGAMe$
  & $\FWCDMBAOCMBSYSTZMINOMEGAMf$
\\
Galactic Extinction
  & $\FWCDMBAOCMBSYSTXTMWOMEGAMa$
  & $\FWCDMBAOCMBSYSTXTMWOMEGAMb$
  & $\FWCDMBAOCMBSYSTXTMWOMEGAMc$
  & $\FWCDMBAOCMBSYSTXTMWOMEGAMd$
  & $\FWCDMBAOCMBSYSTXTMWOMEGAMe$
  & $\FWCDMBAOCMBSYSTXTMWOMEGAMf$
\\
\hline  \SALTII\ SN Ia MODEL PARAMETERS    & & & & & \\
 retraining : include SDSS data
  & $\FWCDMBAOCMBSYSTRTDOMEGAMa$
  & $\FWCDMBAOCMBSYSTRTDOMEGAMb$
  & $\FWCDMBAOCMBSYSTRTDOMEGAMc$
  & $\FWCDMBAOCMBSYSTRTDOMEGAMd$
  & $\FWCDMBAOCMBSYSTRTDOMEGAMe$
  & $\FWCDMBAOCMBSYSTRTDOMEGAMf$
\\
 dispersions of SALT-II surfaces
  & $\FWCDMBAOCMBSYSTDISPOMEGAMa$
  & $\FWCDMBAOCMBSYSTDISPOMEGAMb$
  & $\FWCDMBAOCMBSYSTDISPOMEGAMc$
  & $\FWCDMBAOCMBSYSTDISPOMEGAMd$
  & $\FWCDMBAOCMBSYSTDISPOMEGAMe$
  & $\FWCDMBAOCMBSYSTDISPOMEGAMf$
\\
 $\beta$-variation with redshift
 & $\FWCDMBAOCMBSYSTBETAZOMEGAMa$
 & $\FWCDMBAOCMBSYSTBETAZOMEGAMb$
 & $\FWCDMBAOCMBSYSTBETAZOMEGAMc$
 & $\FWCDMBAOCMBSYSTBETAZOMEGAMd$
 & $\FWCDMBAOCMBSYSTBETAZOMEGAMe$
 & $\FWCDMBAOCMBSYSTBETAZOMEGAMf$
\\
\hline  SELECTION EFFICIENCY    & & & & & \\
 simulated bias
  & $\FWCDMBAOCMBSYSTSImSNLSOMEGAMa$
  & $\FWCDMBAOCMBSYSTSImSNLSOMEGAMb$
  & $\FWCDMBAOCMBSYSTSImSNLSOMEGAMc$
  & $\FWCDMBAOCMBSYSTSImSNLSOMEGAMd$
  & $\FWCDMBAOCMBSYSTSImSNLSOMEGAMe$
  & $\FWCDMBAOCMBSYSTSImSNLSOMEGAMf$
\\
\hline   CALIBRATION    & & & & & \\
 0.01~mag errors in $U,B,V,R,I$
  & $\FWCDMBAOCMBSYSTLOWZZPTOMEGAMa$
  & $\FWCDMBAOCMBSYSTLOWZZPTOMEGAMb$
  & $\FWCDMBAOCMBSYSTLOWZZPTOMEGAMc$
  & $\FWCDMBAOCMBSYSTLOWZZPTOMEGAMd$
  & $\FWCDMBAOCMBSYSTLOWZZPTOMEGAMe$
  & $\FWCDMBAOCMBSYSTLOWZZPTOMEGAMf$
\\
 shifted Bessel90 filters
  & $\FWCDMBAOCMBSYSTBESSOMEGAMa$
  & $\FWCDMBAOCMBSYSTBESSOMEGAMb$
  & $\FWCDMBAOCMBSYSTBESSOMEGAMc$
  & $\FWCDMBAOCMBSYSTBESSOMEGAMd$
  & $\FWCDMBAOCMBSYSTBESSOMEGAMe$
  & $\FWCDMBAOCMBSYSTBESSOMEGAMf$
\\
 vary SDSS AB offsets for $g,r,i$
  & $\FWCDMBAOCMBSYSTABOMEGAMa$
  & $\FWCDMBAOCMBSYSTABOMEGAMb$
  & $\FWCDMBAOCMBSYSTABOMEGAMc$
  & $\FWCDMBAOCMBSYSTABOMEGAMd$
  & $\FWCDMBAOCMBSYSTABOMEGAMe$
  & $\FWCDMBAOCMBSYSTABOMEGAMf$
\\
 vary ESSENCE $R-I$ color zeropoint
  & $\FWCDMBAOCMBSYSTESSENCEZPTOMEGAMa$
  & $\FWCDMBAOCMBSYSTESSENCEZPTOMEGAMb$
  & $\FWCDMBAOCMBSYSTESSENCEZPTOMEGAMc$
  & $\FWCDMBAOCMBSYSTESSENCEZPTOMEGAMd$
  & $\FWCDMBAOCMBSYSTESSENCEZPTOMEGAMe$
  & $\FWCDMBAOCMBSYSTESSENCEZPTOMEGAMf$
\\
 vary SNLS $g,r,i,z$ zeropoints
  & $\FWCDMBAOCMBSYSTSNLSZPTOMEGAMa$
  & $\FWCDMBAOCMBSYSTSNLSZPTOMEGAMb$
  & $\FWCDMBAOCMBSYSTSNLSZPTOMEGAMc$
  & $\FWCDMBAOCMBSYSTSNLSZPTOMEGAMd$
  & $\FWCDMBAOCMBSYSTSNLSZPTOMEGAMe$
  & $\FWCDMBAOCMBSYSTSNLSZPTOMEGAMf$
\\
 vary HST zeropoints
  & $\FWCDMBAOCMBSYSTHSTZPTOMEGAMa$
  & $\FWCDMBAOCMBSYSTHSTZPTOMEGAMb$
  & $\FWCDMBAOCMBSYSTHSTZPTOMEGAMc$
  & $\FWCDMBAOCMBSYSTHSTZPTOMEGAMd$
  & $\FWCDMBAOCMBSYSTHSTZPTOMEGAMe$
  & $\FWCDMBAOCMBSYSTHSTZPTOMEGAMf$
\\
\hline  
 & & & & & & \\
  Total
  & \Large{${}^{+\FWCDMBAOCMBDOMEGAMSYSTPa}_{-\FWCDMBAOCMBDOMEGAMSYSTMa}$}
  & \Large{${}^{+\FWCDMBAOCMBDOMEGAMSYSTPb}_{-\FWCDMBAOCMBDOMEGAMSYSTMb}$}
  & \Large{${}^{+\FWCDMBAOCMBDOMEGAMSYSTPc}_{-\FWCDMBAOCMBDOMEGAMSYSTMc}$}
  & \Large{${}^{+\FWCDMBAOCMBDOMEGAMSYSTPd}_{-\FWCDMBAOCMBDOMEGAMSYSTMd}$}
  & \Large{${}^{+\FWCDMBAOCMBDOMEGAMSYSTPe}_{-\FWCDMBAOCMBDOMEGAMSYSTMe}$}
  & \Large{${}^{+\FWCDMBAOCMBDOMEGAMSYSTPf}_{-\FWCDMBAOCMBDOMEGAMSYSTMf}$}
\\
 & & & & & & \\
\tableline
\tableline
\end{tabular}
%
%\tablenotetext{{\asym}}{ \samplea }
%\tablenotetext{{\bsym}}{ \sampleb }
%\tablenotetext{{\csym}}{ \samplec }
%\tablenotetext{{\dsym}}{ \sampled }
%\tablenotetext{{\esym}}{ \samplee }
%\tablenotetext{{\fsym}}{ \samplef }
%
\end{table*}

% #########################################

\section{Results}
\label{sec:results}

Here we present the Hubble diagram 
and inferred cosmological parameters using the
framework described in \S \ref{sec:wfit}. 
Results based on the \mlcs\ and \SALTII\ methods are presented in
\S \ref{subsec:results_mlcs} and 
\ref{subsec:results_salt2} respectively. We compare 
the results from the two methods in
\S~\ref{sec:results_compare}.

\begin{figure}[h]
  \epsscale{1.1}
  \plotone{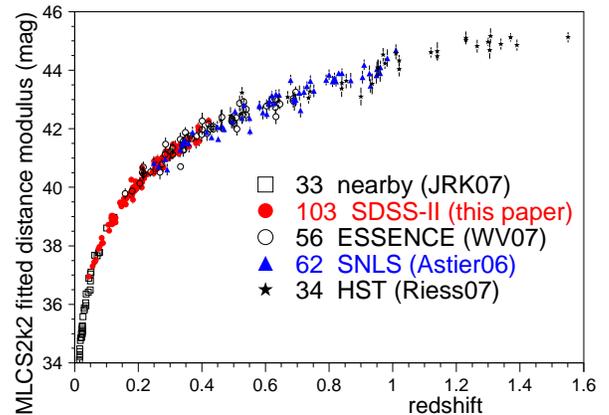}
  \caption{
      Fitted distance modulus (from {\mlcs}) versus redshift for
      the $\NSNTOT$ SNe~Ia from 
      the five samples indicated on the plot. 
        }
  \label{fig:hubble_mlcs}
\end{figure}

The Hubble diagram for the five samples considered in 
this analysis is shown in Fig.~\ref{fig:hubble_mlcs}.
The distance moduli here are obtained from the \mlcs\ 
method described above; the Hubble diagram based on the \SALTII\ method
looks quite similar. Detailed Hubble-residual plots are given 
for each method in \S \ref{subsec:results_mlcs} and 
\ref{subsec:results_salt2}.

% ---------------------------------------
\subsection{Results with {\mlcs}}
\label{subsec:results_mlcs}

Using the \mlcs\ method, we present cosmological results for the 
six sample combinations ({\asym}--{\fsym}) of 
Table~\ref{tb:sample_combos}. 
Table~\ref{tb:mlcs_lcfitres} gives the \specy\ determined redshift 
and marginalized \mlcs\ fit parameters for SNe that pass the 
selection cuts described in \S \ref{sec:sample}.
We use the ensemble of redshifts $z_i$ and estimated 
distance moduli $\mu_i$ for 
each sample combination to fit cosmological 
model parameters, as explained in \S \ref{subsec:wfit_mlcs}.

\begin{table}[hb]
\centering
\caption{  
	Parameters from \mlcs\ light curve fits
	(uncertainties in parentheses). 
	The complete table for all $\NSNTOT$ SNe 
	is given in electronic form in the journal,
	and also at {\wwwTABLES}.
}  % end caption
\tiny
\begin{tabular}{l cccccc}
\tableline 
 SNID  & redshift\tablenotemark{a} & $\mu$ & 
 $A_V$ & $\Delta$ 
       &  MJD$_{\rm peak}$ \\  %%  & $\chi^2/N_{\rm dof}$  \\ 
\tableline  % --------- 
\scriptsize
     762 & $0.1904(.0001)$ & $ 40.05( 0.10)$ & $  0.20( 0.08)$ & $ -0.22( 0.07)$ & $53624.4(  0.4)$ \\ % & $ 34.30/ 48$ \\ 
  1032 & $0.1291(.0002)$ & $ 38.80( 0.10)$ & $  0.07( 0.06)$ & $  0.88( 0.09)$ & $53626.9(  0.2)$ \\ % & $ 54.30/ 52$ \\ 
  1112 & $0.2565(.0002)$ & $ 40.84( 0.18)$ & $  0.13( 0.09)$ & $  0.05( 0.16)$ & $53629.3(  1.0)$ \\ % & $ 46.60/ 53$ \\ 
  1166 & $0.3813(.0005)$ & $ 41.51( 0.18)$ & $  0.16( 0.11)$ & $ -0.25( 0.13)$ & $53630.1(  1.1)$ \\ % & $ 34.60/ 48$ \\ 
  1241 & $0.0858(.0050)$ & $ 38.10( 0.09)$ & $  0.40( 0.06)$ & $  0.02( 0.06)$ & $53634.7(  0.2)$ \\ % & $ 25.50/ 51$ \\ 
  1253 & $0.2609(.0050)$ & $ 40.65( 0.14)$ & $  0.06( 0.05)$ & $  0.08( 0.12)$ & $53634.2(  0.5)$ \\ % & $ 38.90/ 45$ \\ 

\tableline  % ------------------------ 
\end{tabular} 
\tablenotetext{a}{Spectroscopic redshift in CMB frame.}
\label{tb:mlcs_lcfitres}  
\end{table}

% -----------------------------------------------------------------
\subsubsection{Goodness of Fit and Hubble scatter}
\label{subsec:GOF}

Before considering the cosmological parameter results for the various 
combined samples, we examine several measures of fit quality and 
dispersion for each SN~Ia sample treated independently, 
since they provide diagnostic information that is useful to 
consider before combining the samples. 
Table~\ref{tb:mlcs_fitquality} displays these measures:
(i) the $\chisqmu$ statistic for the best-fit \wCDM\ model 
for that sample from Eq.~\ref{eq:mlcs_chisqmu} --- 
in goodness-of-fit tests, 
this statistic is usually compared to 
the number of degrees of freedom, 
given by $N_{\rm dof} = N_{SN}-1$;\footnote{$N_{\rm dof}=$ number of 
SNe minus the number of cosmology parameters 
($H_0$, $w$, $\OM$) + the number of priors (BAO+CMB).
} % end footnote
(ii)
the root-mean-square measure of Hubble scatter,  
$\RMSMU = \sqrt{\sum_i (\mu_i^{\rm fit}-\mu_i)^2}$, 
where $\mu_i$ is the estimated distance modulus 
from the light-curve fit for the $i$'th SN, 
and $\mu_i^{\rm fit}$ is the best-fit \wCDM\ model distance modulus 
at the corresponding redshift $z_i$; 
and (iii) the value of $\sigmuint$ that would be required in 
Eq.~\ref{eq:sigmudef} to make   $\chisqmu = N_{\rm dof}$ in 
Eq.~\ref{eq:mlcs_chisqmu} for that sample. In computing 
the first two measures, we adopt $\sigmuint = \sigmurmsVALUE$, 
the value that yields  $\chisqmu = N_{\rm dof}$ for the nearby SN 
sample and that we use in analyzing the combined samples, 
as explained in \S \ref{subsec:wfit_mlcs}.
The bottom row of Table~\ref{tb:mlcs_fitquality} shows the 
$\chisqmu$ statistic from the SNe in each sample when the 
best-fit \wCDM\ model parameters for the global sample combination 
\samplee\ ({\esym}) are used to determine $\mu_i^{\rm fit}$.  
Compared to the values for the independent fits to each sample,
the $\chisqmu$ values for the global fit are only slightly larger.
For the nearby, ESSENCE, SNLS, and HST samples,
the reduced statistic, $\chisqmu/N_{\rm dof}$, is close to unity, 
and $\RMSMU = \LOWZRMS$ to $\HSTRMS$~mag.
For the \SDSS\ sample, the reduced $\chisqmu/N_{\rm dof} \sim 0.5$,
and $\RMSMU = \SDSSRMS$, 
both substantially smaller than for the other samples.

\begin{table}[hb]
\centering
\caption{  
   	Hubble diagram fit-quality parameters using \mlcs\ distances.	
    }
\begin{tabular}{l | ccccc }
\tableline\tableline
% -------------------------
        
  fit-quality    &  \multicolumn{5}{c}{Result for sample:}  \\
  parameter      & Nearby & SDSS & ESSENCE & SNLS & HST   \\
\tableline
% -----
$\chisqmu$ 
  &  $\LOWZCHISQ$ 
  &  $\SDSSCHISQ$ 
  &  $\ESSECHISQ$ 
  &  $\SNLSCHISQ$ 
  &  $\HSTCHISQ$ 
    \\
$N_{\rm dof}$  
  &  $\LOWZNDOF$ 
  &  $\SDSSNDOF$ 
  &  $\ESSENDOF$ 
  &  $\SNLSNDOF$ 
  &  $\HSTNDOF$ 
    \\
$\RMSMU$
  &  $\LOWZRMS$ 
  &  $\SDSSRMS$ 
  &  $\ESSERMS$ 
  &  $\SNLSRMS$ 
  &  $\HSTRMS$ 
    \\
$\sigmuint$ ($\chisqmu=N_{\rm dof}$)
  &  $\LOWZMLCSDISP$
  &  $\SDSSMLCSDISP$
  &  $\ESSEMLCSDISP$
  &  $\SNLSMLCSDISP$
  &  $\HSTMLCSDISP$
    \\
\tableline
$\chisqmu$ (global fit)
  &  $\LOWZCHISQx$ 
  &  $\SDSSCHISQx$ 
  &  $\ESSECHISQx$ 
  &  $\SNLSCHISQx$ 
  &  $\HSTCHISQx$ 
    \\
\tableline
\end{tabular}
\label{tb:mlcs_fitquality} 
\end{table}

We attribute the smaller scatter and $\chisqmu$ of the \SDSS\
 sample largely to  \spec\ selection effects. 
As described in \S~\ref{sec:survey}, 
when prioritizing candidate SNe for spectroscopic 
observations,
preference was given to those that were far from host-galaxy 
cores and/or that were hosted by redder 
(and presumably less dusty) galaxies. 
In Appendix \ref{app:SDSS_scatter},
this explanation is quantified 
by comparing the Hubble scatter 
and $\chisqmu$ statistic for
the spectroscopic SN~Ia sample to those for the \hostz\   
sample of photometrically identified SNe~Ia that have \spec\ 
redshifts from subsequent host-galaxy observations.
The cosmology analysis accounts for this selection effect 
via the model for the search \eff\
(\S~\ref{subsec:speceff}) in the \mlcs\ fitting prior  
and by including a systematic error that reflects uncertainties 
in the search \eff.

% -----------------------------------------------------------------
\subsubsection{\mlcs\ Hubble Diagrams and Cosmological Parameters}
\label{subsec:MLCS_RESULTS}

\begin{figure}[h]
  \epsscale{1.1}
  \plotone{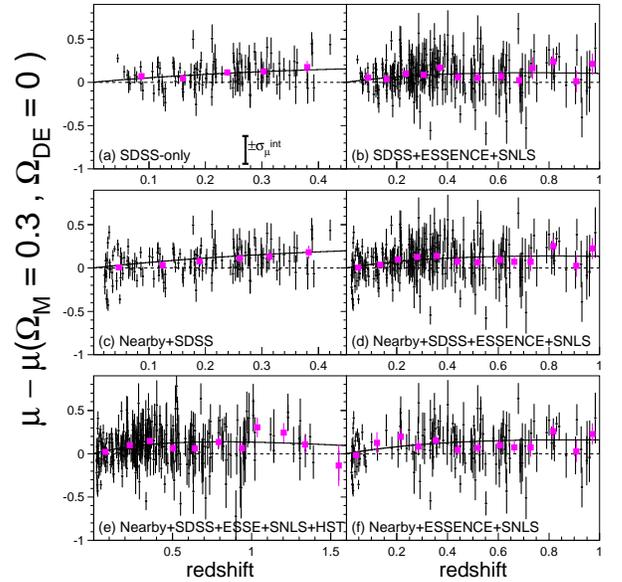}
  \caption{
	Hubble residuals for the \mlcs\ method: 
	differences between measured SN distance moduli and 
	those for an open CDM model ($\OM=0.3$, $\ODE=0$) vs. redshift 
	for the six SN sample combinations. 
	Large, square (pink) points show weighted averages in redshift bins: 
	within each redshift bin, the points are plotted at the 
	weighted mean redshift given by  
	$\bar{z} = (\sum z_i/\sigma_i^2) / (\sum 1/\sigma_i^2)$, 
	where $\sigma_i$ is the distance-modulus \unc. 
	Solid curves show the difference 
	between the best-fit \wCDM\ model distance modulus 
	and that for the open model, normalized 
	to have the same value of the Hubble parameter.  
	The error bars on the data points correspond to the
	distance modulus error $\sigmufit$ from the \mlcs\ 
	light-curve fit (Eq.~\ref{eq:sigmudef}), 
	i.e., they do not include the intrinsic scatter or the 
	effects of redshift and peculiar velocity errors.
	The vertical bar in panel (a) shows the intrinsic \unc,
	$\sigmuint = \sigmurmsVALUE$, included in the cosmology fits
	so that the $\chi^2$ per degree of freedom is unity for 
	the Hubble diagram constructed from the nearby SN~Ia sample alone. 
	}
  \label{fig:muresids_MLCS}
\end{figure}

\begin{figure}[h]
  \epsscale{1.15}
  \plotone{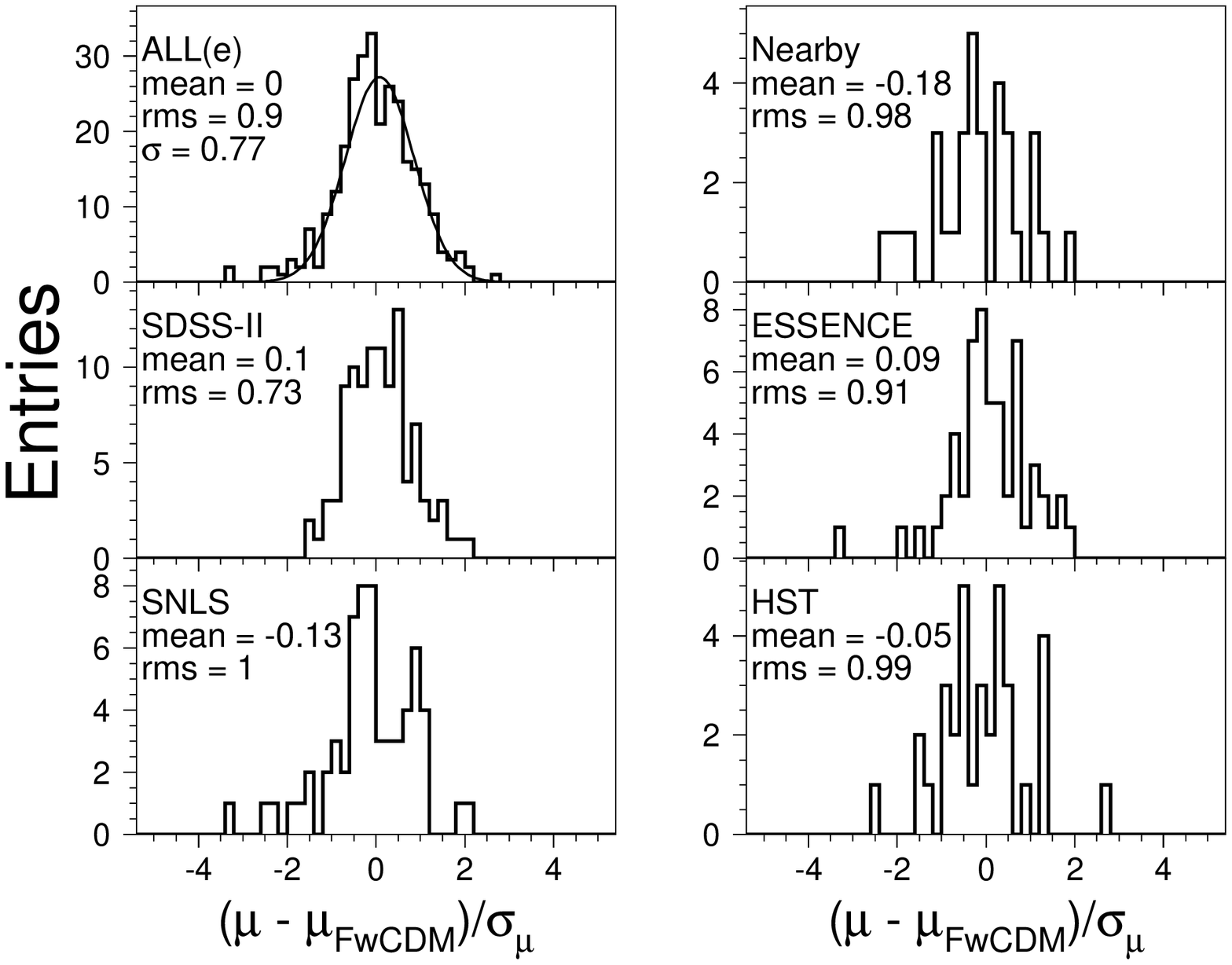}
  \caption{
	Distribution of normalized Hubble residuals (pull) for the 
	\mlcs\ method, for the \samplee\ sample combination {\esym} 
	(upper left, along with Gaussian fit) 
	and for each SN sample indicated on the plots.
	$\mu$ is the measured SN distance modulus, 
	$\MUwCDM$ is the distance modulus from the best-fit
	\wCDM\ model (for sample combination {\esym})
	at the same redshift,
	and $\sigmutot$ is the total \unc\ (Eq.~\ref{eq:sigmudef}). 
	Mean and RMS for each distribution are indicated on each panel.
	}
  \label{fig:muresids_norm_MLCS}
\end{figure}

Figure~\ref{fig:muresids_MLCS} shows the differences between the estimated 
SN distance moduli $\mu_i$ and those for an open CDM model with no 
dark energy ($\OM=0.3$, $\ODE=0$) as a function of redshift, for each 
of the six SN sample combinations; the large square (pink) points show 
weighted averages of these residuals in redshift bins.
Also shown is the Hubble distance-residual curve 
between the best-fit \wCDM\  model for each sample combination (including 
the BAO+CMB prior) 
and that for the open model (solid curves). 
In each panel, the Hubble parameter for the open CDM model has 
been adjusted to agree with that for the best-fit \wCDM\ model, 
so that the \wCDM\ versus open CDM residuals vanish at $z=0$. 
Since the Hubble parameter is not determined in this analysis, 
a constant vertical offset in Fig.~\ref{fig:muresids_MLCS} is irrelevant: 
what is significant are the slope and curvature of the points and the 
best-fit (solid) curves vs. redshift. 
Figure~\ref{fig:muresids_norm_MLCS} shows the distributions of normalized
residuals, $(\mu_i - \MUwCDM)/\sigmutot$,
where $\MUwCDM$ is the distance modulus from the best-fit
\wCDM\ model for sample combination ({\esym}) including the 
BAO+CMB prior,
and $\sigmutot$ is the total \unc\ defined in Eq.~\ref{eq:sigmudef}.
The bulk of the  distribution of all \NSNTOT\ normalized residuals 
(upper left panel of Fig.~\ref{fig:muresids_norm_MLCS}) 
is well fit by a Gaussian with $\sigma = \MLCSMUPULLSIG$; 
outliers increase the rms to $\MLCSMUPULLRMS$.
For the nearby sample, the rms is one as expected, because
$\sigmuint = \sigmurmsVALUE$~mag is determined such that
$\chisqmu/N_{\rm dof}=1$.

\begin{figure}[h]
 \epsscale{1.1}
 \plotone{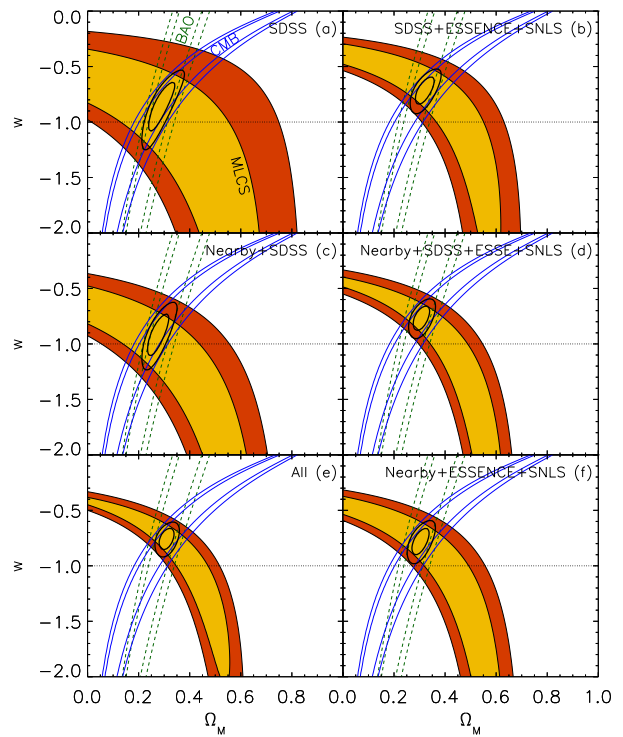}
  \caption{
	For the \wCDM\ model, \mlcs\ statistical-\unc\ contours 
	in the $\OM$-$w$ plane for each of the six SN sample 
	combinations indicated on the plots.
	Shaded regions:  68\% and 95\%
	confidence level regions for the SN data alone; 
	green: corresponding CL contours for SDSS BAO 
	\citep{Eisenstein05}; 
	blue: CL contours for WMAP5 CMB \citep{Komatsu2008}; 
	closed, black contours:  combined constraints from SN+BAO+CMB. 
        }
  \label{fig:wCDM_MLCS_statcont}
\end{figure}

\begin{figure}[h]
 \epsscale{1.1}
 \plotone{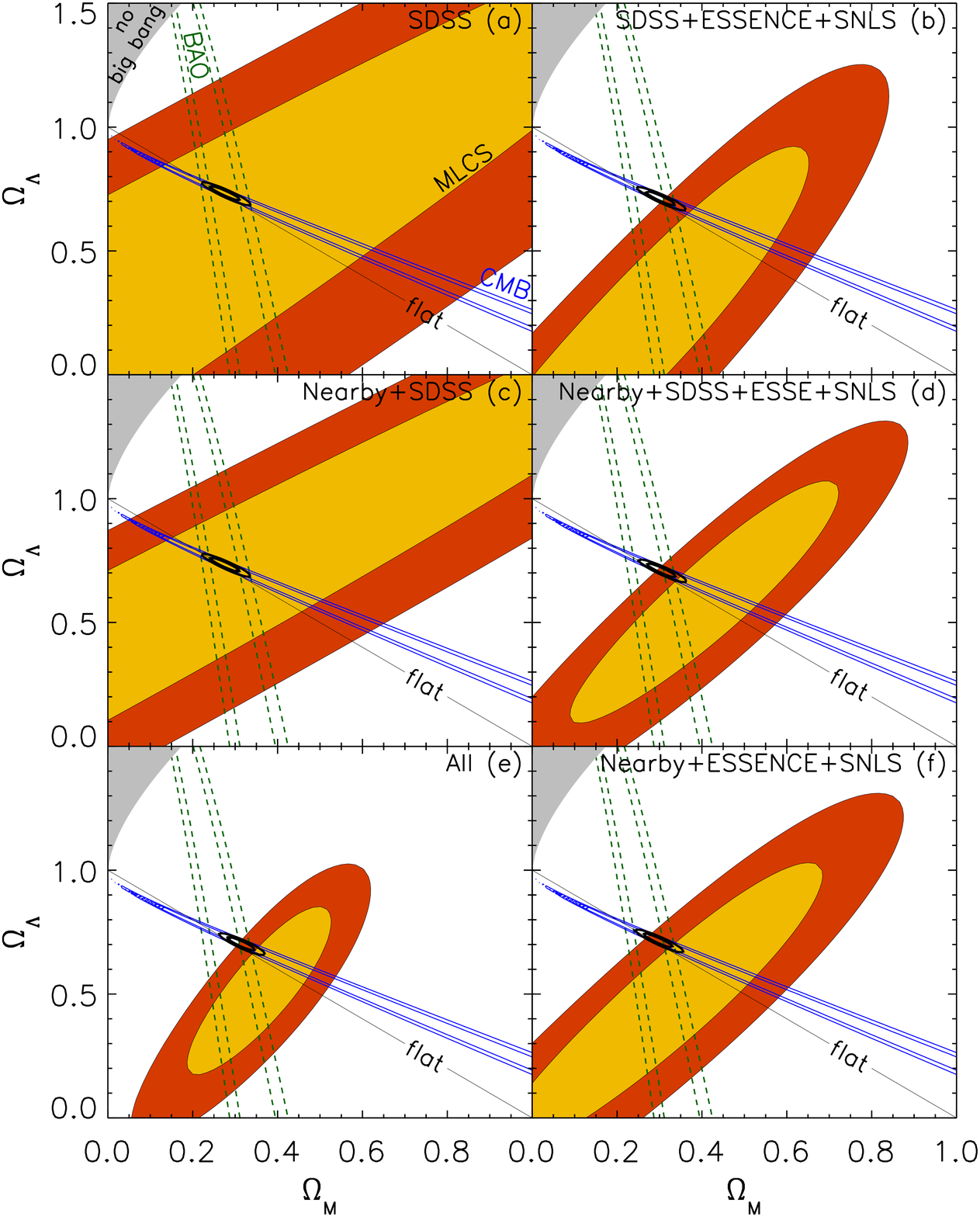}
  \caption{
 	For the \LCDM\ model, \mlcs\ statistical-\unc\ contours 
	in the $\OM$-$\OLAM$ plane for each of the six SN sample 
	combinations 
	indicated on the plots.
	Shaded regions:  68\% and 95\% confidence 
	level regions for the SN data alone; 
	green: corresponding CL contours for SDSS BAO; 
	blue: CL contours for WMAP5 CMB; 
	closed, black contours: combined constraints from SN+BAO+CMB. 
	Grey region indicates models with no Big Bang, i.e., with 
	a bounce at finite value of the FRW scale factor. Solid 
	diagonal line indicates a spatially flat Universe, with
	$\OM+\OLAM=1$. 
        }
  \label{fig:LCDM_MLCS_statcont}
\end{figure}

\begin{figure}[h]
 \epsscale{1.1}
 \plotone{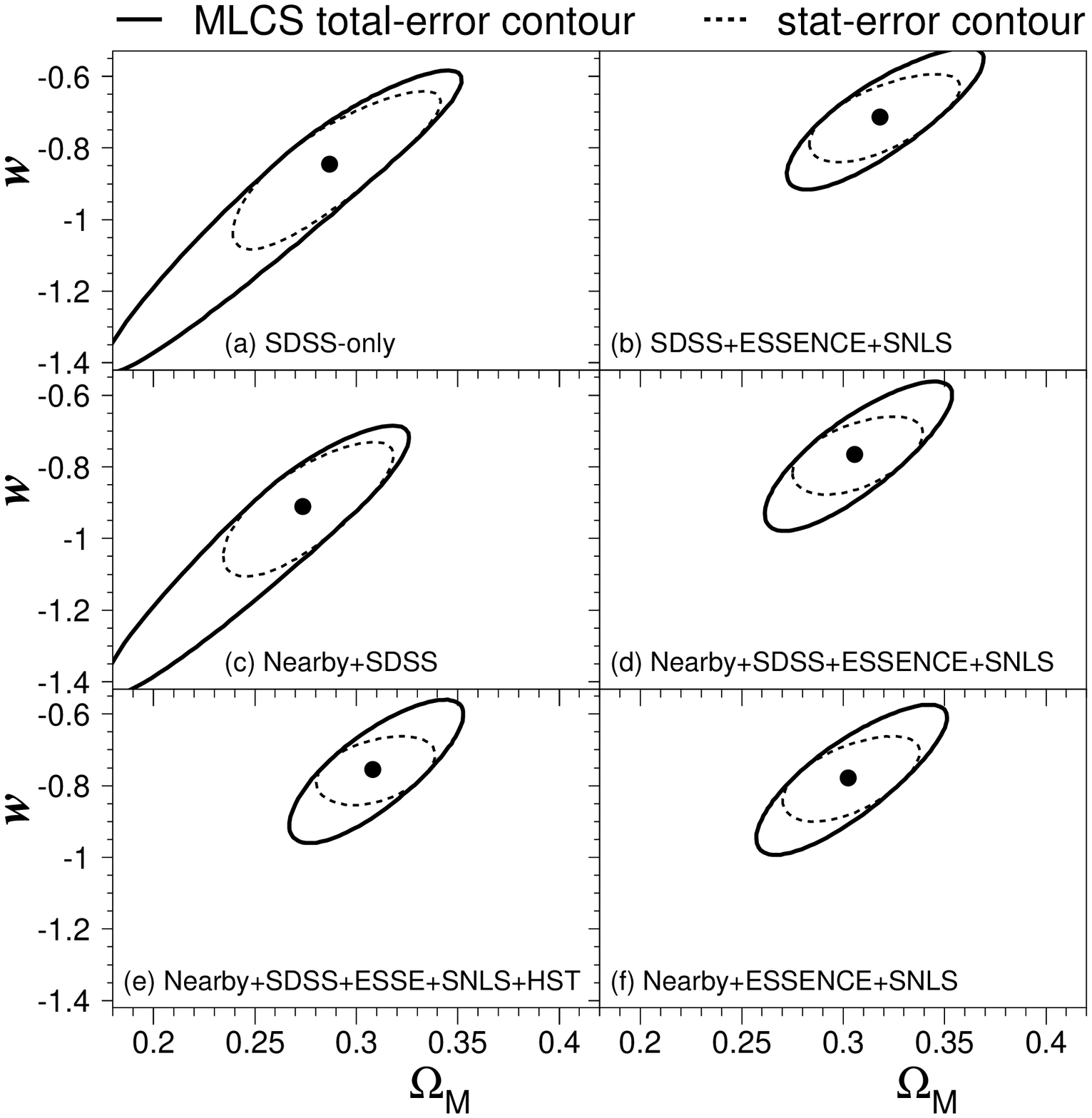}
  \caption{
	For \mlcs\ and the \wCDM\ model, 68\% CL contours in the 
	$\OM$-$w$ plane for each of the six SN sample combinations, 
	using the combined SN+BAO+CMB constraints. 
	Solid contours are total (statistical+systematic) \unc;
	dashed contours are statistical only.
	Systematic errors have been included using the prescription in 
	Appendix~\ref{app:contours}.
        }
  \label{fig:wCDM_MLCS_etotcont}
\end{figure}

\begin{figure}[h]
 \epsscale{1.1}
  \plotone{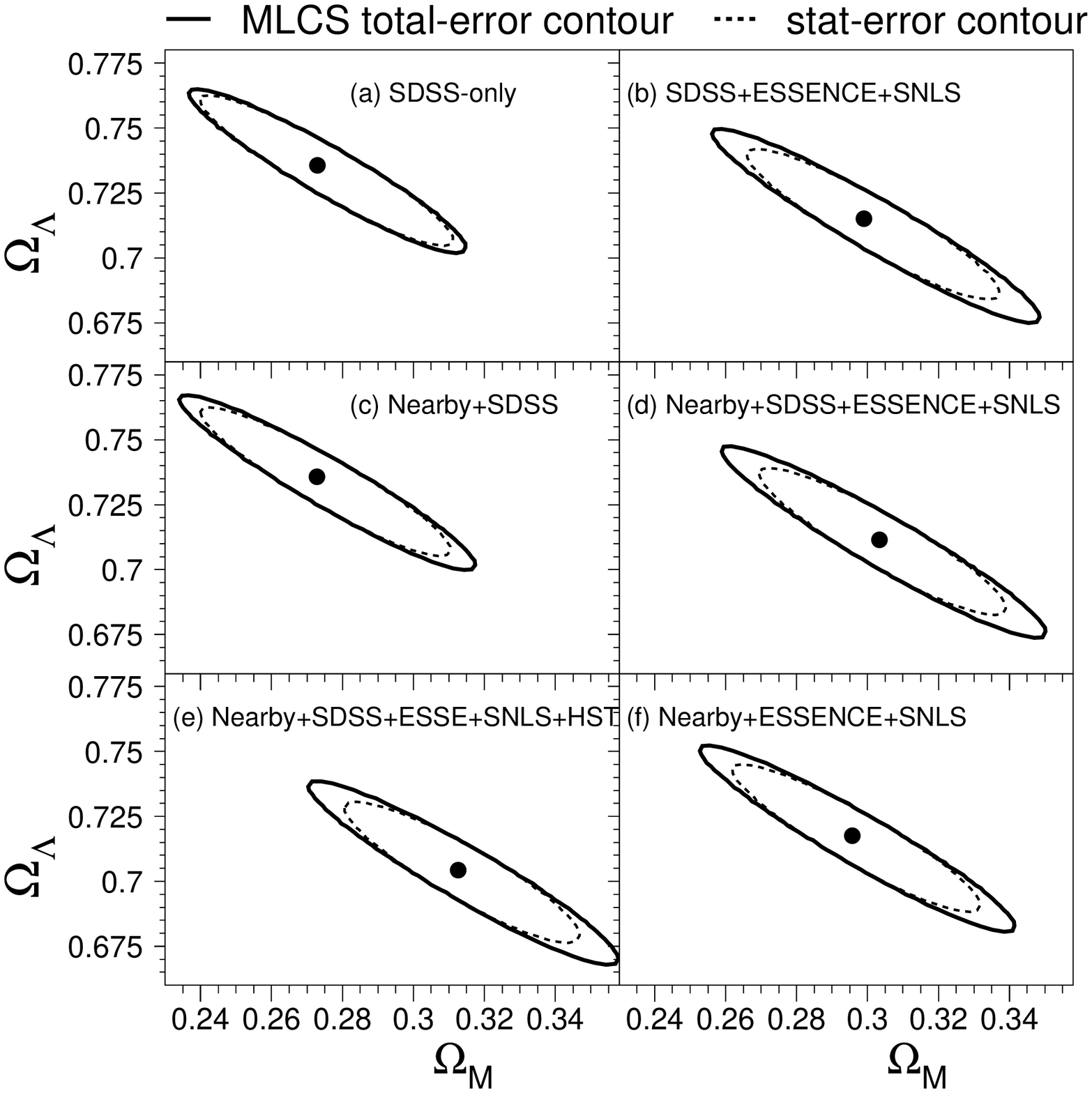}
  \caption{
	For \mlcs\ and the \LCDM\ model, 68\% CL contours in the 
	$\OM$-$\OLAM$ plane for each of the six SN sample combinations, 
	using the combined SN+BAO+CMB constraints. 
	Solid contours are total (statistical+systematic) \unc;
	dashed contours are statistical only.
	Systematic errors have been included using the prescription in 
	Appendix~\ref{app:contours}.
        }
  \label{fig:LCDM_MLCS_etotcont}
\end{figure}

Figs.~\ref{fig:wCDM_MLCS_statcont} and \ref{fig:LCDM_MLCS_statcont}
show the \mlcs\ statistical-\unc\ contours for the 
\wCDM\ and \LCDM\ models: for each of the six sample combinations,
the SN~Ia, BAO, and CMB contours are shown 
along with the combined constraints.
For the combined SN+BAO+CMB results, we include systematic \uncs\ to  
derive total (statistical plus systematic) \unc\ contours,  
as explained in Appendix~\ref{app:contours}, with results shown in 
Figs.~\ref{fig:wCDM_MLCS_etotcont} and \ref{fig:LCDM_MLCS_etotcont} 
for the \wCDM\ and \LCDM\ models 
(note the zoomed axis scales compared to the previous figures).
The best-fit cosmological parameter values 
and uncertainties, 
including the BAO and CMB priors, are given in 
Tables~\ref{tb:results_MLCS_FWCDM} and \ref{tb:results_MLCS_LCDM} 
for the \wCDM\ and \LCDM\ models. 
The distance modulus vs. redshift curve for the best-fit \wCDM\ 
cosmological parameter values (relative to that of an 
open CDM model) are shown as the solid curves in the 
Hubble residual plots in Fig.~\ref{fig:muresids_MLCS}.

Among the six SN sample combinations, 
the best-fit values of the dark energy equation of state 
parameter $w$ fall roughly into two groups.
In the first group, the highest-redshift sample is
from \SDSS:
for the  \samplea\ sample ({\asym}) and \samplec\ sample combination ({\csym}),
$w= \MLCSWRESa$ and $\MLCSWRESc$.
The agreement between these values is consistent with the 
expected RMS spread of 0.07 based on simulations.
The second group comprises the other four sample combinations, 
which include the higher-redshift ESSENCE \& SNLS samples:
$w= \MLCSWRESb,\MLCSWRESd,\MLCSWRESe,\MLCSWRESf$
for sample combinations \bsym,\dsym,\esym,\fsym.
Simulations predict an RMS spread in $w$ of $\sim 0.1$
between the results from these two groups;
the observed difference is therefore 
not statistically significant but may nevertheless be an indicator 
of systematic effects.

Table~\ref{tb:results_MLCS_FWCDM} also shows that 
the statistical and systematic errors in $w$ and $\OM$ 
for sample combinations ({\bsym}) and ({\fsym}) are very similar. 
Since these two sample combinations differ only in the 
substitution of the nearby SN~Ia sample with the \SDSS\  sample, 
this indicates that the first-season \SDSS\ SN sample 
anchors the Hubble diagram with comparable constraining 
power to the nearby sample. 

Using the \samplee\ sample combination ({\esym}), which covers the 
widest redshift range, we obtain 
$w=\MLCSWRESe\pm\MLCSWSTATERRe({\rm stat}) 
   \pm \MLCSWSYSTERRPe({\rm syst})$ 
and 
$\OM =\MLCSOMRESe\pm\MLCSOMSTATERRe({\rm stat}) 
   \pm\MLCSOMSYSTERRPe({\rm syst})$. 
Although this value for $w$ is higher than that obtained from other 
recent SN measurements, we stress that the difference is {\it not} 
due to inclusion of the \SDSS\ SN data: as Table~\ref{tb:results_MLCS_FWCDM} 
shows, we infer nearly identical parameter values for 
sample combination ({\fsym}), 
which excludes the \SDSS\ data. By contrast, 
WV07 inferred $w=-1.07 \pm 0.09({\rm stat}) \pm 0.13({\rm syst})$, 
$\OM=0.267^{+0.028}_{-0.018}({\rm stat})$ 
using \mlcs\ for a sample combination nearly identical to 
({\fsym}) and including the BAO (but not CMB) constraints. 
In \S~\ref{subsec:WV07_compare}, we trace the 
differences between the WV07 result and ours to changes in \mlcs\ model 
parameters and assumptions.

For the \LCDM\ model, in comparison with the \wCDM\ model, 
the SN results carry less weight 
relative to the combined BAO and CMB results in constraining  
the parameters. In particular, Fig.~\ref{fig:LCDM_MLCS_statcont} and 
Table \ref{tb:results_MLCS_LCDM} show that 
the \samplea\ ({\asym}) and \samplec\ ({\csym}) SN samples 
have almost no impact on the maximum likelihood parameter values and 
uncertainties. For the other four SN sample combinations, there is 
some tension between the SN and BAO results: the SN results 
pull the maximum likelihood parameter values along the CMB contour, away 
from the BAO contours. Since the BAO contours are in all cases narrower 
than those for the SNe, however, this shift is small, corresponding 
to $\delta \ODE \simeq -0.03$, $\delta \OM \simeq 0.04$ or less.

%
% Cosmology results from MLCS method, using FWCDM model
% (float w and OM, fix O_tot =1)
%

\begin{table}[hb]
\centering
\caption{  
	For the \wCDM\ model, constraints on $w$ and $\OM$ 
	from \mlcs\ SN distances combined with SDSS BAO and 
 	WMAP5 CMB results. 
    }  % end caption
\begin{tabular}{l | cccccc }
\tableline\tableline
% -------------------------
                   &  \multicolumn{5}{c}{Result for sample combination:}  \\
      & {\asym}
      & {\bsym}
      & {\csym}
      & {\dsym}
      & {\esym}
      & {\fsym}
         \\
\tableline
 $\chisqmu$
  & $\MLCSWCHISQa$
  & $\MLCSWCHISQb$
  & $\MLCSWCHISQc$
  & $\MLCSWCHISQd$
  & $\MLCSWCHISQe$
  & $\MLCSWCHISQf$
    \\ 
 $N_{dof}$
  & $\WNDOFa$
  & $\WNDOFb$
  & $\WNDOFc$
  & $\WNDOFd$
  & $\WNDOFe$
  & $\WNDOFf$
    \\ 
  $\RMSMU$ 
  & $\MLCSWRMSa$
  & $\MLCSWRMSb$
  & $\MLCSWRMSc$
  & $\MLCSWRMSd$
  & $\MLCSWRMSe$
  & $\MLCSWRMSf$
    \\ 
% -------------------
  & & & & \\
 $w$ 
  & $\MLCSWRESa$  
  & $\MLCSWRESb$
  & $\MLCSWRESc$
  & $\MLCSWRESd$
  & $\MLCSWRESe$
  & $\MLCSWRESf$
    \\ 
 $\wstatsym$
  & $\MLCSWSTATERRa$
  & $\MLCSWSTATERRb$
  & $\MLCSWSTATERRc$
  & $\MLCSWSTATERRd$
  & $\MLCSWSTATERRe$
  & $\MLCSWSTATERRf$
    \\ 
\\
 $\wsystsym$
  & {\large ${}^{+\MLCSWSYSTERRPa}_{-\MLCSWSYSTERRMa}$}
  & $\MLCSWSYSTERRPb$
  & {\large ${}^{+\MLCSWSYSTERRPc}_{-\MLCSWSYSTERRMc}$}
  & $\MLCSWSYSTERRPd$
  & $\MLCSWSYSTERRPe$
  & $\MLCSWSYSTERRPf$
    \\
\\ 
 $\wtotsym$
  & {\large ${}^{+\MLCSWTOTERRPa}_{-\MLCSWTOTERRMa}$}
  & $\MLCSWTOTERRPb$
  & {\large ${}^{+\MLCSWTOTERRPc}_{-\MLCSWTOTERRMc}$}
  & $\MLCSWTOTERRPd$
  & $\MLCSWTOTERRPe$
  & $\MLCSWTOTERRPf$
    \\ 
% -------------------
  & & & & \\
 $\OM$ 
  & $\MLCSOMRESa$
  & $\MLCSOMRESb$
  & $\MLCSOMRESc$
  & $\MLCSOMRESd$
  & $\MLCSOMRESe$
  & $\MLCSOMRESf$
    \\ 
 $\OMstatsym$
  & $\MLCSOMSTATERRa$
  & $\MLCSOMSTATERRb$
  & $\MLCSOMSTATERRc$
  & $\MLCSOMSTATERRd$
  & $\MLCSOMSTATERRe$
  & $\MLCSOMSTATERRf$
    \\ \\
 $\OMsystsym$
  & {\large ${}^{+\MLCSOMSYSTERRPa}_{-\MLCSOMSYSTERRMa}$}
  & $\MLCSOMSYSTERRPb$
  & {\large ${}^{+\MLCSOMSYSTERRPc}_{-\MLCSOMSYSTERRMc}$}
  & $\MLCSOMSYSTERRPd$
  & $\MLCSOMSYSTERRPe$
  & $\MLCSOMSYSTERRPf$
    \\  \\
 $\OMtotsym$
  & {\large ${}^{+\MLCSOMTOTERRPa}_{-\MLCSOMTOTERRMa}$}
  & $\MLCSOMTOTERRPb$
  & {\large ${}^{+\MLCSOMTOTERRPc}_{-\MLCSOMTOTERRMc}$}
  & $\MLCSOMTOTERRPd$
  & $\MLCSOMTOTERRPe$
  & $\MLCSOMTOTERRPf$
    \\  
\tableline
\end{tabular}
\tablenotetext{a}{ \samplea } 
\tablenotetext{b}{ \sampleb }
\tablenotetext{c}{ \samplec } 
\tablenotetext{d}{ \sampled }
\tablenotetext{e}{ \samplee }
\tablenotetext{f}{ \samplef }
\label{tb:results_MLCS_FWCDM}
\end{table}

%
% Cosmology results from MLCS method, using LCDM model
% (float Omega_k, fix w = -1)
%

\begin{table}[hb]
\centering
\caption{  
	For the \LCDM\ model, constraints on $\OM$ and $\OLAM$ 
	from \mlcs\ SN distances combined with  SDSS BAO and 
	WMAP5 CMB results.
	}  % end caption
\begin{tabular}{l | rrrrrr }
\tableline\tableline
% -------------------------
                   &  \multicolumn{5}{c}{Result for sample combination:}  \\
      & {\asym}
      & {\bsym}
      & {\csym}
      & {\dsym}
      & {\esym}
      & {\fsym}
         \\
\tableline
 $\chisqmu$
  & $\LCDMCHISQa$
  & $\LCDMCHISQb$
  & $\LCDMCHISQc$
  & $\LCDMCHISQd$
  & $\LCDMCHISQe$
  & $\LCDMCHISQf$
    \\ 
 $N_{dof}$
  & $\LCDMNDOFa$
  & $\LCDMNDOFb$
  & $\LCDMNDOFc$
  & $\LCDMNDOFd$
  & $\LCDMNDOFe$
  & $\LCDMNDOFf$
    \\ 
 $\RMSMU$
  & $\MLCSWRMSa$   % <== use FWCDM RMS ; good to 3rd decimal
  & $\MLCSWRMSb$
  & $\MLCSWRMSc$
  & $\MLCSWRMSd$
  & $\MLCSWRMSe$
  & $\MLCSWRMSf$
% ----------------- 
%  & $\LCDMRMSa$
%  & $\LCDMRMSb$
%  & $\LCDMRMSc$
%  & $\LCDMRMSd$
%  & $\LCDMRMSe$
%  & $\LCDMRMSf$
% ----------------- 
    \\ 
% -------------------
  & & & & \\
 $\OLAM$ 
  & $\LCDMOLa$  
  & $\LCDMOLb$
  & $\LCDMOLc$
  & $\LCDMOLd$
  & $\LCDMOLe$
  & $\LCDMOLf$
    \\ 
 $\sigma_{\OLAM}(stat)$
  & $\LCDMOLSTATERRa$
  & $\LCDMOLSTATERRb$
  & $\LCDMOLSTATERRc$
  & $\LCDMOLSTATERRd$
  & $\LCDMOLSTATERRe$
  & $\LCDMOLSTATERRf$
    \\ 
 $\sigma_{\OLAM}(syst)$
  & $\LCDMOLSYSTERRa$
  & $\LCDMOLSYSTERRb$
  & $\LCDMOLSYSTERRc$
  & $\LCDMOLSYSTERRd$
  & $\LCDMOLSYSTERRe$
  & $\LCDMOLSYSTERRf$
    \\ 
 $\sigma_{\OLAM}(tot)$
  & $\LCDMOLTOTERRa$
  & $\LCDMOLTOTERRb$
  & $\LCDMOLTOTERRc$
  & $\LCDMOLTOTERRd$
  & $\LCDMOLTOTERRe$
  & $\LCDMOLTOTERRf$
    \\ 
% -------------------
  & & & & \\
 $\OM$
  & $\LCDMOMa$
  & $\LCDMOMb$
  & $\LCDMOMc$
  & $\LCDMOMd$
  & $\LCDMOMe$
  & $\LCDMOMf$
    \\ 
 $\OMstatsym$
  & $\LCDMOMSTATERRa$
  & $\LCDMOMSTATERRb$
  & $\LCDMOMSTATERRc$
  & $\LCDMOMSTATERRd$
  & $\LCDMOMSTATERRe$
  & $\LCDMOMSTATERRf$
    \\ 
 $\OMsystsym$
  & $\LCDMOMSYSTERRa$
  & $\LCDMOMSYSTERRb$
  & $\LCDMOMSYSTERRc$
  & $\LCDMOMSYSTERRd$
  & $\LCDMOMSYSTERRe$
  & $\LCDMOMSYSTERRf$
    \\  
 $\OMtotsym$
  & $\LCDMOMTOTERRa$
  & $\LCDMOMTOTERRb$
  & $\LCDMOMTOTERRc$
  & $\LCDMOMTOTERRd$
  & $\LCDMOMTOTERRe$
  & $\LCDMOMTOTERRf$
    \\  
\tableline
\end{tabular}
%
%\tablenotetext{a}{ \samplea } 
%\tablenotetext{b}{ \sampleb }
%\tablenotetext{c}{ \samplec } 
%\tablenotetext{d}{ \sampled }
%\tablenotetext{e}{ \samplee }
%\tablenotetext{f}{ \samplef }
%
\label{tb:results_MLCS_LCDM}
\end{table}

% ------------------------------------------------
  \subsubsection{$U$-Band Anomaly with \mlcs\ }
  \label{subsec:Uanom_MLCS}

\newcommand{\NoULOWZ}{17}
\newcommand{\NoUSDSSugr}{9}
\newcommand{\NoUSDSSgri}{56}
\newcommand{\NoUSNLSgri}{13}

\newcommand{\noUdmu}{\Delta\mu_{noU}}
\newcommand{\noUmu}{\mu_{noU}}
\newcommand{\noBmu}{\mu_{noB}}
% ------------------------------------------------

As noted in \S \ref{subsec:syst_mlcs}, the largest single 
contribution to the systematic error budget comes from 
consideration of the rest-frame $U$-band. Within the \mlcs\ 
framework, the issue is manifest as 
a difference between the nearby sample (JRK07)
and the other SN~Ia samples. 
We have carried out a series of tests in which 
the observer-frame passband corresponding to rest-frame $U$-band 
is excluded from the \mlcs\ light-curve fits. We compare the 
resulting distance modulus estimates, $\noUmu$, with those 
for which rest-frame $U$-band is included, $\mu$,
and define $\noUdmu \equiv \noUmu - \mu$. 
If the \mlcs\ model is a good description of the data, 
we would expect $\noUdmu$ to scatter about zero. 

\begin{figure}[hb!]
  \epsscale{.86}\plotone{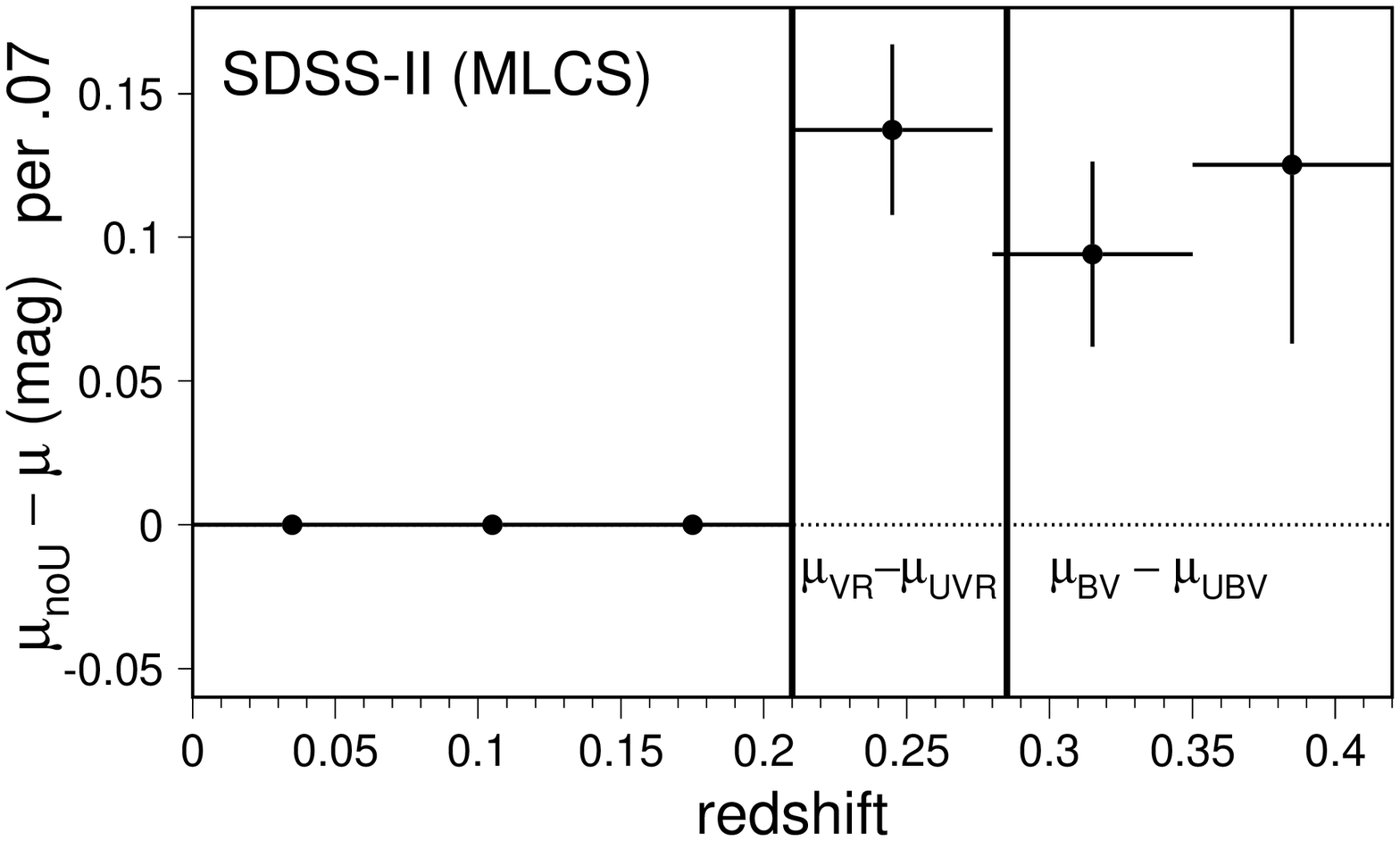}
  \epsscale{.23}\plotone{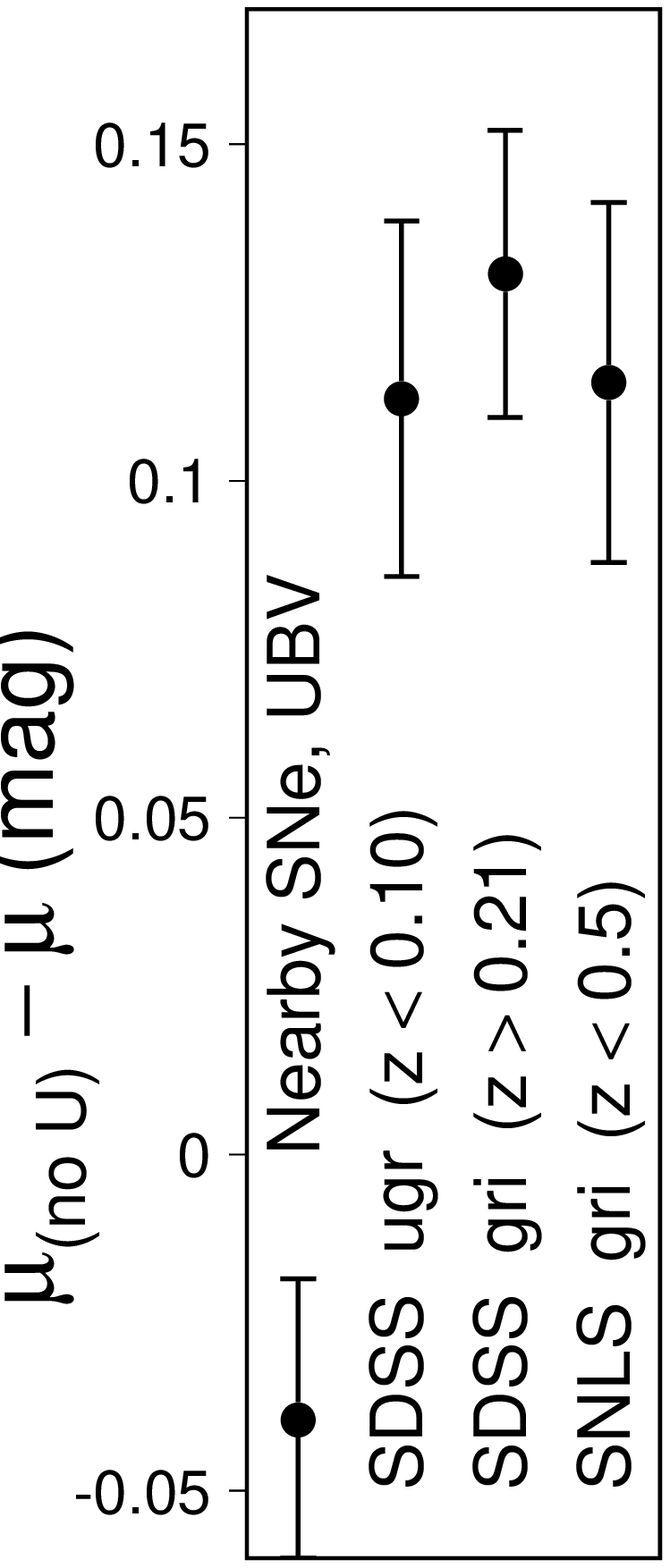}
  \caption{
	Left panel: redshift dependence of the
  	average difference in $\mu$ for \SDSS\ SNe 
	between the nominal \mlcs\ fits and fits in which the 
	observer-frame passband corresponding to rest-frame
	$U$-band is excluded ($g$-band for $z>\ZUSDSS$).
	Labels on the plot indicate the corresponding
	rest-frame $UBV$ \citep{Bessell90} passbands.
	Error bars (rms/$\sqrt{N}$) reflect the statistical \unc\ 
	on the mean $\mu$-difference in each redshift bin.	
	Right panel: shift in average distance modulus when 
	rest-frame $U$-band is excluded from the \mlcs\ fits, 
	for the subsamples discussed in the text.
        }
  \label{fig:noUtest_MLCS}
\end{figure}

The left panel of  Fig.~\ref{fig:noUtest_MLCS} shows the resulting 
change in the \SDSS\ SN Hubble diagram, $\noUdmu$ versus redshift. 
For $z > \ZUSDSS$, observer-frame $g$ corresponds to rest-frame $U$; 
excluding $g$-band in  this redshift range 
results in a shift in the distance modulus of 
$\noUdmu = (\MLCSSDSSnoUdMU)$~mag. 
In the redshift interval $\ZUSDSS <z < 0.285$, 
the remaining observer-frame SDSS passbands $r$ and $i$ 
correspond to rest-frame $V$ and $R$; 
for $z > 0.285$, they correspond to $B$ and $V$. 
In both of those redshift intervals, we see a distance modulus 
shift of $\noUdmu \simeq 0.1 - 0.15$~mag. 
For $z<\ZUSDSS$, the $gri$ filters do not map onto 
rest-frame $U$-band, 
therefore $\noUdmu \equiv 0$ in this redshift range.
Since $\noUdmu$ is determined from strongly correlated
samples, 
i.e., the $r$- and $i$-band data are the same in fits
with and without $U$-band,
its \unc\ is estimated to be ${\rm rms}/\sqrt{N}$, 
where rms is the root-mean-square and $N$ is the number of SNe~Ia.
Applying the same $U$-exclusion test to simulations shows 
that the observed shift has a significance of $\sim 6\sigma$,
consistent with our estimate of the \unc.
The large tilt in the Hubble diagram that results from excluding 
$g$-band at $z>\ZUSDSS$ results in a shift of $\dw \sim 0.3$ for 
the \samplea\ and \samplec\ sample combinations, 
by far the dominant contribution 
to the systematic error budget for those samples.

We further investigate the $U$-band anomaly by comparing the 
fitted distance moduli with and without rest-frame $U$ for 
three subsamples: 
(i) observer-frame $UBV$ vs. $BV$ for \NoULOWZ\ nearby SNe~Ia; 
(ii) observer-frame $ugr$ vs. $gr$ for \NoUSDSSugr\ low-redshift 
\SDSS\ SNe for which the $u$-band signal to noise is sufficient ($z<0.1$), 
and
(iii) observer-frame $gri$ vs. $ri$ for \NoUSNLSgri\ SNLS SNe 
with $\ZUSDSS < z<0.5$ 
and that have at least one $g$-band measurement within 
$\pm 10$ days of maximum brightness.
In each case, the subsamples are chosen such that the {\lc s}
pass the selection criteria of \S~\ref{sec:sample} and include 
three observer passbands, one of which maps onto rest-frame $U$-band.
The differences in distance modulus ($\noUdmu$)
between the two- and three-band fits 
(without and with rest-frame $U$) are shown in the right panel
of Fig.~\ref{fig:noUtest_MLCS}. 
The \SDSS\ \& SNLS subsamples show a consistent shift of about 0.1~mag. 
For comparison, the right panel of Fig.~\ref{fig:noUtest_MLCS}
also shows the average shift for the points in the left-panel test 
at $z>0.21$, again showing consistency. 
By contrast, the nearby subsample is consistent with no shift 
or a slightly negative shift.
Since the \mlcs\ model is trained on a superset of the nearby SN~Ia data, 
we would expect no significant shift for the nearby subsample. 
The redshift range $z<0.1$ in the left panel of 
Fig.~\ref{fig:noUtest_MLCS} is not the same as the 
SDSS-$ugr$ ($z<0.1$) test in the right panel: 
the former is based on observer-frame $gri$ and does not map onto 
rest-frame $U$-band,  while the latter is based on $ugr$ vs. $gr$ 
in order to test excluding rest-frame $U$-band.

Since the $U$-band is particularly sensitive to host-galaxy extinction, 
it is worth exploring whether this anomaly might be an artifact of the 
assumed extinction law. We have repeated the test above, replacing 
the CCM89 color law nominally used in \mlcs\ with 
the empirically determined \SALTII\ color law, 
$CL(\lambda)$ in Eq.~\ref{eq:SALTII_flux_rest}.
The \SALTII\ color law results in a $U$-band extinction that 
is 0.07~mag larger than that from using CCM89 with $R_V = \RV$ 
and the mean extinction value, $A_V = 0.35$.
The differences in the extinction for the other passbands 
are much smaller: $0.014$, $0$, $0.007$, $-0.002$ for $B,V,R,I$.
Repeating the $U$-band exclusion tests with the \SALTII\ color law
results in distance-modulus offsets for the subsamples 
in the right panel of Fig.~\ref{fig:noUtest_MLCS}
that are $\sim 20\%$ smaller than for the nominal test
using the CCM89 color law.

Although excluding rest-frame $U$-band reveals a problem, 
it does not definitively indicate that rest-frame $U$, 
as opposed to one of the other passbands, 
is the source of the problem. 
To study this in more detail, we carry out a similar test 
in which observations in passbands corresponding 
to rest-frame $B$-band are excluded. 
For \SDSS\ SNe with redshifts $z< \ZUSDSS$, this $B$-exclusion 
test corresponds to comparing distance moduli from observer-frame 
$gri$ (rest-frame $BVR$) with those from just 
$r$ and $i$ (rest-frame $VR$). 
The difference in the average distance modulus, 
$\mu_{VR} - \mu_{BVR}$, is $-0.01 \pm 0.02$~mag, 
consistent with no shift. This test suggests that
rest-frame $B$ is not the source of the anomaly
and strengthens the circumstantial evidence 
that rest-frame $U$ is the source of the anomaly.

\begin{figure}[h]
  \epsscale{1.1}
  \plotone{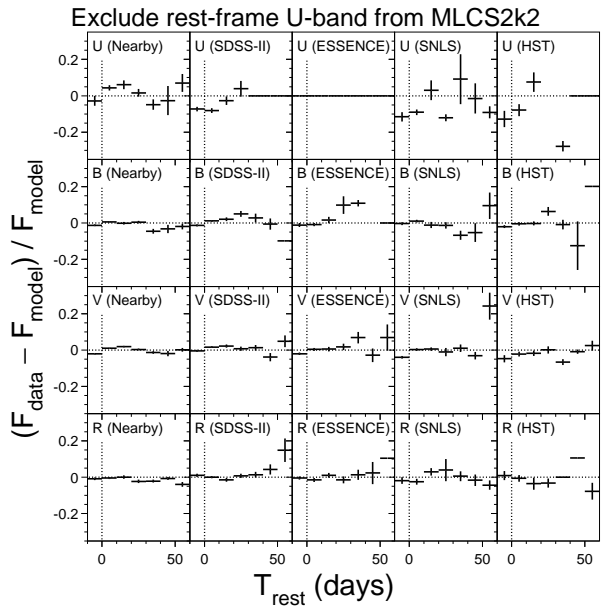}
  \caption{
	Data-model fractional residuals as a function of rest-frame
        epoch in 5-day bins for \mlcs\ light-curve fits.
	Same as Fig.~\ref{fig:lcresid_MLCS}, except fits exclude
	observer-frame filter corresponding to 
 	rest-frame $U$-band
        }
  \label{fig:lcresid_MLCS_noU}
\end{figure}

To further diagnose the $U$-band anomaly, 
we compare the light-curve data and the \mlcs\ model 
light-curve fits as a function of epoch for 
the different rest-frame passbands and the five different SN samples. 
Figure~\ref{fig:lcresid_MLCS} shows the data-model residuals 
for the nominal fits.
Figure~\ref{fig:lcresid_MLCS_noU} shows the residuals 
when the filter corresponding to rest-frame $U$-band 
is excluded from the fit;
to see the $U$-band residuals,
we must use the nominal \mlcs\ model parameters
that include $U$-band in the training. 
When $U$-band is excluded for the nearby SNe~Ia, 
there is a negligible change in the $U$-band residuals; 
this is expected 
because
\mlcs\ is trained on the nearby data. 
For the \SDSS, excluding $U$-band from the fits results in a 
$\sim 0.05$~mag shift in the $U$-band residuals
near the time of peak brightness ($\Trest=0$). 
The ESSENCE survey used only two passbands, $R$ and $I$, and only
a few of their SNe~Ia probe rest-frame $U$-band, 
so we cannot use this sample to probe the problem.
For the SNLS sample,
the residuals have been plotted for those
SNe~Ia that have observer-frame $g$-band measurements
mapping into rest-frame $U$-band (most of the SNe with $z<0.48$);
i.e., the same subset used in Fig.~\ref{fig:noUtest_MLCS}.
When rest-frame $U$-band is excluded, the corresponding shift in the $U$-band
residuals is consistent with the shift seen in the \SDSS.
We have also examined the higher-redshift SNLS subset for  
which observer-frame $r$-band maps onto rest-frame $U$-band; 
the results of those tests are consistent with the tests
based on $g$-band, but with much larger \uncs.
Note that the SNLS subset with excluded $g$-band has 
a \spec\ \eff\ $\simeffspec=1$ (see \S \ref{subec:searcheff}), 
while the \SDSS\ \spec\ \eff\ is significantly smaller than 1.
Although the corresponding \mlcs\ priors are quite different,
the $U$-band anomaly is consistent between them, indicating that it is 
not caused by errors in $\simeffspec$.

Since the \mlcs\ model is trained with the nearby sample, 
the $U$-band anomaly in the other SN~Ia samples results 
in significant systematic \uncs\ that limit the precision
of cosmological parameters obtained with the current implementation
of the \mlcs\ method. There are a number of possible causes for 
the $U$-band discrepancy:
(i) the SN~Ia $U$-band flux could be redshift dependent; 
(ii) selection effects for the nearby sample could result in
a $U$-band flux distribution that is not representative
of the true SN~Ia population; 
(iii) there is a problem with the \mlcs\ model;
(iv) the observer-frame $U$-band flux for the nearby sample 
is not properly translated into the Landolt system;
(v) the SN spectral energy distribution in the UV region 
is not adequately constrained, leading to errors in the \Kcor s.

The first possibility, a redshift-dependent flux in the $U$-band,
is unlikely, given our test based on the $ugr$ passbands for
\NoUSDSSugr\ \SDSS\ SN~Ia with $0.04 < z < 0.09$ 
(right panel in Fig.~\ref{fig:noUtest_MLCS}). Although this 
redshift range is still slightly higher than that for 
the nearby sample ($\bar{z} \sim 0.03$),  
a very rapid redshift evolution of SN~Ia properties 
would be needed to account for the discrepancy.
The second possibility is motivated by the very low selection 
\eff\ for the nearby sample, 
as indicated by the data-simulation comparison of the 
redshift distribution in the upper-left panel 
of Fig.~\ref{fig:ovzdatasim}.
However, the nearby sample shows good
agreement with the other SN~Ia samples in the $B,V,R$ passbands,
so one would have to postulate a selection effect that
biases the $U$-band more than the other bands.
The third possibility, of a problem with the light-curve model,
is difficult to exclude, but the residuals in the nearby sample
(``Nearby'' column of the left panel of Fig.~\ref{fig:lcresid_MLCS}) 
look reasonable; that rules out obvious problems, 
thus narrowing potential problems to the extrapolation 
to higher redshifts and to different passbands.

The fourth possibility seems at first sight unlikely, 
since calibration errors are typically quoted at the level 
of 0.01-0.02~mag, but it could be 
that those errors have been underestimated. 
A mis-calibration of more than 0.1 mag in $U$-band
would be needed to account for the observed anomaly.
JRK07 report that the light-curve residuals are 40\% larger
in $U$-band than in the other bands, but that does not
necessarily point to a calibration problem.
The $U$-band residuals for the nearby sample
(upper-left panel in Fig.~\ref{fig:lcresid_MLCS})
vary with epoch as the SN becomes redder,
suggesting a problem with the definition of the
$U$-band filter.

The last possibility, a difference
in the UV region of the SN spectral energy distribution, 
has been suggested in previous works 
\citep{Foley08,Sullivan09}.
Both \mlcs\ and \SALTII\ models assign larger \uncs\ 
in the UV region compared to the redder bands.
\citet{Ellis08} found that maximum-light SN~Ia SEDs varied 
significantly more at wavelengths $\lambda < 4000$~\AA\
than for redder wavelengths and concluded that the
additional dispersion is not due to extinction from
host-galaxy dust.

Since the $U$-band anomaly appears in multiple samples
(\SDSS\ and SNLS) 
as well as in multiple redshift ranges for the \SDSS\ sample
($z<0.1$ and $z>0.21$ in right panel of Fig.~\ref{fig:noUtest_MLCS})
we believe that the problem lies within the nearby SN sample, 
i.e., with observations in observer-frame $U$-band.
The most likely source of the problem is 
either the training procedure or the 
translation into the Landolt system.
We will address this issue in the future by retraining 
light-curve models with \SDSS\ SNe; the modeling of rest-frame 
$U$-band will make use of $u$-band measurements at $z < 0.1$ and 
$g$-band measurements at $z > \ZUSDSS$.

% ------------------------------------------------
  \subsubsection{ Comparison with WV07 Result}
  \label{subsec:WV07_compare}
% ------------------------------------------------

The \samplef\ sample combination ({\fsym}) 
corresponds to one of the combinations analyzed by the 
ESSENCE collaboration in WV07 (with the exceptions of a 
different minimum redshift cut and different light-curve 
selection criteria). To compare with those results, 
we repeat the sample ({\fsym}) analysis using 
the same BAO prior as in WV07, i.e., without the CMB prior.
We find $w = -0.75 \pm 0.11 ({\rm stat})$,
which differs from the WV07 value by  
$0.32$, or about $3\sigma_{stat}$. 
Since the two analyses are based on the same data,
the statistical significance of the discrepancy is
much larger than $3\sigma_{stat}$. 
If we add the systematic \uncs\ in quadrature
($\dw_{syst} = 0.18$), which is clearly an overestimate,
the discrepancy is still fairly significant ($1.8\sigma$).
Both analyses are based on the \mlcs\ method, but our changes 
in \mlcs\ parameters and priors result in systematic differences 
that are explained below.

\begin{figure}[h]
  \epsscale{1.0}
  \plotone{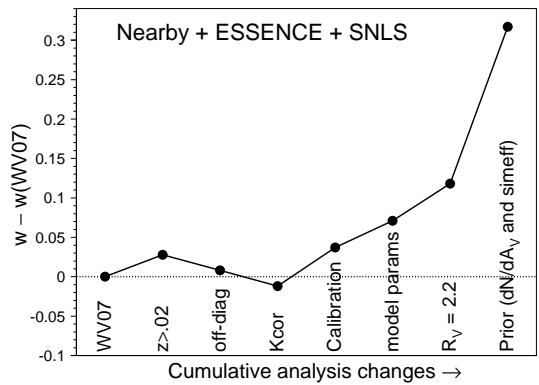}
  \caption{
	Changes in $w$ between the WV07 analysis and the analysis 
	presented here, for the {\samplef} sample combination. 
	Changes are implemented sequentially from 
	left to right, and the cumulative differences 
 	in $w$ are shown.
	The changes are described in the text.
	}
  \label{fig:WV07_evolve}
\end{figure}

We run the \mlcs\ fitter in 
``WV07 mode'' (\S~\ref{subsec:MLCS2k2}), using the WV07 parameter choices 
and reproducing their result for $w$, 
and then make cumulative changes 
in the analysis 
that evolve it toward the 
parameter inputs 
and \lc\ fitting code
used in our fiducial analysis.
The resulting shifts in $w$ relative to the WV07 value 
are shown in Fig.~\ref{fig:WV07_evolve}. 
The changes from WV07 that we implement sequentially are: 
change the minimum redshift for the nearby sample from 
0.015 to 0.02 ($z>0.02$); 
set off-diagonal model covariances to zero (off-diag); 
implement \Kcor\ improvements 
in item 2 of \S~\ref{subsec:MLCS2k2} (Kcor);
replace the Bessell filter shifts introduced by \citet{Astier06} 
and adopted by WV07 to the color transformation method used 
in our analysis 
and simultaneously 
change the primary standard from Vega to BD+17 (Calibration);
use \mlcs\ model parameters $\Mmlcs$, $p$, $q$ 
(Eq.~\ref{eq:MLCS2k2model}) corresponding to $R_V=\RV$ (\S~\ref{subsec:RV}),
but still use $R_V=\RVMW$ and WV07 priors in the \lc\ fits;
use $R_V=\RV$ with WV07 priors in the \lc\ fits;
use the $A_V$ prior and efficiency from this analysis and remove the 
WV07 requirement that each observation has 
SNR$>5$ (prior).

The largest 
source of 
change in $w$ ($\sim 0.25$) results from our 
different assumptions about 
host-galaxy dust and 
the \mlcs\ fitting prior.
Our fitting prior is based on a measurement of 
$R_V$ and of the $A_V$ distribution
using the SDSS dust sample (\S~\ref{sec:dust}) along with 
a comprehensive model of survey efficiencies using 
Monte Carlo simulations. 
In contrast to our analysis, the fitting prior used in WV07 is 
based on the assumption that $R_V=\RVMW$,
that the $A_V$ distribution is represented by the
``galactic line-of-sight'' (glos) distribution,
and  that the \spec\ targeting \eff\ is unity.

% -----------------------------------

\subsection{Results with {\SALTII} }
\label{subsec:results_salt2}

Using the \SALTII\ method (\S \ref{subsec:wfit_SALTII}),
we present cosmological results for the six
sample combinations ({\asym}--{\fsym}) of Table \ref{tb:sample_combos}. 
Table \ref{tb:salt2_lcfitres} gives the spectroscopic redshift 
and derived \SALTII\ fit parameters for each supernova that 
passes the selection cuts of \S \ref{sec:sample}. Recall that
the fit parameters $x_1$ and $c$ are estimated from the 
light-curve fits for each object. For each sample combination, 
those fit parameters are 
used to estimate the cosmological parameters, the global 
parameters $\alpha$ and $\beta$ in Eq. \ref{eq:MUSALTII} 
(see Tables \ref{tb:results_salt2_FWCDM_BAOCMB} and 
\ref{tb:results_salt2_LCDM_BAOCMB}), 
and the distance moduli, 
by minimizing the scatter in the Hubble diagram. 
Since the distance modulus estimate for each SN  
depends upon the sample combination in which the SN is included,  
upon the cosmological model parametrization, and upon the BAO and 
CMB priors, 
we provide tables for each of the six sample combinations and 
for both the \wCDM\ and \LCDM\ models.
Although the cosmological parameters have been 
corrected for the selection bias using simulations
(Table~\ref{tb:SALT2_simcor}), the distance moduli
in Table~\ref{tb:salt2_lcfitres}
do not include any bias corrections.
The entries in this Table should therefore not be used to derive 
cosmological constraints.

\begin{table}[hb]
\centering
\caption{  
	\SALTII\ light curve fit parameters including BAO+CMB priors.
	(uncertainties in parentheses). 
	The complete set of tables for each sample-combination 
	({\asym}--{\fsym}) is given in electronic form in the journal,
        and also at {\wwwTABLES}.
}
\tiny
\begin{tabular}{l cccccc }
\tableline\tableline
 SNID  & redshift\tablenotemark{a} & $\mu$ & 
 $c$   & $x_1$ 
       &  MJD$_{\rm peak}$ \\ %%% & $\chi^2/{\rm dof}$  \\ 
\tableline  % --------- 
   
%  Table of SDSS light curve fit-results for salt2             
% 
   762 & $0.1904(.0000)$ & $ 39.91( 0.09)$ & $ -0.01( 0.03)$ & $  0.73( 0.33)$ & $53625.2(  0.4)$ \\ %    & $ 34.84/ 41$ \\ 
  1032 & $0.1291(.0000)$ & $ 38.75( 0.17)$ & $  0.14( 0.06)$ & $ -3.09( 0.39)$ & $53626.6(  0.4)$ \\ %    & $ 29.06/ 40$ \\ 
  1112 & $0.2565(.0000)$ & $ 40.53( 0.16)$ & $  0.00( 0.04)$ & $ -1.06( 0.69)$ & $53630.2(  0.7)$ \\ %    & $ 40.78/ 43$ \\ 
  1166 & $0.3813(.0000)$ & $ 41.23( 0.25)$ & $  0.03( 0.07)$ & $  1.10( 1.12)$ & $53631.9(  1.2)$ \\ %    & $ 21.11/ 38$ \\ 
  1241 & $0.0858(.0000)$ & $ 37.91( 0.09)$ & $  0.09( 0.03)$ & $ -0.90( 0.17)$ & $53635.3(  0.2)$ \\ %    & $ 13.37/ 41$ \\ 

\tableline  % ------------------------ 
\end{tabular} 
\tablenotetext{a}{Spectroscopic redshift in CMB frame.}
\label{tb:salt2_lcfitres} 
\end{table}

% -----------------------------------------------
\subsubsection{\SALTII\ Hubble Dispersion}
\label{subsec:SALT-disp}
% -----------------------------------------------

In the \SALTII\ method, an intrinsic dispersion ($\sigmuint$)
is added in quadrature to the distance modulus \unc\
(Eq.~\ref{eq:sigmudef}) such that the resulting 
Hubble diagram $\chi^2_{\mu}/N_{\rm dof}$ is equal to one.
Using the \wCDM\  model parametrization and the BAO+CMB prior,
Table~\ref{tb:salt2_sigmuint} gives the $\sigmuint$ values 
obtained from fitting each SN sample independently and 
setting $\chisqmu=N_{\rm dof}$. 
For the nearby, {\SDSS}, and ESSENCE samples,
the \SALTII\ values are similar to those from \mlcs\ in the 
fourth line of Table \ref{tb:mlcs_fitquality}.
For the SNLS sample, the \SALTII\ dispersion is
significantly smaller than that from \mlcs,
while for the HST sample the \SALTII\ dispersion is larger.
The smaller dispersion for SNLS may derive in part from the fact 
that the \SALTII\ model was partially trained on SNLS data.
For sample combinations {\asym}--{\fsym},
the $\sigmuint$ values are in Tables 
\ref{tb:results_salt2_FWCDM_BAOCMB} and 
\ref{tb:results_salt2_LCDM_BAOCMB}.

\begin{table}[hb]
\centering
\caption{  
	Intrinsic dispersion ($\sigmuint$) required for $\chi^2_\mu=N_{\rm dof}$ for 
	each SN~Ia sample fit 
	separately with the \SALTII\ method. 
    }
\begin{tabular}{ l | ccccc }
\tableline\tableline
% -------------------------
        
 &  \multicolumn{5}{c}{$\sigmuint$ for sample:}  \\
 &   Nearby & SDSS-II & ESSENCE & SNLS & HST   \\
\tableline
% ----------
 independent fits &
  $\LOWZSALTDISP$ &
  $\SDSSSALTDISP$ &
  $\ESSESALTDISP$ &
  $\SNLSSALTDISP$ &
  $\HSTSALTDISP$  \\
% -------------
\tableline
\end{tabular}
\label{tb:salt2_sigmuint} 
\end{table}

% ------------------------------------------------------------
\subsubsection{\SALTII\ Hubble Diagrams and Cosmological Parameters}
\label{subsec:SALT2_RESULTS}
% ------------------------------------------------------------

\begin{figure}[h]
  \epsscale{1.1}
  \plotone{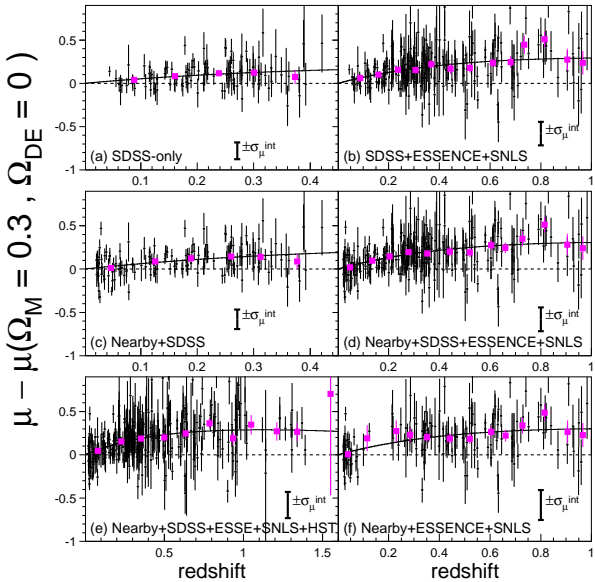}
  \caption{
	Hubble residuals for the \SALTII\ method: 
	differences between measured SN distance moduli and 
	those for an open CDM model ($\OM=0.3$, $\ODE=0$) vs. redshift 
	for the six SN sample combinations. 
	Large, square (pink) points show weighted averages in 
	weighted redshift bins (see Fig.~\ref{fig:muresids_MLCS} caption).
	Solid curves show the difference  between the best-fit 
	\wCDM\ model distance modulus for that sample combination 
	and that for the open model, normalized 
	to have the same value of the Hubble parameter.   
	The error bars on the data points correspond to the
	distance modulus error $\sigmufit$ from the \SALTII\ 
	light-curve fit (Eq.~\ref{eq:sigmudef}), 
	i.e., they do not include the intrinsic scatter or the 
	effects of redshift and peculiar velocity errors.
	The vertical bars show the values of the intrinsic \unc,
	$\sigmuint$, included in the cosmology fits
	so that the $\chi^2$ per degree of freedom is unity for 
        each Hubble diagram.
        }
  \label{fig:muresids_SALT2}
\end{figure}

\begin{figure}[h]
  \epsscale{1.15}
  \plotone{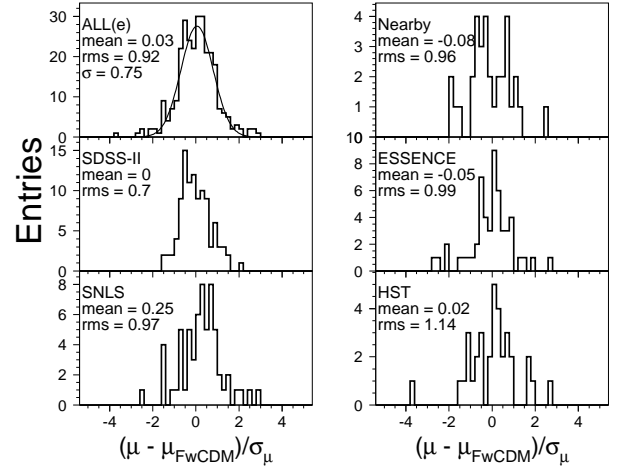}
  \caption{
	Distribution of normalized Hubble residuals (pull) for the 
	\SALTII\ method, for sample combination ({\esym}) comprising 
	all five samples (upper left, along with Gaussian fit),
	and for each SN sample indicated on the panels.
	$\mu$ is the measured SN distance modulus,
	$\MUwCDM$ is the distance modulus from the best-fit
	\wCDM\ model for sample combination ({\esym}),
	and $\sigmutot$ is the total \unc\ (Eq.~\ref{eq:sigmudef}).
	}
  \label{fig:muresids_norm_SALT2}
\end{figure}

Figure \ref{fig:muresids_SALT2} shows the differences between the 
\SALTII\ estimated SN distance moduli $\mu_i$ and those for an open 
CDM model with no dark energy ($\OM=0.3$, $\ODE=0$) as a function 
of redshift.
Figure~\ref{fig:muresids_norm_SALT2} shows the distribution of 
normalized residuals, $(\mu_i - \MUwCDM)/\sigmutot$,
where $\MUwCDM$ is the distance modulus from the best-fit
\wCDM\ model for sample combination (\esym), 
and $\sigmutot$ is the total \unc\ defined in Eq.~\ref{eq:sigmudef}.
The bulk of the distribution of all \NSNTOT\ normalized residuals 
(upper left panel of Fig.~\ref{fig:muresids_norm_SALT2})
is well fit by a Gaussian with $\sigma=\SALTMUPULLSIG$;
outliers increase the rms to $\SALTMUPULLRMS$.
The rms for combination ({\esym}) is slightly smaller than one
because covariances from the fit are 
not included in the calculation of $\sigmutot$.

\begin{figure}[h]
  \epsscale{1.1}
  \plotone{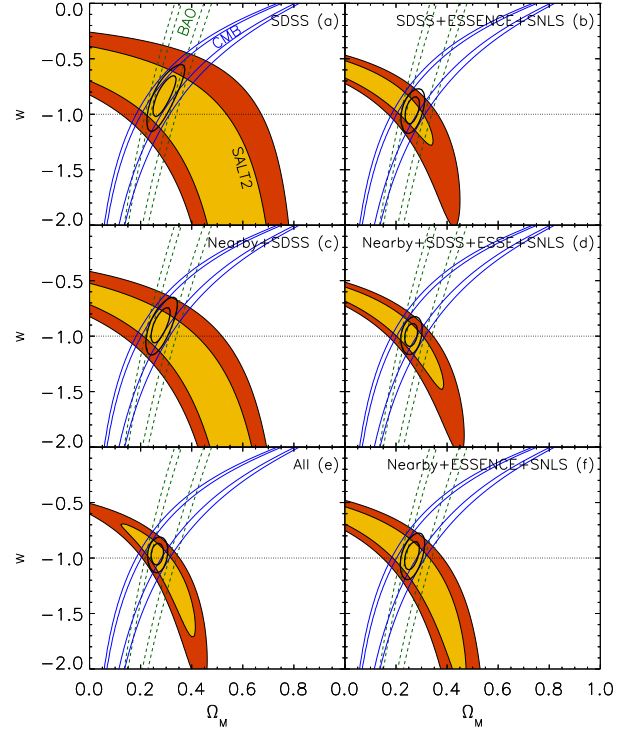}
  \caption{
 	For the \wCDM\ model, \SALTII\ statistical-\unc\ contours 
	in the $\OM$-$w$ plane for each of the six SN sample 
	combinations indicated on the plots.
	Long, black contours:  68\%, 95\%, and 99\% 
	confidence level regions for the SN data alone; 
	green contours: corresponding CL regions for SDSS BAO 
	\citep{Eisenstein05}; 
	blue contours: CL regions for WMAP5 CMB \citep{Komatsu2008}; 
	closed, black contours:  combined constraints from SN+BAO+CMB. 
% -----------
        }
  \label{fig:FWCDM_SALT2_statcont}
\end{figure}

\begin{figure}[h]
  \epsscale{1.1}
  \plotone{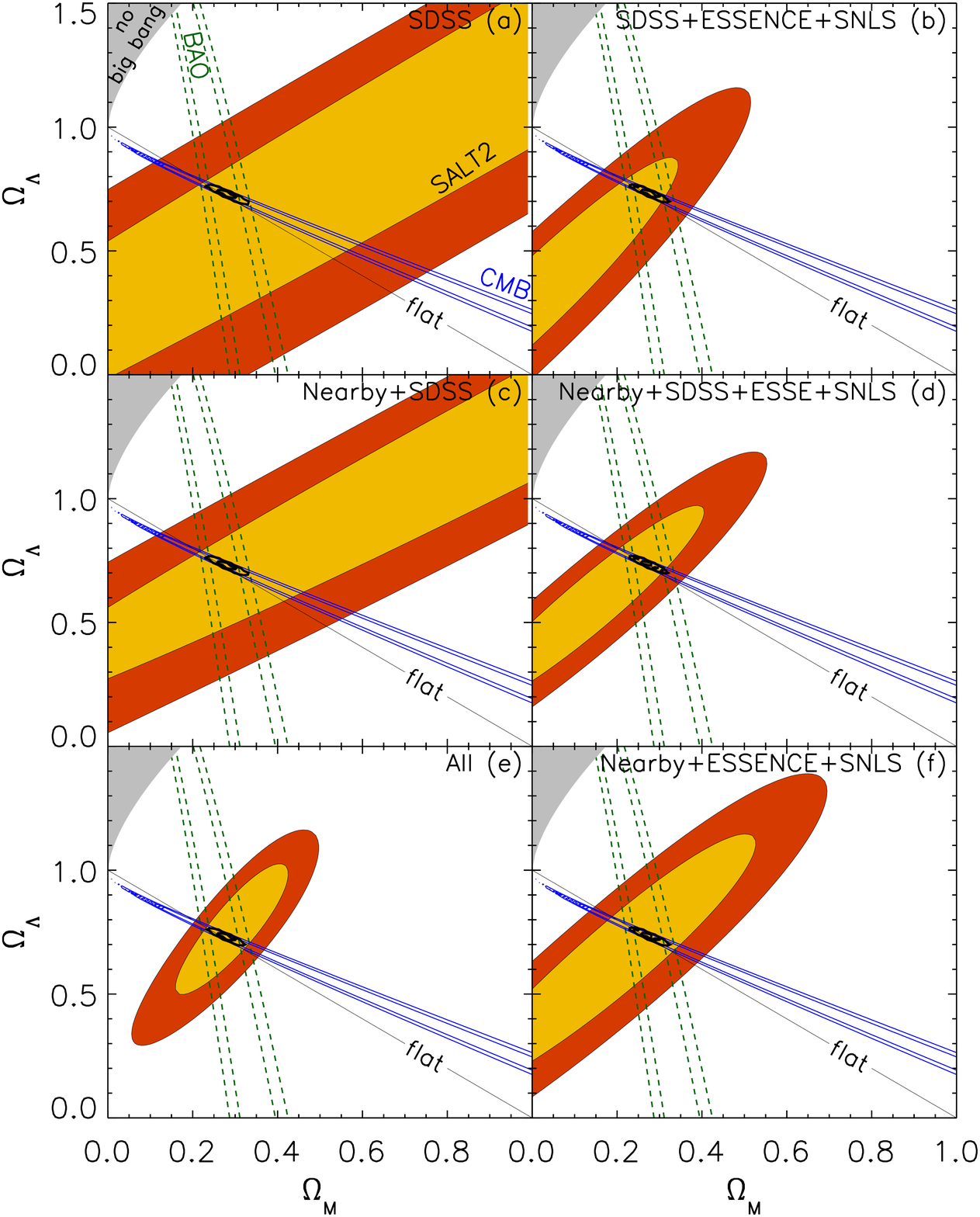}
  \caption{
 	For the \LCDM\ model, \SALTII\ statistical-\unc\ contours 
	in the $\OM$-$\OLAM$ plane for each of the six SN sample 
	combinations indicated on the plots.
	Long, black contours:  68\%, 95\%, and 99\% 
	confidence level regions for the SN data alone; 
	green contours: corresponding CL regions for SDSS BAO 
	\citep{Eisenstein05}; 
	blue contours: CL regions for WMAP5 CMB \citep{Komatsu2008}; 
	closed, red contours:  combined constraints from SN+BAO+CMB. 
% -----------
        }
  \label{fig:LCDM_SALT2_statcont}
\end{figure}

\begin{figure}[h]
  \epsscale{1.1}
  \plotone{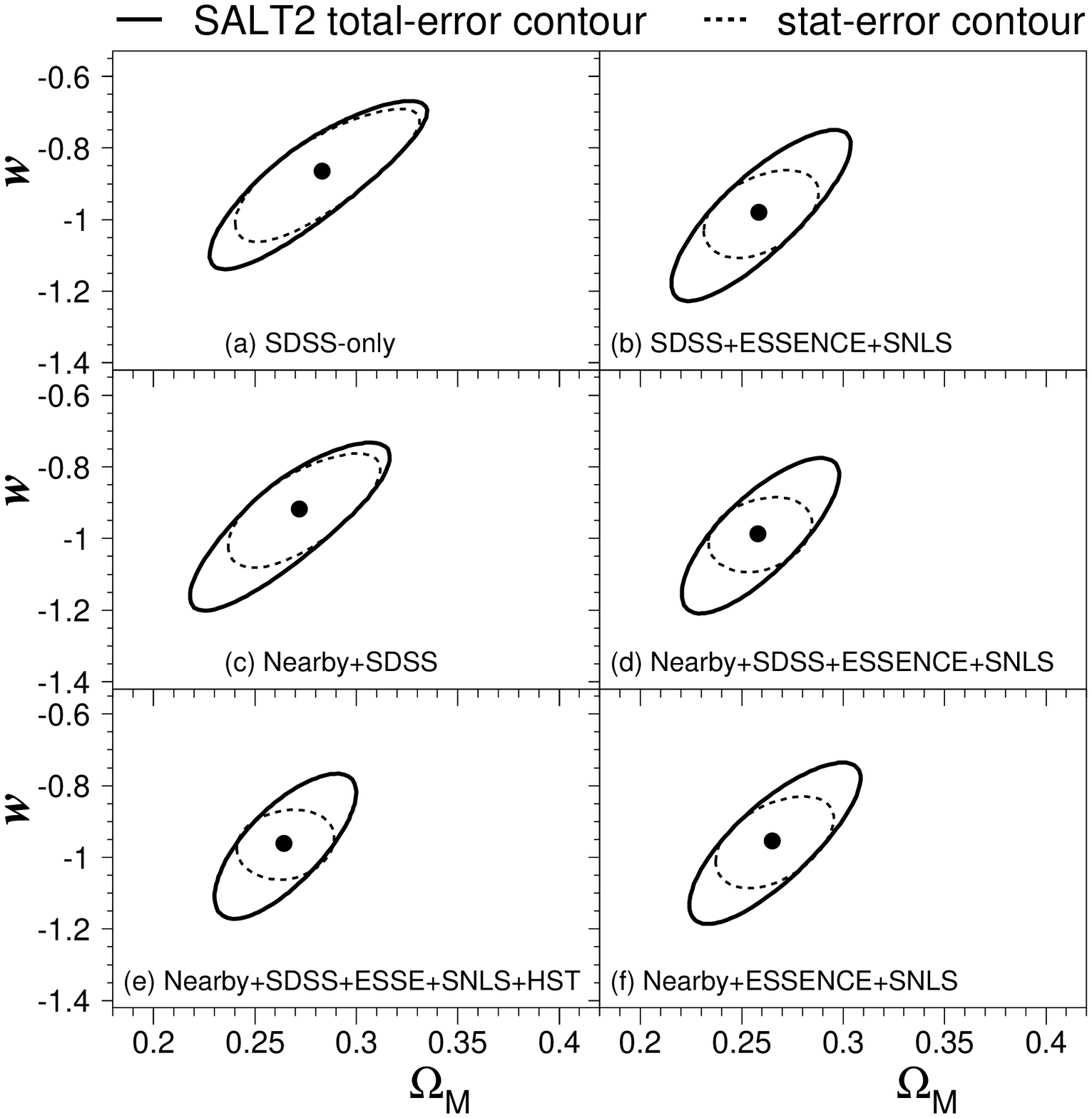}
  \caption{
	For \SALTII\ and the \wCDM\ model, 68\% CL contours in the 
	$\OM$-$w$ plane for each of the six SN sample combinations, 
	using the combined SN+BAO+CMB constraints. 
	Solid contours are total (statistical+systematic) \unc;
	dashed contours are statistical only.
        Systematic errors have been included using the prescription in 
        Appendix~\ref{app:contours}.
        }
  \label{fig:etot_contours_FWCDM_SALT2}
\end{figure}

\begin{figure}[h]
  \epsscale{1.1}
 \plotone{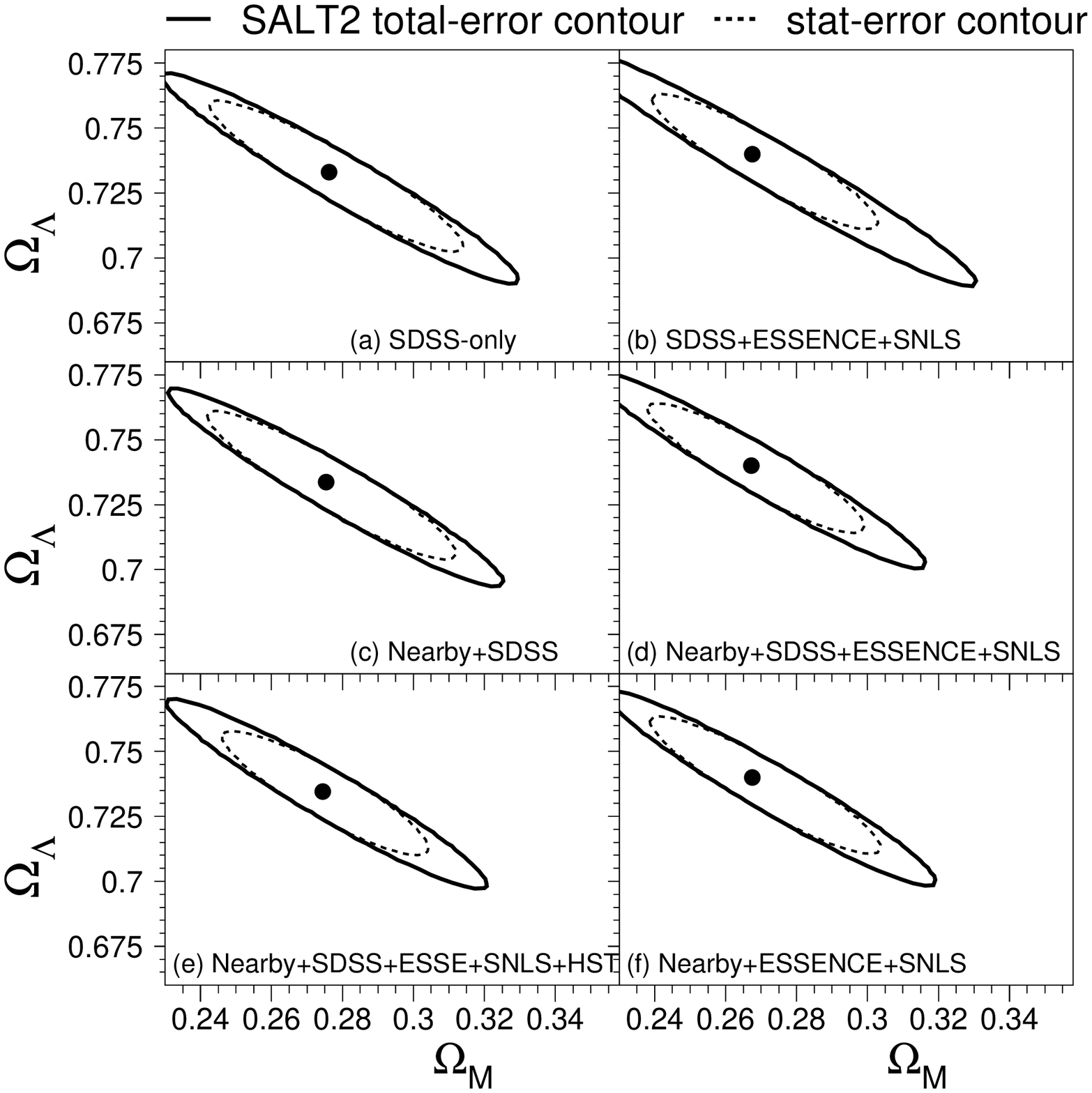}
  \caption{
	For \SALTII\ and the \LCDM\ model, 68\% CL contours in the 
	$\OM$-$\OLAM$ plane for each of the six SN sample combinations, 
	using the combined SN+BAO+CMB constraints. 
	Solid contours are total (statistical+systematic) \unc;
	dashed contours are statistical only.
        Systematic errors have been included using the prescription in 
        Appendix~\ref{app:contours}.
        }
  \label{fig:etot_contours_LCDM_SALT2}
\end{figure}

Figures \ref{fig:FWCDM_SALT2_statcont} and \ref{fig:LCDM_SALT2_statcont} 
show the \SALTII\ statistical-\unc\ contours for the \wCDM\ and 
\LCDM\ models; for the latter, the \SALTII\ SN contours are more 
consistent with the BAO+CMB constraints than the \mlcs\ contours were.
For the combined SN+BAO+CMB results, 
the total uncertainty contours, including systematic errors, 
are shown in Figs. \ref{fig:etot_contours_FWCDM_SALT2} 
and \ref{fig:etot_contours_LCDM_SALT2}. 
The best-fit cosmological parameter values and uncertainties, 
marginalizing over $H_0$ and 
incorporating the bias corrections of Table \ref{tb:SALT2_simcor},
are given in Tables \ref{tb:results_salt2_FWCDM_BAOCMB} and 
\ref{tb:results_salt2_LCDM_BAOCMB}.
The \SALTII\ statistical 
errors on cosmological parameters are consistent with those 
from {\mlcs}. The systematic errors for the two methods
are similar for sample combinations (\bsym) and (\dsym-\fsym), 
but the \SALTII\ systematic \unc\ is significantly smaller
for the sample combinations in which \SDSS\ is the 
high-redshift sample (\asym\ and \csym).
The large difference in the systematic \unc\ is driven by 
the $U$-band anonaly.

Among the six SN sample combinations, the best-fit values of 
$w$ for the \wCDM\ model again fall into two groups, 
as for the \mlcs\ results, but with the opposite trend in 
$w$ compared to {\mlcs}. 
For sample combinations (\asym) and (\csym), 
in which \SDSS\ is the high-redshift sample, 
the \SALTII\ method results in 
$w = \FWCDMBAOCMBWa$ and $\FWCDMBAOCMBWc$, 
comparable to the \mlcs\ results of  $\MLCSWRESa$ and $\MLCSWRESc$ 
for the same sample combinations. 
However, for the other four sample combinations, 
which include higher-redshift SNe, \SALTII\ 
yields $w = \FWCDMBAOCMBWb$ to $\FWCDMBAOCMBWf$, 
compared with the \mlcs\ result of $w = \MLCSWRESb$ to $\MLCSWRESe$
for the same sample combinations.
Based on studies with simulated sample combinations,
the observed difference between these two groups of 
\SALTII\ results is 
not statistically significant.
However, the difference between the \mlcs\ and \SALTII\ results 
for the higher-redshift samples appears to be significant. 
We discuss these differences further in \S~\ref{sec:results_compare}.

% ===========================

\begin{table}
\centering
\caption{
   For the \wCDM\ model, constraints on $w$ and $\OM$ from 
   \SALTII\ SN distances combined with SDSS LRG BAO and WMAP5 CMB results. 
    }
\label{tb:results_salt2_FWCDM_BAOCMB}
\begin{tabular}{l | cccccc}
\tableline\tableline
% -------------------------
                   &  \multicolumn{5}{c}{Result for sample:}  \\
      & {{\asym}}
      & {{\bsym}}
      & {{\csym}}
      & {{\dsym}}
      & {{\esym}}
      & {{\fsym}}
         \\
\tableline
 $\sigmuint$
  & $\FWCDMBAOCMBDISPa$
  & $\FWCDMBAOCMBDISPb$
  & $\FWCDMBAOCMBDISPc$
  & $\FWCDMBAOCMBDISPd$
  & $\FWCDMBAOCMBDISPe$
  & $\FWCDMBAOCMBDISPf$
\\
 $\RMSMU$
  & $\FWCDMBAOCMBRMSa$
  & $\FWCDMBAOCMBRMSb$
  & $\FWCDMBAOCMBRMSc$
  & $\FWCDMBAOCMBRMSd$
  & $\FWCDMBAOCMBRMSe$
  & $\FWCDMBAOCMBRMSf$
\\
\\
 $w$ 
  & $\FWCDMBAOCMBWa$
  & $\FWCDMBAOCMBWb$
  & $\FWCDMBAOCMBWc$
  & $\FWCDMBAOCMBWd$
  & $\FWCDMBAOCMBWe$
  & $\FWCDMBAOCMBWf$
\\
 $\wstatsym$ 
  & $\FWCDMBAOCMBDWSTATa$
  & $\FWCDMBAOCMBDWSTATb$
  & $\FWCDMBAOCMBDWSTATc$
  & $\FWCDMBAOCMBDWSTATd$
  & $\FWCDMBAOCMBDWSTATe$
  & $\FWCDMBAOCMBDWSTATf$
\\ \\
 $\wsystsym$
  & {\large ${}^{+\FWCDMBAOCMBDWSYSTPa}_{-\FWCDMBAOCMBDWSYSTMa}$}
  & $\FWCDMBAOCMBDWSYSTPb$
  & {\large ${}^{+\FWCDMBAOCMBDWSYSTPc}_{-\FWCDMBAOCMBDWSYSTMc}$}
  & $\FWCDMBAOCMBDWSYSTPd$
  & $\FWCDMBAOCMBDWSYSTPe$
  & $\FWCDMBAOCMBDWSYSTPf$
\\  \\
 $\wtotsym$
  & {\large ${}^{+\FWCDMBAOCMBDWTOTPa}_{-\FWCDMBAOCMBDWTOTMa}$}
  & $\FWCDMBAOCMBDWTOTPb$
  & {\large ${}^{+\FWCDMBAOCMBDWTOTPc}_{-\FWCDMBAOCMBDWTOTMc}$}
  & $\FWCDMBAOCMBDWTOTPd$
  & $\FWCDMBAOCMBDWTOTPe$
  & $\FWCDMBAOCMBDWTOTPf$
\\  \\
% ERRTOT/ERRSTAT-SALT2-w-P-a = 1.129
% ERRTOT/ERRSTAT-SALT2-w-M-a = 1.392
% ERRTOT/ERRSTAT-SALT2-w-P-b = 1.970
% ERRTOT/ERRSTAT-SALT2-w-M-b = 1.970
% ERRTOT/ERRSTAT-SALT2-w-P-c = 1.198
% ERRTOT/ERRSTAT-SALT2-w-M-c = 1.742
% ERRTOT/ERRSTAT-SALT2-w-P-d = 2.108
% ERRTOT/ERRSTAT-SALT2-w-M-d = 2.108
% ERRTOT/ERRSTAT-SALT2-w-P-e = 2.111
% ERRTOT/ERRSTAT-SALT2-w-M-e = 2.111
% ERRTOT/ERRSTAT-SALT2-w-P-f = 1.772
% ERRTOT/ERRSTAT-SALT2-w-M-f = 1.772
% 
 $\OM$ 
  & $\FWCDMBAOCMBOMEGAMa$
  & $\FWCDMBAOCMBOMEGAMb$
  & $\FWCDMBAOCMBOMEGAMc$
  & $\FWCDMBAOCMBOMEGAMd$
  & $\FWCDMBAOCMBOMEGAMe$
  & $\FWCDMBAOCMBOMEGAMf$
\\
 $\sigma_{\OM}$ (stat)
  & $\FWCDMBAOCMBDOMEGAMSTATa$
  & $\FWCDMBAOCMBDOMEGAMSTATb$
  & $\FWCDMBAOCMBDOMEGAMSTATc$
  & $\FWCDMBAOCMBDOMEGAMSTATd$
  & $\FWCDMBAOCMBDOMEGAMSTATe$
  & $\FWCDMBAOCMBDOMEGAMSTATf$
\\ \\
 $\OMsystsym$ 
  & {\large ${}^{+\FWCDMBAOCMBDOMEGAMSYSTPa}_{-\FWCDMBAOCMBDOMEGAMSYSTMa}$}
  & $\FWCDMBAOCMBDOMEGAMSYSTPb$
  & {\large ${}^{+\FWCDMBAOCMBDOMEGAMSYSTPc}_{-\FWCDMBAOCMBDOMEGAMSYSTMc}$}
  & $\FWCDMBAOCMBDOMEGAMSYSTPd$
  & $\FWCDMBAOCMBDOMEGAMSYSTPe$
  & $\FWCDMBAOCMBDOMEGAMSYSTPf$
\\ \\
 $\OMtotsym$ 
  & {\large ${}^{+\FWCDMBAOCMBDOMEGAMTOTPa}_{-\FWCDMBAOCMBDOMEGAMTOTMa}$}
  & $\FWCDMBAOCMBDOMEGAMTOTPb$
  & {\large ${}^{+\FWCDMBAOCMBDOMEGAMTOTPc}_{-\FWCDMBAOCMBDOMEGAMTOTMc}$}
  & $\FWCDMBAOCMBDOMEGAMTOTPd$
  & $\FWCDMBAOCMBDOMEGAMTOTPe$
  & $\FWCDMBAOCMBDOMEGAMTOTPf$
\\ \\
% ERRTOT/ERRSTAT-SALT2-OM-P-a = 1.121
% ERRTOT/ERRSTAT-SALT2-OM-M-a = 1.313
% ERRTOT/ERRSTAT-SALT2-OM-P-b = 1.842
% ERRTOT/ERRSTAT-SALT2-OM-M-b = 1.842
% ERRTOT/ERRSTAT-SALT2-OM-P-c = 1.173
% ERRTOT/ERRSTAT-SALT2-OM-M-c = 1.532
% ERRTOT/ERRSTAT-SALT2-OM-P-d = 1.879
% ERRTOT/ERRSTAT-SALT2-OM-M-d = 1.879
% ERRTOT/ERRSTAT-SALT2-OM-P-e = 1.884
% ERRTOT/ERRSTAT-SALT2-OM-M-e = 1.884
% ERRTOT/ERRSTAT-SALT2-OM-P-f = 1.680
% ERRTOT/ERRSTAT-SALT2-OM-M-f = 1.680
\hline 
 $\alpha$ 
  & $\FWCDMBAOCMBALPHAa$
  & $\FWCDMBAOCMBALPHAb$
  & $\FWCDMBAOCMBALPHAc$
  & $\FWCDMBAOCMBALPHAd$
  & $\FWCDMBAOCMBALPHAe$
  & $\FWCDMBAOCMBALPHAf$
\\
 $\sigma_\alpha(stat)$
  & $\FWCDMBAOCMBDALPHASTATa$
  & $\FWCDMBAOCMBDALPHASTATb$
  & $\FWCDMBAOCMBDALPHASTATc$
  & $\FWCDMBAOCMBDALPHASTATd$
  & $\FWCDMBAOCMBDALPHASTATe$
  & $\FWCDMBAOCMBDALPHASTATf$
\\
 $\sigma_\alpha(syst)$
  & $\FWCDMBAOCMBDALPHASYSTPa$
  & $\FWCDMBAOCMBDALPHASYSTPb$
  & $\FWCDMBAOCMBDALPHASYSTPc$
  & $\FWCDMBAOCMBDALPHASYSTPd$
  & $\FWCDMBAOCMBDALPHASYSTPe$
  & $\FWCDMBAOCMBDALPHASYSTPf$
\\
 $\sigma_\alpha(tot)$ 
  & $\FWCDMBAOCMBDALPHATOTPa$
  & $\FWCDMBAOCMBDALPHATOTPb$
  & $\FWCDMBAOCMBDALPHATOTPc$
  & $\FWCDMBAOCMBDALPHATOTPd$
  & $\FWCDMBAOCMBDALPHATOTPe$
  & $\FWCDMBAOCMBDALPHATOTPf$
\\
 $\beta$ 
  & $\FWCDMBAOCMBBETAa$
  & $\FWCDMBAOCMBBETAb$
  & $\FWCDMBAOCMBBETAc$
  & $\FWCDMBAOCMBBETAd$
  & $\FWCDMBAOCMBBETAe$
  & $\FWCDMBAOCMBBETAf$
\\
 $\sigma_\beta(stat)$ 
  & $\FWCDMBAOCMBDBETASTATa$
  & $\FWCDMBAOCMBDBETASTATb$
  & $\FWCDMBAOCMBDBETASTATc$
  & $\FWCDMBAOCMBDBETASTATd$
  & $\FWCDMBAOCMBDBETASTATe$
  & $\FWCDMBAOCMBDBETASTATf$
\\
 $\sigma_\beta (syst)$ 
  & $\FWCDMBAOCMBDBETASYSTPa$
  & $\FWCDMBAOCMBDBETASYSTPb$
  & $\FWCDMBAOCMBDBETASYSTPc$
  & $\FWCDMBAOCMBDBETASYSTPd$
  & $\FWCDMBAOCMBDBETASYSTPe$
  & $\FWCDMBAOCMBDBETASYSTPf$
\\
 $\sigma_\beta (tot)$ 
  & $\FWCDMBAOCMBDBETATOTPa$
  & $\FWCDMBAOCMBDBETATOTPb$
  & $\FWCDMBAOCMBDBETATOTPc$
  & $\FWCDMBAOCMBDBETATOTPd$
  & $\FWCDMBAOCMBDBETATOTPe$
  & $\FWCDMBAOCMBDBETATOTPf$
\\
\tableline
\tableline
\end{tabular}
\tablenotetext{{\asym}}{ \samplea }
\tablenotetext{{\bsym}}{ \sampleb }
\tablenotetext{{\csym}}{ \samplec }
\tablenotetext{{\dsym}}{ \sampled }
\tablenotetext{{\esym}}{ \samplee }
\tablenotetext{{\fsym}}{ \samplef }
\end{table}

\begin{table}
\centering
\caption{
  For the \LCDM\ model, constraints on $\OM$ and $\OLAM$ 
  from \SALTII\ SN distances combined with BAO and CMB 
  results. 
    }
\label{tb:results_salt2_LCDM_BAOCMB}
\begin{tabular}{l | cccccc}
\tableline\tableline
% -------------------------
                   &  \multicolumn{5}{c}{Result for sample:}  \\
      & {{\asym}}
      & {{\bsym}}
      & {{\csym}}
      & {{\dsym}}
      & {{\esym}}
      & {{\fsym}}
         \\
\tableline
 $\sigmuint$
  & $\LCDMBAOCMBDISPa$
  & $\LCDMBAOCMBDISPb$
  & $\LCDMBAOCMBDISPc$
  & $\LCDMBAOCMBDISPd$
  & $\LCDMBAOCMBDISPe$
  & $\LCDMBAOCMBDISPf$
\\
 $\RMSMU$
  & $\LCDMBAOCMBRMSa$
  & $\LCDMBAOCMBRMSb$
  & $\LCDMBAOCMBRMSc$
  & $\LCDMBAOCMBRMSd$
  & $\LCDMBAOCMBRMSe$
  & $\LCDMBAOCMBRMSf$
\\
\\
 $\OL$ 
  & $\LCDMBAOCMBOMEGALa$
  & $\LCDMBAOCMBOMEGALb$
  & $\LCDMBAOCMBOMEGALc$
  & $\LCDMBAOCMBOMEGALd$
  & $\LCDMBAOCMBOMEGALe$
  & $\LCDMBAOCMBOMEGALf$
\\
 $\sigma_{\OL}(stat)$ 
  & $\LCDMBAOCMBDOMEGALSTATa$
  & $\LCDMBAOCMBDOMEGALSTATb$
  & $\LCDMBAOCMBDOMEGALSTATc$
  & $\LCDMBAOCMBDOMEGALSTATd$
  & $\LCDMBAOCMBDOMEGALSTATe$
  & $\LCDMBAOCMBDOMEGALSTATf$
\\
 $\sigma_{\OL}(syst)$ 
  & $\LCDMBAOCMBDOMEGALSYSTPa$
  & $\LCDMBAOCMBDOMEGALSYSTPb$
  & $\LCDMBAOCMBDOMEGALSYSTPc$
  & $\LCDMBAOCMBDOMEGALSYSTPd$
  & $\LCDMBAOCMBDOMEGALSYSTPe$
  & $\LCDMBAOCMBDOMEGALSYSTPf$
\\
 $\sigma_{\OL}(tot)$ 
  & $\LCDMBAOCMBDOMEGALTOTPa$
  & $\LCDMBAOCMBDOMEGALTOTPb$
  & $\LCDMBAOCMBDOMEGALTOTPc$
  & $\LCDMBAOCMBDOMEGALTOTPd$
  & $\LCDMBAOCMBDOMEGALTOTPe$
  & $\LCDMBAOCMBDOMEGALTOTPf$
\\
% ERRTOT/ERRSTAT-SALT2-OL-P-a = 1.422
% ERRTOT/ERRSTAT-SALT2-OL-M-a = 1.404
% ERRTOT/ERRSTAT-SALT2-OL-P-b = 1.801
% ERRTOT/ERRSTAT-SALT2-OL-M-b = 1.801
% ERRTOT/ERRSTAT-SALT2-OL-P-c = 1.369
% ERRTOT/ERRSTAT-SALT2-OL-M-c = 1.335
% ERRTOT/ERRSTAT-SALT2-OL-P-d = 1.557
% ERRTOT/ERRSTAT-SALT2-OL-M-d = 1.557
% ERRTOT/ERRSTAT-SALT2-OL-P-e = 1.570
% ERRTOT/ERRSTAT-SALT2-OL-M-e = 1.570
% ERRTOT/ERRSTAT-SALT2-OL-P-f = 1.441
% ERRTOT/ERRSTAT-SALT2-OL-M-f = 1.441
\\
 $\OM$ 
  & $\LCDMBAOCMBOMEGAMa$
  & $\LCDMBAOCMBOMEGAMb$
  & $\LCDMBAOCMBOMEGAMc$
  & $\LCDMBAOCMBOMEGAMd$
  & $\LCDMBAOCMBOMEGAMe$
  & $\LCDMBAOCMBOMEGAMf$
\\
 $\sigma_{\OM}(stat)$
  & $\LCDMBAOCMBDOMEGAMSTATa$
  & $\LCDMBAOCMBDOMEGAMSTATb$
  & $\LCDMBAOCMBDOMEGAMSTATc$
  & $\LCDMBAOCMBDOMEGAMSTATd$
  & $\LCDMBAOCMBDOMEGAMSTATe$
  & $\LCDMBAOCMBDOMEGAMSTATf$
\\
 $\sigma_{\OM}(syst)$ 
  & $\LCDMBAOCMBDOMEGAMSYSTPa$
  & $\LCDMBAOCMBDOMEGAMSYSTPb$
  & $\LCDMBAOCMBDOMEGAMSYSTPc$
  & $\LCDMBAOCMBDOMEGAMSYSTPd$
  & $\LCDMBAOCMBDOMEGAMSYSTPe$
  & $\LCDMBAOCMBDOMEGAMSYSTPf$
\\
 $\sigma_{\OM}(tot)$ 
  & $\LCDMBAOCMBDOMEGAMTOTPa$
  & $\LCDMBAOCMBDOMEGAMTOTPb$
  & $\LCDMBAOCMBDOMEGAMTOTPc$
  & $\LCDMBAOCMBDOMEGAMTOTPd$
  & $\LCDMBAOCMBDOMEGAMTOTPe$
  & $\LCDMBAOCMBDOMEGAMTOTPf$
\\
% ERRTOT/ERRSTAT-SALT2-OM-P-a = 1.158
% ERRTOT/ERRSTAT-SALT2-OM-M-a = 1.180
% ERRTOT/ERRSTAT-SALT2-OM-P-b = 1.341
% ERRTOT/ERRSTAT-SALT2-OM-M-b = 1.341
% ERRTOT/ERRSTAT-SALT2-OM-P-c = 1.150
% ERRTOT/ERRSTAT-SALT2-OM-M-c = 1.189
% ERRTOT/ERRSTAT-SALT2-OM-P-d = 1.350
% ERRTOT/ERRSTAT-SALT2-OM-M-d = 1.350
% ERRTOT/ERRSTAT-SALT2-OM-P-e = 1.346
% ERRTOT/ERRSTAT-SALT2-OM-M-e = 1.346
% ERRTOT/ERRSTAT-SALT2-OM-P-f = 1.241
% ERRTOT/ERRSTAT-SALT2-OM-M-f = 1.241
\\
 $\alpha$ 
  & $\LCDMBAOCMBALPHAa$
  & $\LCDMBAOCMBALPHAb$
  & $\LCDMBAOCMBALPHAc$
  & $\LCDMBAOCMBALPHAd$
  & $\LCDMBAOCMBALPHAe$
  & $\LCDMBAOCMBALPHAf$
\\
 $\sigma_\alpha(stat)$ 
  & $\LCDMBAOCMBDALPHASTATa$
  & $\LCDMBAOCMBDALPHASTATb$
  & $\LCDMBAOCMBDALPHASTATc$
  & $\LCDMBAOCMBDALPHASTATd$
  & $\LCDMBAOCMBDALPHASTATe$
  & $\LCDMBAOCMBDALPHASTATf$
\\
 $\sigma_\alpha (syst)$ 
  & $\LCDMBAOCMBDALPHASYSTPa$
  & $\LCDMBAOCMBDALPHASYSTPb$
  & $\LCDMBAOCMBDALPHASYSTPc$
  & $\LCDMBAOCMBDALPHASYSTPd$
  & $\LCDMBAOCMBDALPHASYSTPe$
  & $\LCDMBAOCMBDALPHASYSTPf$
\\
 $\sigma_\alpha (tot)$ 
  & $\LCDMBAOCMBDALPHATOTPa$
  & $\LCDMBAOCMBDALPHATOTPb$
  & $\LCDMBAOCMBDALPHATOTPc$
  & $\LCDMBAOCMBDALPHATOTPd$
  & $\LCDMBAOCMBDALPHATOTPe$
  & $\LCDMBAOCMBDALPHATOTPf$
\\ \\
 $\beta$
  & $\LCDMBAOCMBBETAa$
  & $\LCDMBAOCMBBETAb$
  & $\LCDMBAOCMBBETAc$
  & $\LCDMBAOCMBBETAd$
  & $\LCDMBAOCMBBETAe$
  & $\LCDMBAOCMBBETAf$
\\
 $\sigma_\beta(stat)$ 
  & $\LCDMBAOCMBDBETASTATa$
  & $\LCDMBAOCMBDBETASTATb$
  & $\LCDMBAOCMBDBETASTATc$
  & $\LCDMBAOCMBDBETASTATd$
  & $\LCDMBAOCMBDBETASTATe$
  & $\LCDMBAOCMBDBETASTATf$
\\
 $\sigma_\beta(syst)$ 
  & $\LCDMBAOCMBDBETASYSTMa$
  & $\LCDMBAOCMBDBETASYSTMb$
  & $\LCDMBAOCMBDBETASYSTMc$
  & $\LCDMBAOCMBDBETASYSTMd$
  & $\LCDMBAOCMBDBETASYSTMe$
  & $\LCDMBAOCMBDBETASYSTMf$
\\
 $\sigma_\beta (tot)$ 
  & $\LCDMBAOCMBDBETATOTMa$
  & $\LCDMBAOCMBDBETATOTMb$
  & $\LCDMBAOCMBDBETATOTMc$
  & $\LCDMBAOCMBDBETATOTMd$
  & $\LCDMBAOCMBDBETATOTMe$
  & $\LCDMBAOCMBDBETATOTMf$
\\
\tableline
\tableline
\end{tabular}
%
%\tablenotetext{{\asym}}{ \samplea }
%\tablenotetext{{\bsym}}{ \sampleb }
%\tablenotetext{{\csym}}{ \samplec }
%\tablenotetext{{\dsym}}{ \sampled }
%\tablenotetext{{\esym}}{ \samplee }
%\tablenotetext{{\fsym}}{ \samplef }
%
\end{table}

% ------------------------------------------------
  \subsubsection{Redshift Evolution of \SALTII\ Parameters}
  \label{subsec:SALT2z}
% ------------------------------------------------

\begin{figure}[hb]
  \epsscale{1.1}
  \plottwo{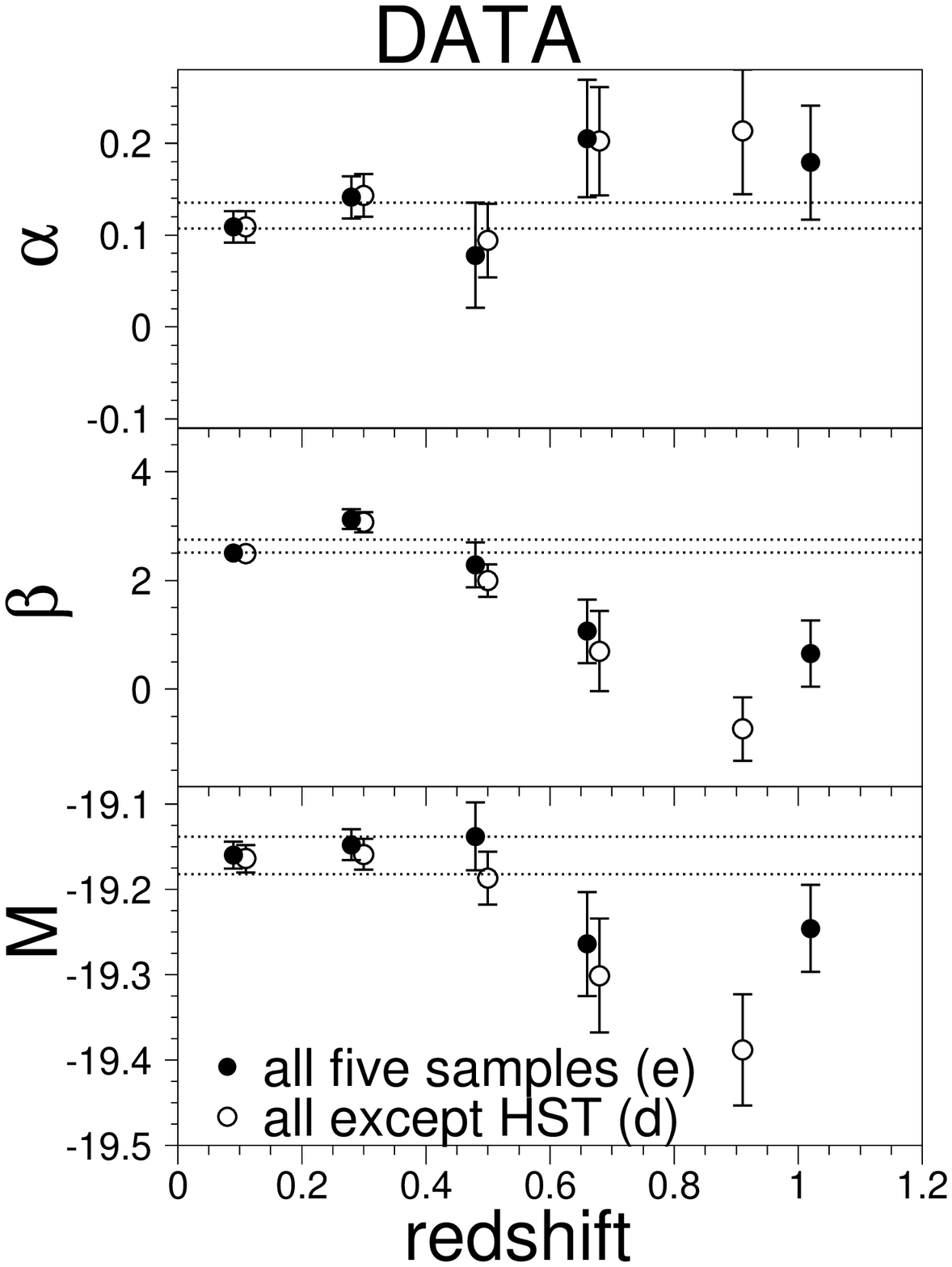}{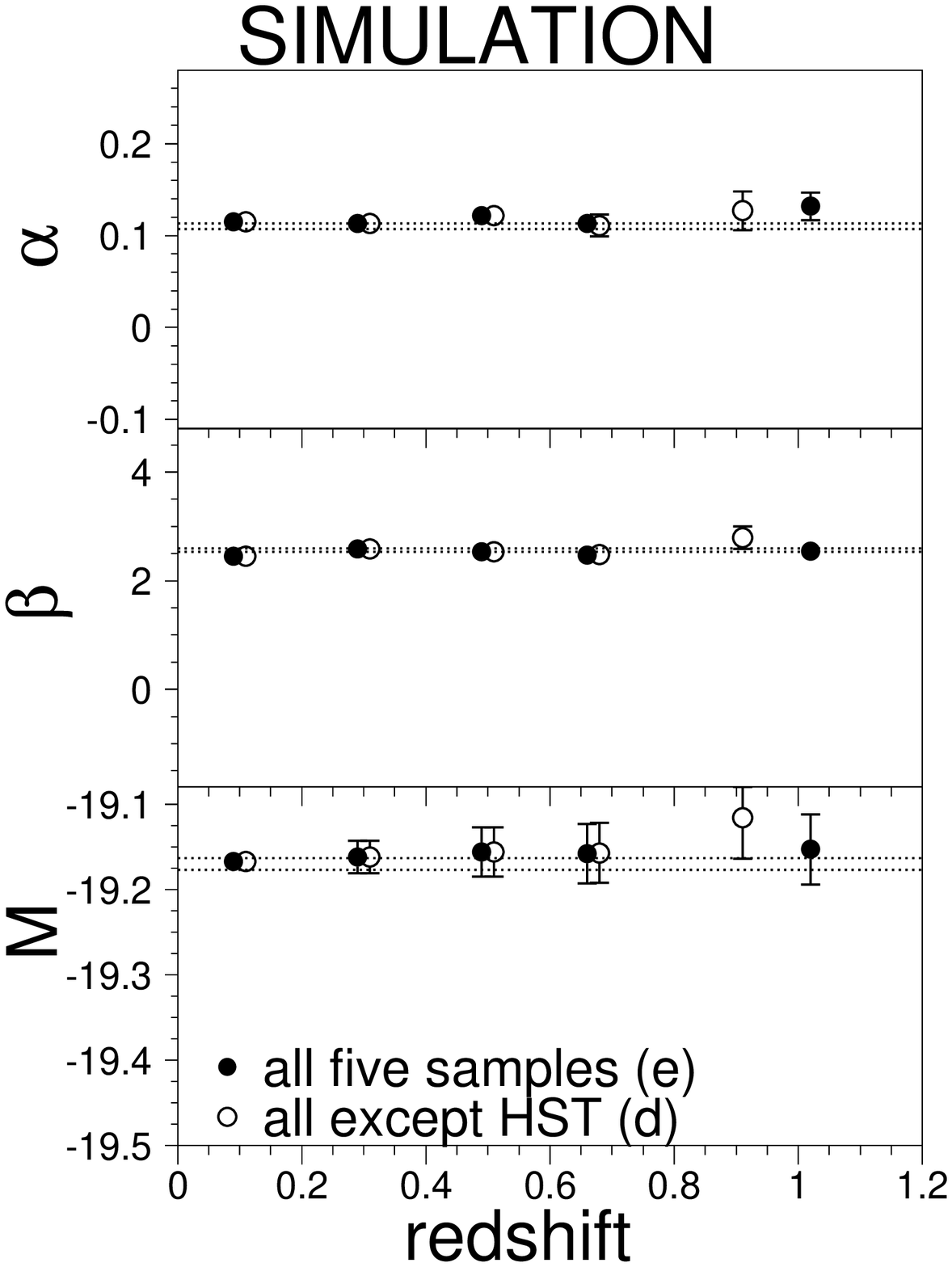}
  \caption{
	\SALTII\ fitted values and \uncs\ for 
	$\alpha$, $\beta$, and $M$,
	evaluated independently in redshift bins for the 
	\samplee\ sample combination (\esym) (solid dots) and for
	the \sampled\ combination (\dsym) (open circles).
	Left is for data; right is for a simulation with
	$10$ times the number of SNe as in the data.
	For each redshift bin, the cosmological parameters 
	$w$ and $\OM$ are fixed to the values from the global 
	combination ({\dsym} or {\esym}) fit.
	The redshift bins are $\Delta z=0.2$ for $z<0.8$; 
	the highest-redshift bin includes all SNe~Ia with $z>0.8$.
	The dashed lines show the $\pm 1\sigma$ statistical error band
	based on the global fit to combination (\esym).
        } % end caption
  \label{fig:SALT2par_zbins}
\end{figure}

In the \SALTII\ model fits, $\alpha$, $\beta$, and $M$ 
(see Eq.~\ref{eq:MUSALTII})
are global parameters that are assumed to be
independent of redshift. 
To test the consistency of this assumption, 
we have carried out \SALTII\ fits separately in five 
redshift bins for sample combinations ({\dsym}) and ({\esym}). 
For the fit in each redshift bin, the cosmological parameters 
$w$ and $\OM$ in the \wCDM\ model 
are fixed to the values from the sample ({\esym}) fit, and the 
BAO+CMB prior is applied. The redshift dependence of the 
best-fit results for $\alpha$, $\beta$, and $M$ are shown 
in the left panels of Fig.~\ref{fig:SALT2par_zbins}.
There is evidence for redshift evolution of the parameters, 
particularly for the color parameter $\beta$, which 
falls with increasing redshift above $z\sim 0.6$. 
This trend is more evident without the HST sample,
suggesting that the $\beta$-variation is driven primarily 
by the SNLS sample. 
We have performed 
this test with high-statistics simulations 
(right panels of Fig.~\ref{fig:SALT2par_zbins})  
and find that the fitted \SALTII\ parameters are 
consistent
across all redshift bins. Since the simulation accounts for 
Malmquist bias, selection effects, and measurement errors, 
the redshift dependence favored by the data is likely due 
to some other effect.

% ------------------------------------------------
  \subsubsection{$U$-Band Anomaly with \SALTII\ }
  \label{subsec:Uanom_SALT2}
% ------------------------------------------------

\begin{figure}[hb!]
 \epsscale{1.1}
  \plotone{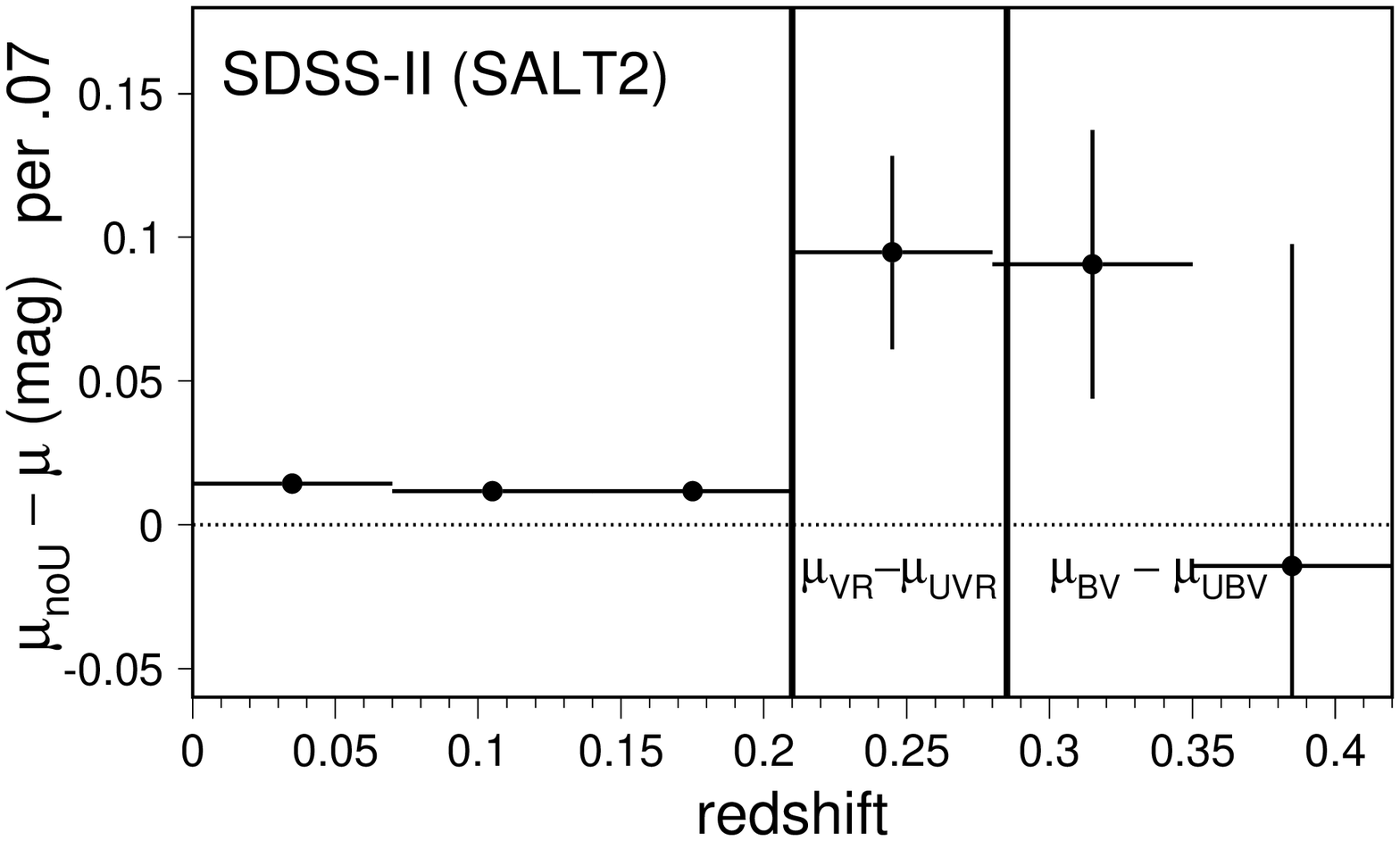}
  \caption{
        Redshift dependence of the
        average difference in distance modulus for \SDSS\ SNe
        between the nominal \SALTII\ light-curve fits and fits in which 
        the observer-frame passband corresponding to rest-frame
        $U$-band is excluded ($g$-band for $z>\ZUSDSS$).
        Labels on the plot indicate the corresponding
        rest-frame $UBVR$ \citep{Bessell90} passbands.
        Error bars (rms/$\sqrt{N}$) reflect the statistical \unc\ 
	on the mean $\mu$-difference in each redshift bin.
        }
  \label{fig:noUtest_SALT2}
\end{figure}

Here we study how the $U$-band anomaly is manifest in the \SALTII\ method.
Fig.~\ref{fig:noUtest_SALT2} shows the change in the binned \SDSS\ SN  
Hubble diagram when the \SALTII\ fit excludes data from the observer-frame 
passband corresponding to rest-frame $U$-band, i.e., excluding 
$g$-band data for $z> \ZUSDSS$. Although 
this change only affects the 
light-curve fits for $z> \ZUSDSS$, it can alter the estimated 
distance moduli at all redshifts since they are derived from 
a global fit to the Hubble diagram. 
For $z<\ZUSDSS$, the average distance modulus shift is 
$(\SALTSDSSnoUdmu)$~mag; 
for $z>\ZUSDSS$ the mean shift is 
$(\SALTSDSSnoUdMU)$~mag relative to the $z<0.21$ shift. 
This relative shift is
about half of that for {\mlcs}, 
but it is still significant: for the \samplea\ sample, 
the exclusion of $U$-band  in the \SALTII\ fits
results in a shift in $w$ of 
$\FWCDMBAOCMBSYSTUBANDWa$ 
that is included as a systematic \unc.

\begin{figure}[h]
  \epsscale{1.1}
  \plotone{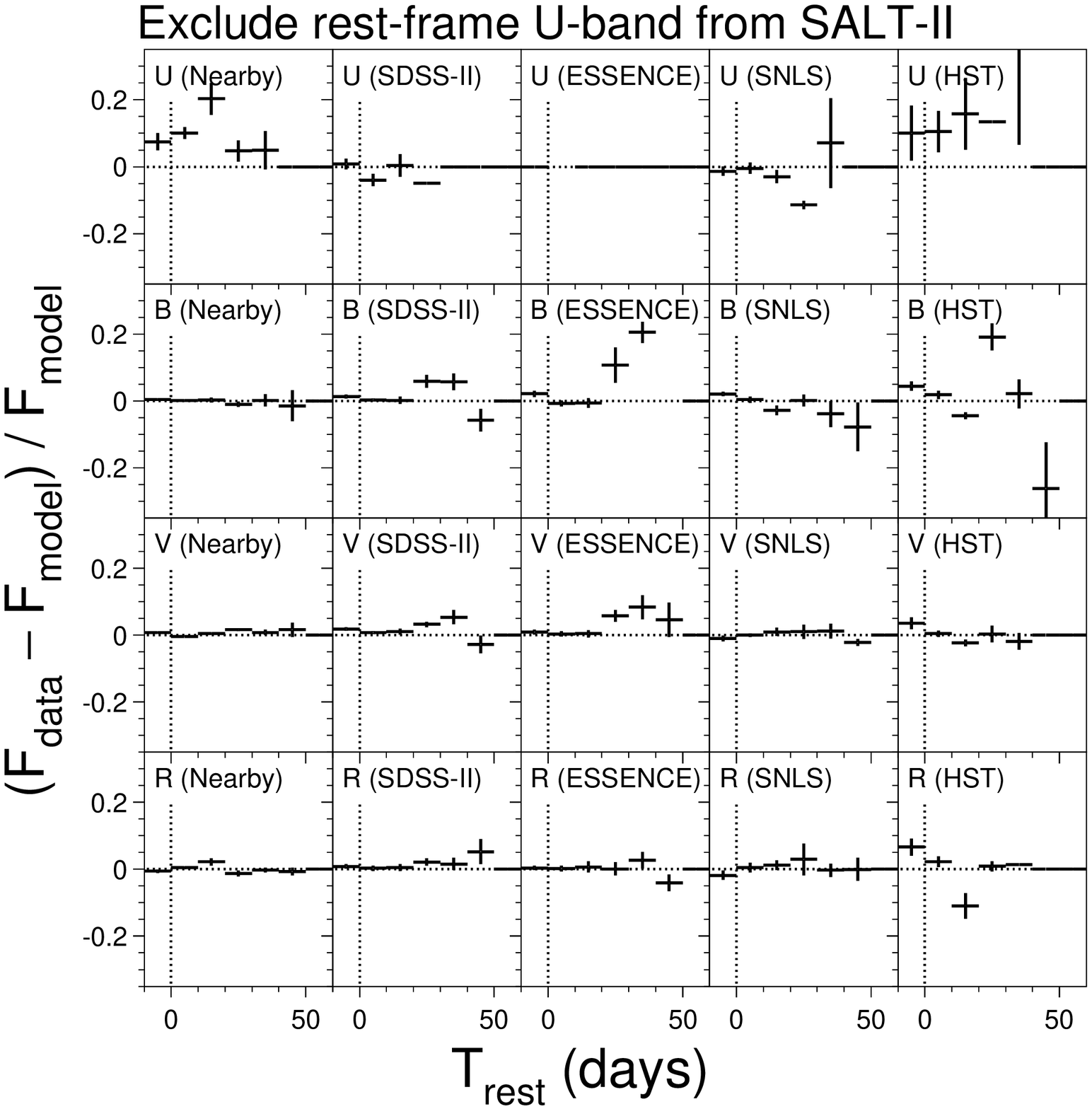}
  \caption{
	Data-model fractional residuals as a function of rest-frame
        epoch in 5-day bins for \SALTII\ light-curve fits.
        Same as Fig.~\ref{fig:lcresid_SALT2}, except fits exclude 
	observer-frame filter corresponding to rest-frame $U$-band.
        }
  \label{fig:lcresid_SALT2_noU}
\end{figure}

The data-model residuals for the \SALTII\ light-curve fits
are shown in Fig.~\ref{fig:lcresid_SALT2} for all 
of the SN~Ia samples in all of the rest-frame passbands.
Fig.~\ref{fig:lcresid_SALT2_noU} shows the residuals 
when the filter corresponding to rest-frame $U$-band 
is excluded from the fit. In both cases, there are systematic 
discrepancies between the model and the data in rest-frame 
$U$-band for the Nearby SN~Ia sample at all epochs. Since the 
\SALTII\ rest-frame $U$-band model was trained primarily on 
higher-redshift SNe, i.e., downweighting observer-frame $U$-band 
data from nearby SNe, this points to a systematic offset between the 
nearby, observer-frame $U$-band data and the higher-redshift, 
rest-frame $U$-band data, which was also qualitatively seen in 
the \mlcs\ fits.

% ------------------------------------

\section{Comparison Between \mlcs\ and \SALTII\ Results }
\label{sec:results_compare}
% TOP
% Comparing SALT-II vs MLCS results 
%

\newcommand{\dwsym}{\mbox{$\Delta w$}}
\newcommand{\dwsymstat}{\mbox{$\dwsym_{stat}$}}
\newcommand{\dwsymsyst}{\mbox{$\dwsym_{syst}$}}

% define misc numbers used in discussion

\newcommand{\dwval}{0.2}      % approx MLCS-SALT2 discrepancy for b,d,e,f
\newcommand{\dwvalsim}{0.05}  % dwval for sim
\newcommand{\dwLOWZSDSS}{0.04}  % w-agreement for LOWZ, SDSS
\newcommand{\LOWZmudifrms}{0.10}   % rms on mu(SALT2)-mu(MLCS)
\newcommand{\SDSSmudifrms}{0.15}   % rms on mu(SALT2)-mu(MLCS)
\newcommand{\ESSEmudifrms}{0.16}   % rms on mu(SALT2)-mu(MLCS)
\newcommand{\ESSEmudifavg}{0.04}   % rms on mu(SALT2)-mu(MLCS)
\newcommand{\dwSDSS}{0.06}         % SALT2-MLCS w-diff, nominal
\newcommand{\dwSDSSnoU}{0.2}      % idem without U-band
\newcommand{\dwmlcsSALTII}{0.2}    % w-shift with salted mlcs
\newcommand{\dwSETg}{0.05}   % w(SALT2)-w(MLCS) for Nearby+SDSS+ESSENCE (set g)
\newcommand{\wMLCSg}{-0.84}  % w(MLCS) for Nearby+SDSS+ESSENCE (set g)
\newcommand{\wSALTg}{-0.89}  % w(SALT2) for Nearby+SDSS+ESSENCE (set g)

% ===========================

As noted in \S \ref{subsec:results_salt2}, the best-fit 
values of $w$ for the \wCDM\ model agree between 
\SALTII\ and \mlcs\ to within \dwLOWZSDSS\ 
for the \samplea\ and \samplec\ sample combinations. 
However, as indicated in 
Tables~\ref{tb:results_MLCS_FWCDM} and \ref{tb:results_salt2_FWCDM_BAOCMB}, 
when the ESSENCE and SNLS data are included 
(sample combinations {\bsym},{\dsym},{\esym}, and {\fsym}), 
the $w$-values increase by $\sim 0.1$ for the
\mlcs\ method, while decreasing by nearly $0.1$ for the 
\SALTII\ method, leading to a discrepancy of 
$\dwsym \sim \dwval$ between the two methods. 
To estimate the statistical significance of this discrepancy,
we have run \mlcs\ and \SALTII\ fits on 
a set of
ten simulated sample combinations generated with the \mlcs\ model, 
each with the same statistics as the data.
The predicted rms-spread on \dwsym\ for combination ({\esym})
(all five samples) is $\dwsymstat \simeq \dwvalsim$, so the 
observed discrepancy appears to be statistically significant.

\begin{figure}[h]
  \epsscale{1.1}
  \plotone{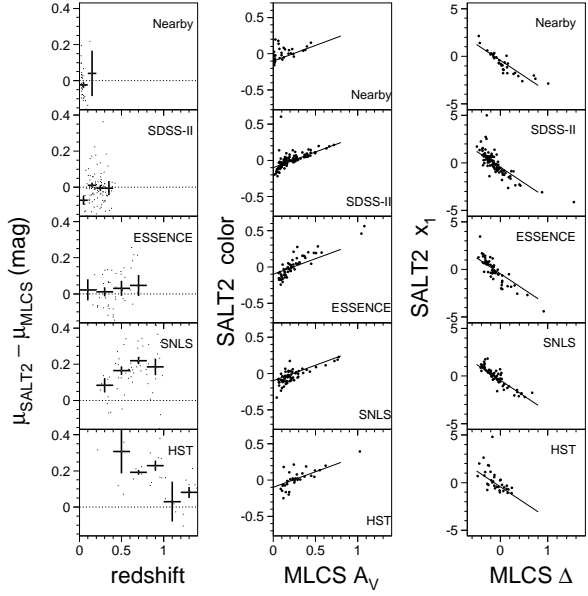}
  \caption{
	Comparisons between \mlcs\ and {\SALTII} light-curve fit 
	parameters for the five SN samples: difference in
	distance modulus vs. redshift (left), 
	color $c$ vs. $A_V$ (middle), and
	$x_1$ vs. $\Delta$ (right). 
	Crosses in left panels show average \& \unc\
	in redshift bins.
	The solid straight lines (middle \& right columns) 
	are the same within a column to guide the eye, and they
	are derived from a fit to all of the samples except for ESSENCE, 
	as explained in the text.
        }
  \label{fig:compare_fitpars}
\end{figure}

To help diagnose this discrepancy, 
we compare the \mlcs\ and \SALTII\ fitted parameters in
Fig.~\ref{fig:compare_fitpars} for each of the SN samples. 
In this figure, 
the \SALTII\ parameters are based on the fit to 
sample combination ({\esym}), including the BAO+CMB prior, 
and the \SALTII\ distances have not been corrected for 
selection bias.
The left panels show the mean difference in distance modulus,
$\MUDIF \equiv \mu_{\rm SALT2} - \mu_{\rm MLCS}$,
as a function of redshift. The values of the Hubble parameter 
have been relatively adjusted so that the values of $\mu$ 
for the best-fit \mlcs\ and \SALTII\ models for \wCDM\ agree 
at $z \rightarrow 0$. 
The two methods yield consistent distance 
estimates for the Nearby, {\SDSS}, and ESSENCE 
samples over the redshift ranges they cover.
Moreover, the scatter in $\MUDIF$ 
is comparable to the intrinsic scatter:
$\sigma(\MUDIF)=\LOWZmudifrms$, $\SDSSmudifrms$ and
$\ESSEmudifrms$~mag for these three samples, respectively.
Analyzing the Nearby+SDSS+ESSENCE 
combination (not one of our standard combinations), 
$w = \wMLCSg$ for \mlcs\ and $w=\wSALTg$ for \SALTII,
indicating good agreement.
The situation changes dramatically when we include the SNLS data, 
for which there is a clear trend of increasing $\MUDIF$ 
with redshift: this is the primary cause of the difference in 
$w$ between \mlcs\ and \SALTII\ for sample combinations 
{\bsym}, {\dsym}, {\esym}, and {\fsym}.

The middle and right panels of Fig.~\ref{fig:compare_fitpars} 
show the correlations between the \mlcs\ and \SALTII\ light-curve 
fit parameters, 
\SALTII\ color $c$ versus \mlcs\ 
extinction 
$A_V$ 
and \SALTII\ versus \mlcs\ shape-luminosity 
parameters $x_1$ versus $\Delta$. 
The fitted slopes, $dc/dA_V$ and $dx_1/d\Delta$,
are consistent among the SN samples, 
except for ESSENCE which has somewhat larger slopes. The latter 
could be related to the fact that this sample has only two
observer-frame passbands and hence less reliable color 
information. 

\begin{figure}[h]
  \epsscale{1.}
  \plotone{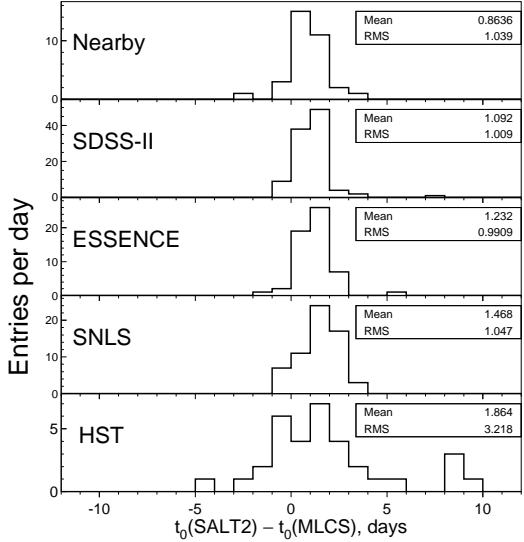}
  \caption{
	Comparison of \SALTII\ and \mlcs\ fitted epoch of 
	maximum $B$-band light for the five SN samples:
	shown are the distributions of $t_0$(\SALTII)$- t_0$(\mlcs). 
	Insets show mean and rms for each sample.
        }
  \label{fig:compare_t0}
\end{figure}

The final light-curve fit parameter to compare is the 
epoch of maximum light in rest-frame $B$-band ($t_0$).
Fig.~\ref{fig:compare_t0} shows the distributions of 
$t_0$(\SALTII)$- t_0$(\mlcs) for the five SN samples. 
On average, the \SALTII\ epoch of peak light is about 
$1-1.5$ days later than that for {\mlcs}, 
with a dispersion of about one day for the ground-based samples.
For the HST sample, the offset is $\sim 2$ days and the
dispersion is larger (3 days) because of a handful of SNe
with large $t_0$-discrepancies. 
Fitting simulated SN samples shows that for the poorly sampled
HST light curves, the fitted $t_0$ from \SALTII\ is expected 
to be more discrepant from the input value than for \mlcs.

\begin{figure}[h]
  \epsscale{1.1}
  \plotone{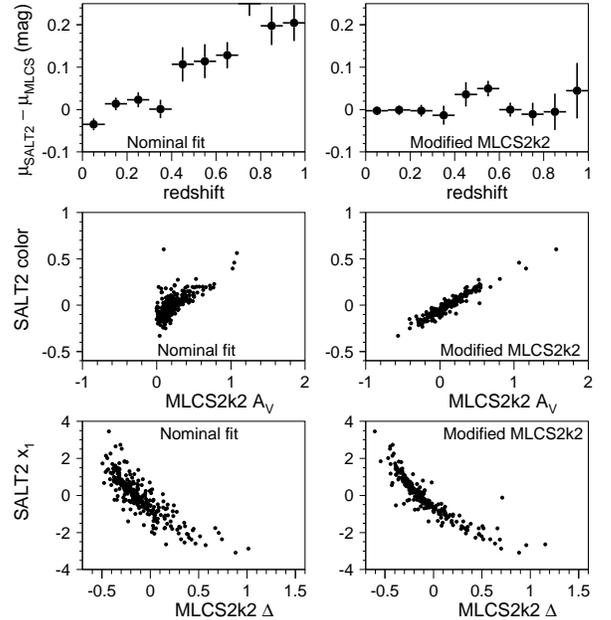}
  \caption{
	For all five samples (combination {\esym}),
	comparison of \SALTII\ fit parameters with those from
	the nominal \mlcs\ fits (left) and from the 
	modified \mlcs\ fits (right) as described in the text.
	}
  \label{fig:compare_saltymlcs}
\end{figure}

To gain some insight into the discrepancy between the
two fitting methods, we have investigated modifications
to the \mlcs\ model that are designed to partially replicate 
the features of the \SALTII\ model. 
The modifications are: 
(i) change the \mlcs\ model vectors so that the $UBVRI$ 
light-curve templates match those of the synthesized \SALTII\ model, 
as described in Appendix \ref{app:saltymlcs}; 
(ii) use a flat prior, $P_{\rm prior}(z,A_V,\Delta)=1$
in Eq.~\ref{eq:prior_master}, allowing negative values of $A_V$; 
(iii) use the \SALTII\ color law $CL(\lambda)$ in place of the 
CCM89 model in the extinction term in Eq. \ref{eq:MLCS2k2model}; 
and
(iv) exclude measurements for which the mean filter 
wavelength exceeds $7000\cdot (1+z)$~\AA. Using this 
modified {\mlcs}
model to fit the light curves, the resulting values of 
$w$ decrease by \dwmlcsSALTII\ for sample combinations 
($\dsym$) and ($\esym$) and are in good 
agreement with the \SALTII\ results.
Fig.~\ref{fig:compare_saltymlcs} compares the \SALTII\ parameters
to those from the nominal \mlcs\ fits and to those from 
the modified \mlcs\ model. For the latter, 
the distance moduli show good agreement over the
entire redshift range, and the correlations between the
shape-color parameters ($c$ vs. $A_V$ and $x_1$ vs. $\Delta$)
are stronger than for the nominal \mlcs\ fits.

\begin{figure}[h]
  \epsscale{1.1}
  \plotone{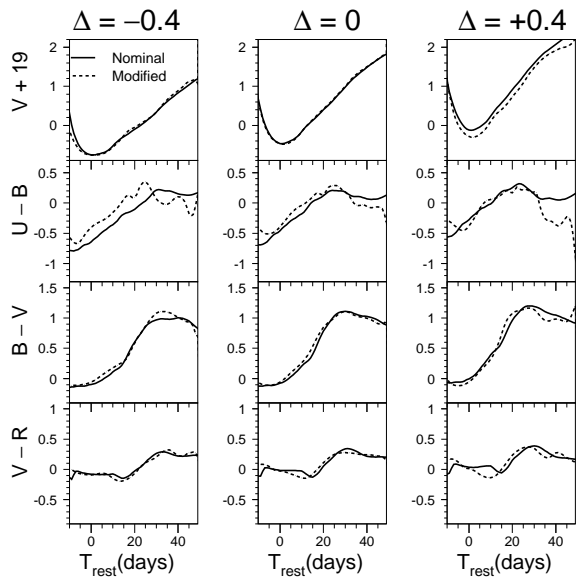}
  \caption{
	For \mlcs\ model, rest-frame $V$-band magnitude (top row)
	\& colors (lower 3 rows) versus epoch.
	Solid curves are for the nominal \mlcs\ model,
	and dashed are for the modified \mlcs\ model. 
	The latter was constructed to reproduce the \SALTII\ 
	flux and color evolution. 
	Each column corresponds to a different value of	
	\mlcs\ parameter $\Delta=-0.4,0,0.4$.
	More than 70\% of SNe~Ia have $\Delta < 0$,
	and 5\% have $\Delta>0.4$.
	}
  \label{fig:compare_mlcs_models}
\end{figure}

Of the four modifications to \mlcs, the last two 
(use of the \SALTII\ color law and a 7000~\AA\ cutoff)
have a negligible effect on the cosmology analysis.
The change in $w$ is mainly due to matching the template light curves 
and to using a flat $A_V$ prior. 
Implementing either of these changes alone results in a 
$w$-shift of $-0.07$ or less; both changes are needed to obtain 
the full $w$-shift of $-\dwmlcsSALTII$.
This exercise shows that the $w$-discrepancy between
\mlcs\ and \SALTII\ is related in part to the training
procedure that determines the spectral and light-curve templates 
and in part to the different assumptions about SN color variations.
The model $V$-band light curve and color evolution for \mlcs\ 
and modified \mlcs\ are compared in
Fig.~\ref{fig:compare_mlcs_models} for different values
of the shape-luminosity parameter $\Delta$. 
The main discrepancy is in $U-B$, which differs
by about 0.1--0.2~mag between \mlcs\ and modified \mlcs.
Since the modified \mlcs\ colors agree well with the 
\SALTII\ colors (not shown), the $U-B$ discrepancy in
Fig.~\ref{fig:compare_mlcs_models} can be interpreted
as the discrepancy between the \mlcs\ and \SALTII\ models.
In particular, since the models agree in $B$-band, 
this plot illustrates the differences in rest-frame $U$-band.

As noted earlier, 
there is strong evidence of systematic discrepancies in 
rest-frame $U$-band between the nearby and higher-redshift samples. 
These discrepancies are reflected in the differences between the 
\mlcs\ and \SALTII\ $U$-band models, differences that account for 
part of the cosmological parameter disagreement between the two 
models. The other major contributor to the cosmological disagreement 
is the differing treatment of SN color variation in the two models. 
There is a trend toward negative apparent \SALTII\ color at 
high-redshift within the SNLS sample. 
\SALTII\ and \mlcs\ with a flat-$A_V$ prior assign these 
blue events large intrinsic luminosities and therefore large 
distance moduli. By contrast, \mlcs\ with the nominal $A_V$ 
prior identifies these events as having $A_V \sim 0$ and assigns 
them lower luminosities and distances. 
As illustrated in Fig.~\ref{fig:ovdatasim_avnoprior},
the nominal \mlcs\ interpretation of these events is consistent with 
the observed color distributions, so it is not obvious which 
model is correct.

In \S~\ref{sec:results} we have considered the approach 
of excluding data corresponding to rest-frame $U$-band 
altogether.
In that case,  
for sample combinations that include ESSENCE and SNLS  
({\bsym}, {\dsym}, {\esym}, {\fsym}), 
the \SALTII\ - \mlcs\ discrepancy of $\dwsym \sim \dwval$ 
is reduced by nearly 50\%. By contrast, 
the same test applied to the \samplea\ sample increases 
$\dwsym$ from \dwSDSS\ to \dwSDSSnoU\ when $U$-band is excluded. 
Moreover, in the nominal fits (including $U$-band), 
the \SALTII\ and \mlcs\ Hubble diagrams appear continuous 
around $z=0.2$, and there is good agreement between them 
across the \SDSS\ redshift range. 
Removing $U$-band 
introduces a noticeable step 
in both Hubble diagrams at $z =0.2$, suggesting that both models 
display unphysical behavior when this information is removed completely.
Such a sharp feature in the Hubble diagram is not well 
captured by a model with constant $w$, so the 
changes in $\dwsym$ quoted above should be interpreted 
with extreme caution.

% #########################################

\section{ Conclusions }
\label{sec:conclude}

% ==============================================
% Conclusion for 1st SDSS-II SN Cosmology Paper
% ==============================================

We have presented measurements of the Hubble diagram for 
type Ia supernovae discovered during Sept.--Nov. 2005 by the 
\SDSS\ Supernova Survey and combined them with other SN~Ia 
data and with BAO and CMB results to derive 
cosmological constraints.

For the \SDSS\ sample, based on stringent light-curve
sampling and signal-to-noise criteria, 
we selected $\NSNSDSS$ SNe~Ia with redshifts 
$\ZMINSDSS < z < \ZMAXSDSS $ 
for analysis from a parent sample of $\NSNSDSSCONF$ \specy\
confirmed SNe~Ia. In an effort to make the analysis as 
homogeneous as possible among the SN data sets,
we have used similar light-curve selection criteria 
and applied the same analysis techniques to all samples.
We have estimated distances for these SNe using both the 
\mlcs\ and \SALTII\  methods,
which differ from each other in a number of respects 
(\S~\ref{subsec:compare}).
The analysis includes significant improvements in the 
implementation of the \mlcs\ method.

To determine the efficiency functions used in the \mlcs\ priors 
and the biases in the \SALTII\ results,
we have carried out detailed Monte Carlo simulations for each of
the SN~Ia data sets, making use of recorded observing conditions. 
The simulation accurately models the 
light curves (fluxes \& errors) and the
photometric selection effects; we have also incorporated
a quantitative model for the \spec\ selection \eff\ 
based on comparing observed and simulated distributions 
in SN redshift and mean extinction.
The simulation provides an excellent description of the data,
as illustrated by the data-simulation comparison
of the flux distribution for each SN sample
(Fig.~\ref{fig:ovdatasim_flux}).

Due to selection effects, supernovae in spectroscopic samples 
tend to be brighter and less extinguished, on average, 
than those of the parent population. For {\SDSS}, 
we have augmented the spectroscopic sample with 
a larger sample that includes photometrically 
identified SNe~Ia with host-galaxy redshift measurements. 
We used this \SDSS\ ``dust'' sample 
in conjunction with the Monte Carlo simulations to 
determine the mean host-galaxy reddening parameter $R_V$ and 
the distributions of extinction and light-curve shape parameter, 
$P(A_V)$ and $P(\Delta)$, for the underlying SN~Ia population. 
By matching observed and predicted SN~Ia colors 
in {\mlcs}, we find $\RVRESULT$, consistent with the recent 
trend toward values lower than the canonical Milky Way average of 
$3.1$. For comparison, the \SALTII\ analysis of the combined SN~Ia 
sample (\esym) yields 
 $ \beta=\FWCDMBAOCMBBETAe \pm \FWCDMBAOCMBDBETASTATe ({\rm stat})
                           \pm \FWCDMBAOCMBDBETASYSTPe$ (syst). 
If all SN~Ia color variation 
(beyond that associated with light-curve shape) 
were due to dust extinction, we would expect $\beta - R_V = 1$.
In this analysis we find $\beta - R_V = 0.5 \pm 0.5$,
consistent with the expectation for extinction from dust,
but also suggesting that there may be additional sources of 
SN color variation.
We find that the underlying $A_V$ distribution is well 
described by an exponential function;
this distribution is marginally consistent with both
the galactic line of sight model used by WV07
and the exponential distribution found 
for the nearby sample by JRK07.

For both the \mlcs\ and \SALTII\ methods, we have carried out 
extensive studies of systematic errors, varying a large number 
of model parameters and assumptions, and we have included  
the resulting cosmological parameter variations 
in the systematic error budget. 
For each method, the largest source of systematic uncertainty 
is associated with the rest-frame $U$-band anomaly.
The anomaly is manifest in discrepancies between the light-curve 
data and the models, indicating the need for improved training of the 
rest-frame $U$-band models.  
When data corresponding to rest-frame $U$-band are excluded 
from the light-curve fits, the estimated distance modulus shifts by 
$\MLCSSDSSnoUdMUval$~mag for {\mlcs} and by 
$\SALTSDSSnoUdMUval$ mag for {\SALTII}. 
This shift occurs in particular redshift ranges, 
leading to abrupt features in the Hubble diagram;  
the effect is most severe for the {\SDSS} SNe, 
since the $U$-band shift occurs at the median redshift of the sample.
For this reason, and because dropping $U$-band 
leads to larger uncertainties from the significantly reduced 
color constraints, we have chosen not to exclude $U$-band for 
our nominal analysis but to include the corresponding changes
as part of the systematic \unc.

We have combined the SN Hubble diagram with BAO and CMB 
measurements to estimate the cosmological parameters. 
For the \wCDM\ model and the combined sample of 
$\NSNTOT$ SNe~Ia from all five surveys, we find 
$w   = \MLCSWRESe \pm \MLCSWSTATERRe ({\rm stat}) 
  \pm \MLCSWSYSTERRPe$ (syst), 
$\OM = \MLCSOMRESe \pm \MLCSOMSTATERRe ({\rm stat}) 
  \pm \MLCSOMSYSTERRPe$ (syst) 
using \mlcs\ and 
$w=\FWCDMBAOCMBWe \pm \FWCDMBAOCMBDWSTATe ({\rm stat})
   \pm \FWCDMBAOCMBDWSYSTPe$ (syst), 
$\OM = \FWCDMBAOCMBOMEGAMe \pm \FWCDMBAOCMBDOMEGAMSTATe ({\rm stat})
  \pm \FWCDMBAOCMBDOMEGAMSYSTPe$ (syst) 
for {\SALTII}. 
This discrepancy of $\dwval$ in $w$ between the two analysis methods 
is {\it not} due to inclusion of the \SDSS\ data: 
for the \samplef\ (\fsym) sample combination, which excludes {\SDSS}, 
we find the same difference in $w$ between the two methods.
Our {\mlcs} results for this sample 
combination (\fsym) differ substantially from those of WV07
(\S~\ref{subsec:WV07_compare}), 
who used an earlier version of {\mlcs} to analyze a nearly 
identical sample. 
We have traced the differences primarily to the different priors 
used, i.e.,  to the different values of $R_V$, to the different 
$A_V$ distributions, and to the inclusion of \spec\ targeting 
\eff\ in the prior.

We have traced the \mlcs\ vs. \SALTII\  discrepancy to the SN model parameters 
determined in the training, particularly the rest-frame $U$-band model 
(Fig.~\ref{fig:compare_mlcs_models}),  
and to the different assumptions about the source of color 
variations in \mlcs\ and \SALTII.
If we restrict the analysis to the 136 SNe~Ia in the 
{\samplec} combination, we find much better 
agreement between the two analysis methods, but with larger 
uncertainties,
$w = \MLCSWRESc \pm \MLCSWSTATERRc ({\rm stat})
  {}^{+\MLCSWSYSTERRPc}_{-\MLCSWSYSTERRMc}$ (syst) 
for \mlcs\ and 
$ w = \FWCDMBAOCMBWc \pm \FWCDMBAOCMBDWSTATc ({\rm stat}) 
{}^{+\FWCDMBAOCMBDWSYSTPc}_{-\FWCDMBAOCMBDWSYSTMc}$ (syst) 
for {\SALTII}. 
We also note that the cosmological parameter uncertainties for the 
\sampleb\ sample combination are similar to those for {\samplef}, 
i.e., the first-season \SDSS\ data sample anchors 
the Hubble diagram with comparable power to the 
Nearby sample.

The \mlcs\ vs. \SALTII\ discrepancy 
for the higher-redshift samples raises the question of 
which method, if either, is more reliable.
Since the \mlcs\ $U$-band model relies entirely on 
$U$-band measurements at low redshift,
while \SALTII\ uses a combination of low and high-redshift data, 
some at redder observer-frame passbands than $U$, 
the \SALTII\ results may be less biased by the $U$-band anomaly
if the problem is related to the observer-frame $U$-band measurements.
On the other hand, the \SALTII\ $\beta$ parameter has a notable
redshift dependence (Fig.~\ref{fig:SALT2par_zbins}) that is
inconsistent with the underyling assumption that it is constant.
Concerning the interpretation of SN color variations,
both models are consistent with the data; in particular,
a supernova that is bluer than the templates could be extremely bright,  
as indicated by \SALTII, or it could be due to measurement 
uncertainty and intrinsic color fluctuations, as interpreted by 
\mlcs\ (Fig.~\ref{fig:ovdatasim_avnoprior}).
Since we cannot definitively determine from the current 
data that either method is better
or incorrect, the overall conclusion is that the cosmological
parameter $w$ conservatively lies between $-1.1$ and $-0.7$~.
This result reflects the fact that present SN Ia samples have 
reached the systematic limits of current SN model fitters. 
Although this result is disappointing, we are optimistic 
that this situation is temporary.
The full three-season \SDSS\ sample, 
with its homogeneous and well-modeled selection function, 
will serve both to anchor the Hubble diagram and 
to retrain the light-curve models. 
Since the $U$-band anomaly is most likely associated with 
the published nearby SN sample, use of the full \SDSS\ data as well 
as updated low-redshift samples from the CfA \citep{Hicken09a}, the 
Carnegie Supernova Project, and the Nearby Supernova Factory,   
should significantly reduce 
this major source of systematic uncertainty.

Finally, our use of photometrically identified SNe~Ia 
to measure host-galaxy dust properties is an important
step toward including these SNe in the Hubble diagram, which 
will increase the statistical power of the data. 
This will be a growing trend in the future, as large surveys, 
including PanSTARRS, the Dark Energy Survey, and LSST,
will discover vastly more SNe than can be confirmed
with available \spec\ resources.

% ======= !!!! FINITO !!! =========

% acknowledgements

\medskip

All software used in this analysis is publicly
available from our website \citep{SNANA09}.
We wish to thank Julien Guy for retraining the \SALTII\ program and 
for consulting on its use and results.
We thank Armin Rest for modifying the ESSENCE SN-search pipeline 
for use in the \SDSS\ SN survey.
We gratefully acknowledge support from 
the Kavli Institute for Cosmological Physics at the University of Chicago,
the National Science Foundation at Wayne State, 
the Japan Society for the Promotion of Science (JSPS), and
the Department of Energy at Fermilab, the University of Chicago, 
and Rutgers University.
R.J.F. is supported by a Clay Fellowship.
Y.~Ihara and T.~Morokuma are supported by a JSPS Fellowship.
A.V.F. is grateful for the support of NSF grant AST-0607485 and
DOE grant DE-FG02-08ER41563.
Funding for the creation and distribution of the SDSS and SDSS-II
has been provided by the Alfred P. Sloan Foundation,
the Participating Institutions,
the National Science Foundation,
the U.S. Department of Energy,
the National Aeronautics and Space Administration,
the Japanese Monbukagakusho,
the Max Planck Society, and 
the Higher Education Funding Council for England.
The SDSS Web site \hbox{is {\tt \wwwSDSS}.}

The SDSS is managed by the Astrophysical Research Consortium
for the Participating Institutions.  The Participating Institutions are
the American Museum of Natural History,
Astrophysical Institute Potsdam,
University of Basel,
Cambridge University,
Case Western Reserve University,
University of Chicago,
Drexel University,
Fermilab,
the Institute for Advanced Study,
the Japan Participation Group,
Johns Hopkins University,
the Joint Institute for Nuclear Astrophysics,
the Kavli Institute for Particle Astrophysics and Cosmology,
the Korean Scientist Group,
the Chinese Academy of Sciences (LAMOST),
Los Alamos National Laboratory,
the Max-Planck-Institute for Astronomy (MPA),
the Max-Planck-Institute for Astrophysics (MPiA), 
New Mexico State University, 
Ohio State University,
University of Pittsburgh,
University of Portsmouth,
Princeton University,
the United States Naval Observatory,
and the University of Washington.

This work is based in part on observations made at the 
following telescopes.
The Hobby-Eberly Telescope (HET) is a joint project of the 
University of Texas at Austin,
the Pennsylvania State University,  Stanford University,
Ludwig-Maximillians-Universit\"at M\"unchen, and 
Georg-August-Universit\"at G\"ottingen.  
The HET is named in honor of its principal benefactors,
William P. Hobby and Robert E. Eberly.  The Marcario Low-Resolution
Spectrograph is named for Mike Marcario of High Lonesome Optics, who
fabricated several optical elements 
for the instrument but died before its completion;
it is a joint project of the Hobby-Eberly Telescope partnership and the
Instituto de Astronom\'{\i}a de la Universidad Nacional Aut\'onoma de M\'exico.
The Apache Point Observatory 3.5 m telescope is owned and operated by 
the Astrophysical Research Consortium. We thank the observatory 
director, Suzanne Hawley, and site manager, Bruce Gillespie, for 
their support of this project. The Subaru Telescope is operated by the 
National Astronomical Observatory of Japan. The William Herschel 
Telescope is operated by the Isaac Newton Group on the island of 
La Palma in the Spanish Observatorio del Roque 
de los Muchachos of the Instituto de Astrofisica de 
Canarias. The W.M. Keck Observatory is operated as a scientific partnership 
among the California Institute of Technology, the University of 
California, and the National Aeronautics and Space Administration. 
The Observatory was made possible by the generous financial support 
of the W. M. Keck Foundation.

% #########################################
%
%            APPENDICES 
%
% #########################################

% \appendix
\begin{appendix}

\section{Spectral Warping for K-Corrections in the \mlcs\ Light Curve Model}
\label{app:kcor}

The \mlcs\ model makes predictions about rest-frame light curves 
in $UBRVI$; K-corrections are used to translate the model light curves so 
that they can be used to fit measurements at non-zero 
redshift in a variety of passbands. This translation requires 
knowledge of the supernova spectral energy distribution (SED) time sequence. 
Since the SN~Ia population exhibits intrinsic color variations, a 
single time sequence will not provide an adequate model. The 
standard practice is therefore to ``warp'' the model SEDs so that 
they match the colors of the photometric model for a given object.
Here we describe our procedure for the spectral warping
used in determining {\Kcor s} within the 
\mlcs\ framework (\S~\ref{subsec:MLCS2k2}).

The procedure begins by computing the rest-frame \mlcs\ model 
magnitudes for the assumed values of extinction ($A_V$), 
time of peak brightness ($t_0$), 
and shape-luminosity parameter ($\Delta$). 
These assumed values are typically determined iteratively as 
part of the $\chi^2$-minimization (Eq.~\ref{eq:mlcs_chi2def}).
For each iteration,
an epoch-dependent SN~Ia SED from \citet*{Hsiao07} is warped 
so that the synthetic colors of the warped SED 
match the {\mlcs} model colors at the corresponding rest-frame epoch. 

In detail, for each epoch and passband, the rest-frame 
model magnitude is first computed as the sum of the first 
four terms on the right-hand side of  Eq.~\ref{eq:MLCS2k2model}, 
where the host-galaxy extinction $\Xhost$ has been determined 
from the unwarped SED using the values of $A_V$ and $R_V$. 
Although the \Kcor\ depends strongly on how the SED is warped,
the value of $\Xhost$ is only weakly dependent on the warping, so  
$\Xhost$ can be determined to good approximation from the unwarped SED. 
As an illustration, when the SED is severely warped to modify 
the $V-B$ color (at peak brightness) by 0.5~mag, 
the estimate of the $V$-band extinction changes by less than 1\%.
Next, the SED is warped by multiplying it with the 
CCM89 galactic extinction law, following JRK07. 
This usage of the CCM89 law to warp is purely  
a mathematical convenience---it has nothing to do with physical 
extinction. 
The ``$A_V$-warp'' is the value of ``$A_V$'' in the warp factor 
for which the synthetic SN~Ia 
color matches the color of the rest-frame model magnitude computed above.
The redshift-dependent K-correction to observer-frame magnitude 
is then determined from this $A_V$-warped SED, using the redshift and 
knowledge of the rest-frame and observer-frame passbands. 
This method is model-independent and can 
therefore be applied to any light-curve model.
The SED 
is warped
independently at each epoch and locally in wavelength near the 
passband of interest; we do not do a global fit to match all 
colors simultaneously. 
There are two potential limitations in this procedure.
First, brightness-dependent spectral features are ignored, i.e., 
we use a single, $\Delta$-independent composite SED at each epoch. 
Second, for rest-frame $U$-band ($I$-band) there is no
constraint for warping the SED blueward (redward) of the
central wavelength.

Compared to the treatment in JRK07, we have made a slight
improvement to the spectral warping used for {\Kcor s}.
The rest-frame filter, $f'$, is chosen as the 
one that covers the equivalent rest-frame wavelength,  
$\lamrest = \lamobs/(1+z)$, where $\lamobs$ is the
central wavelength of the observed passband $f$.
For spectral warping, JRK07 used a fixed 
rest-frame color for a given rest-frame filter; 
e.g., $B-V$ color was used when $f'=B$. 
In this example, the new code uses either $B-V$ or $B-U$ 
for the warping, depending on whether the value of $\lamrest$ 
is closer to the central wavelength of $V$ or $U$.

\section{Filter Set for the Nearby SN Ia Sample}
\label{app:lowzfilters}

The Nearby SN~Ia sample serves as the
low-redshift anchor for the Hubble diagram. A superset of the 
nearby sample was also used to derive (``train'') model parameters for
the \mlcs\ method.
Within the framework of \mlcs, here 
we discuss the filter-response functions and color terms
needed for the \Kcor s that transform from rest-frame
model magnitudes ($UBVRI$) to observer-frame magnitudes.

The nearby SN~Ia sample is a heterogeneous data set for which 
the filter response functions vary.  While some of the filter 
curves have been published, the applicability of the filter curves 
to precision photometry is uncertain. However, all of the 
nearby SN~Ia magnitudes
have been transformed into the Landolt system \citep{Landolt07a},
even if the color-terms are not always available.
Although the Landolt magnitude system is well defined, 
there are no standard filter responses for this system
and hence one cannot compute {\Kcor s}. 
In order to compute {\Kcor s}, 
we determine color transformations from the Landolt system
to the standard $UBVRI$ filter response functions defined 
by \citet*{Bessell90}.
While previous analyses with SNe~Ia used Vega as the 
primary reference, we note that Vega has not been
measured by Landolt. 
Instead of using Vega, we define the primary reference
from the Landolt network, \BDFULL\ (hereafter \BD),
which happens to be the primary reference for SDSS photometry 
and also has a precisely measured HST spectrum.

In order to determine the Bessell-Landolt transformation, 
we use Landolt standards that have excellent spectrophotometric
data and compare the observed Landolt magnitudes to synthetic  
Bessell magnitudes based upon the spectrophotometric data and 
knowledge of the Bessell response functions. We use spectra
from HST CALSPEC 2006\footnote{\tt \wwwCALSPEC}
\citep{CALSPEC07}
because of its availability, high quality,
and consistent calibration. 
For the Landolt standards that overlap with HST~CALSPEC, 
the Landolt measurements are given in Table~\ref{tb:Landolt_mags}
and the synthetic magnitudes in the Bessell system  
in Table~\ref{tb:Landolt_synth}.

To proceed, we define 
a synthetic Landolt-magnitude in passband ``X'' by 
\begin{equation}
  X_{synth}^{\rm Landolt} =  X_{\rm synth}^{\rm Bess} + \Delta X_{\rm synth} ~~,
\end{equation}
where $X_{\rm synth}^{\rm Bess}$ is the synthetic magnitude constructed 
from the source spectrum and the filter response from 
\citet*{Bessell90}, and $\Delta X_{\rm synth}$
is a correction based on color transformations as follows,
\begin{eqnarray}
  \Delta V_{\rm synth}     & = &  k_0 [ (B-V)_{\rm obs} - (B-V)_{\BD} ] \nonumber  \\
  \Delta (B-V)_{\rm synth} & = &  k_1 [ (B-V)_{\rm obs} - (B-V)_{\BD} ] \nonumber  \\
  \Delta (U-B)_{\rm synth} & = &  k_2 [ (U-B)_{\rm obs} - (U-B)_{\BD} ] \nonumber \\  
  \Delta (V-R)_{\rm synth} & = &  k_3 [ (V-R)_{\rm obs} - (V-R)_{\BD} ] \nonumber \\  
  \Delta (R-I)_{\rm synth} & = &  k_4 [ (R-I)_{\rm obs} - (R-I)_{\BD} ] ~.\nonumber \\  
                       &   &  \label{eq:kdef}
\end{eqnarray}
The subscript ``obs'' refers to observed (instrumental) magnitudes 
for a standard star or supernova, and the \BD\ subscript indicates a 
Landolt measurement. 
These color transformations are defined so that there is no correction
for the reference star, \BD. 
The color coefficients ($k_{i=0,4}$) are determined by fitting 
Eq.~\ref{eq:kdef} with the Landolt measurements and synthetic 
Bessell magnitudes
in Tables~\ref{tb:Landolt_mags}-\ref{tb:Landolt_synth}, 
with Vega excluded from the fit.
The resulting $k_i$ values are shown in Table~\ref{tb:color}; 
typical values are in the few percent range. 
We have also calculated the color terms 
determined with Vega as the primary reference, with results given in the 
last column of Table \ref{tb:color}.
The use of color terms is an approximation that can lead to 
significant errors, but the small size of the color terms 
indicate that this error should be negligible.

As an alternative to using color transformations, 
we follow \citet*{Astier06} and define 
a modified set of $UBVRI$ Bessell filter response functions
in which the central wavelength of each filter passband  
is shifted from that of \citet*{Bessell90} but the shape of 
the response curve is unchanged. 
The shifts are defined such that the color terms,
as defined in Eq.~\ref{eq:kdef}, are zero.
The corresponding wavelength shifts relative to the 
filter responses defined by \citet*{Bessell90} are
given in Table~\ref{tb:lamshift} under the column
``HST standards.''
The shifts used in \citet*{Astier06} are also given in 
Table~\ref{tb:lamshift} for comparison.
The differences in wavelength shifts are likely due to the
different choices of spectral standards: 
we use HST standards (Table~\ref{tb:Landolt_mags}),
while \citet*{Astier06} used ground-based spectra from 
\citet{Hamuy92,Hamuy94}.

The recipe for {\Kcor s} is as follows.
The $UBVRI$ model magnitudes for {\mlcs} are assumed to be in the 
Landolt system.
The inverse of Eq.~\ref{eq:kdef} is used to convert these Landolt 
magnitudes into magnitudes in the \citet*{Bessell90} system, 
and then a \Kcor\ is applied
in the usual manner. For the SDSS, ESSENCE, and SNLS surveys,
the filter response functions are well understood, and therefore
the {\Kcor s} are well-defined. For the nearby SN sample,
the  \Kcor\ transforms into a hypothetical observer-frame $UBVRI$ system
with a filter response described by \citet*{Bessell90}; 
in this case, Eq.~\ref{eq:kdef} is applied again to transform back 
to Landolt magnitudes.

\begin{table}[hb]
\centering
\caption{  
    Measurements from Landolt 2007 that overlap with HST CALSPEC data.
    $N_{obs}$ is the number of Landolt observations.
    }
\begin{tabular}{lrrrrrrrr}
\tableline\tableline
   star & $N_{obs}$ & $V$ & $B-V$ & $U-B$ & $V-R$ & $R-I$ & $V-I$ \\
\tableline
BD+17\arcdeg
4708 \tablenotemark{e}     &  28 &      9.464 &     0.443 &    -0.183 &     0.298 &     0.320 &     0.618 \\
--                         &-       &$\pm$0.0026&$\pm$0.0015&$\pm$0.0021&$\pm$0.0011&$\pm$0.0009&$\pm$0.0013\\
Vega\tablenotemark{a}      &-       &     0.017 &    -0.002 &    -0.004 &    -0.007 &     0.004 &    -0.003 \\
--                         &-       & x & x & x & x & x  & x  \\
G191B2B\tablenotemark{b}   &  48 &     11.773 &    -0.326 &    -1.205 &    -0.149 &    -0.181 &     -0.327\\
--                         &-    &$\pm$0.0028&$\pm$0.0014&$\pm$0.0026&$\pm$0.0016&$\pm$0.0017&$\pm$0.0025\\
GD71\tablenotemark{c}      & 104 &     13.032 &    -0.249 &    -1.107 &    -0.137 &    -0.164 &     -0.320\\
--                         &-      &$\pm$0.0015&$\pm$0.0014&$\pm$0.0024&$\pm$0.0015&$\pm$0.0022&$\pm$0.0028\\
GD153\tablenotemark{d}     &   4 &      13.346 &    -0.286 &    -1.169 &    -0.138 &    -0.180 &     -0.319\\
--                         &-      & $\pm$0.004& $\pm$0.004& $\pm$0.005& $\pm$0.006& $\pm$0.008& $\pm$0.002\\
AGK+81\arcdeg
4211 \tablenotemark{e}     &  39 &      11.936&    -0.340 &    -1.204 &     -0.154&    -0.191 &     -0.345\\
--                         &-      &$\pm$0.0024&$\pm$0.0013&$\pm$0.0030&$\pm$0.0013&$\pm$0.0021&$\pm$0.0019\\
BD+28\arcdeg
4211 \tablenotemark{e}     &  32 &      10.509&    -0.341 &    -1.246 &    -0.147 &    -0.176 &     -0.322\\
--                         &-    &$\pm$0.0027&$\pm$0.0018&$\pm$0.0039&$\pm$0.0011&$\pm$0.0012&$\pm$0.0018\\
BD+75\arcdeg
325 \tablenotemark{e}      &  37 &      9.548 &     -0.334&    -1.212 &    -0.150 &     -0.187&    -0.336 \\
--                         &-    &$\pm$0.0018&$\pm$0.0010&$\pm$0.0020&$\pm$0.0008&$\pm$0.0018&$\pm$0.0018\\
Feige 110 \tablenotemark{e}& 26 &      11.832 &    -0.305 &    -1.167 &    -0.138 &     -0.180&    -0.313 \\
--                         &-     &$\pm$0.0018&$\pm$0.0010&$\pm$0.0033&$\pm$0.0012&$\pm$0.0022&$\pm$0.0020\\
Feige 34 \tablenotemark{e} &  31 &     11.181 &    -0.343 &    -1.225 &    -0.138 &    -0.144 &    -0.283 \\
--                         &-     &$\pm$0.0025&$\pm$0.0011&$\pm$0.0041&$\pm$0.0013&$\pm$0.0018&$\pm$0.0018\\
GRW+70\arcdeg
325 \tablenotemark{e}      &  36 &     12.773 &    -0.091 &    -0.875 &    -0.100 &    -0.104 &    -0.206 \\
--                         &-      &$\pm$0.0027&$\pm$0.0017&$\pm$0.0022&$\pm$0.0013&$\pm$0.0017&$\pm$0.0020\\
HZ21 \tablenotemark{e}      &  40 &    14.688 &    -0.327 &    -1.236 &    -0.149 &    -0.201 &    -0.350 \\
--                         &-     &$\pm$0.0022&$\pm$0.0016&$\pm$0.0033&$\pm$0.0022&$\pm$0.0043&$\pm$0.0049\\
HZ44 \tablenotemark{e}      &  40 &    11.673 &    -0.291 &    -1.196 &    -0.141 &    -0.181 &    -0.322 \\
--                         &-     &$\pm$0.0016&$\pm$0.0011&$\pm$0.0027&$\pm$0.0009&$\pm$0.0011&$\pm$0.0014\\
HZ4 \tablenotemark{e}      &  51 &     14.506 &     0.086 &    -0.675 &    -0.074 &    -0.060 &    -0.136 \\
--                         &-     &$\pm$0.0027&$\pm$0.0017&$\pm$0.0022&$\pm$0.0013&$\pm$0.0017&$\pm$0.0020\\
\tableline
\end{tabular}
\tablenotetext{a}{
	Vega measurements are not from Landolt.  
	The $V$-magnitude is from the HST analysis of \citet*{Bohlin06}, 
	and the colors are from \citet*{Bessell98}. 
	Vega error estimates are not given in the references.
  } 
\tablenotetext{b}{
	The reported $V$ magnitude was adjusted by \citet{Bohlin00}. 
    	The magnitude reported by \citet{Landolt07b} is 11.781.     
    	The errors and the colors are from Landolt \textit{loc cit.}  
  } 
\tablenotetext{c}{\citet{Landolt06}} 
\tablenotetext{d}{
 	Private communication from \citet{Landolt06} reports more 
	observations but the same values as published 
	previously \citep{Landolt92}.
  } 
\tablenotetext{e}{\citet{Landolt07b}.} 
  \label{tb:Landolt_mags}
\end{table}

\begin{table}[hb]
\centering
\caption{  
   For Landolt stars in Table~\ref{tb:Landolt_mags},
   synthetic magnitudes are computed using HST CALSPEC spectra and
   filters defined by \citet*{Bessell90}.
    }
\begin{tabular}{lrrrrr}
\tableline\tableline
    star & $V$ & $B-V$ & $U-B$ & $V-R$ & $R-I$ \\
\tableline
BD+17\arcdeg 4708  &   9.4640 &  0.4430 & -0.1830 &  0.2980 &   0.3200 \\
Vega               &   0.0169 &  0.0048 &  0.0213 & -0.0114 &  -0.0086 \\
\hline
G191B2B            &  11.7777 & -0.3021 & -1.2475 & -0.1535 &  -0.2016 \\
GD71               &  13.0371 & -0.2265 & -1.1476 & -0.1438 &  -0.1840 \\
GD153              &  13.3509 & -0.2583 & -1.1933 & -0.1477 &  -0.1916 \\
AGK+81\arcdeg 4211 &  11.9226 & -0.3221 & -1.2573 & -0.1583 &  -0.2051 \\
BD+28\arcdeg 4211  &  10.5076 & -0.3202 & -1.2735 & -0.1581 &  -0.2122 \\
BD+75\arcdeg 325   &   9.5301 & -0.3136 & -1.2672 & -0.1509 &   0.1987 \\
Feige 110          &  11.8295 & -0.2969 & -1.2064 & -0.1475 &  -0.1697 \\
Feige 34           &  11.1731 & -0.3312 & -1.2726 & -0.1334 &  -0.1563 \\
GRW+70\arcdeg 5824 &  12.7515 & -0.0605 & -0.8913 & -0.1118 &  -0.1287 \\
HZ21               &  14.6880 & -0.3270 & -1.2677 & -0.1346 &  -0.1969 \\
HZ44               &  11.6606 & -0.2643 & -1.2213 & -0.1385 &  -0.2001 \\
HZ4                &  14.4818 &  0.1210 & -0.6627 & -0.0879 &  -0.0758 \\
\tableline
\end{tabular}
  \label{tb:Landolt_synth}
\end{table}

\begin{table}[hb]
\centering
\caption{  
   Calculated color coefficients for \BD\ and for Vega.
    }
\begin{tabular}{lccc}
\tableline\tableline
      &              & value for  & value for  \\
Band  & coefficient  & {\BDFULL}  & Vega       \\
\tableline
$V$   & $k_0$   &  $-0.010 \pm 0.004$  &  $-0.015 \pm 0.004$ \\
$B-V$ & $k_1$   &  $+0.027 \pm 0.005$  &  $+0.033 \pm 0.005$ \\
$U-B$ & $k_2$   &  $-0.035 \pm 0.004$  &  $-0.050 \pm 0.004$ \\
$V-R$ & $k_3$   &  $-0.010 \pm 0.005$  &  $+0.004 \pm 0.005$ \\
$R-I$ & $k_4$   &  $-0.029 \pm 0.007$  &  $-0.010 \pm 0.007$ \\
\tableline
\end{tabular}
  \label{tb:color}
\end{table}

\begin{table}[hb]
\centering
\caption{  
    Wavelength shifts for the \citet*{Bessell90} filters.
     }
\begin{tabular}{ccc}
\tableline\tableline
           & \multicolumn{2}{c}{filter shift in {\AA} for:} \\
   Bessell & HST         & Astier et al.     \\
   filter  & standards   &  (2006)           \\
\tableline  % ------------------------
  $U$  & $+13 \pm 4$  & --     \\
  $B$  & $-15 \pm 4$  & $-41$  \\
  $V$  & $+12 \pm 6$  & $-27$  \\
  $R$  & $~+7 \pm 9$  & $-21$  \\
  $I$  & $-45 \pm 21$ & $-25$  \\
\tableline  % ------------------------
\end{tabular}
  \label{tb:lamshift}  
\end{table}

%% \clearpage
\section{Primary Magnitudes for \BD\ and Vega}
\label{app:primarymags}
%
% Jan 2009: explain interpolated Landolt mags for BD17
%

\newcommand{\MVEGALAND}{M_{\rm Vega}^{\rm Land}}
\newcommand{\MVEGASYNTH}{M_{\rm Vega}^{\rm synth}}
\newcommand{\MVEGA}{M_{\rm Vega}}
\newcommand{\ZPSYNTH}{{\rm ZP}^{\rm synth}}
\newcommand{\MBD}{M_{\rm BD17}}
\newcommand{\MBDLAND}{M_{\rm BD17}^{\rm Land}}
\newcommand{\MBDSYNTH}{M_{\rm BD17}^{\rm synth}}

For the \mlcs\ method, 
the primary magnitudes for each filter system are 
given in Table~\ref{tb:primary_mags} 
for \BD\ and Vega.
The magnitudes for $UBVRI$ are taken from
Landolt measurements, and the SDSS $gri$ magnitudes are
given in the AB system.
The primary magnitudes in the other filter systems 
are obtained by the interpolation method described below.

\begin{table*}[!h]
%\centering
\caption{Primary Landolt magnitudes used for \Kcor s. }
\begin{center}
\leavevmode
\begin{tabular}{lll}
\tableline\tableline % --------------------------------------------
 filter system   & primary  &  magnitudes  \\
\tableline % -------------------------------------------------
\tablenotemark{a} Landolt $U,B,V,R,I$ 
    & \BD\ & 9.724,  9.907,  9.464,  9.166,  8.846  \\
    & Vega & 0.017,  0.021,  0.023,  0.030,  0.026  \\
\hline % ----------------
  SDSS~2.5m $g,r,i$   
                     &  \BD\  & 9.644,  9.350,  9.256           \\
 (AB system)         &  Vega  & $-0.106$, $0.142$,  $0.356$     \\
\hline  % ---------
\tablenotemark{b} CTIO(ESSENCE) $R,I$      
  &  \BD\  & 9.152,  8.855   \\
  &  Vega  & 0.024,  0.020   \\
\hline  % --------
\tablenotemark{b}  CFHT(SNLS) $g,r,i,z$  
   &  \BD\  & 9.720,  9.222,  8.911,  8.756   \\  % NEW (minor fix)
   &  Vega  & 0.016,  0.022,  0.021,  0.016   \\
\hline  % ---------
\tablenotemark{b} 
 HST (F110W, F160W,     &  \BD\   & 8.558, 8.141, 9.337, 8.898, 8.746   \\
 F606W, F775W, F850LP)  &  Vega   & 0.003, 0.000, 0.020, 0.021, 0.015   \\
\tableline % -------------------------------------------------
\end{tabular}
\tablenotetext{a}{Measured by Landolt.}
\tablenotetext{b}{Interpolated as described in 
                    the text.}
\label{tb:primary_mags}
\end{center}
\end{table*}

We compute the \BD\ magnitudes by first interpolating 
$UBVRI$ magnitudes from Vega and using these
magnitudes to determine offsets that are used to correct
the synthetic magnitudes (from HST spectra) for \BD.
The Vega interpolation is based solely on the central
wavelength of each passband and does not depend
on the detailed shape of the transmission curve.
This approach is reasonable since the Vega magnitudes
and colors are all small, and the Vega spectrum is smooth.

Since we use the published magnitudes for the SN data, 
using our own analysis of the \BD\ magnitudes  means that 
we have effectively adjusted the photometry of the published data.  
The implicit assumption is that the original photometry is correct
relative to Vega, but that Vega itself has a slightly different 
value than was assumed previously.  
The Vega and \BD\ magnitudes do not agree exactly with the difference 
expected for the synthetic magnitudes computed from the HST spectra,
so Vega and \BD\ define slightly different photometric systems.  
While we believe that our use of \BD\ provides an 
accurate and more consistent description of all the photometric data, 
it should be emphasized that our assumed magnitudes for Vega differ 
by amounts that are small compared to the zeropoint errors quoted 
in the original publications and the differences are well within our quoted 
systematic errors.

As an example, we illustrate the determination of the primary 
\BD\ magnitudes for the SNLS $griz$ passbands.
For the interpolation, we need Landolt $UBVRI$ magnitudes 
referenced to \BD,
$\MVEGALAND = \MVEGASYNTH + (\MBDLAND - \MBDSYNTH)$.
For $V$-band, $\MVEGALAND(V) = 0.003 + (9.464 - 9.450) = 0.017$~mag.
The results for all passbands are
$\MVEGALAND(UBVRI) = $0.026, 0.022, 0.017, 0.028, 0.037~mag.
We interpolate these $UBVRI$ magnitudes to the central 
wavelengths of the SNLS $griz$ filters to get interpolated 
Vega magnitudes ($\MVEGA^{\rm interp}$),
with the results shown in the first row of 
Table~\ref{tb:BD17_for_SNLS}. The zeropoint offests
for synthetic magnitudes ($\ZPSYNTH$) are the differences
between the interpolated and synthetic magnitudes for Vega.
The \BD\ magnitudes in the Landolt system are then given by
$\MBDLAND = \MBDSYNTH + \ZPSYNTH$, and the results are given
in the seventh row of Table \ref{tb:primary_mags}.
%~\ref{tb:BD17_for_SNLS}. 

\begin{table}[hb]
%\centering
\caption{  
  Magnitudes used to compute Landolt \BD\ magnitudes for SNLS $griz$.
    }
\begin{center}
\leavevmode
\begin{tabular}{l r r r r}
\tableline\tableline
type of    & \multicolumn{4}{c}{Magnitude for:} \\
magnitude  & g & r & i & z \\
\tableline
$\MVEGA^{\rm interp}$   
      &  $0.020$ &  $0.025$ &  $0.035$ &  $0.030$  \\
%      &          &          &          &        \\
%
$\MVEGA^{\rm synth}$   
      &  $-0.099$ &  $0.149$ & $0.376$ &  $0.513$  \\
%      &          &          &          &        \\
%
$\ZPSYNTH$
      &   $0.119$ &  $-0.124$ & $-0.341$ &  $-0.483$  \\
%      &          &          &          &        \\
%
$\MBD^{\rm synth}$
      & $ 9.601$ & $9.346$ & $9.253$ & $9.240$  \\
%      &          &          &          &        \\
%
%$\MBD^{\rm Land}$
%      & $9.720$ & $9.222$ & $8.911$ & $8.756$ \\
\tableline
\end{tabular}
  \label{tb:BD17_for_SNLS}
\end{center}
\end{table}

\section{ Determining the Underlying Distribution of 
          Extinction ($A_V$) and Light-Curve Shape ($\Delta$) }
\label{app:unfold}
%
% Defined parameters
%

\newcommand{\iAV}{i_{A_V}}
\newcommand{\iDelta}{i_{\Delta}}
\newcommand{\iz}{i_z}

\newcommand{\ivecstar}{\vec{x}^{\ast}}
\newcommand{\ivecobs}{\vec{x}}

\newcommand{\iAVGEN}{i_{A_V}^{\ast}}
\newcommand{\iDeltaGEN}{i_{\Delta}^{\ast}}
\newcommand{\izGEN}{i_z^{\ast}}

\newcommand{\AVGEN}{A_V^{\ast}}
\newcommand{\DeltaGEN}{\Delta^{\ast}}
\newcommand{\zGEN}{z^{\ast}}

Within the framework of the \mlcs\ model (\S~\ref{subsec:MLCS2k2}),
we extract the underlying $A_V$ and $\Delta$ distributions
from the SDSS dust sample (\S~\ref{subsec:dustsample})
by making use of simulated light curves processed in
exactly the same way as the observed light curves.  
The underlying distributions of $A_V$ and $\Delta$ are defined
such that when these underlying distributions are input into 
the simulation, the fitted distributions from the simulated
\lcs\ match the data distributions.
We assume that these underlying distributions
are independent of redshift, and we fix the CCM89 extinction law
parameter to $R_V = \RV$, following the analysis of \S~\ref{subsec:RV}.

Here we use the Bayesian unfolding method of \citet{Agostini_95},
and tests with simulated samples shows that one iteration 
is adequate with our statistics.
For this discussion, we use an asterisk superscript to indicate 
the underlying true value for a parameter;
the lack of an asterisk indicates a measured value obtained
from fitting with the \mlcs\ method.
Simulated \lcs\  are generated with a 
flat distribution in $\AVGEN$ (over the range 0 to 4),
a flat distribution in $\DeltaGEN$ (from $-0.6$ to $+2$),
and a flat distribution in redshift (from 0 to 0.4).
For each simulated light curve that passes the 
selection criteria of \S~\ref{sec:sample},
the \mlcs\ model is used with a flat, non-informative 
prior on $A_V$ and $\Delta$
to extract a fitted value of $A_V$ and $\Delta$.  
These fitted values, in general, are different from the underlying 
$\AVGEN$ and $\DeltaGEN$. 
The accuracy of the inferred extinction, $A_V - \AVGEN$, 
has a typical root-mean-square (rms) of 0.2, with 
little dependence on redshift. 
The accuracy of the fitted shape-luminosity parameter, 
$\Delta-\DeltaGEN$, has an rms of a few hundredths for $z < 0.1$,
and the rms increases to about 0.2 at $z \sim 0.3$.

We use the simulation to calculate the conditional probability
of extracting fitted values of $A_V$ and $\Delta$ 
from a light curve generated with values 
$\AVGEN$ and $\DeltaGEN$, as a function of redshift.
Since the generated, fitted, and observed values are drawn from 
continuous distributions, we bin these quantities with bin-sizes
0.2 for $A_V$, 0.1 for $\Delta$, and 0.05 for redshift. 
Each bin is identified by an index:
$\ivecstar \equiv \iAVGEN,\iDeltaGEN,\izGEN$
for generated quantities, and 
$\ivecobs \equiv \iAV,\iDelta,\iz$
for fitted quantities.
Since redshifts are \specy\ determined with high precision,
we always have $\iz = \izGEN$.
From the simulation, we calculate
the conditional probability distribution,
\begin{equation}
   P_{\rm sim}(\ivecobs | \ivecstar) = 
   {n(\ivecobs)}/{N(\ivecstar)}~,
\end{equation}
where $n(\ivecobs)$ is the number of fits producing 
values of $(A_V,\Delta,z)$ that lie in the specified bins, and
$N(\ivecstar)$ is the number of light curves 
generated with $(\AVGEN,\DeltaGEN,\zGEN)$ that lie in the
specified bins. 
The underlying two-dimensional distribution of 
$\AVGEN$ and $\DeltaGEN$ is then obtained from the \SDSS\ data by
\begin{equation}
   P(\AVGEN,\DeltaGEN) =     
    \sum_{\ivecobs}
    \left[
      \frac{ P_{\rm sim}(\ivecobs | \ivecstar)}
           { \simeff(\ivecstar) 
             \sum_{\ivecstar} [P_{\rm sim}(\ivecobs | \ivecstar) ] }
        \times N_{\rm data}(\ivecobs)
     \right]~, 
\end{equation}
where $\simeff = \simeffpipe\times\simeffcuts$
is the simulated \eff\ that includes the combined effects
of the image-subtraction pipeline (\S~\ref{subec:searcheff})
and selection cuts (\S~\ref{sec:sample}),
$N_{\rm data}(\ivecobs)$ is the number
of observed SNe~Ia in redshift bin $\iz$ with fitted 
$A_V,\Delta$ that lie in the specified bins,
and the summation is over the three-dimensional grid
of observed $A_V$, $\Delta$, and redshift.
The corresponding one--dimensional distributions,
$P(\AVGEN)$ and $P(\DeltaGEN)$, are obtained
by marginalizing the two--dimensional distribution 
over the other variable.
We have extensively tested
this procedure for extracting the true distributions of
$\AVGEN$ and $\DeltaGEN$ on simulated
data samples, and the technique gives excellent agreement 
between the true (generated) distributions
and the distributions extracted from the simulated data.

\begin{figure}[hb]
  \epsscale{0.9}
  \plotone{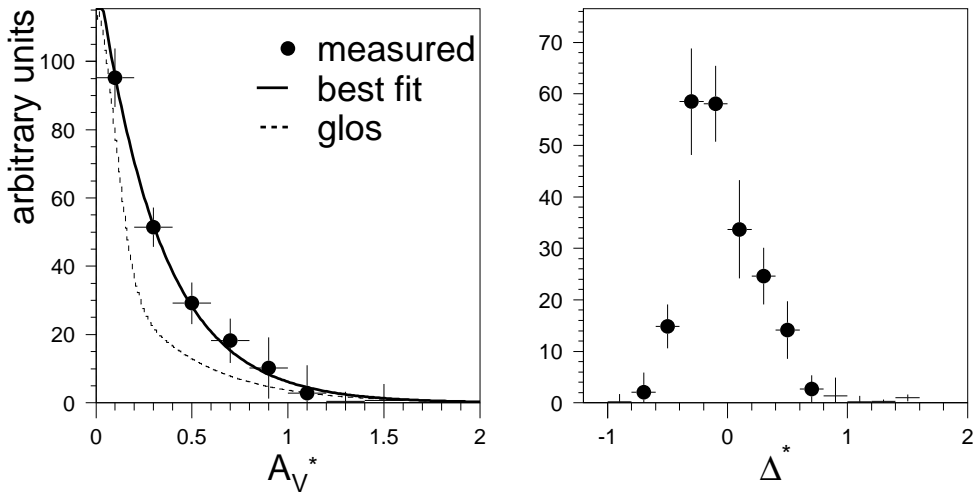}
  \caption{
	The distributions of $\AVGEN$ and $\DeltaGEN$ based on the 
 	\SDSS\ dust sample and the procedure described in the text.
	The $\AVGEN$ distribution includes overlays of the best-fit
	exponential function (solid curve) and the glos prior (dashed).
	}
  \label{fig:avdeltrue}
\end{figure}

We apply this technique to the \SDSS\ dust sample; the extracted
$\AVGEN$ and $\DeltaGEN$ distributions are shown in 
Figure~\ref{fig:avdeltrue}.
The error bars reflect the statistics of our data sample 
as well as the statistical \uncs\ in the simulated \eff\
and $P_{\rm sim}$ distributions.
We fit these distributions to analytic functions that
can be easily computed for the \mlcs\ fitting prior.
The underlying $\AVGEN$ distribution is accurately fit by 
an exponential function,
\begin{equation}
   dN/d\AVGEN = \exp(-\AVGEN/\TAUV)~.
\end{equation}
The $\DeltaGEN$ distribution is described by a bifurcated Gaussian 
with peak position $\Delta_0$ and
different positive side and negative side widths, 
$\sigma+$ and $\sigma-$ respectively.  We find
\begin{eqnarray} 
\TAUV    & = & \TAUAV \pm \TAUAVERRSTAT_{\rm stat} \pm \TAUAVERRSYST_{\rm syst} \\
\Delta_0 & = & \DELTAPEAK     \pm 0.029_{\rm stat} \pm 0.013_{\rm syst}  \\
\sigma-  & = & \DELTASIGMINUS \pm 0.046_{\rm stat} \pm 0.022_{\rm syst} \\
\sigma+  & = & \DELTASIGPLUS  \pm 0.029_{\rm stat} \pm 0.015_{\rm syst} ~,
\end{eqnarray}
where the uncertainties are statistical and systematic respectively.
These distributions are the basis for the {\mlcs} priors shown in 
Figure~\ref{fig:ovdatasim_avdelta}.  

In addition to the statistical errors, we have considered a number
of systematic effects in the determination of these distributions.  
We vary $R_V$ by $\pm 1\sigma$ (see \S~\ref{subsec:RV});  
this variation has the largest effect on the inferred $\AVGEN$ distribution.   
We vary the maximum redshift for the SN~Ia light
curves included in the data sample from the nominal 0.30 to 0.25.
We also consider the difference between the two different methods for
modeling SN~Ia intrinsic luminosity variations 
(see \S~\ref{subec:hubblescat}).  
Numerous minor variations in the analysis procedure, 
such as changing 
bin sizes and the use of binned or smoothed efficiencies,
have a negligible effect.  
The \uncs\ are summarized in Table~\ref{tb:avdelsys} and are summed
in quadrature to obtain the total \unc.  
\begin{table}[ht]
\centering
\caption{  
	Uncertainties in the parameters that describe
      	the distributions of $\AVGEN$ ($\TAUV$) and 
	$\DeltaGEN$ ($\Delta_0,\sigma+,\sigma-$).
        }
\begin{tabular}{lcccc}
\tableline\tableline % --------------------------------------------
source of                & \multicolumn{4}{c}{Uncertainty on:} \\
uncertainty              & $\TAUV$                    
                         & $\Delta_0$  & $\sigma+$   & $\sigma-$ \\
\tableline % -------------------------------------------------
statistical              & $\TAUAVERRSTAT$ 
                         & $0.054$ & $0.051$ & $0.033$ \\
% --------
$R_V=\RV\pm\RVERRTOT$    & $0.050 $        
                         & $0.005$ & $0.004$ & $0.008$ \\
% -----
redshift range           & $0.040$        
                         & $0.012$ & $0.014$ & $0.004$ \\
% -------
simulated color smearing & $0.030$        
                         & $0.003$ & $0.001$ & $0.006$ \\
% ------
analysis details         & $0.012$        
                         & $0.001$ & $0.001$ & $0.002$ \\
\tableline % -------------------------------------------------
Total Systematic         & $\TAUAVERRSYST$ & $0.013$ & $0.015$ & $0.011$ \\
\tableline % -------------------------------------------------
\tableline % -------------------------------------------------
Total Uncertainty        & $\TAUAVERRTOT$  & $0.056$ & $0.053$ & $0.035$ \\
\tableline % -------------------------------------------------
\end{tabular}
   \label{tb:avdelsys}
\end{table}

As noted above, an exponential shape for the 
distribution of $\AVGEN$ agrees very well with the data, and 
we adopt this shape in the {\mlcs} prior. Nevertheless, it is useful 
to check how well other proposed distributions match the data. In particular, 
WV07 considered a ``galaxy line of sight'' (glos) 
distribution for $\AVGEN$.  
The glos model includes an exponential $\AVGEN$ distribution plus
a narrow Gaussian with zero mean
and small width that is meant 
to represent SNe~Ia that would be expected to be observed with negligible
host galaxy dust extinction on half of the observed lines of sight, given 
a random distribution of observed host-galaxy orientations.  
When we fit the inferred $\AVGEN$ distribution with the glos model,
we find poor agreement 
between the best-fit model and the data, as shown in Fig.~\ref{fig:avdeltrue}.  
Even allowing the most extreme of the systematic variations 
(e.g., varying $R_V$), the best agreement between the data and
the glos model gives a confidence level of only $\sim 2$\%, 
calculated from the $\chi^2$ between the data and the model.  
If we allow the relative amplitude of the narrow Gaussian component of 
the glos model to vary, 
the fit returns an amplitude consistent with zero.  
The \unc\ on $\TAUV$, the slope of the exponential, 
therefore accurately describes the \unc\ in the functional shape of the 
$\AVGEN$ distribution.

\section{Discussion of Hubble Scatter for the SDSS Sample }
\label{app:SDSS_scatter}
%
% TOP
% Discuss hostz chisqmu & scatter.
% Rule out suspected reasons for small Hubble scatter.
%

As noted in \S \ref{subsec:results_mlcs}, 
the Hubble scatter and $\chisqmu$ statistic are significantly smaller 
for the \specy\ confirmed \SDSS\ SN~Ia sample
than for the other SN samples (see Table~\ref{tb:mlcs_fitquality}).
Here we investigate this anomaly by analyzing the 
\SDSS\ \hostz\ sample, described in \S \ref{subsec:dustsample},
comprising $\NHOSTZALLZ$ photometrically identified SNe~Ia 
with \specy\ determined host-galaxy redshifts.
Recall that we required $z<0.3$ for the SN sample used
to measure host-galaxy dust properties; 
for the discussion below, we do not impose a redshift cut,
thereby adding \NHOSTZGTZCUT\ \hostz\ SNe~Ia with $z>0.3$.

For the \hostz\ sample, the rest-frame magnitudes at peak brightness
are nearly $0.2$ mag fainter on average than those for the 
\specy\ confirmed SNe at the same redshift. 
The mean inferred extinction, $\AVMNSYMBOL$, is nearly 0.2 mag 
larger for the \hostz\ sample as well, indicating that the 
\spec\ sample is not complete for intrinsically underluminous or 
extinguished events, as already inferred in \S \ref{sec:sim}.
Performing a cosmological fit to the \hostz\ sample alone results in
$\chisqmu/N_{\rm dof} = \HOSTZCHISQDOF$ and
$\RMSMU \sim \HOSTZSCATTER$~mag 
(where, as before, we set $\sigmuint=0.16$), 
both of which are significantly larger than the corresponding 
values for the  \specy\ confirmed SNe~Ia,  
$\chisqmu/N_{\rm dof} = \SDSSCHISQ/\SDSSNDOF$ and $\RMSMU = \SDSSRMS$~mag.
A cosmological fit to the confirmed {\it plus} \hostz\ \SDSS\ sample 
results in $\chisqmu/N_{\rm dof} = \DUSTCHISQDOF$ 
($\sim \DUSTFITPROB$\% probability)
and $\RMSMU \sim \DUSTSCATTER$~mag,
consistent with the fit-quality parameters
for the other SN~Ia samples in Table~\ref{tb:mlcs_fitquality}. 
We therefore conclude that the lower  $\chisqmu$ and Hubble scatter 
for the \specy\ confirmed \SDSS\ sample are largely caused by 
spectroscopic selection effects.
As a crosscheck on the simulation, we have also analyzed simulated 
\spec\ and \hostz\ samples, and find the corresponding $\RMSMU$ 
values to be in excellent agreement with the data.

For completeness, we consider and exclude several other possibilities
for the smaller scatter in the \SDSS\ sample.
\begin{enumerate}
 \item	One possibility is that the 
Scene Model Photometry method (SMP, \S \ref{subsec:SMP}) overestimates the 
	flux uncertainties. Another is that SMP somehow provides 
	a dramatic improvement in accuracy compared to the photometry
	methods used in other surveys. To test these possibilities,
	we have processed the \SDSS\ \spec\ sample photometry with the
	image-subtraction pipeline that was used for preliminary 
	photometric measurements during the SN survey. 
	The resulting cosmological fit yields very 
	similar $\chisqmu$ and Hubble scatter, indicating that SMP 
	is not the cause. 
 \item 	To test if the \mlcs\ fitter overestimates distance-modulus errors
	($\sigmufit$ in Eq. \ref{eq:sigmudef}) for the \SDSS\ sample, 
	we have compared the average $\sigmufit$ values for each of the 
	SN samples
	at the mean redshifts of the samples. 
	For the Nearby, \SDSS, ESSENCE, and SNLS samples, the mean 
	redshifts are $0.035$, 0.22, 0.42, and 0.63.
        For SN sub-samples within small redshift windows 
        centered on the mean redshifts,
	the average $\sigmufit$ values are 0.07, 0.11, 0.19, and 0.18~mag, 
	with an \unc\ of $\sim 0.01$ on the average.
	For the \SDSS, the corresponding average 
        $\sigmufit = 0.11$, smaller than for 
	the ESSENCE \& SNLS surveys. Therefore, it appears that an 
	overestimate of 
	the distance-modulus \unc\ is not the cause of the
	smaller $\chisqmu$ for the \SDSS\ sample.
 \item 	We have split the \SDSS\ \spec\ sample into lower- 
	and higher-redshift halves at the median redshift 
	$z = 0.22$. Both halves have
	consistent values of $\chisqmu$.
\end{enumerate}

We conclude that the small $\chisqmu$ and Hubble
scatter for the \specy\ confirmed \SDSS\ sample are primarily effects  
of the survey selection function, particularly the \spec\ follow-up. 
To reduce biases in the cosmological analysis, it is important
to either include the \hostz\ sample in the analysis or to model 
the selection effects for the spectroscopic sample. In this paper, 
we have followed the second course.

\section{Total-Uncertainty Contours in the $w$-$\OM$ Plane}
\label{app:contours}

For the combined SN+BAO+CMB cosmology results,
we describe a simple method to generate total-\unc\ contours,  
i.e, contours that include statistical and systematic errors,
in the plane of $w$ versus $\OM$ for the \wCDM\ model, and 
in the plane of $\OM$ versus $\ODE$ for the \LCDM\ model.
Figures~\ref{fig:wCDM_MLCS_etotcont}-\ref{fig:LCDM_MLCS_etotcont}
show the statistical and total error contours for the \wCDM\ and 
\LCDM\ models using the \mlcs\ method, and 
Figures~\ref{fig:etot_contours_FWCDM_SALT2}-\ref{fig:etot_contours_LCDM_SALT2}
show the analogous contours using the \SALTII\ method.

A first-principles treatment of systematic errors would 
include all the systematic-error parameters and variations 
as nuisance parameters, evaluating the likelihood function 
on a multi-dimensional grid and then marginalizing over the 
nuisance parameters to obtain the likelihood for the 
cosmological parameters. For the large number of systematic 
effects we have considered, this approach would be 
computationally expensive. Instead, we take advantage of the 
empirical fact that the best-fit cosmological parameter 
results from the numerous systematic tests described in 
\S~\ref{sec:syst} lie very close to a straight line 
defined by the BAO+CMB prior, with slope $dw/d\OM \simeq 5$ 
for the \wCDM\ model, and $d\ODE/d\OM \simeq -0.8$ for the \LCDM\ model.
For {\wCDM} systematic tests in which the $w$-variations are within 
$0.1$, the root-mean-square $w$-scatter about this line is $\sim 0.002$.
For larger $w$-variations, the curvature of the
BAO+CMB prior becomes more noticable; for the largest
$w$-variation of $-0.3$, the value of $w$ lies 0.06 away
from the straight line approximation. 
For the purposes of illustrating contours,  
this linear approximation is adequate.
To incorporate the systematic errors, we stretch the 
statistical \unc\ contour along the line defined by the BAO+CMB 
prior, where the stretch factor is given by the ratio of 
total-to-statistical \uncs\ on the cosmological parameter $w$:
$\wtotsym/\wstatsym$.

Note that this approach is not valid in general: 
it depends on the relative precision of the SN results
and the BAO+CMB constraints. For example, using a simulated 
SN sample with three times the data statistics of sample {\esym}, 
the scatter about the BAO+CMB line increases by a factor of several.

\section{Modification of \mlcs\ Light-curve Templates to Match {\SALTII}}
\label{app:saltymlcs}
% =============================================
%
% Feb 27, 2009
% Translation of SALT2 SEDs into MLCS vectors
% for salty-MLCS vectors that reproduce SALT2
% cosmology. Referenced from Sec 11 
% (mlcs-SALT2 comparisons)
%
% =============================================

For the comparison in \S \ref{sec:results_compare}, we modified 
the \mlcs\ model vectors so that the light-curve templates 
match synthetic light curves derived from the \SALTII\ spectral 
surfaces. 
This translation of the \SALTII\ model begins
with the spectral surface as a function of rest-frame
epoch and $x_1$, with $c=0$ (Eq.~\ref{eq:SALTII_flux_rest}).
Synthetic \SALTII\ $UBVRI$ magnitudes are then calculated on a grid
of epochs and $x_1$ values, and $\alpha\cdot x_1$ is added 
to each synthetic magnitude (with $\alpha=0.12$).
The $x_1$-grid is now relabeled with the \mlcs\ parameter
$\Delta$ using the relation obtained from the data samples,
\begin{equation}
\Delta = -0.1799 - 0.1902 x_1 + 0.038447 x_1^2 - 0.0043656 x_1^3 ~.
\end{equation}
Next, an overall magnitude adjustment is made so that
the peak $V$-band magnitude for $\Delta=0$ matches that
of the nominal \mlcs\ model. 
For the final step, a quadratic fit of magnitude versus $\Delta$
is done for each epoch and each $UBVRI$ filter.
The resulting quadratic parameters
${\Mmlcs}^{e,f'}$, $p^{e,f'}$, and $q^{e,f'}$,
where $e$ is an epoch index and $f'$ is a filter index,
define the {\SALTII}-modified \mlcs\ model.
The $\Delta$-vs-magnitude fits were done in the
interval $-0.55 < \Delta < +1.1$, and the rms-scatter
varies between 0.03 and 0.08 mag.

The $UBVRI$ model-magnitude errors are estimated from 
Fig.~6 of \cite{Guy07}.
Compared to the \mlcs\ model errors, the \SALTII\ errors are 
smaller near peak brightness and larger at later epochs.

\end{appendix}

% ==============================================================
% BIBLIOGRAPHY

\bibliographystyle{apj}
\bibliography{sdss_sncosm09}

% ####################################
  \end{document}